\documentclass{article}

\usepackage{PRIMEarxiv}

\usepackage{setspace}
\usepackage[utf8]{inputenc} 
\usepackage[T1]{fontenc}    
\usepackage[hidelinks]{hyperref}       
\usepackage{url}            
\usepackage{booktabs}       
\usepackage{amsfonts}       
\usepackage{nicefrac}       
\usepackage{microtype}      
\usepackage{lipsum}
\usepackage{fancyhdr}       
\usepackage{graphicx}       

\usepackage{caption}

\usepackage{xcolor}

\usepackage{float}
\usepackage{amsmath,amssymb}

\usepackage{placeins}

\usepackage[ruled,vlined]{algorithm2e}

\newcommand{\revision}{}

\usepackage{caption}
\usepackage{subcaption}

\usepackage{grffile}

\usepackage{bbm}

\usepackage[round]{natbib}

\title{Identifying microbial drivers in biological phenotypes with a Bayesian Network Regression model
}

\author{
  Samuel Ozminkowski \\
  Department of Statistics \\
  Wisconsin Institute for Discovery \\
  University of Wisconsin-Madison\\
  Madison, WI\\
   \\
   \And
  Claudia Sol\'{i}s-Lemus \\
  Department of Plant Pathology \\
  Wisconsin Institute for Discovery \\
  University of Wisconsin-Madison\\
  Madison, WI\\
  \texttt{solislemus@wisc.edu} \\
}

\begin{document}
\maketitle

\captionsetup[figure]{list=no}
\captionsetup[table]{list=no}
\pagebreak

\begin{abstract}
1. In Bayesian Network Regression models, networks are considered the predictors of continuous responses.
These models have been successfully used in brain research to identify regions in the brain that are associated with specific human traits, yet their potential to elucidate microbial drivers in biological phenotypes for microbiome research remains unknown.
In particular, microbial networks are challenging due to their high-dimension and high sparsity compared to brain networks. Furthermore, unlike in brain connectome research, in microbiome research, it is usually expected that the presence of microbes have an effect on the response (main effects), not just the interactions.

2. Here, we develop the first thorough investigation of whether Bayesian Network Regression models are suitable for microbial datasets on a variety of synthetic and real data under diverse biological scenarios. We test whether the Bayesian Network Regression model that accounts only for interaction effects (edges in the network) is able to identify key drivers (microbes) in phenotypic variability.

3. We show that this model is indeed able to identify influential nodes and edges in the microbial networks that drive changes in the phenotype for most biological settings, but we also identify scenarios where this method performs poorly which allows us to provide practical advice for domain scientists aiming to apply these tools to their datasets.

4. BNR models provide a framework for microbiome researchers to identify connections between microbes and measured phenotypes. We allow the use of this statistical model by providing an easy-to-use implementation which is publicly available Julia package at \url{https://github.com/solislemuslab/BayesianNetworkRegression.jl}.
\end{abstract}



\section*{Introduction}

Microbial communities are among the main driving forces of biogeochemical processes in the biosphere. For one, many critical soil processes such as mineral weathering and soil cycling of mineral-sorbed organic matter are governed by mineral-associated microbes \citep{Fierer2012, Whitman2018, Cates2019, Kranz2019, Whitman2019}. Additionally, plant and soil microbiome drive phenotypic variation related to plant health and crop production \citep{Allsup2019, Rioux2019, Lankau2022, Lankau2020b}. Lastly, the human gut microbiome plays a key role in the regulation of human health and behaviour \citep{young2017role, dupont2020intestinal, siddiqui2021gut} and similar host-microbe associations have been studied for lung \citep{sulaiman2020perspectives, stavropoulou2021unraveling} or skin microbiome \citep{callewaert2020skin}.
Understanding the composition of microbial communities and how these compositions shape specific biological phenotypes is crucial to comprehend complex biological processes in soil, plants and humans alike.

Standard approaches to study the connection between microbial communities and biological phenotypes rely on abundance matrices to represent the microbial compositions \citep{Kunin2008, Holmes2012, Sankaran2018, Holmes2019, martin2020modeling,Minot2019, Sankaran2019, Williamson2019}. Different experimental settings are defined and then microbial compositions are measured (as abundances) on each experimental setting. Next, the abundance matrices are used as input in a regression-type (or machine-learning) analysis to relate the microbial community to  phenotypes of interest.

This standard pipeline, however, has limitations to find real connections between microbes and phenotypes.
Many times, these standard approaches focus on the relationship between a single microbial \revision{operational taxonomic unit (OTU)} and the phenotype,
adjusting for possible confounders like soil mineral characteristics. 
\revision{An OTU is a group of closely related organisms, an abstraction of the classic system of biological classification which allows us to specify a given level of relatedness between species in each group.}
This univariate procedure has
multiple assumptions that are violated by the complexity of the microbiome
and can lead to elevated type-I error rates as well as reduced
power.  For example, univariate analyses of individual OTUs ignore
correlations and interactions among the microbial communities, 
which could lead to power loss if the tested OTU is weakly
correlated with an unmeasured relevant OTU. In contrast, high-dimensional regression models allow the inclusion of multiple microbial OTUs simultaneously \citep{Xia2013-he, Lin2014-ii, Tang2019-xt, Grantham2020-as, Subedi2020-ay}, yet these models can be complicated by multicollinearity \citep{Zou2005} or instability
due to overfitting. 
Furthermore, standard regression analyses rarely account for interactions among microbes or ignore
potential
epistasis \citep{Cordell2009,McKinney2009,Fan2011,McKinney2012} among
genes across the microbiome that can reduce power if not properly
modeled \citep{Kraft2007,Broadaway2015}.
Finally, microbial OTUs are usually represented as relative abundances (compositional data) which is restricted to sum to 1 and this affects how proportions behave in different experimental settings (e.g. changes in proportions in the microbial composition does not necessarily reflect actual biological changes in the interactions \citep{blanchet2020co}).

Given that relative abundances only provide a snapshot of the composition of the community at the specific time of sampling and do not account for correlations between microbes, microbial interaction networks have been recently preferred to represent microbial communities \citep{hsu2019microbial, zhou2020identification, zhang2020florfenicol, benidire2020phytobeneficial}. Yet models to connect a microbial network to a biological phenotype remain unknown. 
On one side, recent years have seen an explosion of methods to infer microbial networks from a variety of data types \citep{Peixoto2019-qy, Fang2017-oh, Kurtz2015-rk, Friedman2012-lu}, \revision{including the increasingly popular Bayesian Networks \citep{sazal2020inferring, ramazi2021exploiting}}. However, these methods aim to reconstruct \textit{one} microbial network and do not attempt to connect this network to any biological phenotypes.
On the other side, novel statistical theory has been developed to study samples of networks \citep{durante2017nonparametric, durante2018bayesian, guha2021bayesianB}, yet again, these methods do not aim to understand the connection between the networks and a phenotype of interest. There has only been a handful of new methods that aim to identify associations between a sample of networks (predictors) and a phenotype (response) via a regression framework \citep{ma2019semi, wang2019symmetric, guha2021bayesian}. These methods, however, have only been studied for brain connectome networks which, unlike microbial networks, are intrinsically dense, and thus, methods to find associations between a sample of microbial networks and a biological phenotype remain unknown.

\begin{figure}[ht]
\centering
\includegraphics[scale=0.35]{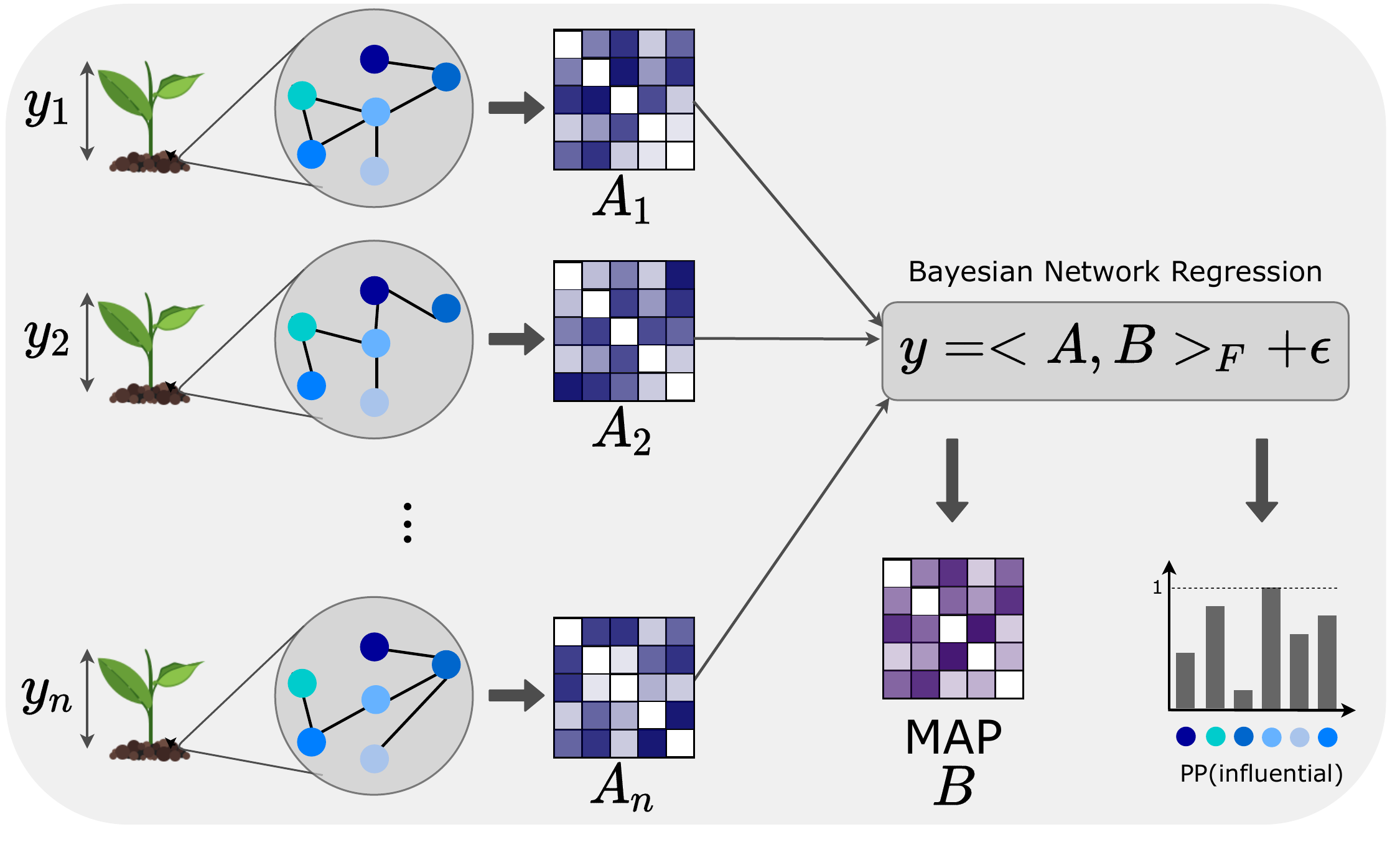}
\caption[Graphical abstract]{{\bf Graphical abstract.}
Samples contain a measured phenotype (e.g. height in plants) $y_i$ and a microbial network as predictor which is converted into its adjacency matrix $\mathbf A_i$. The Bayesian Network Regression model infers the regression coefficient matrix $\mathbf B$ with its maximum \textit{a posteriori} (MAP) and the posterior probability of being an influential node for every node.}
\label{fig:ga}
\end{figure}

In this paper, we introduce a Bayesian Network Regression (BNR) model that uses the microbial network as the predictor of a biological phenotype (Fig. \ref{fig:ga}). This model intrinsically accounts for the interactions among microbes and is able to identify influential edges (interactions) and influential nodes (microbes) that drive the phenotypic variability. While the model itself is not new \citep{ma2019semi, wang2019symmetric, guha2021bayesian}, it has only been studied for brain connectome networks, and thus, its applicability to microbial networks which are inherently more high-dimensional and sparser has not been studied.
Here, we test the BNR model on a variety of simulated scenarios with varying degrees of sparsity and effect sizes, as well as different biological assumptions on the effect of the microbes on the phenotype such as additive effects, interactions effects or functional redundancy \citep{rosenfeld2002functional}. We show that this model is able to identify influential nodes and edges in the microbial networks that drive changes in the phenotype for most biological settings, but we also identify scenarios where this method performs poorly which allows us to provide practical advice for domain scientists aiming to apply these tools to their datasets.
In addition, we implement the method in an open-source publicly available and easy-to-use new Julia package (\texttt{BayesianNetworkRegression.jl}) with online documentation and step-by-step tutorial which will allow scientists to easily apply this model on their own data. The computational speed and efficiency of the package makes it suitable to meet the needs of large datasets.

\noindent \textbf{Main contributions.} The Bayesian Network Regression (BNR) model is not new \citep{ma2019semi, wang2019symmetric, guha2021bayesian}, yet its applicability to microbial datasets has never been explored. Here, we develop the first thorough investigation of whether BNR models are suitable for microbial datasets on a variety of synthetic data that was generated under realistic biological scenarios. We also test the BNR model on real soil microbiome data and validate the findings with the published results \citep{wagg2019fungal}.
In addition, we introduce a novel Julia package (\texttt{BayesianNetworkRegression.jl}) with extensive documentation that implements the BNR model and has broad applicability for the microbiome research community.

\section*{Materials and methods}

\subsection*{Model and priors}

We use the Bayesian Network Regression model initially defined in \citet{guha2021bayesian} to elucidate associations between microbial drivers and biological phenotypes. We present below the theoretical details of the model and priors for the sake of completeness.

Let $y_i$ denote the scalar continuous phenotype for sample $i$ and let $\mathcal N_i$ be the network that represents the microbial community in sample $i$. In this work, we assume that these networks have already been estimated and will be assumed to be known without error. That is, at this stage, we do not propagate statistical error in the inference of the microbial network (but see Discussion). For each microbial network, we can compute its adjacency matrix $\mathbf A_i \in \mathbb R^{V \times V}$ where $V$ represents the number of nodes in the microbial network. \revision{The diagonal of the adjacency matrix $\mathbf A_i$ describes the nodes themselves - either the quantity of the node in the sample or presence/absence indicators.} While \citet{guha2021bayesian} assume that all networks must have the same nodes (as is the case for brain connectome networks), here we allow different networks to have different nodes so that $V$ represents the total number of nodes that appear in at least one network.

The network regression model is then defined as 
\begin{align*}
    y_i = \mu + 
    \langle \mathbf A_i, \mathbf B \rangle_F + \epsilon_i
\end{align*}
where $\mu \in \mathbb{R}$ is the overall mean effect, $\epsilon_i \sim N(0,\tau^2)$ is the error term, $\mathbf B \in \mathbb R ^{V \times V}$ is the symmetric network coefficient matrix, and $\langle \cdot \rangle_F$ represents the Frobenius inner product. 

Given the symmetric structure of the predictor $\mathbf A_i$, we can rewrite this model with a design matrix $\mathbf{X}$ where the $i$th row of $\mathbf X$ is set to be the upper triangle of the $i$th adjacency matrix $\mathbf{A}_i$ (including the diagonal) so that $\mathbf{X} \in \mathbb{R}^{n \times q}$ with $q=\frac{V(V+1)}{2}$\revision{, where $q$ is the number of nodes and edges in the network (that is, the number of elements in the upper triangle of the matrix, including the diagonal)}:
\begin{align*}
    \textbf{A}_i &= \begin{bmatrix} a_{i,1,1} & a_{i,1,2} & ... & a_{i,1,V} \\ \vdots & \ddots & ... & \vdots \\ a_{i,V,1} & a_{i,V,2} & ... & a_{i,V,V} \end{bmatrix}_{V\times V}
    \\ \textbf{X} &= \begin{bmatrix} a_{1,1,2} & a_{1,1,3} & ... & a_{1,1,V} & a_{1,2,3} & ... & a_{1,2,V} & ... & a_{1,V+1,V} \\
    \vdots & \vdots & \vdots & \vdots & \vdots & \vdots & \vdots & \vdots & \vdots \\
    a_{n,1,2} & a_{n,1,3} & ... & a_{n,1,V} & a_{n,2,3} & ... & a_{n,2,V} & ... & a_{n,V+1,V}
    \end{bmatrix}_{n \times q}.
\end{align*}

The responses $\textbf{y} \in \mathbb{R}^n$ then follow a Normal distribution $\textbf{y} \sim N\left(\mu + \textbf{X}\gamma, \tau^2\mathbb{I}_{n}\right)$ with an overall mean $\mu \in \mathbb{R}$, regression coefficients $\gamma \in \mathbb{R}^q$ and error variance $\tau^2 \in \mathbb{R}$. Here, $\mathbb{I}_n$ represents the identity matrix of dimension $n$. We note that the regression coefficients of this model ($\gamma_{kl}$) represent the effect of the edge in the microbial network between node $k$ and node $l$ (interaction effects) in the response and they are connected to the original regression coefficient matrix $\mathbf B$ as $b_{ij} = \gamma_{ij}/2$. \revision{Unlike the original model \citep{guha2021bayesian} that assumes that there are no main effects for the presence of the microbes in the sample, it is not appropriate for microbial research where the phenotype is expected to be affected by both the presence of the microbes and their interactions. Thus, we extend the model to include main effects.}

The prior for the regression coefficients $\gamma \in \mathbb{R}^q$ is given by
\begin{align}
    \gamma_{kl} \sim N(\textbf{u}^T_k \boldsymbol{\Lambda} \textbf{u}_l, \tau^2s_{kl})
\label{bnrmodel}
\end{align}
where $\mathbf u_1,\dots,\mathbf u_V \in \mathbb{R}^R$ are latent variables corresponding to each of the $V$ nodes, $\mathbf{\Lambda} \in \mathbb{R}^{R \times R}$ is equal to $\textrm{diag}(\lambda_1,...,\lambda_R)$ for $\lambda_i \in \{-1,0,1\}$ and $s_{kl} \in \mathbb R$ is a scale parameter. We note that the effect of the interaction between node $k$ and node $l$ on the response  is positive if $\textbf{u}^T_k \boldsymbol{\Lambda} \textbf{u}_l >0$ (or similarly, negative if $\textbf{u}^T_k \boldsymbol{\Lambda} \textbf{u}_l <0$ or zero if $\textbf{u}^T_k \boldsymbol{\Lambda} \textbf{u}_l =0$). \revision{$R \in \mathbb{N}$} is the dimension of the latent variable \revision{$\textbf{u}$ (and therefore the latent dimension of the node space) and is a hyperparameter} chosen by the user. We find in our simulations that it has a strong effect in the floating-point stability of the implementation (\revision{specifically, an $R$ value which is too high will cause floating-point errors such as catastrophic cancellation --} see Simulations).

The matrix $\mathbf \Lambda$ governs which entries in the latent variables $\mathbf u_k \in \mathbb{R}^R$ are informative and we set the following prior:
\begin{align*}
    \lambda_r &\sim \begin{cases}         0 & \textrm{ with probability } \tilde{\pi}_{1r}\\        1 & \textrm{ with probability } \tilde{\pi}_{2r}\\        -1 & \textrm{ with probability } \tilde{\pi}_{3r}    \end{cases}
\end{align*}
\revision{for each $r\in\{1,\dots,R\}$} and with hyper prior $(\tilde{\pi}_{1r}, \tilde{\pi}_{2r}, \tilde{\pi}_{3r}) \sim \textrm{Dirichlet}(r^\eta, 1, 1)$ for $\eta > 1$. 
Note that the probability of 0 in the Dirichlet is governed by the index $r$ (and $\eta$) which is meant to bias inference towards lower dimensional\revision{-representations of the latent variable $\textbf{u}$}.
These $\tilde{\pi}$ parameters control the sparsity of the regression coefficient matrix $\mathbf B$. It is traditionally assumed that only a subset of microbes in the sample are key drivers of the phenotype.

To determine which nodes are influential (non-zero effect on the response), we set a spike-and-slab prior \citep{ishwaran2005spike}:
\begin{align*}
    \textbf{u}_k &\sim \begin{cases}         N(\textbf{0},\textbf{M}) & \textrm{if } \xi_k = 1 \\        \delta_0 & \textrm{if } \xi_k = 0    \end{cases}
\end{align*}
where $\delta_0$ is the Dirac-delta function at $0$, $\mathbf M \in \mathbb{R}^{R \times R}$ is a covariance matrix, $\mathbf 0$ is an $R$-dimensional vector of zeros, and $\xi \in \{0,1\}^V$ is a column vector of dimension $V$ where each value denotes whether that node is influential on the response or not. \revision{Note that $\textbf{u}_k \in \mathbb{R}^{R}$ for each $k \in \{1,...,V\}$.} We assume that $\xi_k \sim \textrm{Bernoulli}(\Delta)$ with hyper priors $\Delta \sim \textrm{Beta}(a_\Delta, b_\Delta)$ for $a_\Delta,b_\Delta \in \mathbb R$ and $\mathbf M \sim \textrm{InverseWishart}(v, \mathbb{I}_R)$ for $v \in \mathbb R$.

Lastly, the prior for the scale parameters ($\mathbf s \in \mathbb{R}^q$) is given by $s_{kl} \sim \textrm{Exp}(\theta/2)$ with hyper prior $\theta \sim \textrm{Gamma}(\zeta, \iota)$ for shape $\zeta \in \mathbb R$ and rate $\iota \in \mathbb R$, and the prior for the overall mean ($\mu$) and error variance ($\tau^2$) is assumed to be non-informative $\pi(\mu,\tau^2) \propto \frac{1}{\tau^2}$.

Table \ref{tab:pars} in the Appendix contains all parameters in the model and their descriptions.

\subsection*{Posteriors}

The posterior distribution of the overall mean and the error variance are given by
\begin{align}
    \mu | \mathbf{y,X},\gamma,\tau^2 &\sim N\left( \frac{\textbf{1}_n^T(\textbf{y} - \textbf{X}\gamma)}{n}, \frac{\tau^2}{n} \right) \label{eq:mu}\\
    \tau^2 | \mathbf{y,X},\mu,\gamma,\mathbf{W,D} &\sim \textrm{InverseGamma}  [( n/2 + V(V+1)/4), \nonumber\\ &\frac{1}{2}(\textbf{y} - \mu\textbf{1}_n - \textbf{X}\gamma)^T (\textbf{y} - \mu\textbf{1}_n - \textbf{X}\gamma) + (\gamma - \textbf{W})^T \textbf{D}^{-1}(\gamma - \textbf{W}) ]
    \label{eq:tau}
\end{align}
where $\mathbf 1_n$ is an $n$-dimensional vector of ones, $\mathbf W \in \mathbf R^{q}$ is a vector given by 
\begin{align*}
    \mathbf{W} &= \begin{bmatrix} \mathbf{u}_1^T\boldsymbol\Lambda\mathbf{u}_2 \\
        \vdots \\ \mathbf{u}_1^T\boldsymbol\Lambda\mathbf{u}_V \\ \vdots \\ \mathbf{u}_{V}^T\boldsymbol\Lambda\mathbf{u}_V \end{bmatrix}
\end{align*}
and $\mathbf D \in \mathbb R^{q \times q}$ is a diagonal matrix with the vector of scale parameters \textbf{s} in the diagonal.

The posterior distributions for the scale parameters ($s_{kl}$) and their hyper parameter ($\theta$) are given by
\begin{align}
    s_{kl} | \gamma_{kl}, \mathbf{u, \Lambda}, \tau^2, \theta &\sim \textrm{GeneralizedInverseGaussian} \left[ \frac{1}{2}, \frac{(\gamma_{kl} - \textbf{u}_k^T \boldsymbol{\Lambda} \textbf{u}_l)^2}{\tau^2, }, \theta \right] \label{eq:s}\\
    \theta | \mathbf{s} &\sim \textrm{Gamma}\left[ \left( \zeta + \frac{V(V+1)}{2} \right), \left( \iota + \sum\limits_{k<l} \frac{s_{kl}}{2} \right) \right]. \label{eq:theta}
\end{align}

The posterior distribution for the regression coefficients ($\gamma$) is given by 
\begin{align}
    \gamma |& \mathbf{y,X,D,W},\mu,\tau^2 \sim \nonumber \\ &N\left( (\textbf{X}^T\textbf{X} + \textbf{D}^{-1})^{-1} (\textbf{X}^T(\textbf{y} - \mu\textbf{1}_n) + \textbf{D}^{-1}\textbf{W}), \hspace{2mm} \tau^2 (\textbf{X}^T\textbf{X} + \textbf{D}^{-1})^{-1}\right). \label{eq:gamma}
\end{align}

Next, for the auxiliary variables, the posterior distribution of the latent variables ($\mathbf u_k$) is given by
\begin{align}
    \mathbf{u}_k | w_{\mathbf{u}_k}, \mathbf{m}_{\mathbf{u}_k}, \mathbf \Sigma_{\mathbf{u}_k} &\sim w_{\mathbf{u}_k} \delta_0(\mathbf{u}_k) + (1 - w_{\mathbf{u}_k})N(\mathbf{u}_k | \mathbf{m}_{\mathbf{u}_k}, \mathbf \Sigma_{\mathbf{u}_k})
    \label{eq:u}
\end{align}
where 
\begin{align}
    w_{\textbf{u}_k} = \frac{(1 - \Delta)N(\boldsymbol{\gamma}_k | \textbf{0}, \tau^2\textbf{H}_k)}{(1 - \Delta)N(\boldsymbol{\gamma}_k | \textbf{0}, \tau^2\textbf{H}_k) + \Delta N(\boldsymbol{\gamma}_k | \textbf{0}, \tau^2\textbf{H}_k + \textbf{U}_k^*\textbf{M}\textbf{U}_k^{*T})}
\label{eq:w_uk}
\end{align}
for $\boldsymbol{\gamma}_{k} = (\gamma_{1k}, ..., \gamma_{k-1, k}, \gamma_{k,k+1}, ..., \gamma_{kV}) \in \mathbb R ^q$, 
and $N(x|m,v)$ corresponding to the Gaussian PDF evaluated on $x$ for mean $m$ and covariance $v$, $\Delta$ given by
\begin{align}\Delta | a_\Delta, b_\Delta, \xi \sim \textrm{Beta}\left[ \left(a_\Delta + \sum_{k=1}^V \xi_k\right), \left(b_\Delta + \sum_{k=1}^V (1 - \xi_k)\right) \right] \label{eq:delta},
\end{align} $\textbf{H}_k = \textrm{diag}(s_{1k}, ..., s_{k-1,k}, s_{k,k+1}, ..., s_{kV}) \in \mathbb R ^{(V-1) \times (V-1)}$, $\textbf{U}_k^* = (\textbf{u}_1 : ... : \textbf{u}_{k-1} : \textbf{u}_{k+1} : ... : \textbf{u}_V)^T \boldsymbol{\Lambda} \in \mathbb R ^{(V-1) \times R}$, $\mathbf 0$ is the $(V-1)$-dimensional vector of zeros, and matrix $\mathbf M$ sampled from the posterior distribution:
\begin{align}
    \textbf{M} | \mathbf{u,\Lambda},v &\sim \textrm{InverseWishart}\left[\left(\mathbb{I}_R + \sum\limits_{k: \textbf{u}_k \neq \textbf{0}} \textbf{u}_k \boldsymbol{\Lambda} \textbf{u}_k^T \right), \left( v + \sum_{k=1}^V \mathbf 1(\textbf{u}_k \neq \textbf{0}) \right) \right] \label{eq:M}
\end{align}
where $\mathbf 1(\cdot)$ is the indicator function and $\mathbb I_R$ is the identity matrix of dimension $R$.

In addition, the posterior mean ($\mathbf m_{\mathbf u_k}$) and posterior covariance matrix ($\mathbf \Sigma_{\mathbf u_k}$) are defined as
\begin{align*}
    \textbf{m}_{\textbf{u}_k} &= \frac{1}{\tau^2}\boldsymbol{\Sigma}_{\textbf{u}_k} \textbf{U}_k^{*T}\textbf{H}_k^{-1}\gamma_k \\
    \boldsymbol{\Sigma}_{\textbf{u}_k} &= \left( \frac{1}{\tau^2}\textbf{U}_h^{*T} \textbf{H}_k^{-1}\textbf{U}_k^* + \textbf{M}^{-1} \right)^{-1}.
\end{align*}

The posterior probability of the vector $\xi$ is given by 
\begin{align}
\xi_k | w_{\mathbf{u}_k} \sim \textrm{Bernoulli}(1 - w_{\textbf{u}_k}) \label{eq:xi}
\end{align}
with the same definition of $w_{\mathbf u_k}$ as in Eq.~\ref{eq:w_uk}. 

The posterior distribution of the $\lambda_r$ values is given by
\begin{align}
    \lambda_r | \gamma, \mathbf{u,\Lambda,D} &\sim \begin{cases}             0 & \textrm{ with probability } p_{1r}\\            1 & \textrm{ with probability } p_{2r}\\            -1 & \textrm{ with probability } p_{3r}            \end{cases}
    \label{eq:lambda}
\end{align}
with
\begin{align*}
    p_{1r} &= \frac{\tilde{\pi}_{1r} N(\gamma | \textbf{W}_0,\tau^2\textbf{D})}{\tilde{\pi}_{1r}N(\gamma | \textbf{W}_0,\tau^2\textbf{D}) + \tilde{\pi}_{2r}N(\gamma | \textbf{W}_1,\tau^2\textbf{D}) + \tilde{\pi}_{3r}N(\gamma | \textbf{W}_{-1},\tau^2\textbf{D})}        
\\ p_{2r} &= \frac{\tilde{\pi}_{2r}N(\gamma | \textbf{W}_1,\tau^2\textbf{D})}{\tilde{\pi}_{1r}N(\gamma | \textbf{W}_0,\tau^2\textbf{D}) + \tilde{\pi}_{2r}N(\gamma | \textbf{W}_1,\tau^2\textbf{D}) + \tilde{\pi}_{3r}N(\gamma | \textbf{W}_{-1},\tau^2\textbf{D})}        
\\ p_{3r} &= 1 - p_{1r} - p_{2r} 
\end{align*}
where $N(x|m,v)$ corresponds to the Gaussian PDF evaluated at $x$ for mean $m$ and covariance $v$ and
\begin{align*}
    \textbf{W}_0 &= \left[ \textbf{u}_1^T\boldsymbol{\Lambda}_0\textbf{u}_2, ..., \textbf{u}_1^T\boldsymbol{\Lambda}_0\textbf{u}_V, ..., \textbf{u}_{V}^T \boldsymbol{\Lambda}_0\textbf{u}_V \right]^T \in \mathbb R^q , \\ \boldsymbol{\Lambda}_0 &= \textrm{diag}(\lambda_1, ..., \lambda_{r-1}, 0, \lambda_{r+1}, ..., \lambda_R) \in \mathbb R^{R \times R}, \\
    \textbf{W}_1 &= \left[ \textbf{u}_1^T\boldsymbol{\Lambda}_1\textbf{u}_2, ..., \textbf{u}_1^T\boldsymbol{\Lambda}_1\textbf{u}_V, ..., \textbf{u}_{V}^T \boldsymbol{\Lambda}_1\textbf{u}_V \right]^T \in \mathbb R^q ,\\
    \boldsymbol{\Lambda}_1 &= \textrm{diag}(\lambda_1, ..., \lambda_{r-1}, 1, \lambda_{r+1}, ..., \lambda_R) \in \mathbb R^{R \times R},\\
    \textbf{W}_{-1} &= \left[ \textbf{u}_1^T\boldsymbol{\Lambda}_{-1}\textbf{u}_2, ..., \textbf{u}_1^T\boldsymbol{\Lambda}_{-1}\textbf{u}_V, ..., \textbf{u}_{V}^T \boldsymbol{\Lambda}_{-1}\textbf{u}_V \right]^T \in \mathbb R^q ,\\
    \boldsymbol{\Lambda}_{-1} &= \textrm{diag}(\lambda_1, ..., \lambda_{r-1}, -1, \lambda_{r+1}, ..., \lambda_R) \in \mathbb R^{R \times R}.
\end{align*}

Lastly, the posterior distribution of the hyper parameters $(\tilde{\pi}_{1r},\tilde{\pi}_{2r},\tilde{\pi}_{3r})$ is given by
\begin{align}
 (\tilde{\pi}_{1r}, \tilde{\pi}_{2r}, \tilde{\pi}_{3r}) | \eta, \mathbf{\Lambda} &\sim \nonumber \\ \textrm{Dirichlet}&\left(r^\eta + \sum_{r=1}^R\mathbf 1(\lambda_r = 0), 1 + \sum_{r=1}^R\mathbf 1(\lambda_r = 1), 1 + \sum_{r=1}^R\mathbf 1(\lambda_r = -1)\right) \label{eq:pi}
\end{align}
where again $\mathbf 1(\cdot)$ represents the indicator function.

The main parameters of interest are the regression coefficients $\gamma$ which represent the effect of the interactions among microbes on the response and the parameters $\xi$ that represent whether each of the $V$ microbes are influential on the response or not. We obtain the posterior probability that a node is influential by taking the mean of $\xi$ over the samples.
Samples of the posterior distributions are obtained using Gibbs sampling as described in the next section.

\subsection*{Gibbs Sampling}

We sample the posterior distributions using Gibbs sampling as described in Algorithm \ref{algo:gibbs}. 

\begin{algorithm}
\caption{Posterior Gibbs Sampling}
\label{algo:gibbs}
\SetAlgoLined
\KwResult{MCMC samples of the posterior distribution of parameters of interest}
Initialization\;
\While{not enough samples}{
  Sample $\tau^2 | \mathbf{y,X},\mu,\gamma,\mathbf{W,D} \sim \textrm{InverseGamma}$ (Eq. \ref{eq:tau})\;
  Sample $\xi | w_{\mathbf{u}_k} \sim \textrm{Binomial}$ (Eq. \ref{eq:xi})\;
  \For{$k$ in $1:V$}{
    Sample $\mathbf{u}_k | w_{\mathbf{u}_k}, \mathbf{m}_{\mathbf{u}_k}, \mathbf \Sigma_{\mathbf{u}_k} \sim \xi_k \times \textrm{Normal}$ (Eq. \ref{eq:u})\;
  }
  Sample $\gamma | \mathbf{y,X,D,W},\mu,\tau^2 \sim \textrm{Normal}$ (Eq. \ref{eq:gamma})\;
  Sample $s_{kl} | \gamma_{kl}, \mathbf{u, \Lambda}, \tau^2, \theta \sim \textrm{GeneralizedInverseGaussian}$ (Eq. \ref{eq:s})\;
  Sample $\theta | \mathbf{s} \sim \textrm{Gamma}$ (Eq. \ref{eq:theta})\;
  Sample $\Delta \sim \textrm{Beta}$ (Eq. \ref{eq:delta})\;
  Sample $\mathbf M | \mathbf{u,\Lambda},v \sim \textrm{InverseWishart}$ (Eq. \ref{eq:M})\;
  Sample $\mu | \mathbf{y,X},\gamma,\tau^2 \sim \textrm{Normal}$ (Eq. \ref{eq:mu})\;
  \For{$r$ in $1:R$}{
    Sample $\lambda_r | \gamma, \mathbf{u,\Lambda,D} \sim [0,1,-1]$ (Eq. \ref{eq:lambda})\;
    Sample $(\tilde{\pi}_{1r}, \tilde{\pi}_{2r}, \tilde{\pi}_{3r}) | \eta, \mathbf{\Lambda}  \sim \textrm{Dirichlet}$ (Eq. \ref{eq:pi})\;
  }
}
\end{algorithm}

\subsection*{Open-source software}

We released a Julia package to perform the sampling scheme which provides posterior estimates and convergence statistics for the Bayesian Network Regression model available as \texttt{BayesianNetworkRegression.jl} at the GitHub repository \url{https://github.com/solislemuslab/BayesianNetworkRegression.jl}. In addition, we provide all reproducible scripts for the simulation study (described in the next section) in the GitHub repository  \url{https://github.com/samozm/bayesian_network_regression_imp}.

\subsection*{Simulations}

One of the main objectives of this manuscript is to test the applicability of the Bayesian Network Regression model when facing sparse data that is ubiquitous in microbiome research. There are two main sources of sparsity: 1) the matrix of regression coefficients $\mathbf B$ is assumed to be sparse (sparsity controlled by $\pi$) which means that there are few microbial drivers that affect the phenotype and 2) the data matrix is sparse (represented by the adjacency matrix $\mathbf A_i$ with sparsity controlled by the number of sampled microbes $k$) because we do not have complete sampling of all microbes. We test different levels of sparsity both on the regression coefficient matrix $\mathbf B$ ($\pi=0.3, 0.8$) and in the adjacency matrices $\mathbf A_i$ ($k=8,15,22$ sampled microbes out of $30$ total). In addition, we test two levels of effect sizes ($\mu=0.8,1.6$) so that the entries in the regression coefficient matrix $\mathbf B$ are distributed $N(\mu,\sigma=1.0)$. For all simulation cases, all possible combinations of parameter values are tested.

\begin{figure}[!ht]
\centering
\includegraphics[scale=0.4]{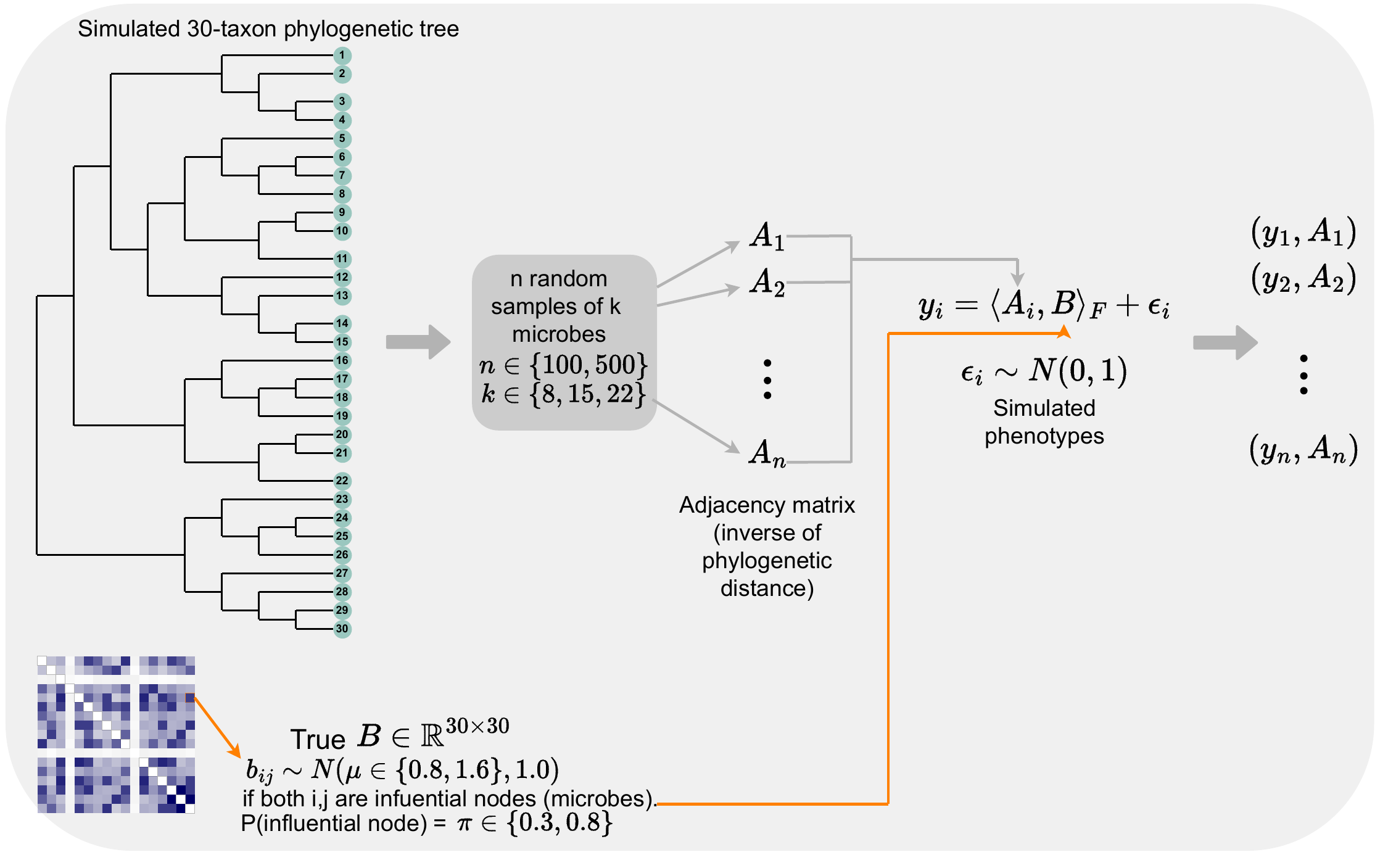}
\caption[Description of theoretical simulations]{{\bf Description of theoretical simulations.} We simulate a 30-taxon phylogenetic tree as the representation of the true microbial community, and then select $k$ microbes per sample with which to build an adjacency matrix ($\mathbf A_i$) per sample. The phenotype $y_i$ is then computed as the Frobenius product of the sample adjacency matrix ($\mathbf A_i$) and the true matrix of regression coefficients $\mathbf B$ plus Gaussian noise.
}
\label{fig:unrealistic}
\end{figure}

\begin{figure}[!ht]
\centering
\includegraphics[scale=0.35]{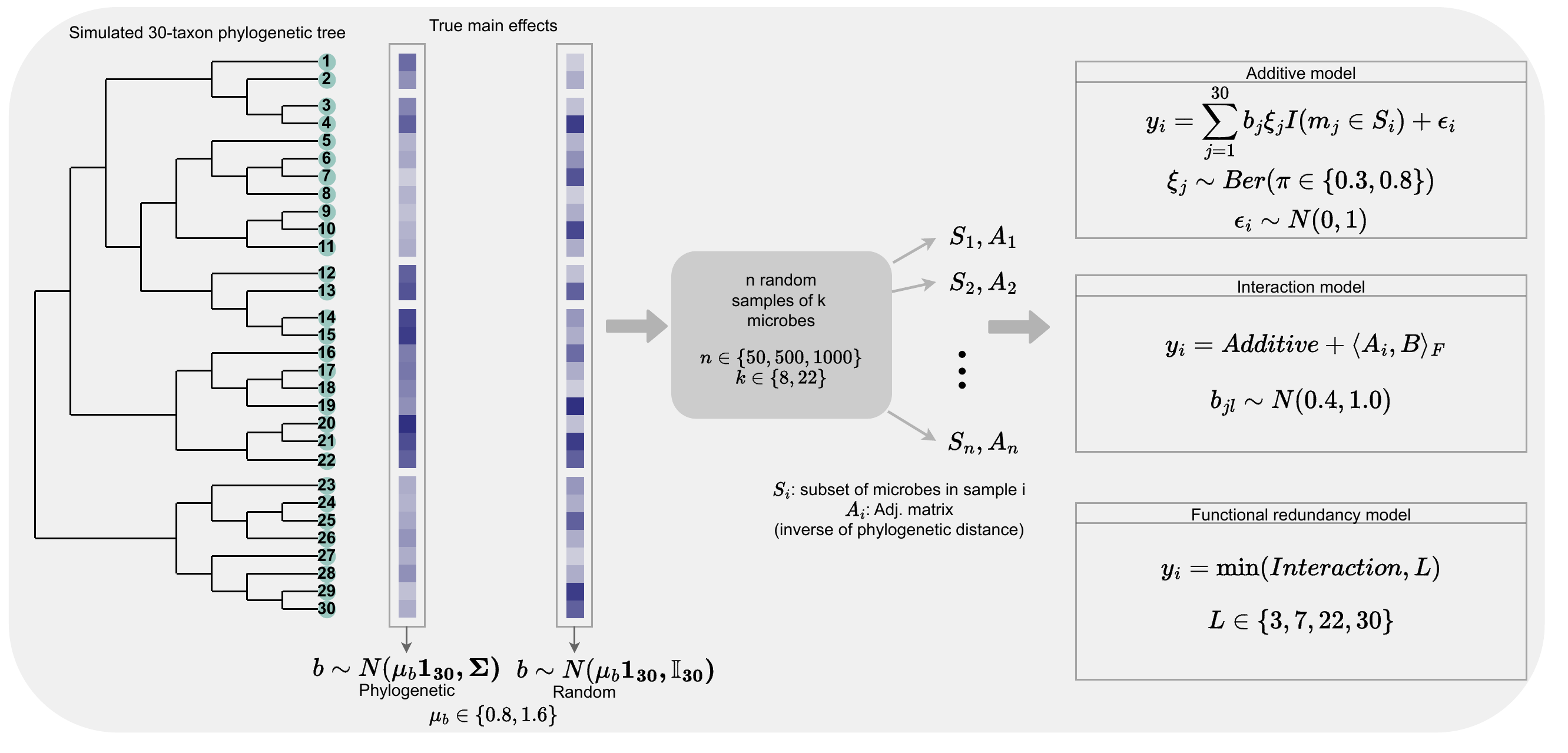}
\caption[Description of realistic simulations]{{\bf Description of realistic simulations.} We simulate a 30-taxon phylogenetic tree as the true microbial community, and then select $k$ microbes per sample \revision{($S_i = \{m_{i_1},...,m_{i_k}\}$)} with which to build an adjacency matrix ($\mathbf A_i$) per sample. The phenotype $y_i$ is computed under three models: additive, interaction or functional redundancy. Within each model, there are two options for the generation of the true microbial effects: randomly sampled independently of other microbes or phylogenetically-informed in which related microbes have similar effects on the phenotype. 
\revision{For the case of $n=50$, we performed data augmentation to reach $n=200$ samples.}
}
\label{fig:realistic}
\end{figure}

We split the simulations into two scenarios: 1) theoretical simulations (graphical description in Fig.~\ref{fig:unrealistic}) and 2) realistic simulations (graphical description in Fig.~\ref{fig:realistic}). We describe both scenarios next.

\vspace{0.1cm}
\noindent \textbf{Theoretical simulations.} We simulate a 30-taxon phylogenetic tree using the \texttt{rtree} function from the R package \texttt{ape} \citep{ape} that randomly splits edges until the desired number of leaves is attained. This tree represents the true community of microbes. We generate the true matrix of regression coefficients ($\mathbf B \in \mathbb R^{30 \times 30}$) by flipping a biased coin for every entry $b_{ij}$ to determine if the edge connecting nodes $i$ and $j$ is an influential edge. We vary the probability of influential node as $\pi=0.3, 0.8$ as already mentioned above. If the edge is indeed set as influential, the entry $b_{ij}$ is sampled from a Normal distribution with mean $\mu = 0.8$ or $1.6$ and variance equal to $1.0$.

For each sample, we randomly select $k$ (set as $8,15$, or $22$) microbes out of the 30 total microbes. We build the adjacency matrix for that sample using the phylogenetic distance between microbes. That is, the entry $a_{ij}$ is equal to the inverse of the phylogenetic distance between microbe $i$ and microbe $j$ if both microbes are present in the sample (and zero otherwise). \revision{Diagonal entries $a_{ii}$ are set to zero - we do not consider the presence-absence of microbes in the theoretical simulations}. Note that we are not estimating the phylogenetic tree for a given sample given that we are not simulating \revision{genetic} sequences, and thus, we are ignoring estimation error in the phylogenetic pipeline at this point. Future work will incorporate this type of error to assess its implications downstream (see Discussion).

For each sample, we calculate the phenotype $y_i$ as the Frobenius \revision{inner} product between $\mathbf B$ and the adjacency matrix for that sample $\mathbf A_i$ plus a Gaussian random error with mean zero and variance of $1.0$. Because the generation of the phenotype follows the same model as the Bayesian Network Regression, we denote this scenario as ``theoretical".

We vary the sample size as $n=100,500$. The whole simulation process is illustrated in Fig.~\ref{fig:unrealistic} and the mathematical details are described in the Appendix.

\vspace{0.1cm}
\noindent \textbf{Realistic simulations.} Again, we simulate a 30-taxon phylogenetic tree which represents the true community of microbes using the \texttt{rtree} function from the R package \texttt{ape} \citep{ape} that randomly splits edges until the desired number of leaves is attained. For each microbe, we simulate its true effect on the phenotype under two settings: 1) random in which each microbe has an effect $b_i$ that is distributed \revision{according to the Normal distribution} with mean $\mu_b=0.8$ or $1.6$ and variance of $1.0$ independently of other microbes, and 2) phylogenetic in which we simulate the whole vector of microbial effects $\mathbf b$ as a Brownian motion on the phylogenetic tree using the Julia package \texttt{PhyloNetworks} \citep{phylonetworks}. That is, $\mathbf b \sim N(\mu_b \mathbf 1_{30}, \mathbf \Sigma)$ where $\mu_b=0.8$ or $1.6$, $\mathbf 1_{30}$ is a 30-dimensional vector of ones, and $\mathbf \Sigma$ is the covariance matrix imposed by the phylogenetic tree. In this last setting, microbes that are closely related have similar effects on the phenotype. In the random setting, the main effects of each microbe on the phenotype are generated without regard to the relationships between them.

The generation of the sample adjacency matrices ($\mathbf A_i$) is the same as in the theoretical scenario\revision{, with the exception that we now include an indicator for whether the microbe is in each sample.} Namely, for each sample, we randomly select $k$ (set as $8$ or $22$) microbes out of the 30 total microbes. We build the adjacency matrix for that sample using the phylogenetic distance between microbes. That is, \revision{for $i \neq j$,} the entry $a_{ij}$ is equal to the inverse of the phylogenetic distance between microbe $i$ and microbe $j$ if both microbes are present in the sample (and zero otherwise). \revision{For each microbe $i$, $a_{ii}$ is set to one if microbe $i$ is in the sample and zero otherwise.}

For the computation of the phenotype, we test three settings:

\vspace{0.1cm}
\noindent \textit{Additive model.} The phenotype $y_i$ is computed as the sum of effects $b_i$ for the influential microbes \revision{($\xi_j = 1$)} that are present in the sample plus Gaussian noise. As before, \revision{each} microbe is considered influential with probability $\pi=0.3$, or $0.8$.

\vspace{0.1cm}
\noindent \textit{Interaction model.} In addition to the main effects already described in the additive model, the phenotype also contains interaction terms $a_{jl}\times b_{jl}$, with $b_{jl} \sim N(0.4,1.0)$ if microbes $j$ and $l$ are both influential and $a_{jl}$ as the inverse of phylogenetic distance between microbes $j$ and $l$ if both are in the sample. Since the interaction term is positive, this model is sometimes referred to as a super-additive model.

\vspace{0.1cm}
\noindent \textit{Functional redundancy model.} Here, we assume that the effect of the microbes on the phenotype is not unbounded. That is, different microbes can have the same function, and thus, the phenotype is not affected by both microbes at the same time. We model this mathematically by imposing a threshold $L$ on the phenotype after it was computed following the interaction model. In this setting, we cannot impose the same threshold on all combinations of $\mu_b$ and $\pi$ because the phenotype values will have different ranges. Therefore, we utilize different thresholds for each setting to try to guarantee that not all response values will be capped: $L=3$ for the $\pi=0.3,\mu=0.8$ case, $L=7$ for the $\pi=0.3,\mu=1.6$ case, $L=22$ for the $\pi=0.8,\mu=0.8$ and $L=30$ for $\pi=0.8, \mu=1.6$. Values of $L$ were chosen so that some but not all responses were capped.


Because the phenotype is not computed using the Frobenius product directly (as in the theoretical simulations), but instead it is generated based on biologically reasonable settings (additive, interaction and functional redundancy models), we denote these scenarios as ``realistic". The whole simulation process is illustrated in Fig. \ref{fig:realistic} and the mathematical details are described in the Appendix.

We vary the sample size as \revision{$n=50,500,1000$. For the case of $n=50$, we perform Gaussian data augmentation to increase sample size to $n=200$. This simulation scenario is included to justify the application of the model on the real dataset which has 50 samples.
For each of the 50 true samples, an augmented sample is created. First, to generate the response, an offset is generated from $N(0,s_P^2/4)$, where $s_P^2$ is the sample variance of the response in the 50 original samples. To generate augmented microbiome data, each microbe is present in the augmented sample with probability 0.9 if it was in the original sample or 0.1 if it was not, independently of other microbes. This augmentation technique preserves the underlying relationship between the graph nodes and edges in the original sample while adding a small amount of noise to both the network and the response.} 


\subsubsection*{MCMC convergence}

For each simulation setting, we run three MCMC chains and assess convergence using the $\hat{R}$ convergence criterion proposed in \citet{vehtari2019rank}. We consider convergence to have been \revision{unambiguously} achieved if \revision{$\hat{R}\leq1.01$} for all of the $\gamma$ and $\xi$ variables as suggested in \citet{vehtari2019rank}. \revision{For all cases, half of generated samples are discarded as burn-in. A minimum of 10,000 total samples and a maximum of 800,000 total samples are generated for the theoretical simulations, while a minimum of 5,000 total generated samples and a maximum of 600,000 generated samples are generated for the realistic simulations. A total of 6 theoretical simulation cases (out of 108) and 14 realistic simulation cases (out of 144) fail to achieve a $\hat{R} \leq 1.01$, but all achieve $\hat{R} \leq 1.1$ (Tables \ref{tab:burn_in_unrealistic} and \ref{tab:_burn_in_realistic}). We accept this as reasonable evidence they have achieved convergence, noting that every dimension of both $\xi$ and $\gamma$ variables achieves this $\hat{R}$ value. } See Results for information on computing times.

\begin{table}[!ht]
\centering
\begin{tabular}{|l|l|l|l|l|l|l|l|}
\hline
 {\bf $\mu$ } & {\bf $\pi$ } & {R} & {$k$} &  {\revision{Data} sample size} & {$\hat{R}(\xi)$ } & {$\hat{R}(\gamma)$}\\ \hline
1.6 & 0.3 & 5 & 22 & 100 & 1.05 & 1.06 \\
1.6 & 0.8 & 5 & 22 & 100 & 1.04 & 1.02 \\
1.6 & 0.3 & 5 & 8  & 500 & 1.00 & 1.02 \\
0.8 & 0.3 & 7 & 15 & 500 & 1.06 & 1.06 \\
0.8 & 0.8 & 7 & 22 & 100 & 1.01 & 1.02 \\
0.8 & 0.3 & 9 & 15 & 500 & 1.07 & 1.01 \\\hline
\end{tabular}
\caption[Maximum $\hat{R}$ values for theoretical simulation cases]{\revision{Maximum $\hat{R}$ values for theoretical simulation cases that did not achieve $\hat{R} \leq 1.01$ for all $\gamma$ and $\xi$ variables.}}
\label{tab:burn_in_unrealistic}
\end{table}

\begin{table}[!ht]
\centering
\begin{tabular}{|l|l|l|l|l|l|l|l|}
\hline
 {\bf $\mu$ } & {\bf $\pi$ } & {R} & {$k$} & {\revision{Data} sample size} & {Simulation type} & {$\hat{R}(\xi)$ } & {$\hat{R}(\gamma)$}\\ \hline
1.6 & 0.3 & 5 & 22 & 50   &  redundant random   & 1.02 & 1.01 \\
0.8 & 0.3 & 7 & 22 & 1000 &  redundant phylo    & 1.06 & 1.06 \\
0.8 & 0.3 & 7 & 22 & 1000 &  redundant random   & 1.02 & 1.03 \\
0.8 & 0.3 & 7 & 22 & 1000 &  interaction random & 1.02 & 1.01 \\
0.8 & 0.3 & 7 & 22 & 1000 &  redundant phylo    & 1.09 & 1.00 \\
0.8 & 0.3 & 7 & 22 & 1000 &  redundant random   & 1.02 & 1.03 \\
0.8 & 0.3 & 7 & 22 & 500  &  redundant phylo    & 1.02 & 1.06 \\
0.8 & 0.3 & 7 & 22 & 500  &  redundant phylo    & 1.03 & 1.01 \\
1.6 & 0.3 & 7 & 22 & 1000 &  redundant random   & 1.10 & 1.04 \\
1.6 & 0.3 & 7 & 22 & 1000 &  interaction phylo  & 1.04 & 1.01 \\
1.6 & 0.3 & 7 & 22 & 1000 &  redundant phylo    & 1.08 & 1.00 \\
1.6 & 0.3 & 7 & 22 & 1000 &  redundant random   & 1.02 & 1.00 \\
1.6 & 0.3 & 7 & 22 & 500  &  redundant random   & 1.05 & 1.03 \\
1.6 & 0.8 & 7 & 22 & 1000 &  additive phylo     & 1.00 & 1.02 \\\hline
\end{tabular}
\caption[Maximum $\hat{R}$ values for realistic simulation cases]{\revision{Maximum $\hat{R}$ values for theoretical simulation cases that did not achieve $\hat{R} \leq 1.01$ for all $\gamma$ and $\xi$ variables.}}
\label{tab:_burn_in_realistic}
\end{table}

\subsection*{Connections of Bacterial Microbiome to Phosphorous Leaching}

We re-analyze the experimental data collected by \citet{wagg2019fungal} to study the fungal-bacterial interactions in soil microbiome and how they are connected to specific biological phenotypes. In this experiment, a soil diversity gradient was produced by filtering inoculum through different meshes from less than 5mm to less than 0.001mm\revision{, in order to create various microbial networks}. In addition, a number of indicators of soil microbiome functions were collected, of which we focus on phosphorous leaching as our response. Following \citet{wagg2019fungal}, we construct the microbiome network for each sample as follows:
\begin{enumerate}
    \item Remove all OTUs occurring fewer than 40 samples in the dataset
    \item Construct a "meta-network" using data from all samples. This is done with the R package Spiec-Easi \citep{kurtz2015spieceasi}.
    \item For each sample, construct a "sample network" from the meta-network by including only edges between OTUs appearing in that sample. An OTU appears in a sample if its relative abundance is different than zero. If an edge does not appear in the sample, its edge value is set to 0. If an OTU appears in the sample its "node value" (edge between the node and itself) is set to 1, and 0 otherwise.
\end{enumerate}

After removal of OTUs than do not appear in at least 40 samples, the resulting meta-network contains 90 OTUs, for a model matrix of dimension $\mathbf{X} \in \mathbb{R}^{50 \times 4005}$ given that the experiment collected 50 samples.
Our simulation results indicate that this number of samples will likely not be enough to achieve the necessary statistical power to detect influential microbes or edges. Therefore, we employ data augmentation to generate \revision{150 more samples}, for a total of \revision{200 samples. We use the same data augmentation technique described previously (Realistic Simulations) with one change. Because phosphorous leaching must always be positive, if the augmented response is negative a new value for the augmented response is generated from $\frac{1}{15}\chi_3^2$ (i.e. one draw from a chi-square distribution with 3 degrees of freedom, divided by 15) to guarantee all response values are positive.} The original phosphorous leaching values range from 0.0040 to 0.6840, with a mean of 0.1593 mg/L. The augmented phosphorous leaching values range from 0.0066 to 0.6760 with a mean of 0.1900 mg/L. For the original data, sample sizes of distinct OTUs range from 52 to 90 with a mean number of OTUs per sample of 82. For the augmented data, samples sizes of distinct OTUs range from 55 to 84 with a mean number of OTUs per sample of 75. Much of the sparsity in the data comes from the relationships between OTUs, rather than the number of OTUs collected in each sample. Only about 2\% of the entries in the meta-matrix are non-zero.

\subsubsection*{MCMC convergence}
 
As in the simulation study, we run three MCMC chains and assess convergence using the $\hat{R}$ convergence criterion proposed in \citet{vehtari2019rank}. \revision{We again aim for $\hat{R}\leq1.01$ for all of the $\gamma$ and $\xi$ variables as a cutoff value that signifies convergence. We are not able to achieve this value, but we find that \revision{100,000} burn-in followed by \revision{100,000} samples (\revision{200,000} total generations) give us $\hat{R}(\xi)\leq  1.02$ and $\hat{R}(\gamma) \leq 1.017$. We accept this as reasonable evidence of convergence. }

\section*{Results}

\subsection*{Theoretical Simulations}

\begin{figure}[!ht]
\centering
\begin{subfigure}[t]{0.49\textwidth}
    \centering
    \includegraphics[scale=0.27]{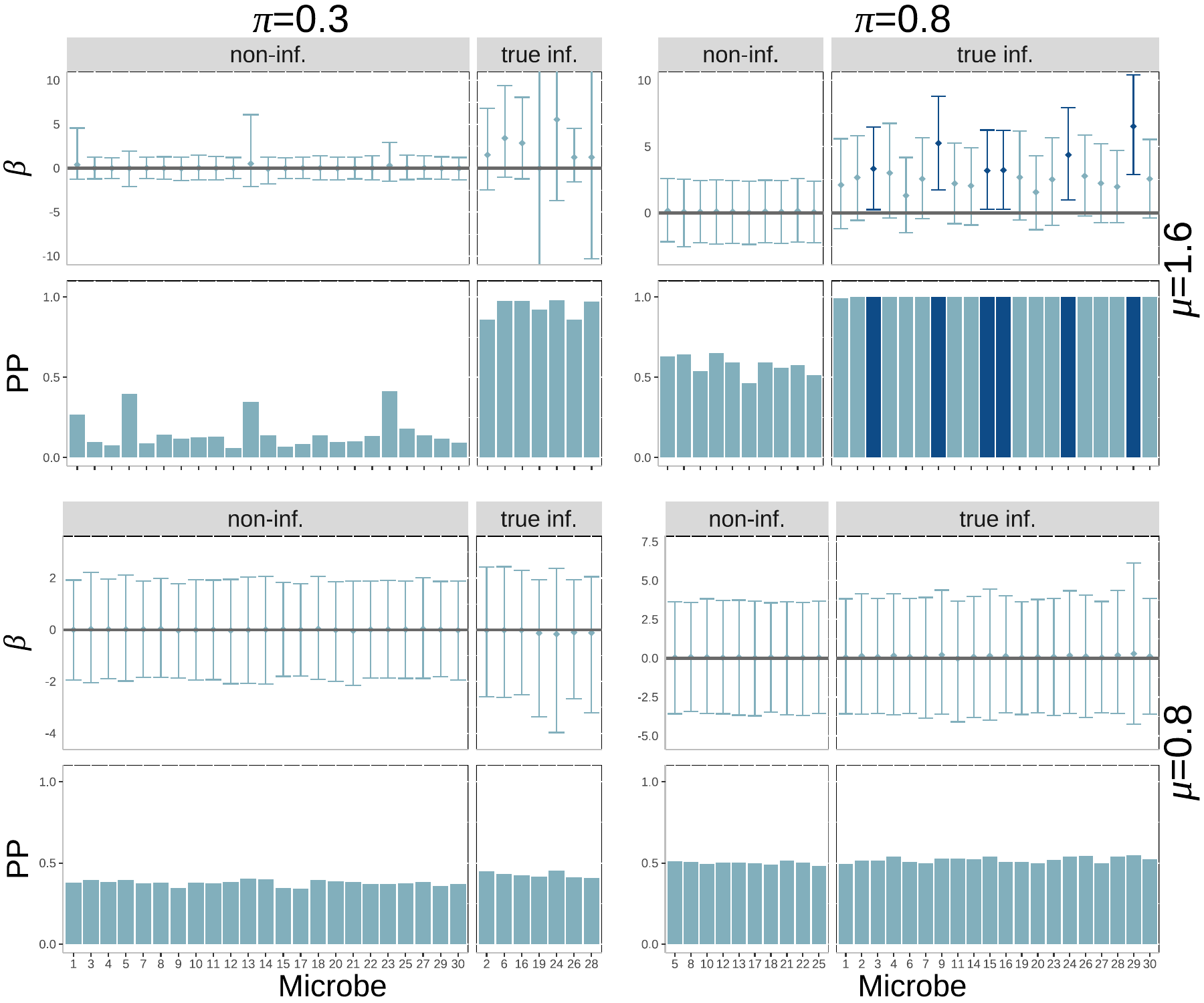}
    \caption{$\mathbf {n=100}$}
\end{subfigure}
\begin{subfigure}[t]{0.49\textwidth}
    \centering
    \includegraphics[scale=0.27]{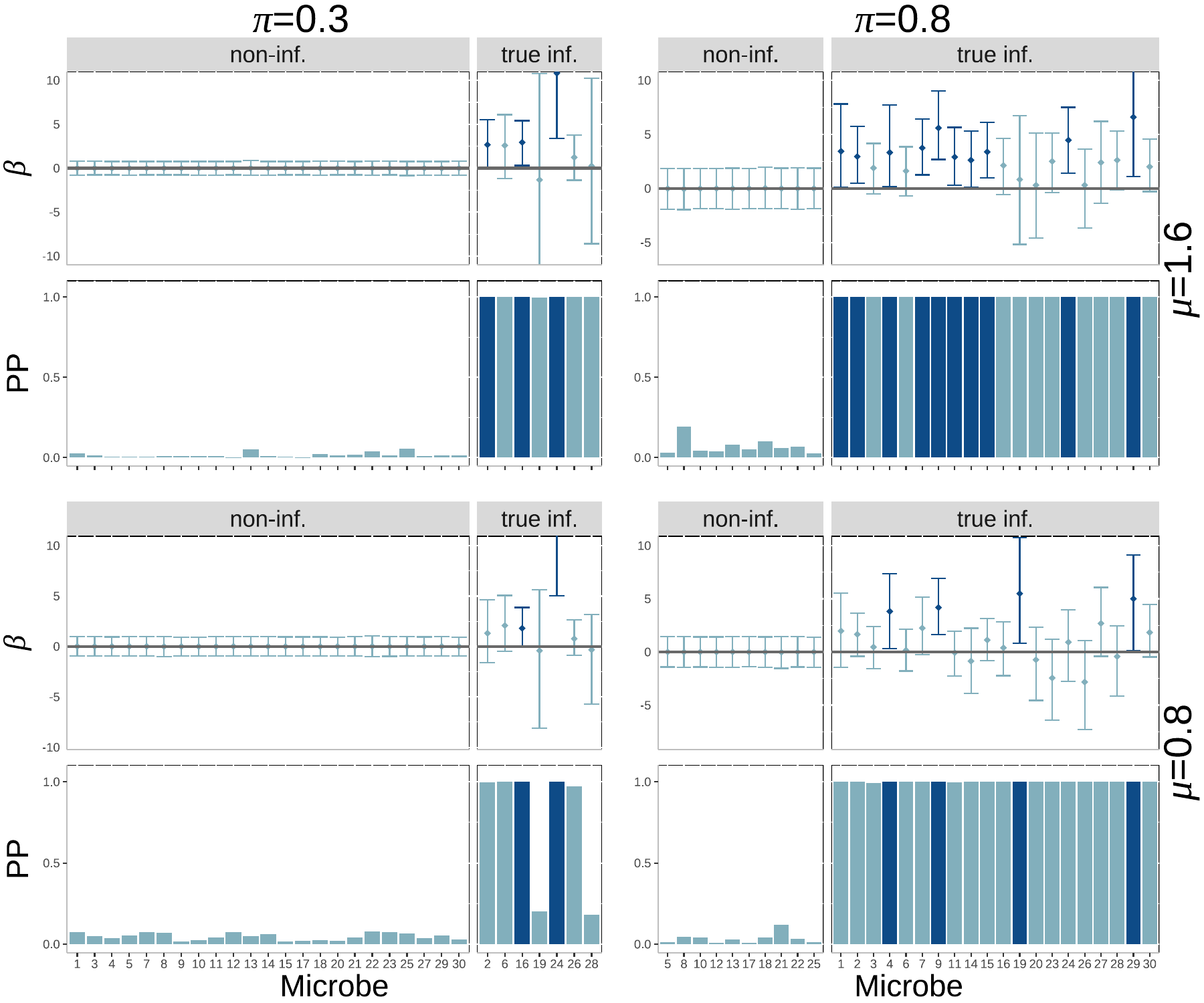}
    \caption{$\mathbf {n=500}$}
\end{subfigure}

\caption[Posterior probability of influential nodes and coefficients for nodes (theoretical simulations)]{{\bf Posterior probability of influential nodes and coefficients for nodes (theoretical simulations).}
All simulations are done with latent dimension $R=7$ and $k=8$ sampled microbes. Different \revision{groups of 4 panels (left side vs. right side)} represent different sample sizes ($n=100,500$). \revision{Within each group, we have four panels corresponding to the two values of edge effect size ($\mu=0.8, 1.6$) and two values of probability of influential node ($\pi=0.3, 0.8$) which controls the sparsity of the regression coefficient matrix ($\mathbf B$). Within each of these panels we have two plots: 95\% credible intervals (top) and posterior probability of influence (bottom - calculated as the mean of the $\xi$ variable for the node across Gibbs samples) for each node.} Each bar corresponds to one node (microbe). \revision{Within each plot the bars and intervals are colored depending on whether the node is found to be influential (dark) or not influential (light) based on the 95\% credible intervals. Each plot is split based on whether the nodes are truly influential (right) or not (left).}
}
\label{fig:nodes}
\end{figure}

Fig. \ref{fig:nodes} shows the posterior probability of being an influential node (key microbe) \revision{(bottom pane in each plot) as well as the credible interval for the estimated main effect of each microbe (top pane of every plot) for (a) $n=100$ and (b) $n=500$. Each plot corresponds to a pair of effect size ($\mu=0.8,1.6$) and sparsity ($\pi=0.3,0.8$) at $k=8$ sampled microbes in the regression coefficient matrix $\mathbf B$ (see Appendix Fig. \ref{fig:nodes-all} for the $k=22$ case) with latent dimension $R=7$. Each plot is also separated in non-influential microbes (left) and true influential microbes (right). Posterior probability for a node is calculated as the mean of the $\xi$ value for that node across retained Gibbs samples. We expect the model to identify true influential microbes by tall bars on the right side of the pane (high estimated posterior probability of influential node) and non-influential microbes as short bars on the left side of the pane (low estimated posterior probability of influential node)}. For smaller sample size ($n=100$), the effect sizes need to be larger ($\mu=1.6$) for the nodes to be accurately detected as influential (tall bars in bottom-right panes in each plot).
For a larger sample size ($n=500$), the model has a high PP for truly influential nodes \revision{(tall bars in bottom-right panes in each plot)} and a low PP for non-influential nodes \revision{(short bars in bottom-left panes in each plot)} regardless of the values of $k,\mu,\pi$. There seem to be no major differences in the performance of the method in terms of regression coefficient sparsity ($\pi$) or adjacency matrix sparsity ($k$) with the exception that for smaller effect sizes ($\mu=0.8$) in small sample size setting ($n=100$), less sparsity in $\mathbf A$ ($k=22$) improves the detection of influential nodes compared to $k=8$. \revision{
The coloring of the intervals (and the corresponding posterior probability bars) indicates whether the nodes are found to have a significant effect on the response by the main coefficient at the 10\% level (credible intervals do not cross 0). In this simulation, using the credible intervals results in a more conservative test that fails to identify many significant microbes (high but light bars in the bottom-right pane).}

We further compare the performance when changing the latent dimension ($R$). In Fig. \ref{fig:nodes}, we have a latent dimension of $R=7$ which produced more accurate results than $R=5$ (Fig. \ref{fig:nodes-sm} in the Appendix). Latent dimension of $R=9$ (Fig. \ref{fig:nodes-sm2} in the Appendix) produces slightly better accuracy compared to $R=7$, but it also creates floating-point instability in the \revision{computations}. As \citet{guha2021bayesian} suggest, we aim to find the smallest $R$ value that produces good performance to guarantee floating-point stability. In our case, we choose a latent dimension of $R=7$ for all the remaining simulations \revision{(aside from the data augmentation simulations, for which we find that a lower value of $R=5$ is required for floating-point stability)}.
Lastly, results for $k=15$ sampled microbes can be found in the Figs. \ref{fig:nodes-sm3-R5}, \ref{fig:nodes-sm3-R7}, \ref{fig:nodes-sm3-R9} in the Appendix with no considerable differences with respect to $R$. 

We conclude that \revision{using posterior probability of influence,} the method is able to detect influential nodes (microbes) for a sufficiently large sample size ($n=500$ here) regardless of the effect size ($\mu$) and sparsity ($\pi$) of the regression coefficient $\mathbf B$ and regardless of the sparsity of the adjacency matrix ($k$). For smaller sample size ($n=100$), either larger effect sizes are needed ($\mu=1.6$) or less sparsity in the adjacency matrix $\mathbf A$ ($k=22$).

\revision{Fig.~\ref{fig:edges8} (Appendix) shows} 95 \% credible intervals for the regression coefficients per edge ordered depending on whether they are truly non-influential edges (top of each panel) or truly influential edges (bottom of each panel) \revision{ for each combination of edge effect size ($\mu=0.8, 1.6$) and probability of influential node ($\pi=0.3, 0.8$). This figure represents the case of $k=8$ sampled microbes}. The color of the intervals depends on whether it intersects zero (light) and hence estimated to be non-influential or does not intersect zero (dark) and hence estimated to be influential by the model. These panels allow us to visualize false positives (dark intervals on the top panel) or false negatives (light intervals on the bottom panel) for all simulation settings ($n,k,\pi,\mu$). Smaller samples size (top $n=100$) has considerably more false negatives compared to larger sample size (bottom $n=500$) as evidenced by the many light intervals in the ``True influential edges" panels. This is true especially for the cases of high sparsity in $\mathbf B$ ($\pi=0.3$ for both $\mu=0.8$ and $1.6$) and low effect size with less sparsity in $\mathbf B$ ($\pi=0.8,\mu=0.8$).
Overall, all simulation settings show controlled false positive rate as evidenced by few dark intervals on the ``True non-influential edges" panels, regardless of sample size ($n$), effect size ($\mu$) and sparsity in $\mathbf B$ ($\pi$).
Fig. \ref{fig:edges22} \revision{(Appendix)} shows the same plot for $k=22$ sampled microbes instead of $k=8$. The conclusions are the same which provides evidence that the identification of influential edges does not depend on the number of microbes in the samples.

\begin{figure}[!ht]
    \centering
    \includegraphics[scale=0.38]{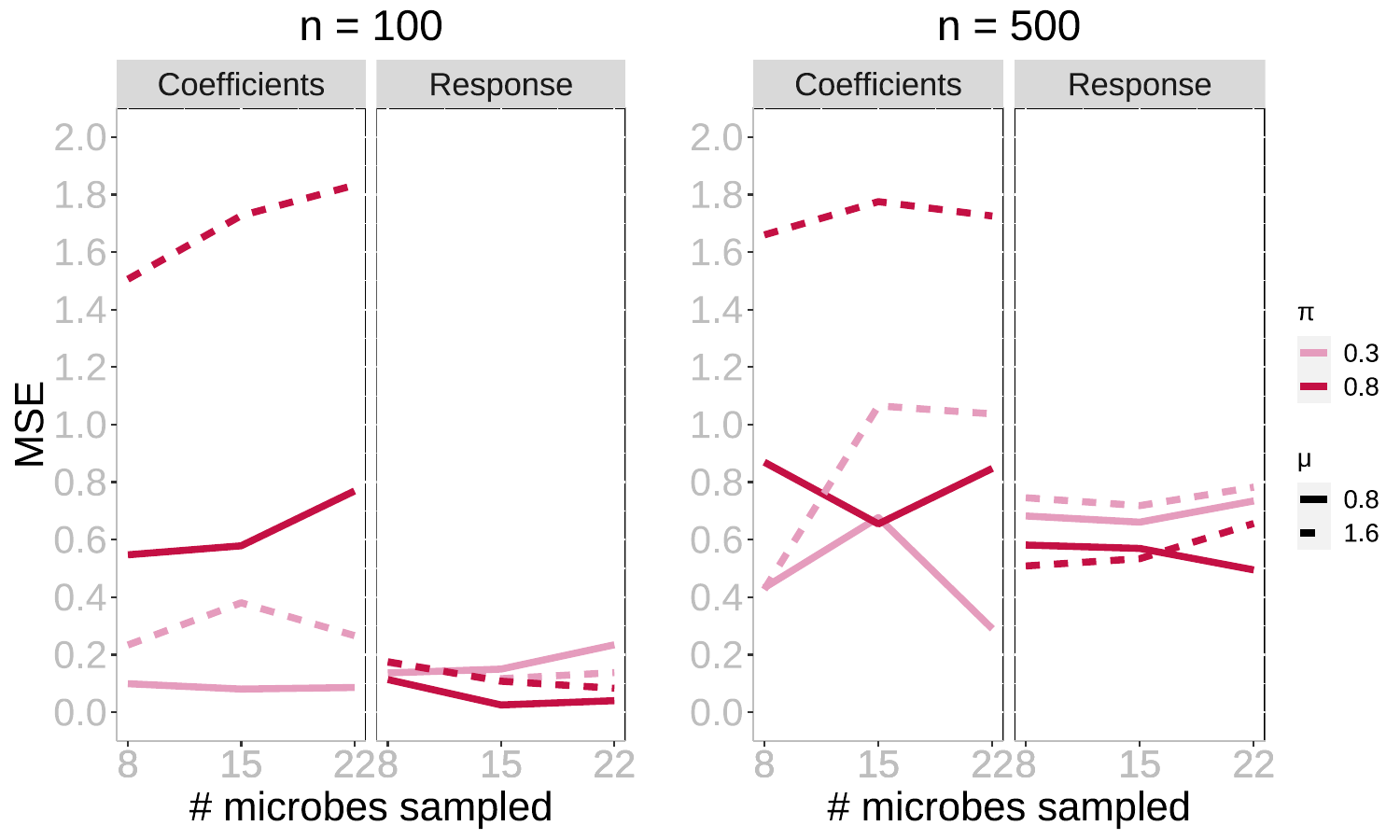}
    \caption[Mean Square Error for regression coefficients and response]{{\bf Mean Square Error for regression coefficients and response \revision{($R=7$)}.} X axis corresponds to the number of sampled nodes (microbes) which relates to the sparsity of the adjacency matrix $\mathbf X$. Dashed lines correspond to different values of the true mean for edge effects ($\mu=0.8, 1.6$) and different colors correspond to different sparsity levels on the regression coefficient matrix $\mathbf B$ ($\pi=0.3,0.8$).}
    \label{fig:mse}
\end{figure}

\revision{Fig. \ref{fig:mse} shows the mean squared error (MSE) (for both coefficients and response) for all 12 cases for sample sizes of $n=100$ and $n=500$. The model appears to generate coefficients which, in aggregate, predict the response well, though the coefficients themselves may not approximate the true coefficients as well. For the coefficients, the model is better at approximating the true coefficients when the edge effect is strong. Interestingly, in the $n=100$ case especially, the model performs better (in terms of MSE of coefficients) for the sparser coefficient matrices - the opposite result observed in node identification. This is likely due to the edge coefficients being biased towards zero. Fig.~\ref{fig:rates} (Appendix) shows false positive and false negative rates for each case (using an arbitrary cutoff for significance of 0.5), and Fig. \ref{fig:roc} (Appendix) shows ROC curves, calculated as the cutoff increases from 0.1 to 1. For both sample sizes ($n=100,500$), the model performs better (in terms of true and false positives and negatives) with stronger edge effects ($\mu=1.6$ vs $\mu=0.8$). Among both edge effect strengths, the model performs better with the larger sample size ($n=500$ vs $n=100$). There are very few false positives for the $n=500$ case. The ROC curves show very good performance for the larger sample size case ($n=500$).}

\subsection*{Realistic simulations}

\subsubsection*{Additive model}

Fig. \ref{fig:nodes_add_rand} (Appendix) and Fig. \ref{fig:nodes_add_phylo} (Appendix) show both the posterior probability of influential nodes \revision{and the coefficients for nodes} for random coefficients and phylogenetic coefficients respectively. Both types of coefficients produce similar results. \revision{The model performs poorly for all cases as it is unable to identify influential nodes regardless of sample size ($n$) and characteristics of $\mathbf B$ ($\pi,\mu$). The model is able to identify a few influential microbes through the node coefficients, but not through posterior probability of influence in the sparse sample ($k=8$) case only.}

\subsubsection*{Interaction model}

\begin{figure}[!ht]
    \centering
    \begin{subfigure}[t]{0.49\textwidth}
        \includegraphics[scale=0.27]{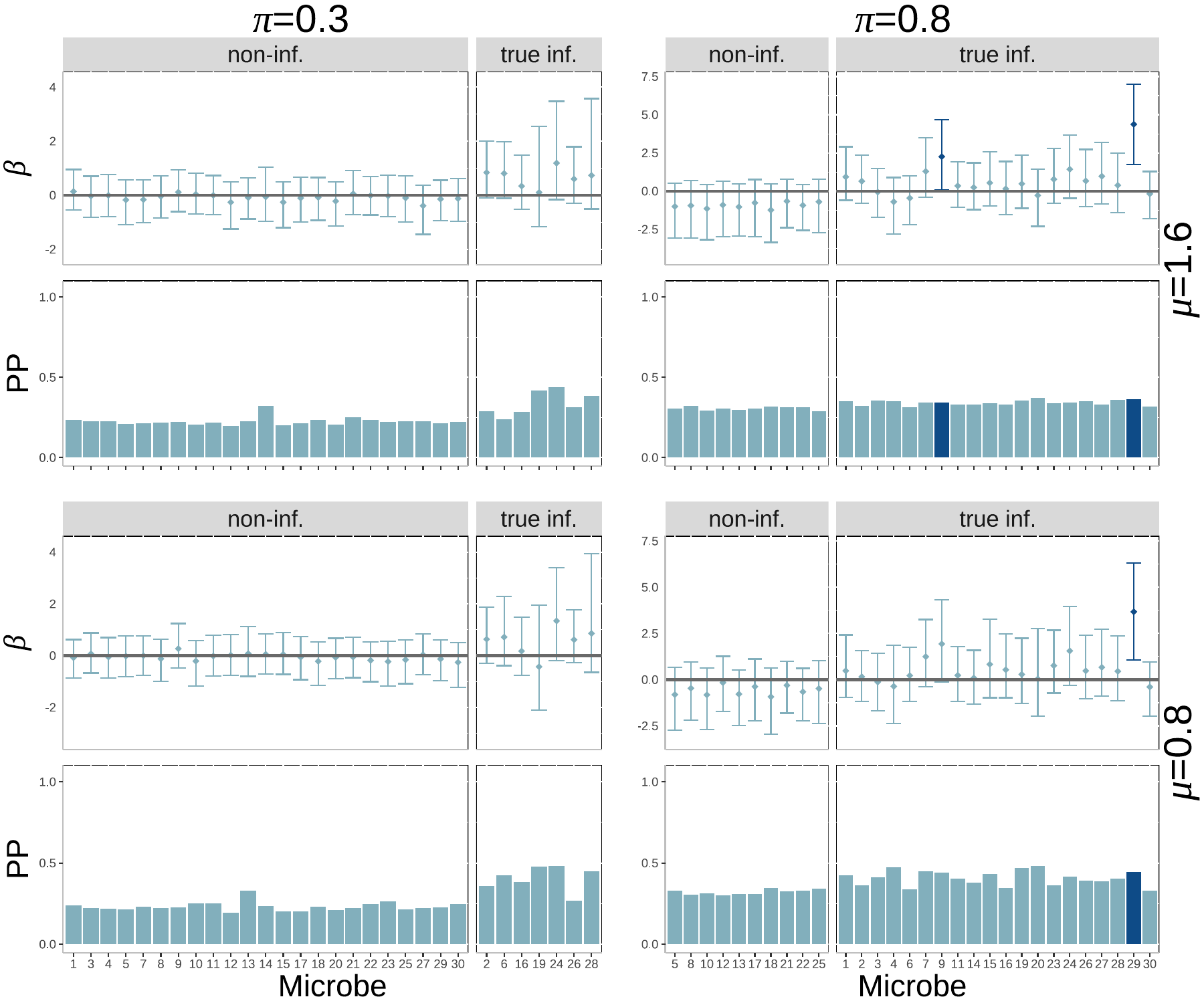}
        \caption{$\mathbf{n=500}$}
    \end{subfigure}
    \begin{subfigure}[t]{0.49\textwidth}
        \includegraphics[scale=0.27]{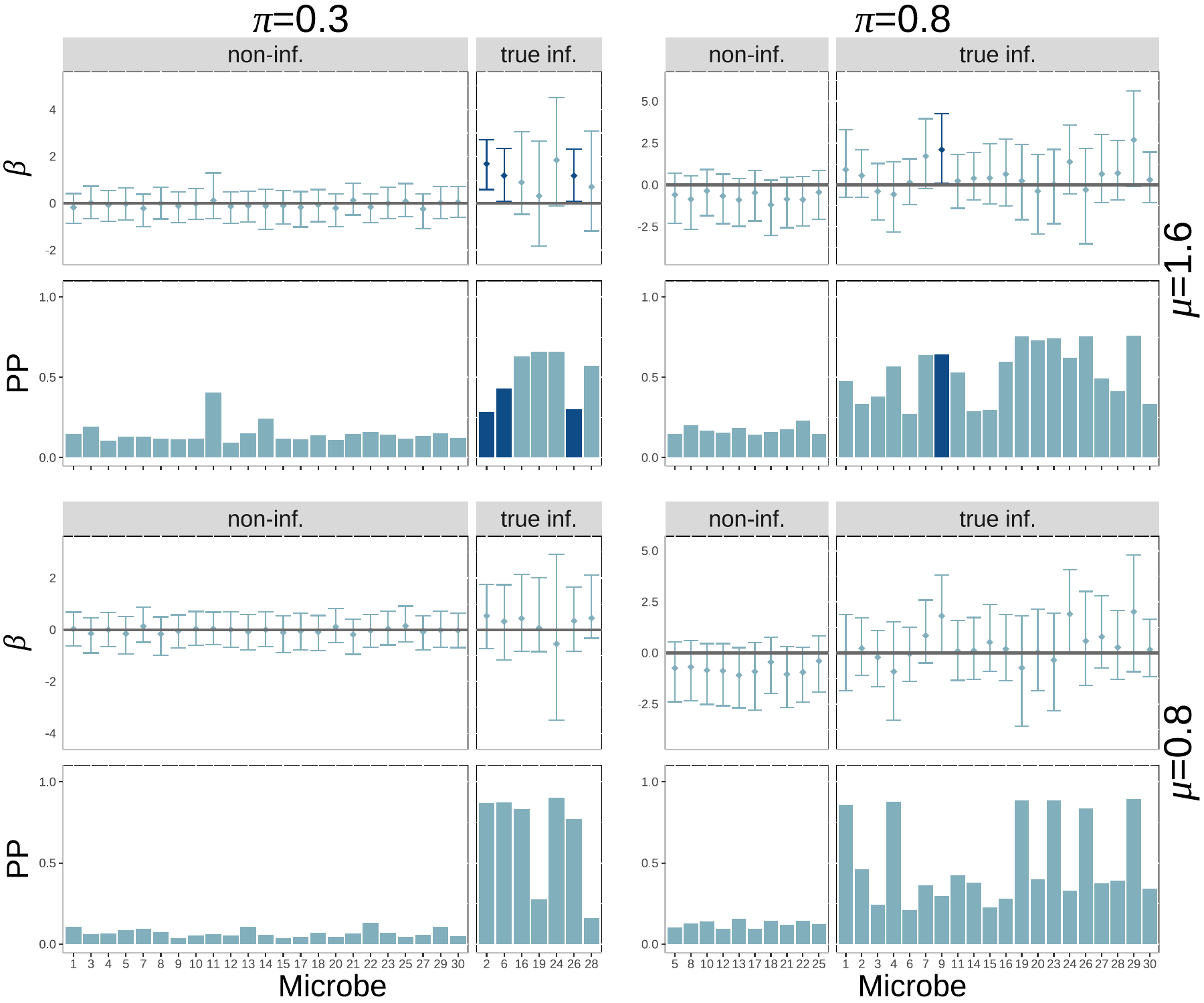}
    \caption{$\mathbf{n=1000}$}
    \end{subfigure}
    \caption[Posterior probability of influential nodes and coefficients for nodes (interaction model with phylogenetic coefficients) ]{{\bf Posterior probability of influential nodes and coefficients for nodes (interaction model with phylogenetic coefficients).}
    All plots shown are for $k=8$ sampled microbes. Different \revision{ groups of four panels (left side vs. right side)} represent different sample sizes ($n=100,500$). \revision{Within each group, we have four panels corresponding to the two values of edge effect size ($\mu=0.8, 1.6$) and two values of probability of influential node ($\pi=0.3, 0.8$) which controls the sparsity of the regression coefficient matrix ($\mathbf B$). Within each of these panels we have two plots: 95\% credible intervals (top) and posterior probability of influence (bottom - calculated as the mean of the $\xi$ variable for the node across Gibbs samples) for each node.} Each bar corresponds to one node (microbe). \revision{Within each plot the bars and intervals are colored depending on whether the node is found to be influential (dark) or not influential (light) based on the 95\% credible intervals. Each plot is split based on whether the nodes are truly influential (right) or not (left).}
    }
    \label{fig:nodes_int_phylo}
\end{figure}

Fig. \ref{fig:nodes_int_phylo} (\revision{$k=8$ - for phylogenetic coefficients}) shows the posterior probabilities of influential nodes \revision{as well as the node coefficient 95\% credible intervals} under the interaction model (Fig. \ref{fig:nodes_int_rand} for random coefficients \revision{and Fig. \ref{fig:nodes_int_phylo_adx} for both node sparsity values ($k=8,22$)} are in the Appendix). Both types of coefficients show similar performance \revision{in terms of posterior probabilties of microbe influence} when there is low sparsity in the $\mathbf B$ matrix ($\pi=0.8$). Namely, \revision{for the larger sample size ($n=1000$)} the method estimates a high PP for truly influential nodes (tall bars in the lower-right panes) and a low PP for non-influential nodes (short bars in the lower left panes) for all cases of low sparsity in $\mathbf B$ ($\pi=0.8$) number of sampled microbes ($k$) or effect size ($\mu$). 

For scenarios of high sparsity in $\mathbf B$ ($\pi=0.3$), both types of coefficients \revision{show good performance with the higher sample size($n=1000$), though performance is better under the phylogenetic regime, where results are similar to those in the low sparsity case ($\pi=0.8$). For both types of coefficients, there does not seem to be much if any difference in model performance between different node coefficient signal strengths ($\mu=0.8,1.6$) or number of microbes sampled ($k=8,22$). However, in terms of using posterior node coefficients for influential node identification (95\% credible intervals in upper panes in each quadrant of Fig. \ref{fig:nodes_int_phylo}), the lower signal strength ($\mu=0.8$) appears to be too weak, with the model finding very few truly influential microbes (dark blue bars/credible intervals on the right pane of each plot) in this case. However, note that there are several cases ($\pi=0.8,\mu=1.6,n=500$ for phylogenetic coefficients, $\pi=0.3,\mu=1.6,n=500$ for random coefficients) in which the posterior node coefficient credible intervals identify truly influential nodes not identified using posterior probabilities of influence. Also note that there are no false positive identifications (dark blue bars/credible intervals in the left pane of each plot) using this method.}

Fig. \ref{fig:edges_int_rand} \revision{(Appendix)} shows the 95 \% credible intervals for edge effects under the interaction model with random coefficients for $k=8$ sampled nodes. The results are very similar for $k=22$ sampled nodes (Fig. \ref{fig:edges_int_rand2} in the Appendix) and for phylogenetic coefficients (Fig. \ref{fig:edges_int_phylo} for $k=8$ and Fig. \ref{fig:edges_int_phylo2} for $k=22$, both in the Appendix). Namely, the model displays a low false positive rate as evidenced by few dark intervals on the ``True non-influential edges" panels for all sample sizes ($n$), sparsity levels ($\pi$) and effect sizes ($\mu$). We highlight that it is expected that there will be few differences when comparing the two effect sizes ($\mu=0.8,1.6$) as these quantities refer to the main (node) effects, not the interaction (edge) effects which was set as 0.4 for all simulations. The purpose of these simulations is to test if changes in the main (node) effects biased the performance of the BNR model to detect influential edges. It appears from these figures that there is no such bias. That is, when the biological phenotype is generated under the interaction model, \revision{as long as the sample size is high enough} the BNR has good performance to identify the influential edges regardless of the number of microbes sampled ($k$), the sparsity level ($\pi$), node effect sizes ($\mu$), and type of coefficient (random vs phylogenetic) with controlled false positive and false negative rates in all settings.

\subsubsection*{Functional redundancy model}

\begin{figure}[!ht]
\centering
\begin{subfigure}[t]{0.49\textwidth}
    \includegraphics[scale=0.27]{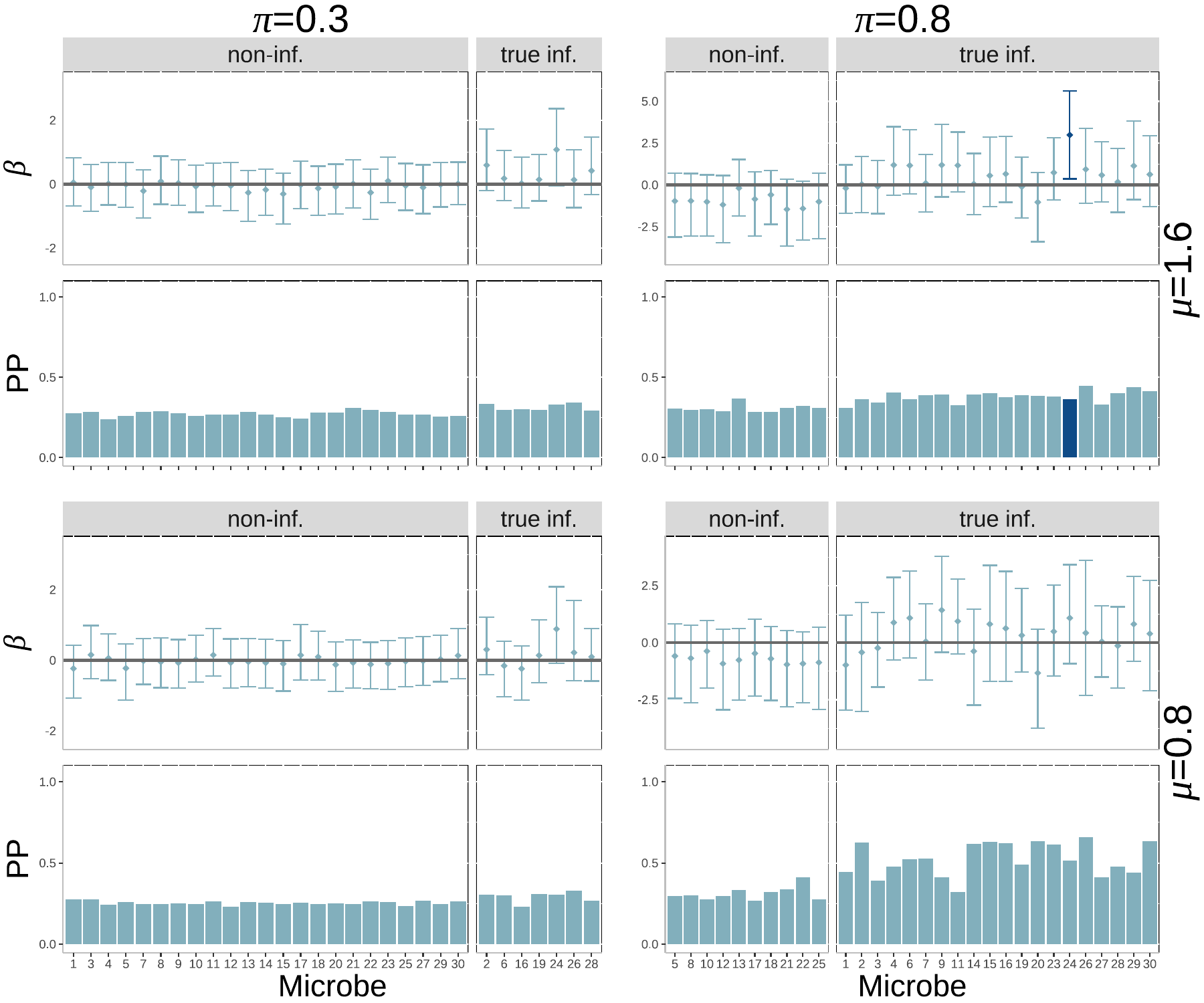}
    \caption{$\mathbf{n=500}$}
\end{subfigure}
\begin{subfigure}[t]{0.49\textwidth}
    \includegraphics[scale=0.27]{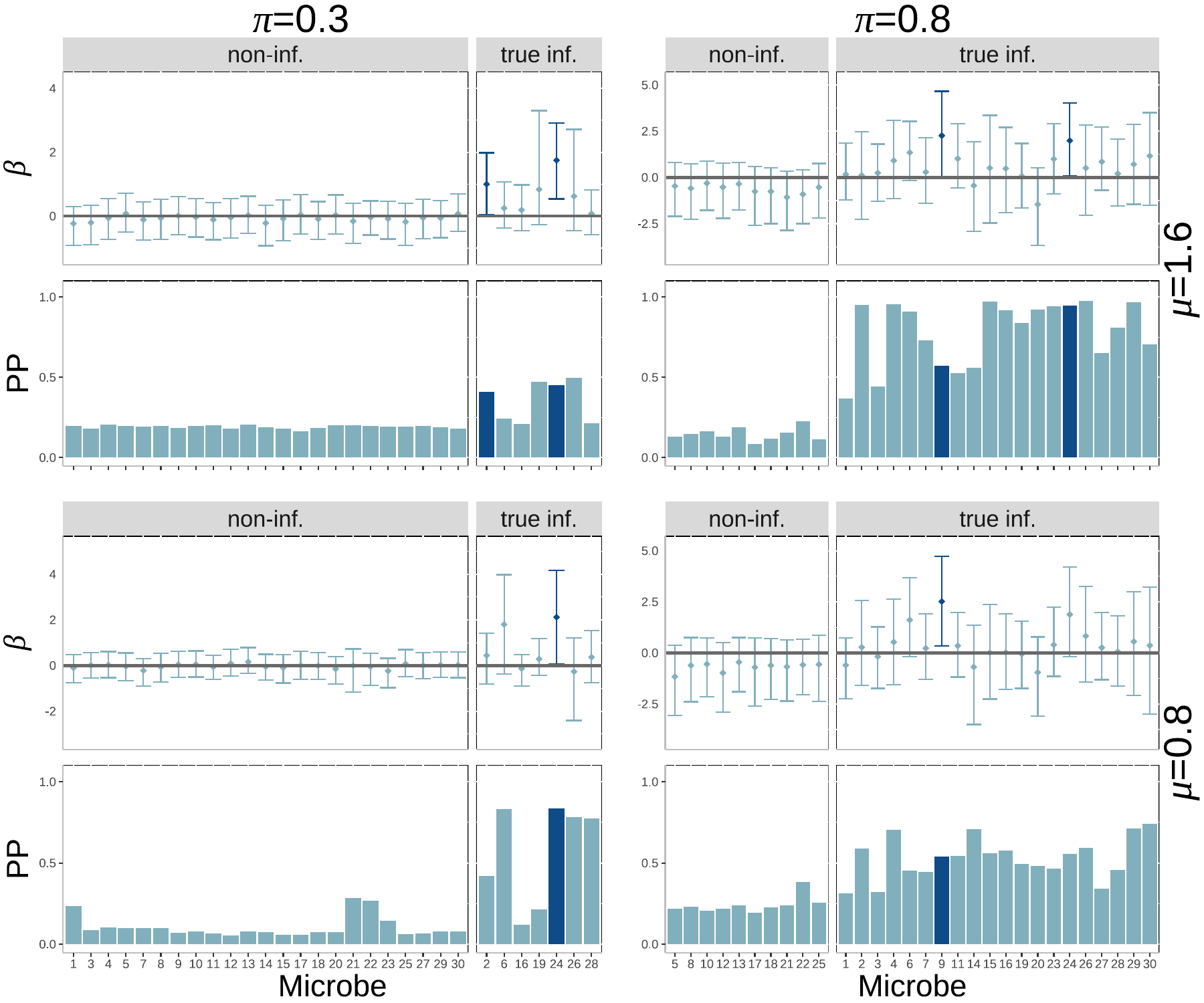}
    \caption{$\mathbf{n=1000}$}
\end{subfigure}
\caption[Posterior probability of influential nodes and coefficients for nodes (functional redundancy model with random coefficients)]{{\bf Posterior probability of influential nodes and coefficients for nodes (functional redundancy model with random coefficients).}
Different \revision{groups of four panels (quadrants) represent different sample sizes ($n=100,500$) and different number of sampled microbes ($k=8$). Within each group, we have four panels corresponding to the two values of edge effect size ($\mu=0.8, 1.6$) and two values of probability of influential node ($\pi=0.3, 0.8$) which controls the sparsity of the regression coefficient matrix ($\mathbf B$). Within each of these panels we have two plots: 95\% credible intervals (top) and posterior probability of influence (bottom - calculated as the mean of the $\xi$ variable for the node across Gibbs samples) for each node. Each bar corresponds to one node (microbe). Within each plot the bars and intervals are colored depending on whether the node is found to be influential (dark) or not influential (light) based on the 95\% credible intervals. Each plot is split based on whether the nodes are truly influential (right) or not (left).}
}
\label{fig:nodes_red_rand}
\end{figure}

Fig. \ref{fig:nodes_red_rand} shows the posterior probability of influential nodes \revision{as well as the node coefficient 95\% credible intervals} under the functional redundancy model with random coefficients \revision{for the case of $k=8$ (Fig. \ref{fig:nodes_red_rand2} in the Appendix for the case of $k=22$)}. Using posterior node probabilities, we observe that there is a high false positive rate \revision{ when there are many nodes sampled ($k=22$) in all sparsity and coefficient strength settings. Also, there is a high false negative rate in the remaining cases for the smaller sample size $n=500$. Good performance is only achieved when few microbes are sampled $k=8$ and the sample size is large $n=1000$}. This behavior is similar with phylogenetic coefficients  (see Figs. \ref{fig:nodes_red_phylo_k8}, \ref{fig:nodes_red_phylo_k22} in the Appendix). This implies that the BNR model is unable to identify influential microbes under a model of functional redundancy unless there are very few sampled microbes ($k=8$). This result could be explained by the fact that multiple sampled microbes could cause the phenotype to reach the threshold more easily and thus, there is less information on the variability of the response to estimate the effects and influential probabilities. \revision{However, using the node coefficient credible intervals, we are able to identify more influential coefficients (dark blue bars/intervals on the right pane of each plot) in the $k=22$ case with fewer false positives (dark blue bars/intervals on the left pane of each plot).}

Fig. \ref{fig:edges_red_rand} \revision{(Appendix)} shows the 95 \% credible intervals for edge effects under the functional redundancy model with random coefficients for $k=8$ sampled nodes. The results are very similar for phylogenetic coefficients (Fig. \ref{fig:edges_red_phylo} for $k=8$ in the Appendix). Namely, there is controlled false positive rate (dark intervals in the ``True non-influential edges" top panels) for the case of small sample size (top: $n=500$) which strangely worsens slightly for larger sample size (bottom: $n=1000$) for all cases of $\pi,\mu$. False negative rate (light intervals in the ``True influential edges" top panels) appears unaffected by $n,\pi,\mu$. When $k=22$ microbes are sampled instead, the model performs worse in all settings (Fig. \ref{fig:edges_red_rand2} for random coefficients for $k=22$ sampled nodes and Fig. \ref{fig:edges_red_phylo2} for phylogenetic coefficients for $k=22$, both in the Appendix). Namely, there are inflated false positive and false negative rates for all settings of $n,\pi,\mu$. 
Similarly to the identification of influential nodes, it seems that the BNR model is unable to accurately identify influential edges under a functional redundancy model when many nodes are sampled ($k=22$). That is, when the biological phenotype is generated under the functional redundancy model, the BNR has good performance to identify influential nodes and influential edges only when there are few microbes sampled ($k=8$).

\subsubsection*{\revision{MSE,} false positive and false negative rates}

We present the plots for false positive and negative rates \revision{ as well as mean squared error} in the Appendix.
Fig. \ref{fig:rates-rand} shows the false positive and false negative rates for edges and nodes for different simulation settings in terms of $n,k,\pi,\mu$ for additive (top), interaction (middle) and functional redundancy (bottom) models with random coefficients. 
An interaction model with low sparsity in $\mathbf B$ ($\pi=0.8$) shows the best performance in terms of controlled false positive and false negative rates for all settings of $n$ and $\mu$. Under the interaction model, there is inflated false negative rate of edges. \revision{Overall, there does not seem to be much effect of number of nodes sampled ($k=8$ vs $k=22$)}. Sample size ($n=500,1000$) does not appear to have an influence given that both columns show similar rate patterns.
False positive rates for edges (dark blue bars) seems to be controlled in all additive and interaction simulation settings showing that the BNR model is accurate in identifying truly non-influential edges, \revision{with the notable exception of the case of $\mu=1.6,pi=0.3$. This case shows very odd results in all aspects (see also Fig. \ref{fig:nodes_int_rand}). We believe this to be a case where the model was affected by computational issues (like catastrophic cancellation or a similar error).}

Under the additive model, false positive rate of nodes (dark purple bars) appear to be the  concern when there are \revision{many} nodes sampled ($k=22$) \revision{ and low coefficient matrix sparsity ($\pi = 0.8$). As noted before, this is likely due to the fact that in this case many of the samples were above the threshold at which the response was capped.}

In general, the BNR model with random coefficients is able to accurately detect influential nodes and edges when there are truly interactions effects (Fig. \ref{fig:rates-rand} middle: interaction model), especially when there is low sparsity of the coefficient matrix $\mathbf B$ ($\pi=0.8$). Under the additive model, the BNR suffers from high false positive rate of nodes.

Fig. \ref{fig:rates-phylo} shows the false positive and false negative rates for edges and nodes for different simulation settings in terms of $n,k,\pi,\mu$ for additive (top), interaction (middle) and functional redundancy (bottom) models with phylogenetic coefficients. Again, an interaction model shows the best performance in terms of controlled false positive and false negative rates for all settings of $\pi,n$ and $\mu$. 

\revision{With a larger sample size ($n=1000$)} there is better controlled false positive rate of nodes (dark purple bars) in the \revision{$\pi=0.8$} case than in the random coefficients case (Fig. \ref{fig:rates-rand}) which means that when the effects of microbes are expected to be phylogenetically-informed, the BNR model is able to accurately identify the influential nodes compared to effects of microbes that are randomly assigned.

The results for the \revision{additive} model (top) are very similar to those with the random coefficients (Fig. \ref{fig:rates-rand}). \revision{For the redundancy model (bottom), the results are very similar with low coefficient sparsity ($\pi=0.8$), while for high sparsity the high false negative rates are mostly replaced with high false positive rates when many nodes are sampled $k=22$.}

\revision{Fig. \ref{fig:mse-phylo-adx} shows the mean squared error for response values and coefficent values for different simulation settings in terms of $n,k,\pi,\mu$ for additive (top), interaction (middle) and functional redundancy (bottom) models with phylogenetic coefficients. Fig. \ref{fig:mse-rand-adx} shows the same for models with random coefficients. In general, MSE is better with higher sample sizes. In many cases, coefficient MSE values for the case of $k=22$ nodes sampled were so high that it was not feasible to include them in the plots. Note, however, that these cases still had relatively good MSE values in terms of the response.}

\revision{\subsection*{The Effect of Data Augmentation}}

\revision{We find that performing data augmentation on samples of size 50 significantly improves model performance. In both coefficients and response, the MSE is significantly lower when the model is run on the augmented data as opposed to the unaugmented data (Fig. \ref{fig:mse_red_rand_aug} Appendix). While the model performance is still relatively poor in terms of identifying truly influential microbes, augmenting the data to \revision{200} samples we are able to identify one significant microbe using the 95\% credible intervals (Figs. \ref{fig:nodes_red_rand_aug_adx_k8}, \ref{fig:nodes_red_rand_aug_adx_k22} Appendix).}

\subsection*{Bacterial drivers of Phosphorous Leaching in Soil}

\begin{figure}[!ht]
\centering
\includegraphics[scale=0.55]{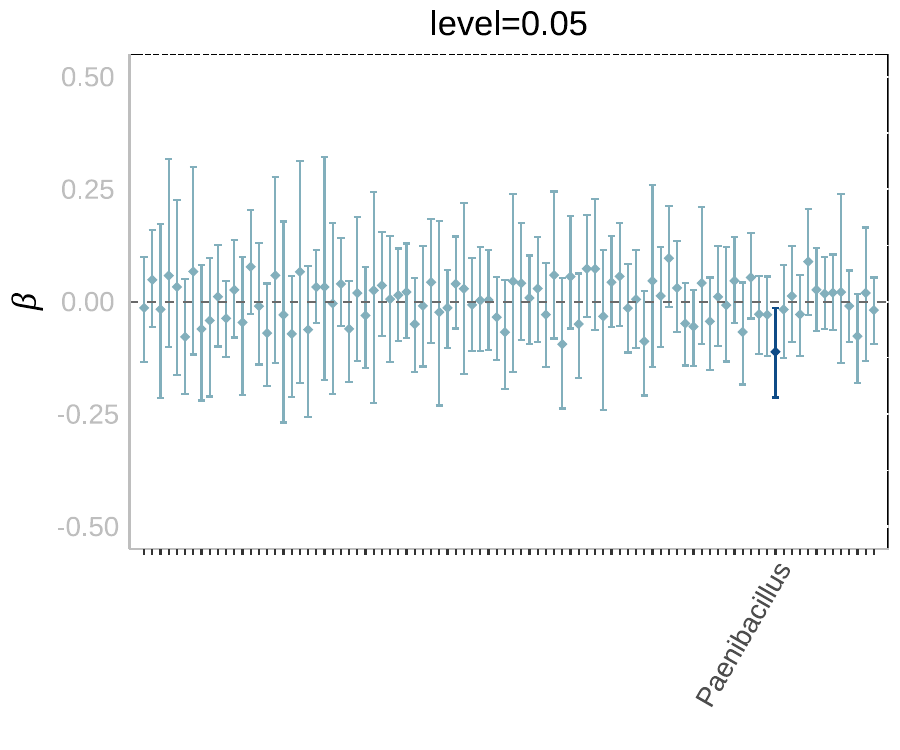}
\includegraphics[scale=0.55]{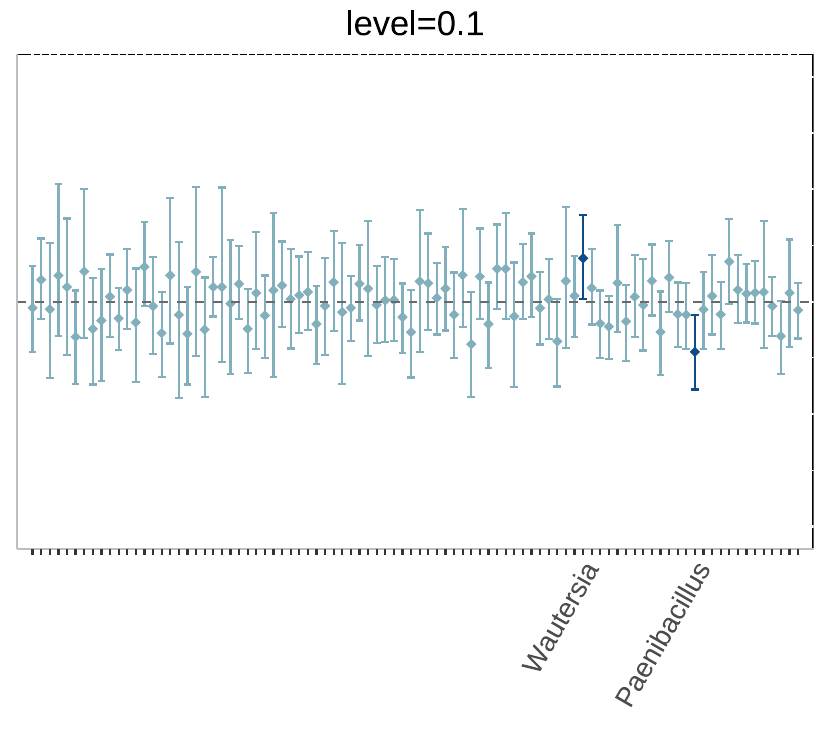}
\caption[Credible intervals for microbial (node) effects on phosphorous leaching]{{\bf Credible intervals for microbial (node) effects on phosphorous leaching.}
\revision{Left: 95\% credible intervals for the estimated effect of each microbe. Right: 90\% credible intervals for the estimated effect of each microbe.  
Within each plot the intervals are colored depending on whether the node is found to be influential (dark) or not influential (light) based on whether the interval intersects zero. 
}
}
\label{fig:nodes-wagg}
\end{figure}

For our analysis of the Wagg data (\citet{wagg2019fungal}) we use a latent variable dimension of $R=5$ as well as $\nu=12$. Fig. \ref{fig:nodes-wagg} shows \revision{node coefficients} for the OTUs in the augmented bacterial microbiome dataset \citep{wagg2019fungal} \revision{(posterior probabilities of influence given in Fig. \ref{fig:nodes-wagg-adx} in the Appendix)}. Likely due to the limited number of samples, the model is not able to accurately differentiate the strength of the effect of different OTUs. \revision{There is no cutoff probability we could reasonably use to determine whether an OTU is identified as influential by the BNR model based on on posterior probabilty. Luckily, using 95\% credible intervals we're able to identify one microbe whose presence } is influential on the amount of phosphorous leaching in the soil (Table \ref{tab:genus}) \revision{(using 90\% credible intervals we identify two microbes)}. We include also the reported association in the original study \citep{wagg2019fungal}. 

\begin{table}[!ht]
    \centering
    \begin{tabular}{|l|l|c|c|}
        \hline
        Genus & Class & \citet{wagg2019fungal} & CI level\\
        \hline
      \textit{Wautersia} & Betaproteobacteria & No association & 0.10 \\
       \textit{Paenibacillus} & Bacilli & Negative & 0.05\\\hline
    \end{tabular}
    \caption[Bacterial genus identified by our BNR model as key drivers of phosphorous leaching in soil (based on 95\% (90\%) credible intervals)]{{\bf Bacterial genus identified by our BNR model as key drivers of phosphorous leaching in soil (\revision{based on 95\% (90\%) credible intervals})}. The original study \citet{wagg2019fungal} reported bacterial classes that were positively or negatively associated with phosphorous leaching (third column). No association means that the original study did not find any connection between the bacterial class and the response. \revision{The ``CI level'' column indicates at what level credible interval the node was significant.}}
    \label{tab:genus}
\end{table}

We unfortunately could not find any influential edges (interactions among bacterial genera) with our model (Fig. \ref{fig:edges-wagg} in the Appendix) as all posterior 95\% credible intervals for the edge effects intersect zero. This is consistent with the simulation study. The model typically requires more than \revision{200} samples to reach the statistical power to identify key interactions.
Furthermore, we note that this phenotype (phosphorous leaching) probably falls under the functional redundancy setting as it is known that several taxa can indeed support the same function \citep{wagg2019fungal}. Our simulation study showed that under a functional redundancy model, the BNR has good performance only when there are few microbes sampled and many are expected to be influential (low sparsity in $\mathbf B$). The data collection setting in \citet{wagg2019fungal} was expected to produce high sampling rates which is in disagreement with the ideal setting for the BNR model which also explains the low statistical power to detect influential nodes. 
However, the requirement of low sampling could be an artifact of flaws in the simulation study. High sampling would artificially create more response values that surpass the threshold, inducing low variability in the response, and thus, restricting the power of the model. 
As a point of comparison, we present the histograms of both the simulated "functional redundancy" response and the actual phosphorous leaching response (Fig. \ref{fig:response_histograms} in the Appendix).
For the simulation case of $k=8$ microbes in each sample 
, the mean response value is 6.9 with a standard deviation of 4.25, and out of the 500 samples in this case, just one is truncated.
For the phosphorous leaching data (actual and augmented data), the mean amount of phosphorous leaching was 0.17 mg/L with a standard deviation of \revision{0.17}. Both histograms show sufficient spread and variability, yet the visual perception can be deceiving and the model could potentially require more samples for accurate estimation of parameters. We conclude that for cases of functional redundancy, presence/absence of microbes does not appear to be a suitable predictor of phenotype variability, and other measures like diversity can be more relevant (as those used in the original study \citep{wagg2019fungal}).

\subsection*{Computational speed}

All simulations were run on Julia 1.7.1 with compiler optimization level set to 0, IEEE math, and bounds-checking on, on servers provided by the Wisconsin Institute for Discovery. These servers run a Linux-based operating system on Intel Xeon Gold 6254 CPUs. 
For the theoretical simulations, the inference took an average of \revision{1 minute 34 seconds per 10,000 Gibbs samples (burn-in and retained)} for sample size of $n=100$ and \revision{4 minutes 31 seconds per 10,000 Gibbs samples (burn-in and retained)} for sample size of $n=500$. \revision{Recall that between 10,000 and 50,0000 Gibbs samples (burn-in and retained) were required to be generated to achieve convergence.}

For the realistic simulations, the inference took an average of \revision{1 minute 6 seconds per 10,000 Gibbs samples (burn-in and retained) for sample size of $n=50$, 1 minute  41 seconds per 10,000 Gibbs samples (burn-in and retained) for sample size of $n=200$, 6 minutes 41 seconds per 10,000 Gibbs samples (burn-in and retained) for sample size of $n=500$, and 17 minutes 14 seconds per 10,000 Gibbs samples (burn-in and retained)} for sample size of $n=1000$. \revision{Note that a minimum of 5,000 and maximum of 600,000 Gibbs samples (burn-in and retained) were required to be generated to achieve convergence. Also note that for $n=50$ and $n=200$ samples an R value of 5 was used while for $R=500$ and $R=1000$ an R value of 7 was used.}

\revision{For the real data analysis, performing 200,000 generations (Gibbs samples) took 8 hours 17 minutes 41 seconds (24 minutes 53 seconds per 10,000 generations). Note the time per generation is much higher for the real data due to the higher dimension of the coefficient matrix.}



\section*{Discussion}


In this work, we present the first deep investigation of the applicability of the Bayesian Network Regression (BNR) model on microbiome data. In addition, we introduce the first user-friendly implementation of the BNR model in an open source well-documented Julia package \texttt{BayesianNetworkRegression.jl} available on GitHub \url{https://github.com/solislemuslab/BayesianNetworkRegression.jl}.

The model performs well under a variety of settings of data sparsity (sparsity in network adjacency matrix $\mathbf A$) and sparsity of influential drivers (sparsity in the coefficient regression matrix $\mathbf B$) when the model generating the simulated data matches the BNR model (denoted theoretical simulations). When the model generating the simulated data did not match the BNR model (realistic simulations), the model performance varied depending on the specific generating model.

The current version of the BNR model (Equation \ref{bnrmodel}) can accurately identify influential nodes (microbes) and influential edges (interactions among microbes) under most realistic biological settings, but it requires large sample sizes (e.g. $n=1000$ tested here). Future work will involve the extension of the model to environmental predictors that can also have an effect on the phenotype of interest. Furthermore, future work will incorporate downstream estimation error of the adjacency matrix which is currently taken as perfectly reconstructed from the phylogenetic tree.

In terms of the real data analysis, we investigated the connections between the bacterial soil microbiome data and phosphorous leaching as in \citet{wagg2019fungal}. Unlike in our simulation study, the sample size was merely 50 samples which we augmented to \revision{200 via Gaussian data augmentation techniques}. The augmented sample was still not sufficient to achieve statistical power to detect influential edges (interactions), yet it was enough to identify influential nodes (microbes). \revision{At the 5\% level, one bacterial OTU was identified a key driver of phosphorous leaching in soil, which had already been identified as significantly associated with the response in the previous study \citep{wagg2019fungal}. Another, which had not been identified in the previous study \citep{wagg2019fungal} was identified at the 10\% level.}


While many existing public datasets do not have equivalent sample sizes to the ones tested in our simulation study, 
we believe that this is a case of ``build it and they will come". Until now, real data analyses of microbiome data involved \textit{one} matrix of relative abundances which is used to estimate \textit{one} co-occurrence microbial network. The assumption behind this one co-occurrence matrix is that correlations represent interactions, and these interactions are \textit{global}. That is, the interactions will appear in all contexts and all samples.

In recent years, \revision{some} microbiome researchers believe that the interactions are context-dependent and that there will be different interactions on different environmental conditions. 
These different interactions would produce different microbial networks, each in turn associated with a specific biological phenotype of interest.
BNR is the ideal model to represent this setting as it requires a sample of networks with edge variability (which violates the global interactions assumption of most public microbiome datasets) and it requires each of the microbial networks to be associated with a phenotype value. The downside is that the model requires hundreds of these network-phenotype samples.
While these large sample conditions appear to be outside the norm of current observational microbiome research, we believe that the BNR model could be quite useful to identify key microbial drivers of biological phenotypes in experimental settings. Indeed, scientists can design experiments with $k$ microbes and then, measure the phenotype of interest. Different samples would correspond to different replicates under the same set (or different) of microbes. This setup actually aligns with our simulation study where we have $k$ microbes per sample and the true phenotype value is computed only using those $k$ microbes. Future work will incorporate other covariates into the model corresponding to replicate, experimental or environmental conditions. Furthermore, in conjunction with data augmentation techniques, the BNR model could be applied on a variety of real datasets, as this will be another line of future work. Despite its limitations, we view the BNR model as one novel tool that microbiome researchers could utilize to identify key microbiome drivers in biological phenotypes of interest given its robustness and accuracy under a variety of biologically relevant conditions.

\section*{Acknowledgments}
This work was supported by the Department of Energy [DE-SC0021016 to CSL] and by the National Institute of Food and Agriculture, United States Department of Agriculture [Hatch project 1023699 to CSL]. 
We thank Sharmistha Guha and Abel Rodriguez for communications to clarify specifics of their model implementation, and 
we thank Richard Lankau for meaningful conversations about microbial communities.

\section*{Conflicts of interest}
The authors declare no conflicts of interest.

\section*{Author contributions}

SO implemented the Julia package, ran the simulations and real data analysis and wrote the first draft of the manuscript. CSL devised the idea of the project, chose the model, and produced the first iteration of figures for the manuscript. All authors contributed critically to the drafts and gave final approval for publication.

\section*{Data availability statement}
Software is publicly available on github: \url{https://github.com/solislemuslab/soil-microbiome-nn}.
We did not generate new data.

\bibliographystyle{abbrvnat}
\bibliography{manuscript}  

\newpage 

\appendix

\section{Appendix}
\captionsetup[figure]{list=yes}
\captionsetup[table]{list=yes}

\singlespacing

\subsection{Table of parameters}

\begin{table}[h]
    \centering
    \caption{Parameters in the model and in the priors}
    \begin{tabular}{|c|l|}
        \hline 
        Parameter & Description  \\
        \hline
        $V$ & number of nodes \\
        \hline
        $q$ & number of edges \\
        \hline
        $\tau^2 \in \mathbb{R}$ & variance of error term \\
        \hline
        $\mathbf{u} \in \mathbb{R}^{R \times V}$ & $R$-dimensional latent variables for each node \\
        \hline 
        $\xi \in \{0,1\}^V$ & binary vector to denote if a node is influential \\
        \hline
        $\gamma \in \mathbb{R}^q$ & regression coefficients for edge effects \\
        \hline 
        $\mathbf{s} \in \mathbb{R}^q$ & scale parameters for the variance of $\gamma$ \\
        \hline
        $\theta \in \mathbb{R}$ & exponential parameter for the scale $\mathbf{s}$ \\
        \hline
        $\Delta \in \mathbb{R}$ & Bernoulli parameter for $\xi$ \\
        \hline 
        $\mathbf M \in \mathbb{R}^{R \times R}$ & covariance matrix for latent variables $\mathbf u$ \\
        \hline
        $\mu \in \mathbb{R}$ & overall mean \\
        \hline
        $\lambda \in \{0,1,-1\}^R$ & vector that governs which entries in the latent variables are informative \\
        \hline 
        $\tilde{\pi} \in \mathbb{R}^{R \times 3}$ & Dirichlet prior probability for $\lambda$ being $0,1,-1$ \\
        \hline
    \end{tabular}
    \label{tab:pars}
\end{table}

\subsection{Mathematical description of theoretical simulations}

Graphical description of the theoretical simulations is in Fig.~\ref{fig:unrealistic}. Here, we list the variables \revision{(see also Table \ref{tab:pars})}:
\begin{itemize}
    \item $n$: sample size ($n=100, 500$)
    \item $d$: number of total microbes ($d=30$)
    \item $k$: number of microbes sampled ($k=8, 15, 22$) which controls the sparsity in $\mathbf A$
    \item $R$: dimension of latent variables $\mathbf u$ ($R=5, 7, 9$)
    \item $\pi$: probability of Bernoulli per node ($\pi= 0.3, 0.8$) which controls the sparsity in $\mathbf B$
    \item $\mu$: mean of effect size in $\mathbf B$ ($\mu=0.8, 1.6$) which controls the magnitude in $\mathbf B$
\end{itemize}

\noindent Then, the simulating algorithm is:
\begin{enumerate}
    \item Simulate a 30-taxon phylogenetic tree using the \texttt{rtree} function from the R package \texttt{ape} \citep{ape}
    \item Simulate whether each node is influential and the true effects of each interaction for $i=1,\dots,d$ and $j=1,\dots,30$
    \begin{align*}
        \xi_i &\sim \textrm{Bernoulli}(\pi)
        \\ \mathbf{B}_{ij} &\sim \xi_i \xi_j  \textrm{Normal}(\mu,1)
    \end{align*}
    \item For every sample $i=1,\dots,n$, calculate the adjacency matrix $\mathbf A_i$ as follows
    \begin{enumerate}
        \item Select $k$ microbes randomly \revision{$S_i = \{m_{i_1},m_{i_2},\dots,m_{i_k}\}$, with indices} $K^*_i = \{i_1,i_2,\dots,i_k\}$
        \item For $p,q \in K^*$, let $d_{p,q}$ be the phylogenetic distance between microbes $p$ and $q$, so that the $(p,q)$ entry in the adjacency matrix is given by
        \begin{align*}
            \mathbf A_{i}[p,q] = \frac{1}{d_{p,q}}
        \end{align*}
    \end{enumerate}
    \item For every sample $i=1,\dots,n$, calculate the response value as follows
    \begin{align*}
        y_i = <\mathbf A_i,\mathbf B>_F + \epsilon, \hspace{3mm} \epsilon \sim \textrm{Normal}(0,1)
    \end{align*}
\end{enumerate}
At the end of the simulating algorithm, we have a collection of pairs $(y_1,\mathbf A_1), \dots, (y_n,\mathbf A_n)$ to be used as data in the BNR implementation.

\subsection{Mathematical description of realistic simulations}

Graphical description of the realistic simulations is in Fig.~\ref{fig:realistic}. Here, we list the variables \revision{(see also Table \ref{tab:pars})}:
\begin{itemize}
    \item $n$: sample size ($n=500,1000$)
    \item $d$: number of total microbes ($d=30$)
    \item $k$: number of microbes sampled ($k=8,22$) which controls the sparsity in $\mathbf A$
    \item $R$: dimension of latent variables $\mathbf u$ ($R=7$)
    \item $\pi$: probability of Bernoulli per node ($\pi=0.3,0.8$) which controls the sparsity in $\mathbf B$
    \item $\mu_b$: mean of main effect size in $\mathbf B$ ($\mu_b=0.8,1.6$) which controls the magnitude in $\mathbf B$
    \item $L$: limit for response $y$ under the functional redundancy model\\
    \begin{center} \begin{tabular}{c|c|c}
         $\mu$ & $\pi$ & $L$ \\\hline
         0.8 & 0.3 & 3 \\
         1.6 & 0.3 & 7 \\
         0.8 & 0.8 & 22 \\
         1.6 & 0.8 & 30
    \end{tabular}\end{center}
\end{itemize}

Next, we describe the simulating algorithm for the three models.

\subsubsection{Additive Simulations}
\begin{enumerate}
    \item Simulate a 30-taxon phylogenetic tree using the \texttt{rtree} function from the R package \texttt{ape} \citep{ape}
    \item Simulate whether a specific node is influential and its true main effect for $i=1,\dots,30$:
    \begin{enumerate}
        \item $\xi_i \sim \textrm{Bernoulli}(\pi)$
        \item $b_i \sim \textrm{Normal}(\mu_b,\Sigma)$ 
        \begin{itemize}
            \item $\Sigma = \mathbb{I}_{30}$ (case of random coefficients)
            \item $\Sigma$ governed by the phylogenetic tree using the Julia package \texttt{PhyloNetworks} \citep{phylonetworks} (case of phylogenetic coefficients)
        \end{itemize}
    \end{enumerate}
    \item For every sample $i=1,\dots,n$ calculate the adjacency matrix ($\mathbf A_i$) as follows
    \begin{enumerate}
        \item Select $k$ microbes randomly \revision{$S_i = \{m_{i_1},m_{i_2},\dots,m_{i_k}\}$, with indices} $K_i^* = \{i_1,\dots,i_k\}$
        \item For $p,q \in K^*_i$, let $d_{p,q}$ be the phylogenetic distance between microbes $p$ and $q$ so that the $(p,q)$ entry in the adjacency matrix is given by
        \begin{align*}
            \mathbf A_{i}[p,q] = \frac{1}{d_{p,q}}
        \end{align*}
    \end{enumerate}
    \item For every sample $i=1,\dots,n$, calculate the response value as follows
    \begin{align*}
        y_i &= \sum_{j=1}^d b_j m_j\xi_j + \epsilon, \hspace{3mm} \epsilon \sim \textrm{Normal}(0,1) \\
        m_j &= \mathbf{1}(j \in K^*_i)
    \end{align*}
    where $\mathbf 1(\cdot)$ is the indicator function.
\end{enumerate} 
At the end of the simulating algorithm, we have a collection of pairs $(y_1,\mathbf A_1), \dots, (y_n,\mathbf A_n)$ to be used as data in the BNR implementation.

\subsubsection{Interaction Simulations}
\begin{enumerate}
    \item Simulate a 30-taxon phylogenetic tree using the \texttt{rtree} function from the R package \texttt{ape} \citep{ape}
    \item Simulate whether a specific node is influential, its true main effect for $i=1,\dots,30$ and all interaction effects $j=1,\dots,30$:
    \begin{enumerate}
        \item $\xi_i \sim \textrm{Bernoulli}(\pi)$
        \item $b_i \sim \textrm{Normal}(\mu_b,\Sigma)$ 
        \begin{itemize}
            \item $\Sigma = \mathbb{I}_{30}$ (case of random coefficients)
            \item $\Sigma$ governed by the phylogenetic tree using the Julia package \texttt{PhyloNetworks} \citep{phylonetworks} (case of phylogenetic coefficients)
        \end{itemize}
        \item $b_{ij} \sim \textrm{Normal}(0.4,1)$
    \end{enumerate}
    \item For every sample $i=1,\dots,n$ calculate the adjacency matrix ($\mathbf A_i$) as follows
    \begin{enumerate}
        \item Select $k$ microbes randomly \revision{$S_i = \{m_{i_1},m_{i_2},\dots,m_{i_k}\}$, with indices} $K^*_i = \{j_1,j_2,\dots,j_k\}$
        \item For $p,q \in K^*$, let $d_{p,q}$ be the phylogenetic distance between microbes $p$ and $q$ so that the $(p,q)$ entry in the adjacency matrix is given by
        \begin{align*}
            \mathbf A_{i}[p,q] = \frac{1}{d_{p,q}}
        \end{align*}
    \end{enumerate}
    \item For every sample $i=1,\dots,n$, calculate the response value as follows:
    \begin{align*}
        y_i &= \sum_{l=1}^d b_l m_l\xi_l + \sum_{l=1}^d \sum_{j=1}^d b_{lj} (\mathbf A_i[l,j]) \xi_l\xi_j + \epsilon, \hspace{3mm} \epsilon \sim \textrm{Normal}(0,1) \\
        m_l &= \mathbf{1}(l \in K^*_i)
    \end{align*}
     where $\mathbf 1(\cdot)$ is the indicator function.
\end{enumerate} 
At the end of the simulating algorithm, we have a collection of pairs $(y_1,\mathbf A_1), \dots, (y_n,\mathbf A_n)$ to be used as data in the BNR implementation.

\subsubsection{Functional Redundancy Simulations}

\begin{enumerate}
    \item Simulate a 30-taxon phylogenetic tree using the \texttt{rtree} function from the R package \texttt{ape} \citep{ape}
    \item Simulate whether a specific node is influential, its true main effect for $i=1,\dots,30$ and all interaction effects $j=1,\dots,30$:
    \begin{enumerate}
        \item $\xi_i \sim \textrm{Bernoulli}(\pi)$
        \item $b_i \sim \textrm{Normal}(\mu_b,\Sigma)$ 
        \begin{itemize}
            \item $\Sigma = \mathbb{I}_{30}$ (case of random coefficients)
            \item $\Sigma$ governed by the phylogenetic tree using the Julia package \texttt{PhyloNetworks} \citep{phylonetworks} (case of phylogenetic coefficients)
        \end{itemize}
        \item $b_{ij} \sim \textrm{Normal}(0.4,1)$
    \end{enumerate}
    \item For every sample $i=1,\dots,n$ calculate the adjacency matrix ($\mathbf A_i$) as follows
    \begin{enumerate}
        \item Select $k$ microbes randomly \revision{$S_i = \{m_{i_1},m_{i_2},\dots,m_{i_k}\}$, with indices} $K^*_i = \{j_1,j_2,\dots,j_k\}$
        \item For $p,q \in K^*$, let $d_{p,q}$ be the phylogenetic distance between microbes $p$ and $q$ so that the $(p,q)$ entry in the adjacency matrix is given by
        \begin{align*}
            \mathbf A_{i}[p,q] = \frac{1}{d_{p,q}}
        \end{align*}
    \end{enumerate}
    \item For every sample $i=1,\dots,n$, calculate the response value as follows:
    \begin{align*}
        y_i &= \min \left\{ \sum_{l=1}^d b_l m_l\xi_l + \sum_{l=1}^d \sum_{j=1}^d b_{lj} (\mathbf A_i[l,j]) \xi_l\xi_j + \epsilon, L \right\}, \hspace{3mm} \epsilon \sim \textrm{Normal}(0,1) \\
        m_l &= \mathbf{1}(l \in K^*_i) 
    \end{align*}
     where $\mathbf 1(\cdot)$ is the indicator function.
\end{enumerate} 
At the end of the simulating algorithm, we have a collection of pairs $(y_1,\mathbf A_1), \dots, (y_n,\mathbf A_n)$ to be used as data in the BNR implementation.

\subsection{Simulation plots}

\listoftables
\listoffigures

\subsubsection{Theoretical case}
\FloatBarrier

\begin{figure}[!ht]
    \centering
    \begin{subfigure}[t]{0.49\textwidth}
        \includegraphics[scale=0.27]{plot-ci-nodes-n100-k8-R7.pdf}
        \caption{$\mathbf{n=100}$, $\mathbf{k=8}$}
    \end{subfigure}
    \begin{subfigure}[t]{0.49\textwidth}
        \includegraphics[scale=0.27]{plot-ci-nodes-n500-k8-R7.pdf}
    \caption{$\mathbf{n=100}$, $\mathbf{k=22}$}
    \end{subfigure}\\
    \begin{subfigure}[t]{0.49\textwidth}
        \includegraphics[scale=0.27]{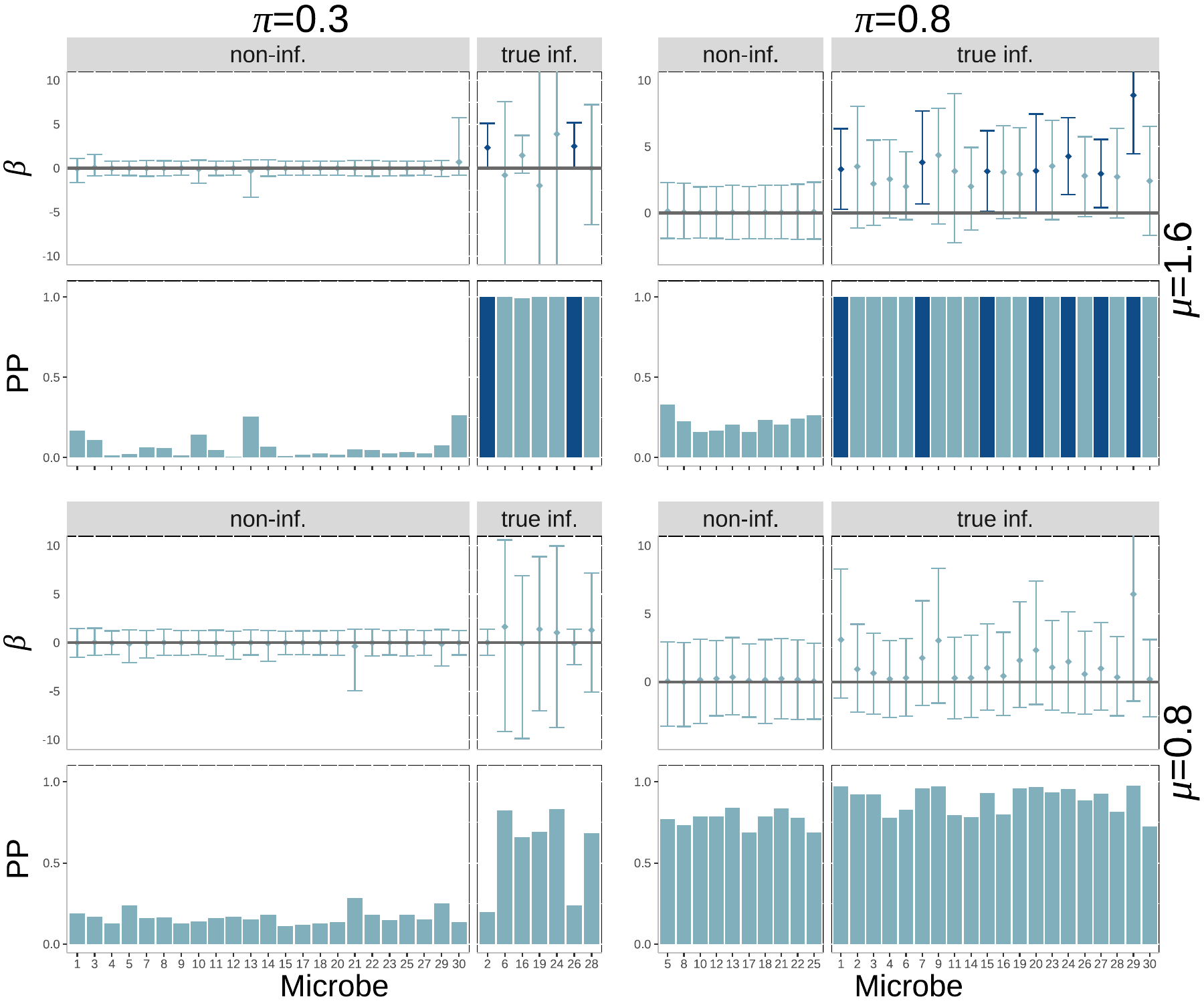}
    \caption{$\mathbf{n=500}$, $\mathbf{k=8}$}
    \end{subfigure}
    \begin{subfigure}[t]{0.49\textwidth}
        \includegraphics[scale=0.27]{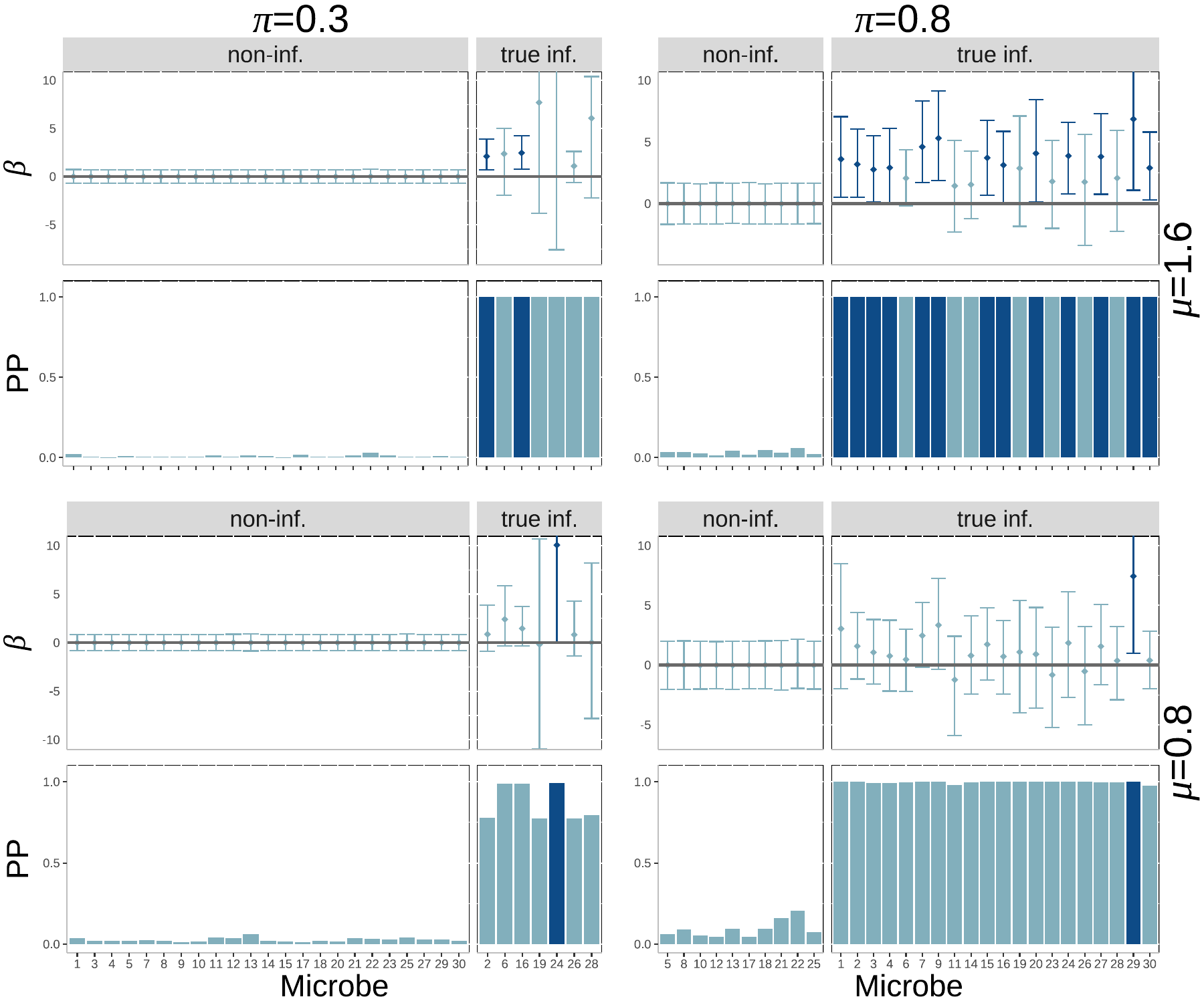}
        \caption{$\mathbf{n=500}$, $\mathbf{k=22}$}
    \end{subfigure}
    \caption[Posterior probability of influential nodes and "edge" coefficients for nodes(theoretical simulations)]{{\bf Posterior probability of influential nodes (theoretical simulations).}
    Different groups of 4 panels represent different number of sampled microbes ($k=8,22$) which controls the sparsity of the adjacency matrix and different sample sizes ($n=100,500$). \revision{Within each group, we have four panels corresponding to the two values of edge effect size ($\mu=0.8, 1.6$) and two values of probability of influential node ($\pi=0.3, 0.8$) which controls the sparsity of the regression coefficient matrix ($\mathbf B$). Within each of these panels we have two plots: 95\% credible intervals (top) and posterior probability of influence (bottom - calculated as the mean of the $\xi$ variable for the node across Gibbs samples) for each node.} Each bar corresponds to one node (microbe). \revision{Within each plot the bars and intervals are colored depending on whether the node is found to be influential (dark) or not influential (light) based on the 95\% credible intervals. Each plot is split based on whether the nodes are truly influential (right) or not (left).}
    }
    \label{fig:nodes-all}
\end{figure}

\begin{figure}[!ht]
\centering
     \begin{subfigure}[b]{0.49\textwidth}
        \centering
        \includegraphics[scale=0.19]{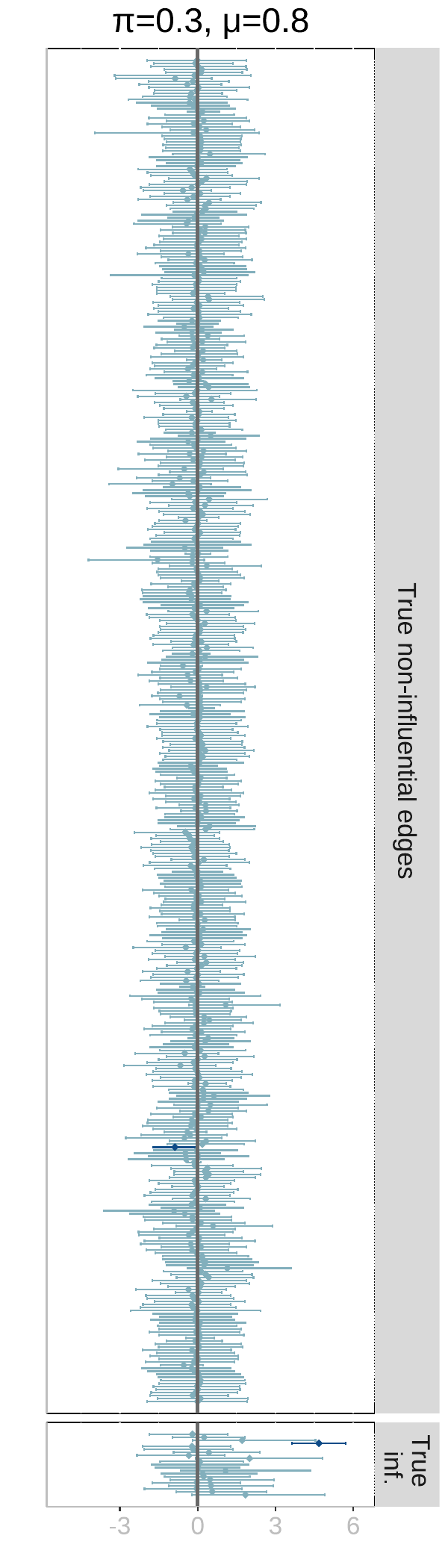}
        \includegraphics[scale=0.19]{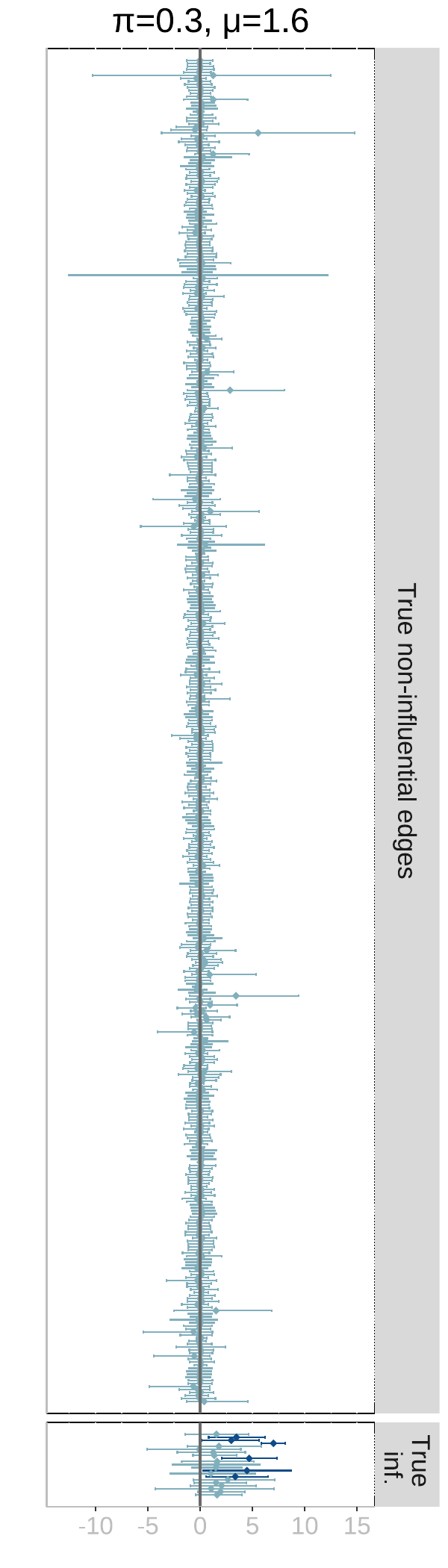}
        \includegraphics[scale=0.19]{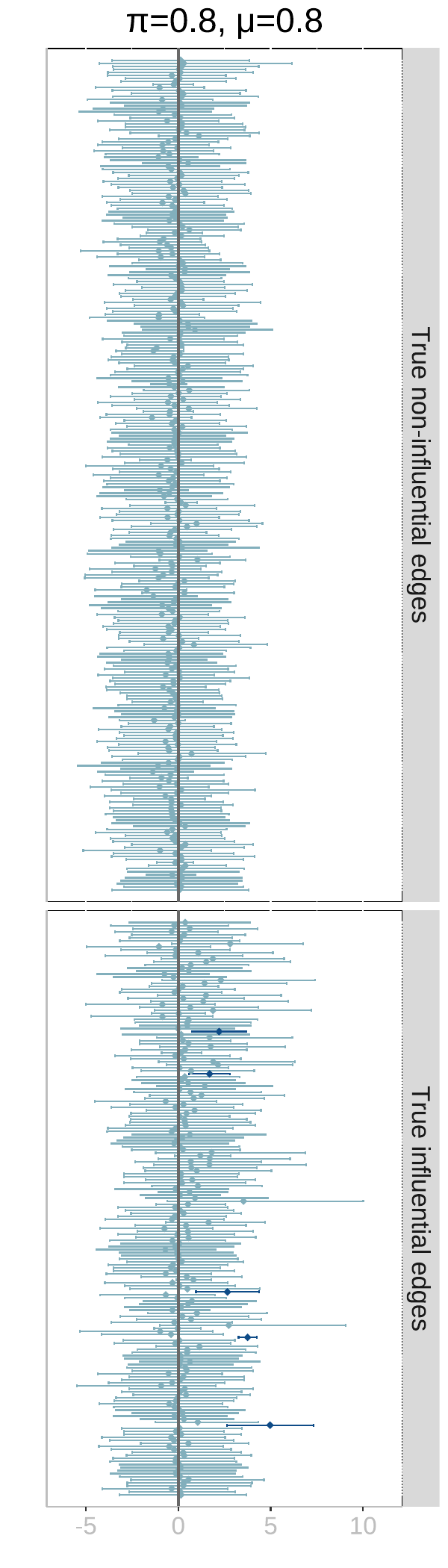}
        \includegraphics[scale=0.19]{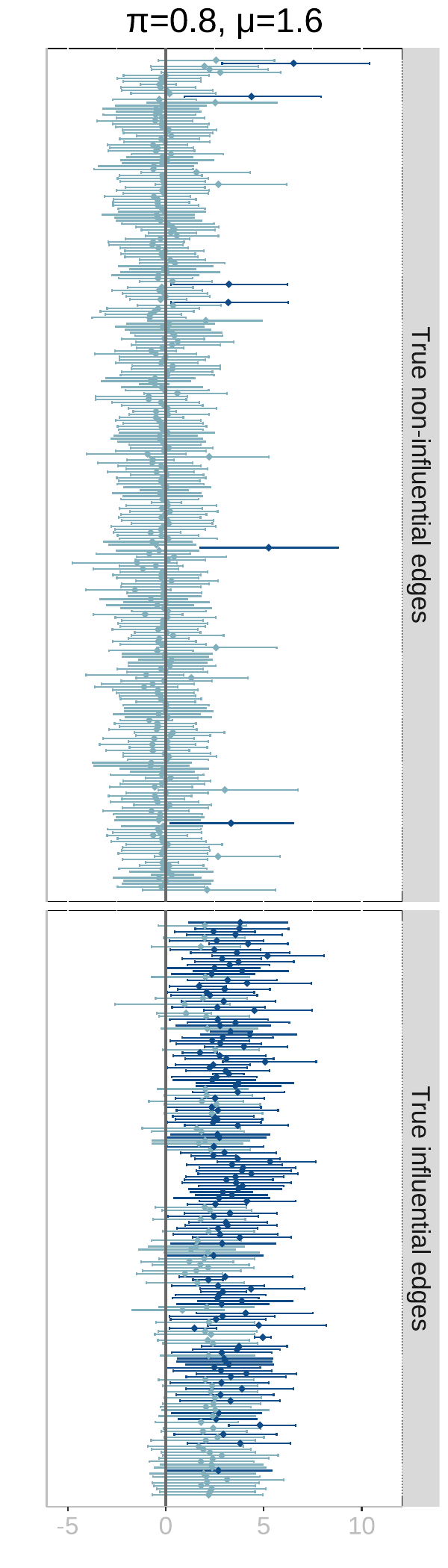}
        \caption{$\mathbf{n=100}$}
     \end{subfigure}
     \begin{subfigure}[b]{0.49\textwidth}
        \centering
        \includegraphics[scale=0.19]{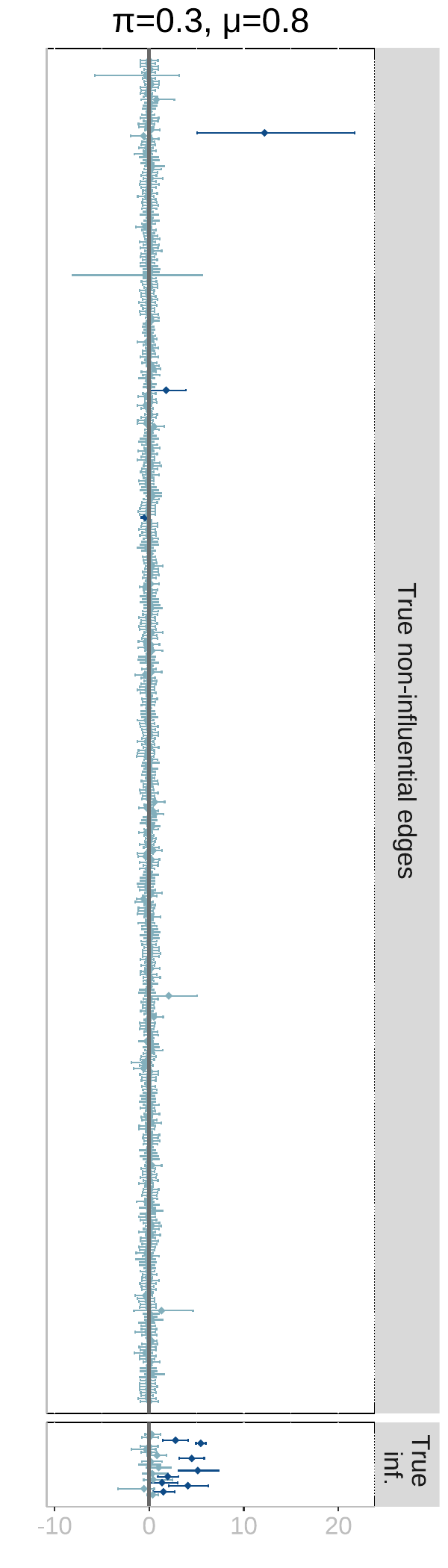}
        \includegraphics[scale=0.19]{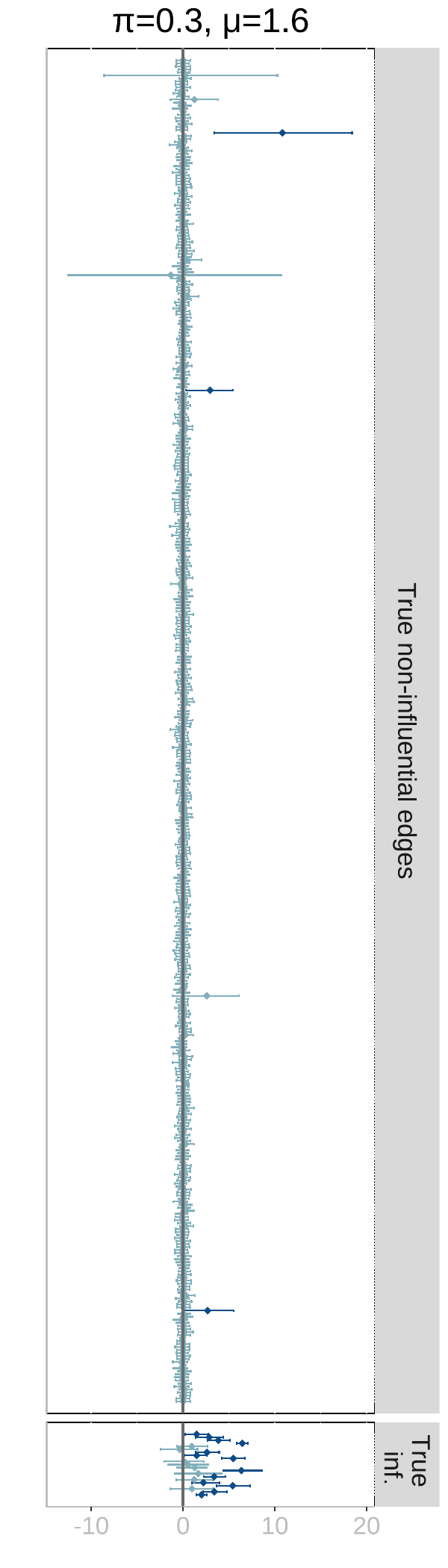}
        \includegraphics[scale=0.19]{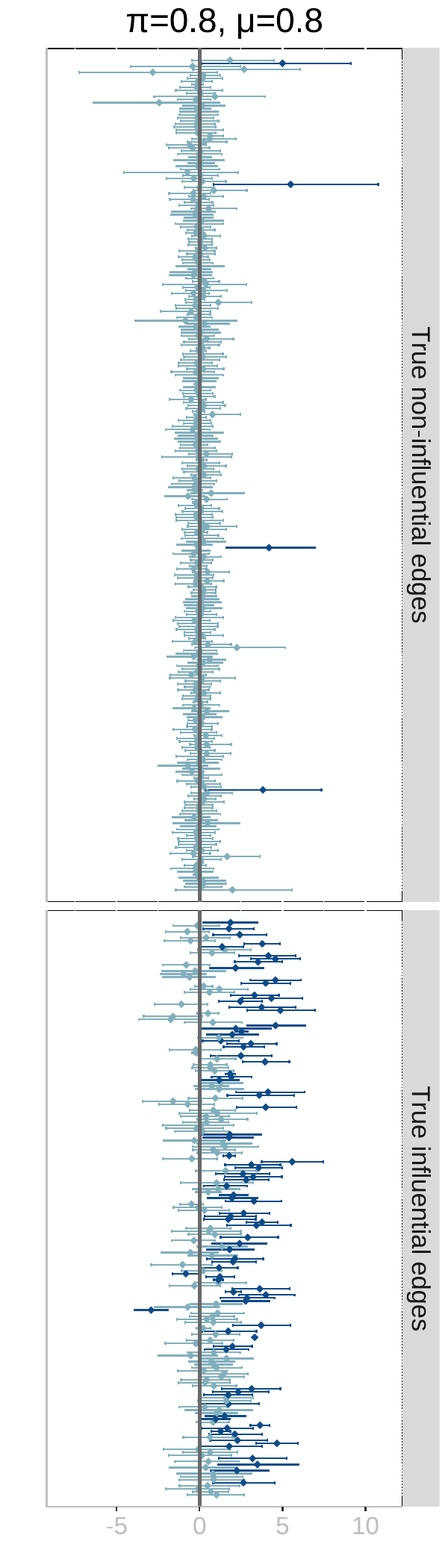}
        \includegraphics[scale=0.19]{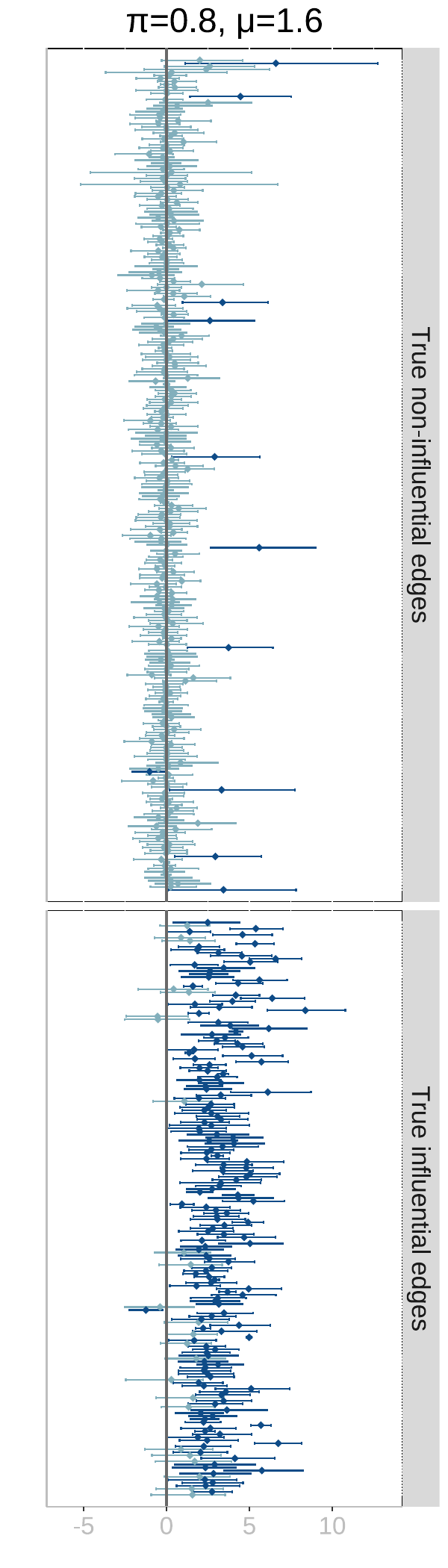}
        \caption{$\mathbf{n=500}$}
     \end{subfigure}
\caption[95 \% credible intervals for edge effects for $k=8$ sampled nodes (theoretical simulations)]{{\bf 95 \% credible intervals for edge effects for $k=8$ sampled nodes (theoretical simulations).} Top (a): Sample size of $n=100$. Bottom (b): Sample size of $n=500$. Each panel corresponds to a scenario of $\pi=0.3, 0.8$ (which controls the sparsity of the regression coefficient matrix $\mathbf B$) and $\mu=0.8,1.6$. We plot the 95 \% credible intervals for the regression coefficients per edge ordered depending on whether they are truly non-influential edges (top of each panel) or truly influential edges (bottom of each panel). The color of the intervals depends on whether it intersects zero (light) and hence estimated to be non-influential or does not intersect zero (dark) and hence estimated to be influential by the model. These panels allow us to visualize false positives (dark intervals on the top panel) or false negatives (light intervals on the bottom panel).
}
\label{fig:edges8}
\end{figure}

\begin{figure}[!ht]
\centering
\begin{subfigure}[b]{0.49\textwidth}
\centering
\includegraphics[scale=0.19]{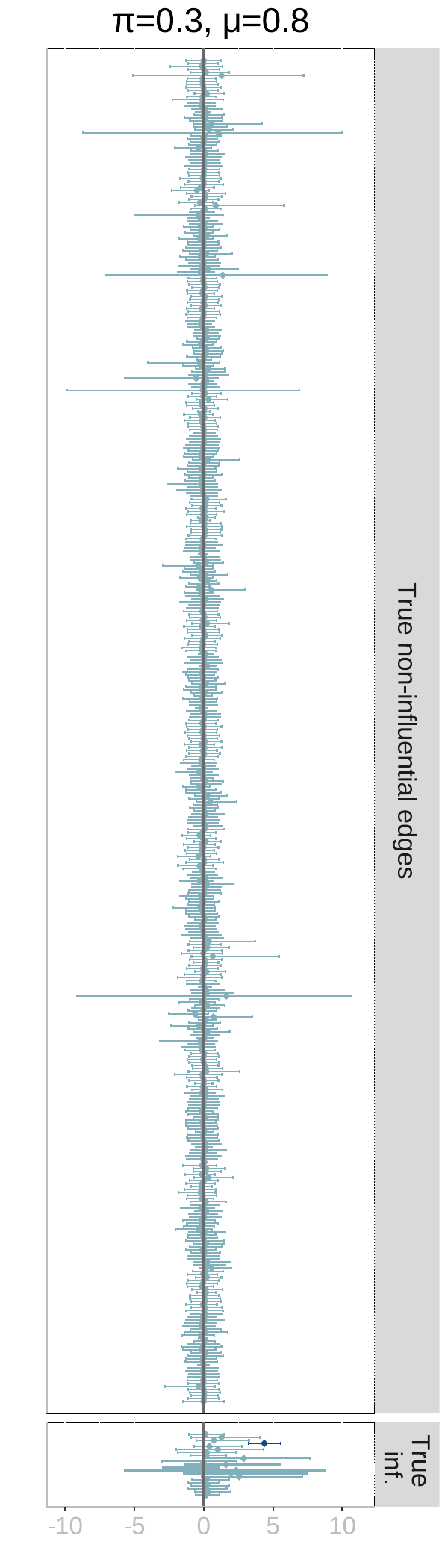}
\includegraphics[scale=0.19]{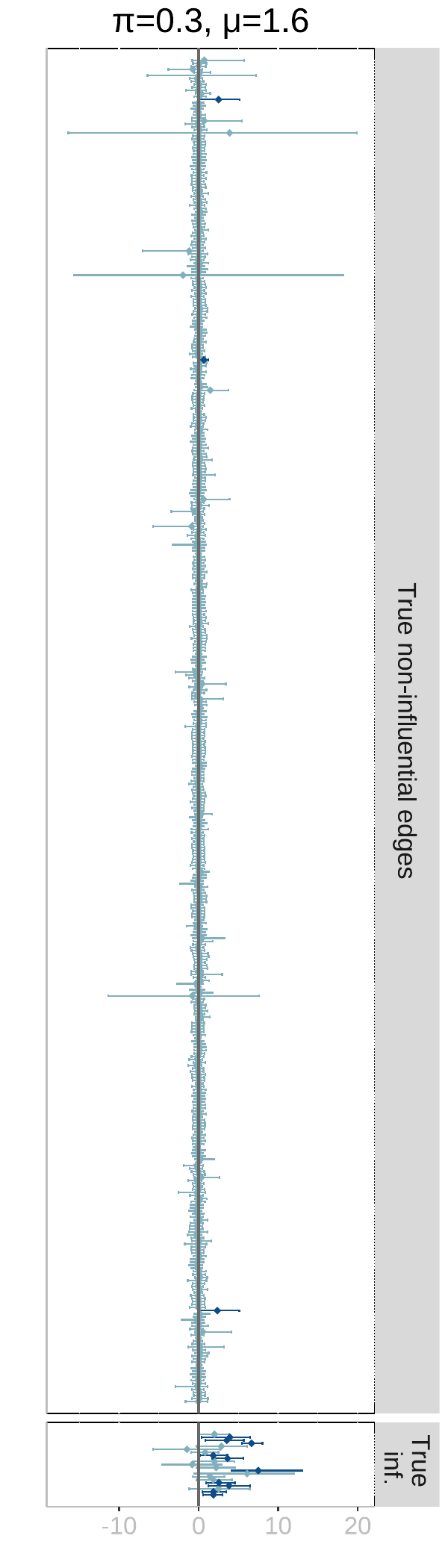}
\includegraphics[scale=0.19]{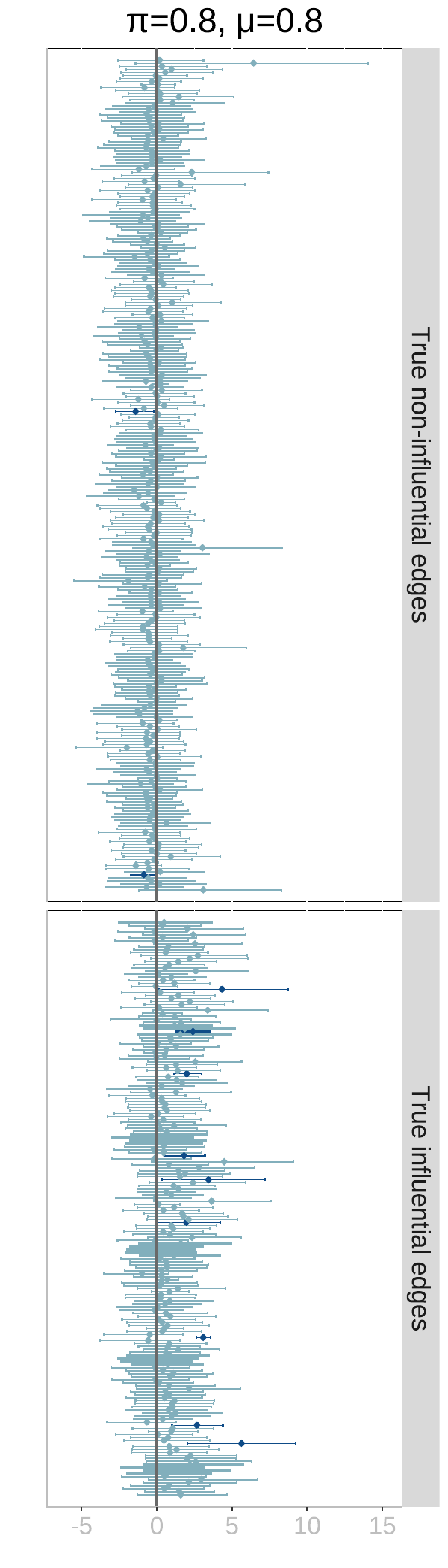}
\includegraphics[scale=0.19]{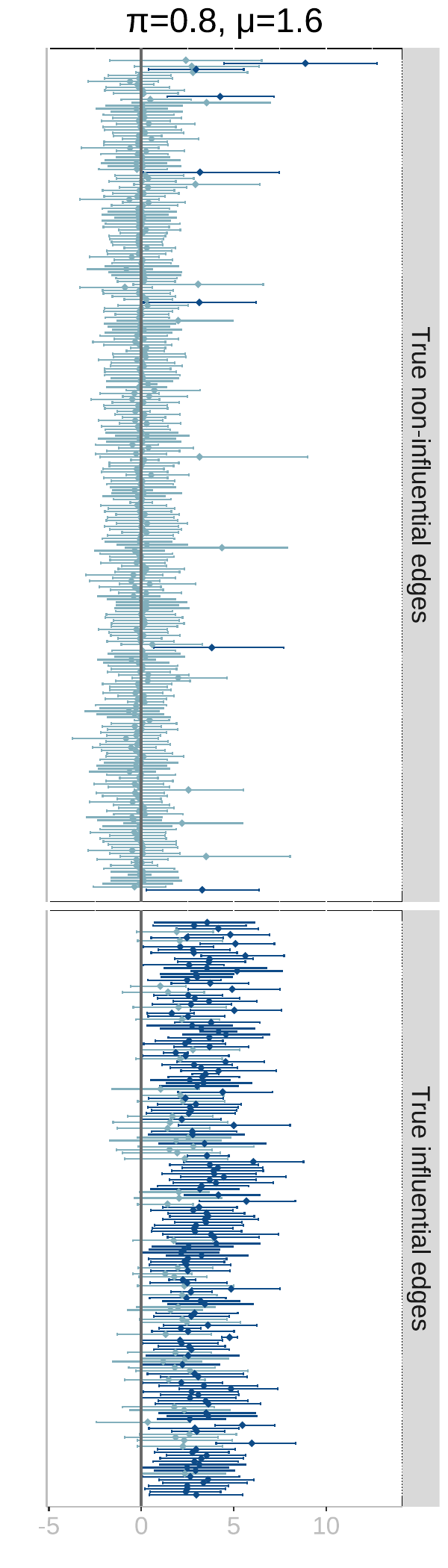}
\caption{$\mathbf{n=100}$}
\end{subfigure}
\begin{subfigure}[b]{0.49\textwidth}
\centering
\includegraphics[scale=0.19]{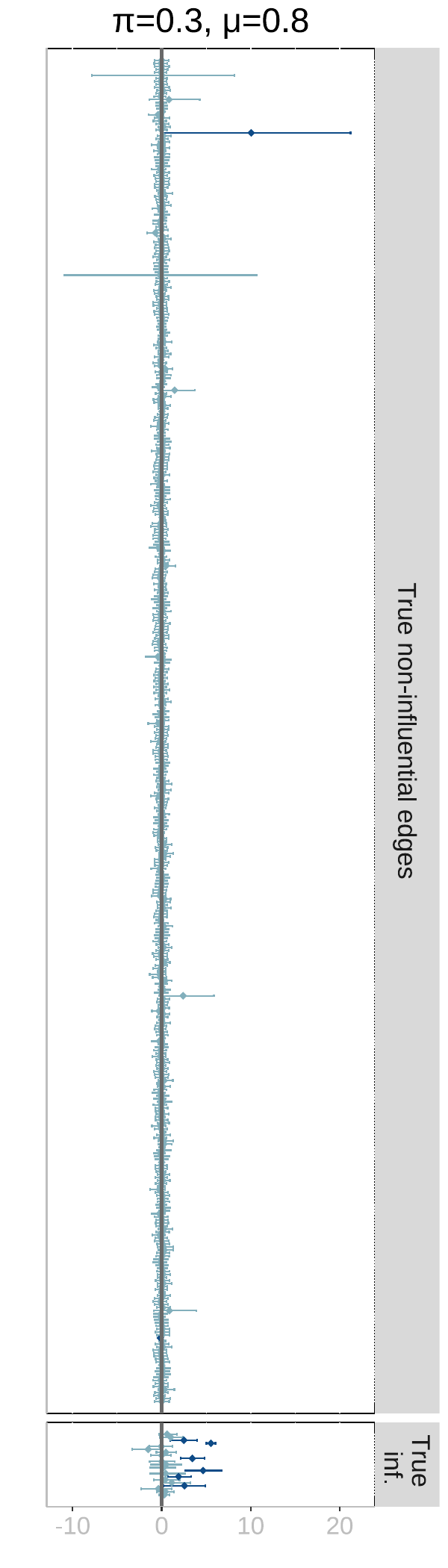}
\includegraphics[scale=0.19]{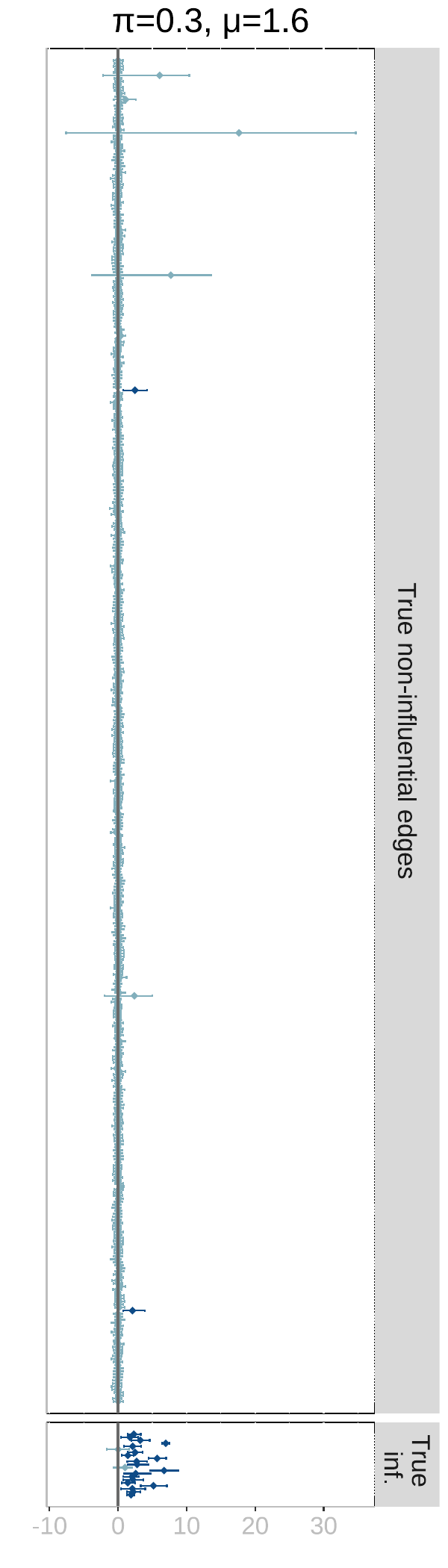}
\includegraphics[scale=0.19]{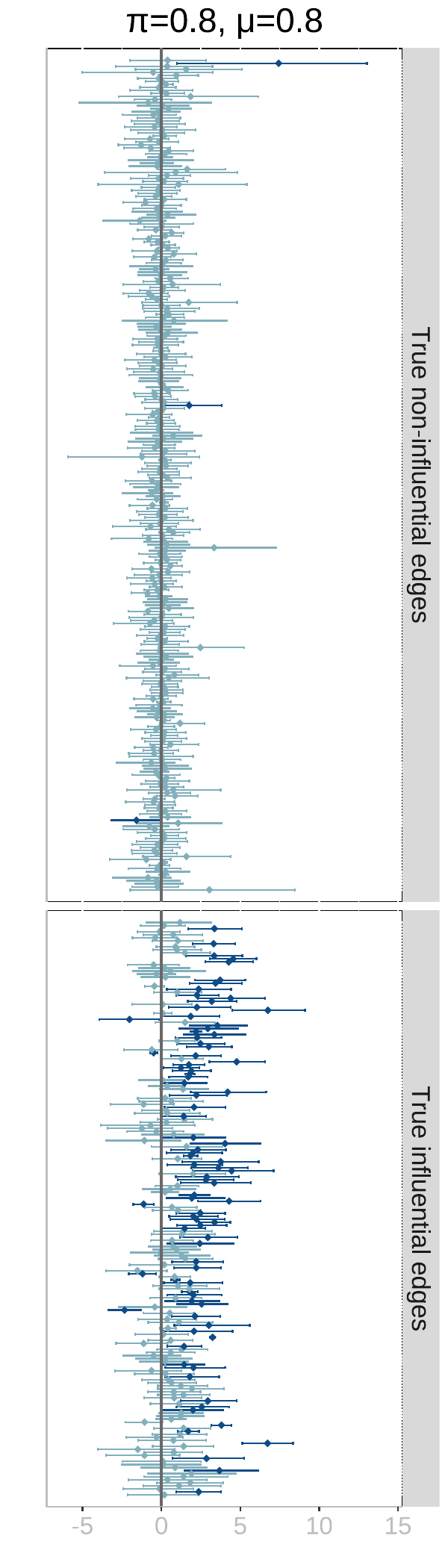}
\includegraphics[scale=0.19]{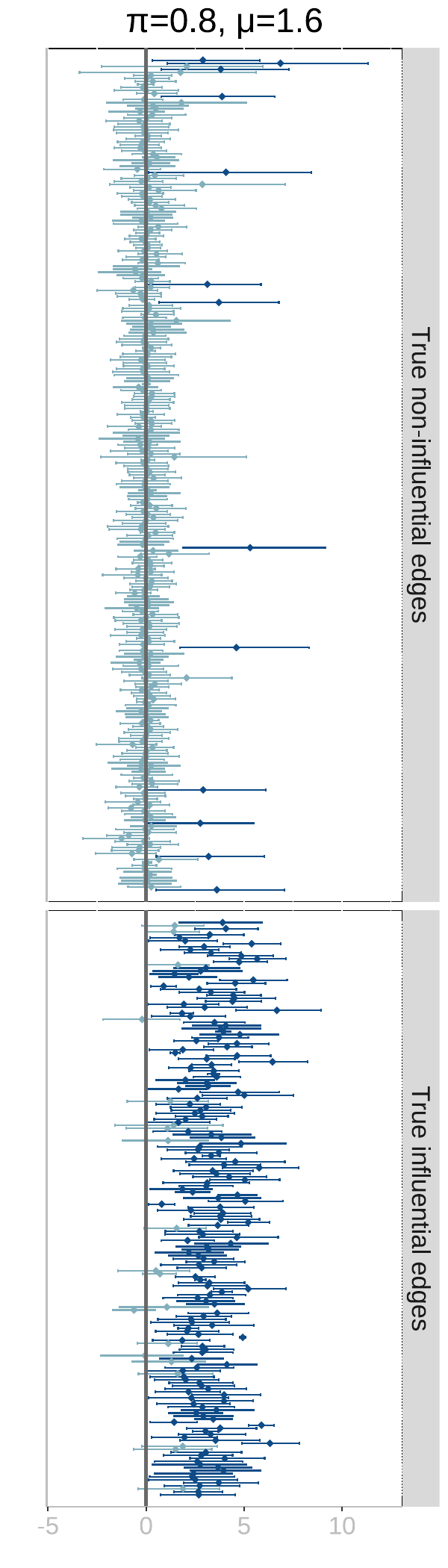}
\caption{$\mathbf{n=500}$}
\end{subfigure}
\caption[95 \% credible intervals for edge effects for $k=22$ sampled nodes (theoretical simulations)]{{\bf 95 \% credible intervals for edge effects for $k=22$ sampled nodes (theoretical simulations).} Top (a): Sample size of $n=100$. Bottom (b): Sample size of $n=500$. Each panel corresponds to a scenario of $\pi=0.3, 0.8$ (which controls the sparsity of the regression coefficient matrix $\mathbf B$) and $\mu=0.8,1.6$. We plot the 95 \% credible intervals for the regression coefficients per edge ordered depending on whether they are truly non-influential edges (top of each panel) or truly influential edges (bottom of each panel). The color of the intervals depends on whether it intersects zero (light) and hence estimated to be non-influential or does not intersect zero (dark) and hence estimated to be influential by the model. These panels allow us to visualize false positives (dark intervals on the top panel) or false negatives (light intervals on the bottom panel).}
\label{fig:edges22}
\end{figure}

\begin{figure}[!ht]
\centering
\begin{subfigure}[t]{0.49\textwidth}
    \centering
    \includegraphics[scale=0.27]{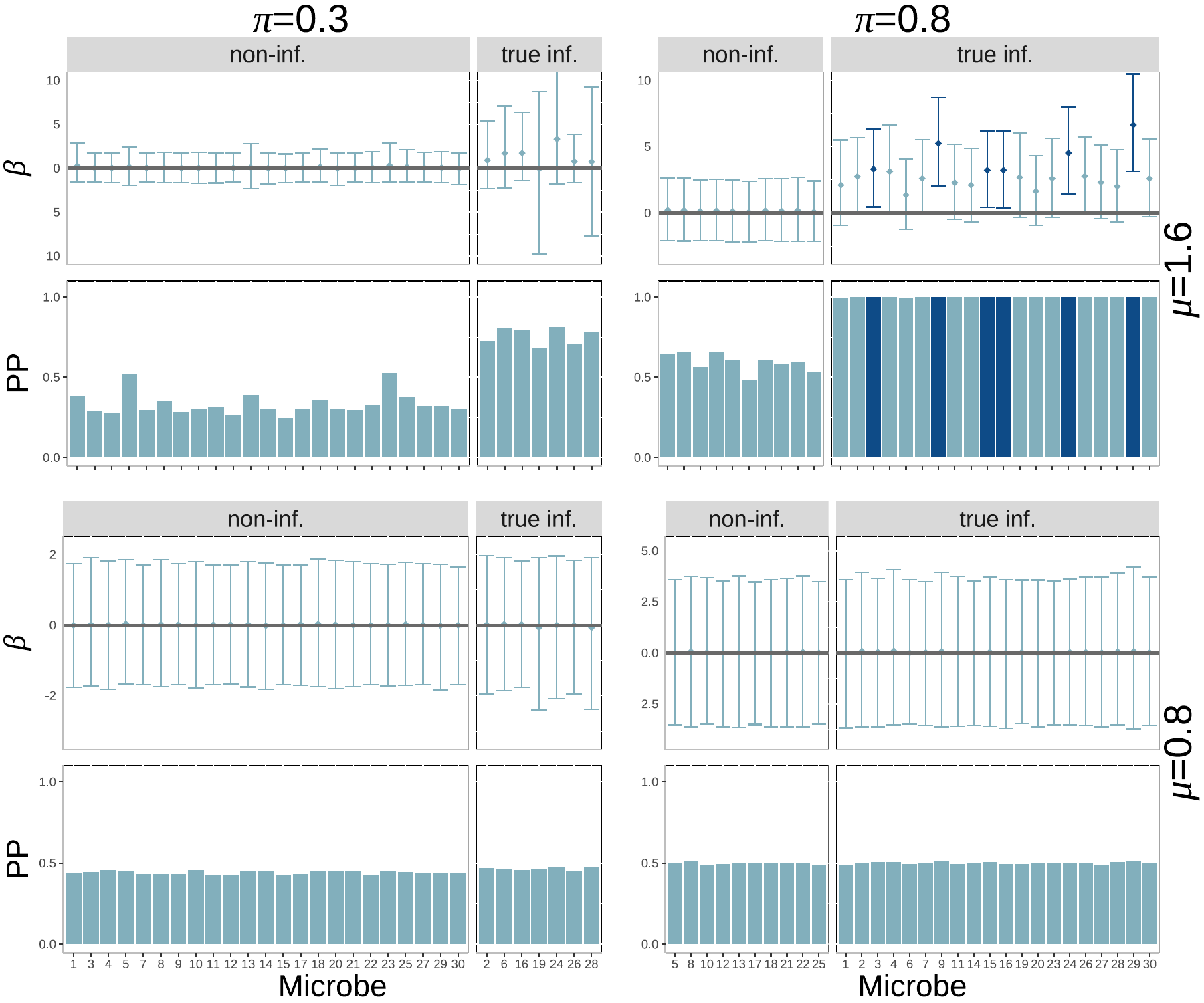}
    \caption{$\mathbf{n=100}$, $\mathbf{k=8}$}
\end{subfigure}
\begin{subfigure}[t]{0.49\textwidth}
    \centering
    \includegraphics[scale=0.27]{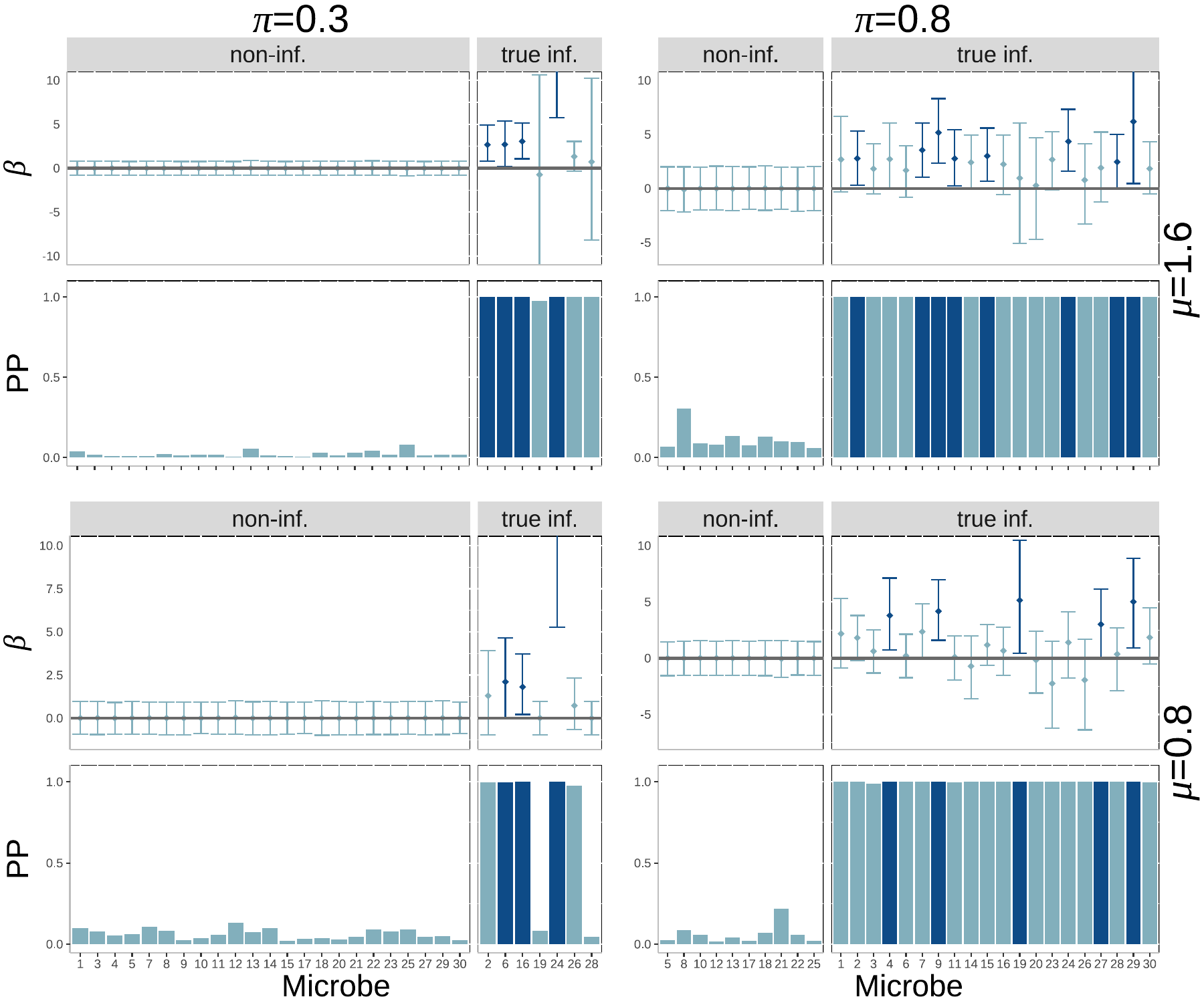}
    \caption{$\mathbf{n=500}$, $\mathbf{k=8}$}
\end{subfigure}\\
\begin{subfigure}[t]{0.49\textwidth}
    \centering
    \includegraphics[scale=0.27]{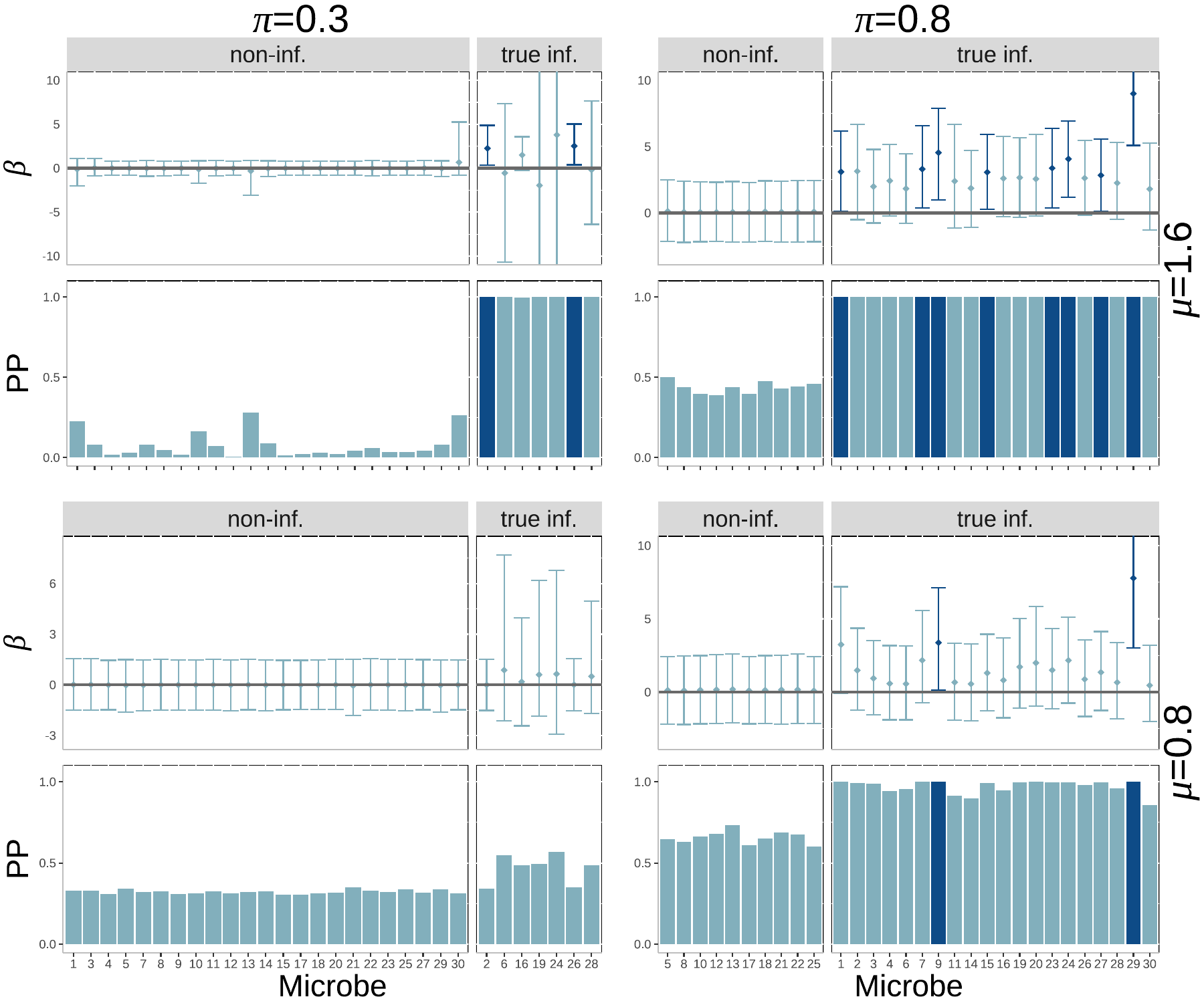}
    \caption{$\mathbf{n=100}$, $\mathbf{k=22}$}
\end{subfigure}
\begin{subfigure}[t]{0.49\textwidth}
    \centering
    \includegraphics[scale=0.27]{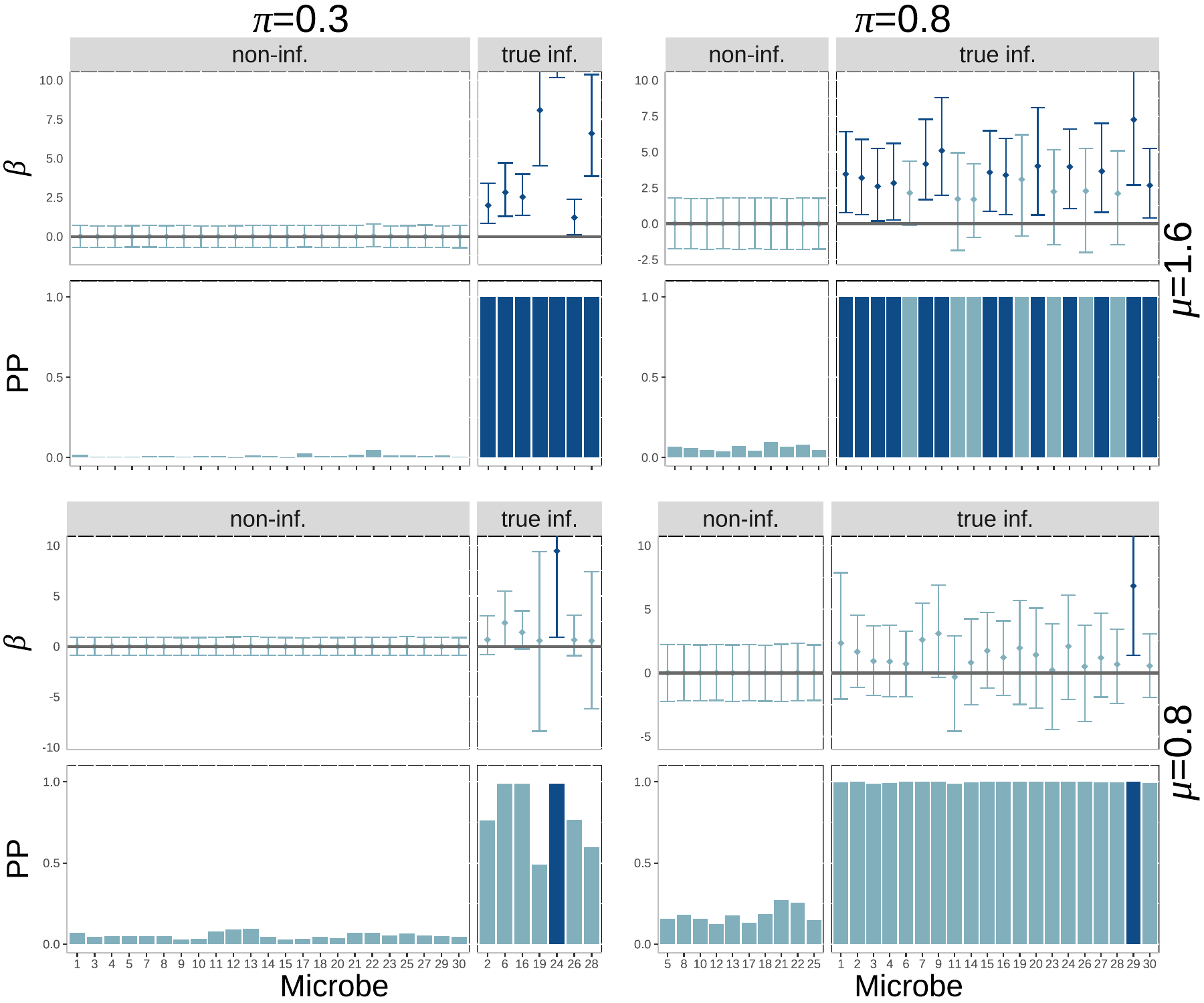}
    \caption{$\mathbf{n=500}$, $\mathbf{k=22}$}
\end{subfigure}
\caption[Posterior probability of influential nodes and coefficients for nodes (theoretical simulations $R=5$)]{{\bf Posterior probability of influential nodes (theoretical simulations $R=5$).}
Different groups of four panels represent different number of sampled microbes ($k=8,22$) which controls the sparsity of the adjacency matrix and different sample sizes ($n=100,500$). \revision{Within each group, we have four panels corresponding to the two values of edge effect size ($\mu=0.8, 1.6$) and two values of probability of influential node ($\pi=0.3, 0.8$) which controls the sparsity of the regression coefficient matrix ($\mathbf B$). Within each of these panels we have two plots: 95\% credible intervals (top) and posterior probability of influence (bottom - calculated as the mean of the $\xi$ variable for the node across Gibbs samples) for each node.} Each bar corresponds to one node (microbe). \revision{Within each plot the bars and intervals are colored depending on whether the node is found to be influential (dark) or not influential (light) based on the 95\% credible intervals. Each plot is split based on whether the nodes are truly influential (right) or not (left).}
}
\label{fig:nodes-sm}
\end{figure}

\begin{figure}[!ht]
\centering
\begin{subfigure}[t]{0.49\textwidth}
    \centering
    \includegraphics[scale=0.27]{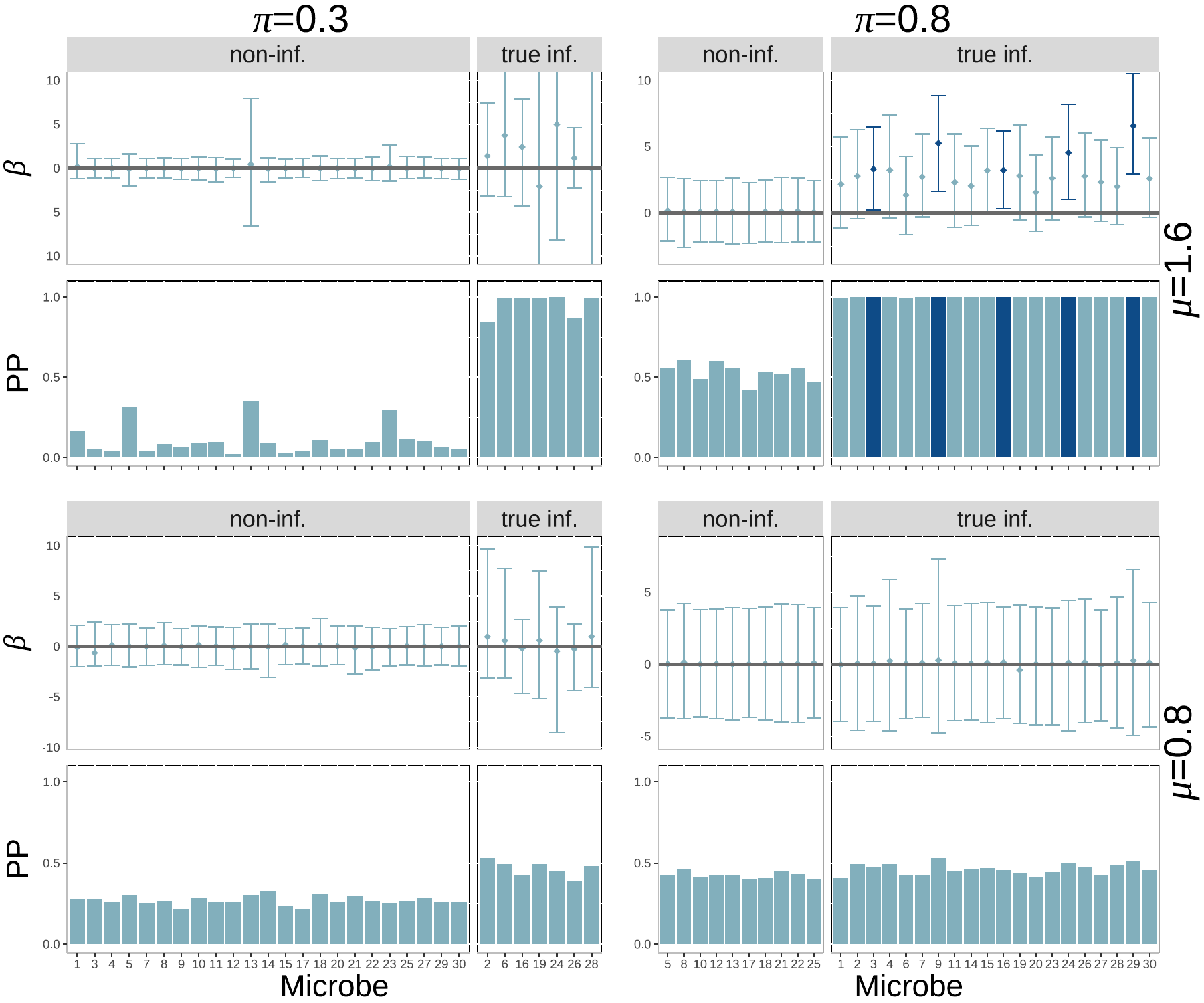}
    \caption{$\mathbf{n=100}$, $\mathbf{k=8}$}
\end{subfigure}
\begin{subfigure}[t]{0.49\textwidth}
    \centering
    \includegraphics[scale=0.27]{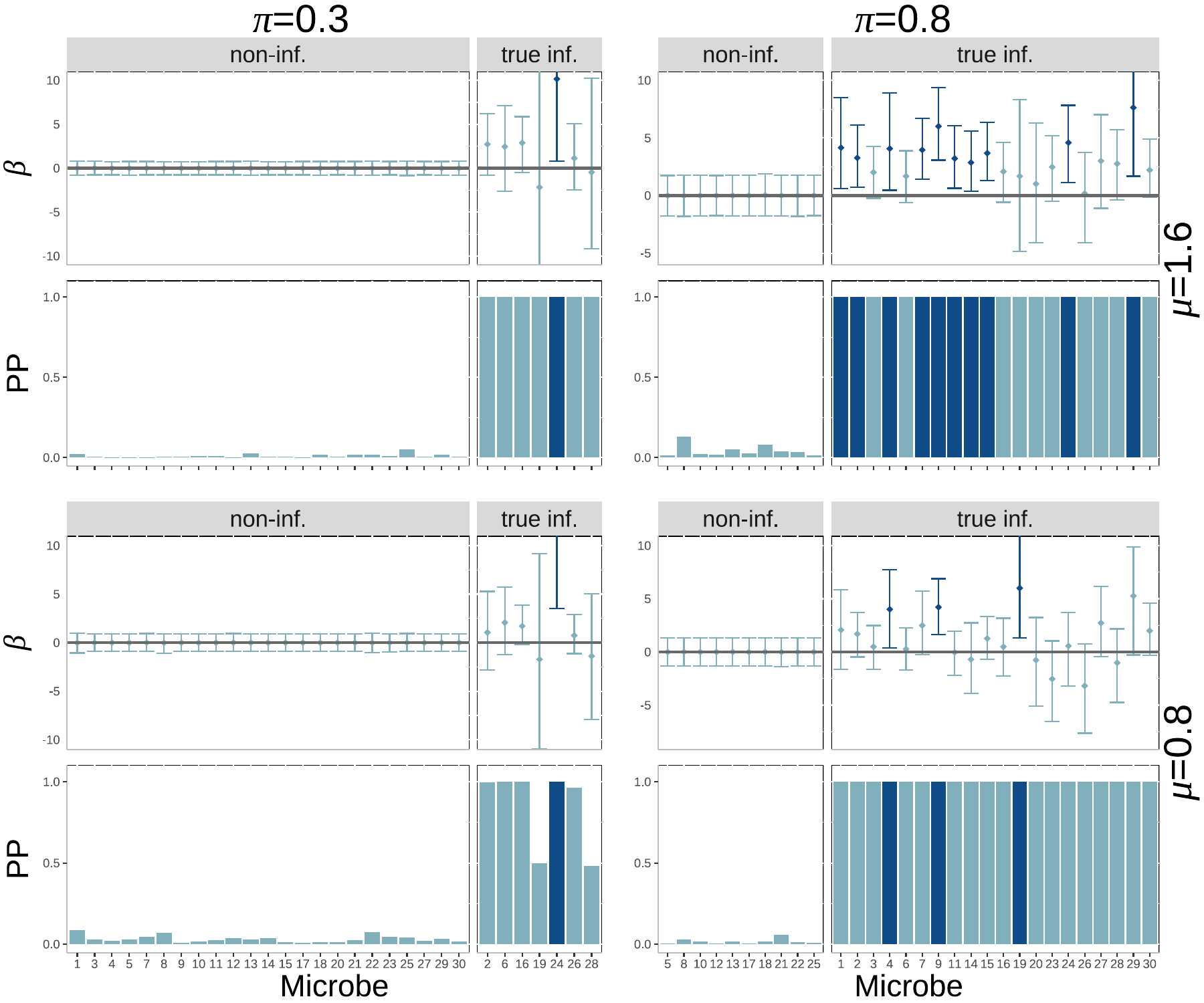}
    \caption{$\mathbf{n=500}$, $\mathbf{k=8}$}
\end{subfigure}\\
\begin{subfigure}[t]{0.49\textwidth}
    \centering
    \includegraphics[scale=0.27]{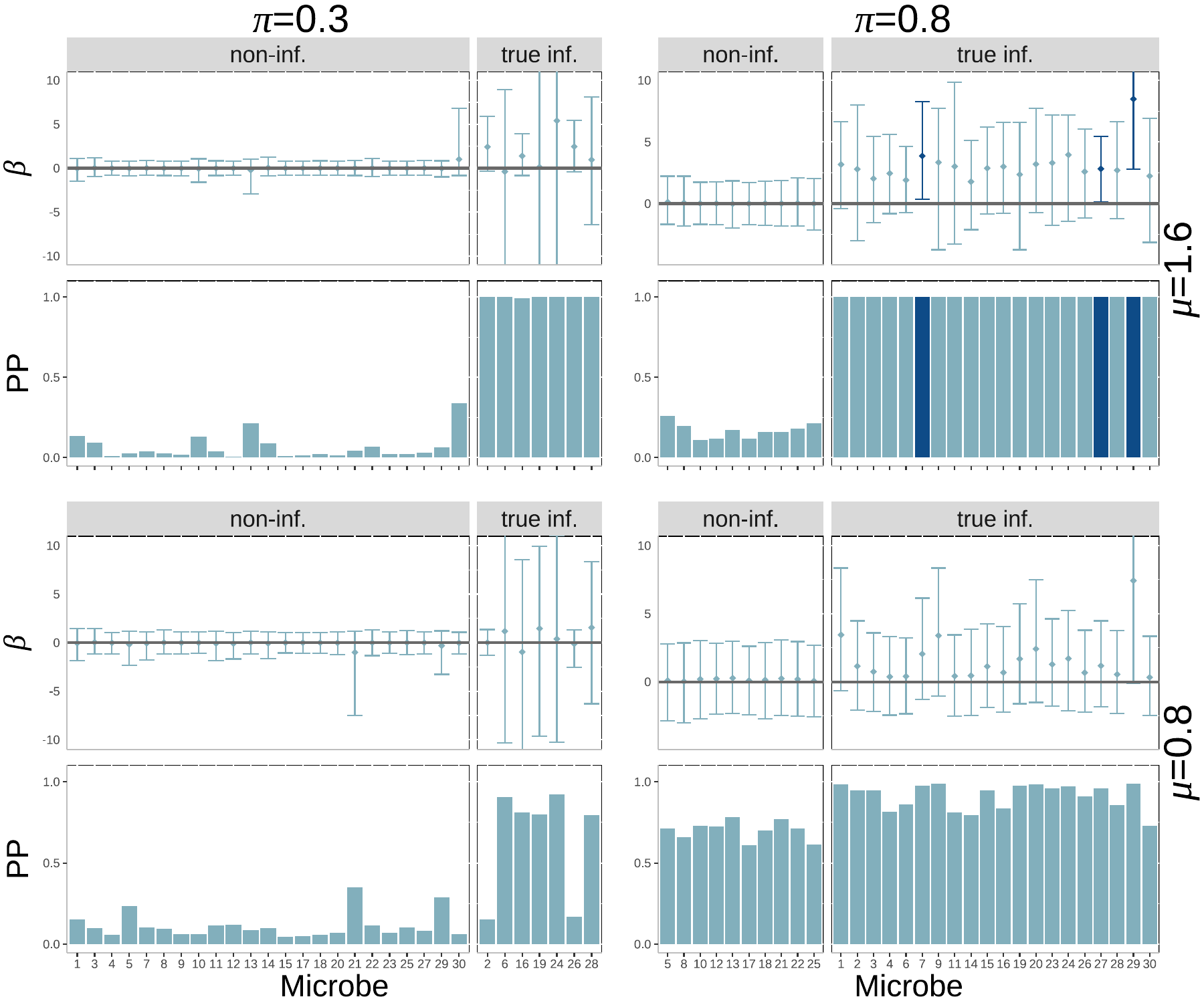}
\caption{$\mathbf{n=100}$, $\mathbf{k=22}$}
\end{subfigure}
\begin{subfigure}[t]{0.49\textwidth}
    \centering
    \includegraphics[scale=0.27]{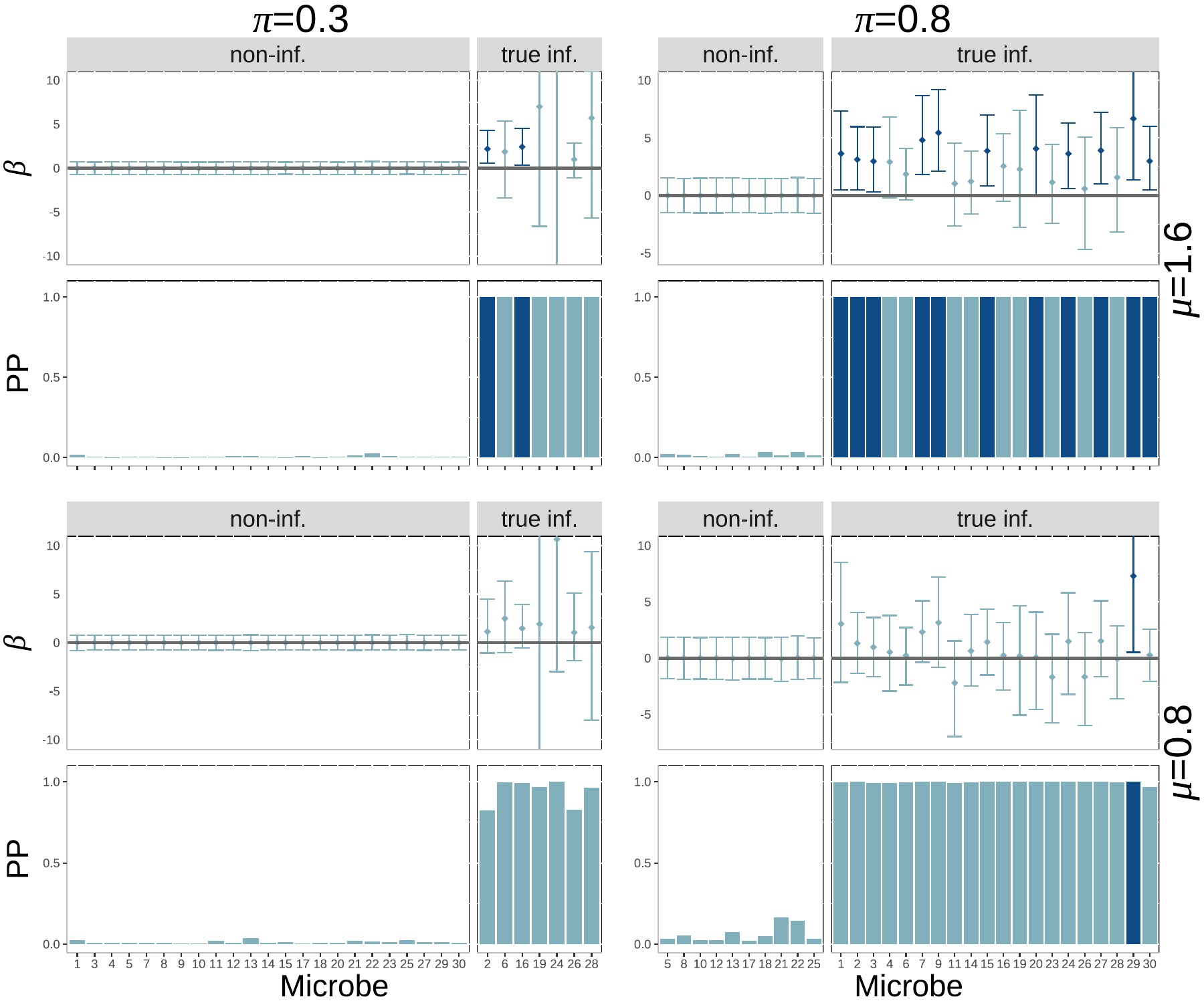}
\caption{$\mathbf{n=500}$, $\mathbf{k=22}$}
\end{subfigure}
\caption[Posterior probability of influential nodes and coefficients for nodes (theoretical simulations $R=9$)]{{\bf Posterior probability of influential nodes (theoretical simulations $R=9$).}
Different groups of four panels represent different number of sampled microbes ($k=8,22$) which controls the sparsity of the adjacency matrix and different sample sizes ($n=100,500$). \revision{Within each group, we have four panels corresponding to the two values of edge effect size ($\mu=0.8, 1.6$) and two values of probability of influential node ($\pi=0.3, 0.8$) which controls the sparsity of the regression coefficient matrix ($\mathbf B$). Within each of these panels we have two plots: 95\% credible intervals (top) and posterior probability of influence (bottom - calculated as the mean of the $\xi$ variable for the node across Gibbs samples) for each node.} Each bar corresponds to one node (microbe). \revision{Within each plot the bars and intervals are colored depending on whether the node is found to be influential (dark) or not influential (light) based on the 95\% credible intervals. Each plot is split based on whether the nodes are truly influential (right) or not (left).}
}
\label{fig:nodes-sm2}
\end{figure}

\begin{figure}[!ht]
\centering
\begin{subfigure}[t]{0.49\textwidth}
    \centering
    \includegraphics[scale=0.27]{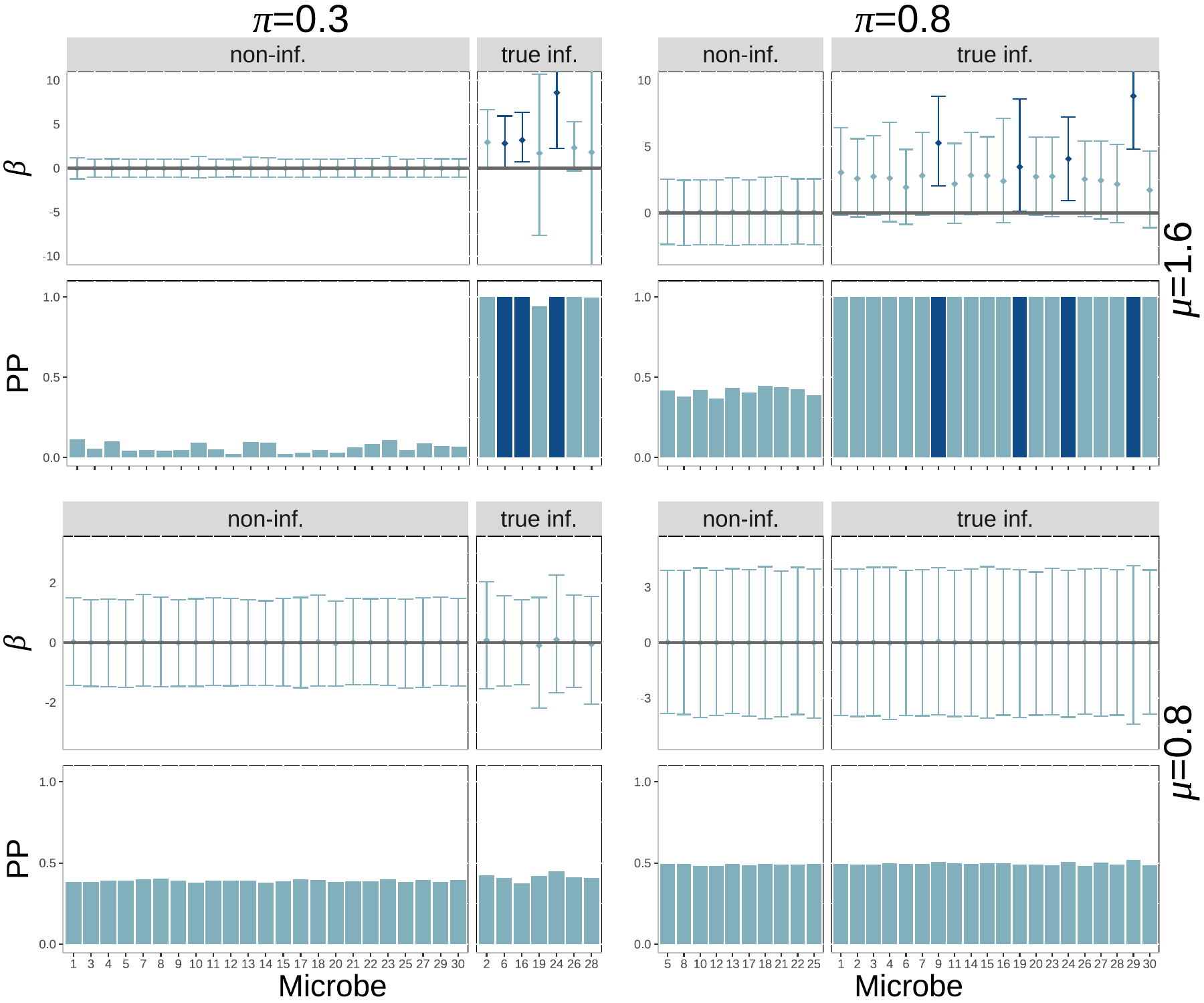}
    \caption{$\mathbf{n=100}$, $\mathbf{R=5}$}
\end{subfigure}
\begin{subfigure}[t]{0.49\textwidth}
    \centering
    \includegraphics[scale=0.27]{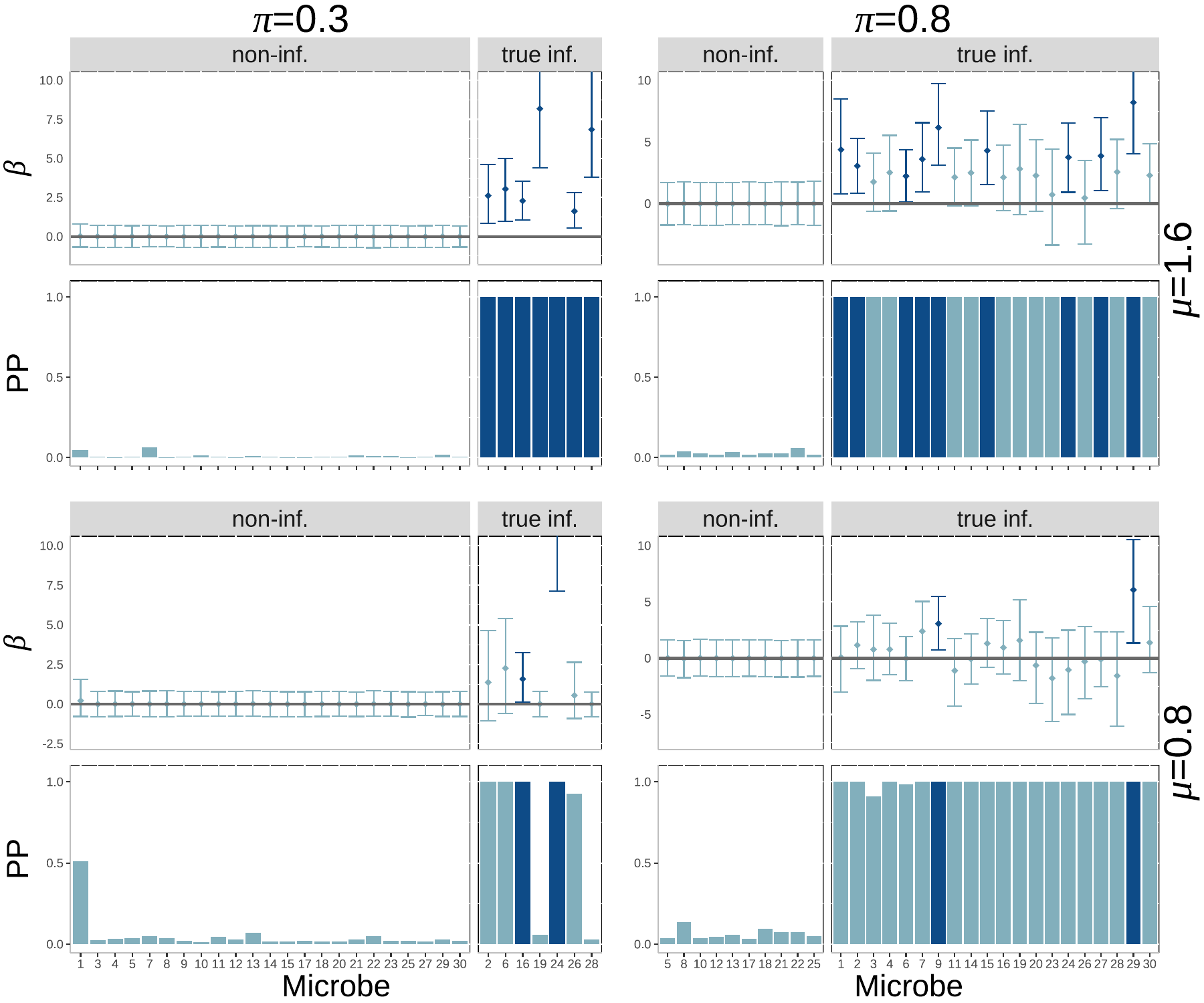}
    \caption{$\mathbf{n=500}$, $\mathbf{R=5}$}
\end{subfigure}
\caption[Posterior probability of influential nodes and coefficients for nodes (theoretical simulations $k=15$, $R=5,7,9$)]{{\bf Posterior probability of influential nodes and coefficients for nodes (theoretical simulations $k=15$, $R=5$).}
Different groups of four panels represent different sample sizes ($n=100,500$). \revision{Within each group, we have four panels corresponding to the two values of edge effect size ($\mu=0.8, 1.6$) and two values of probability of influential node ($\pi=0.3, 0.8$) which controls the sparsity of the regression coefficient matrix ($\mathbf B$). Within each of these panels we have two plots: 95\% credible intervals (top) and posterior probability of influence (bottom - calculated as the mean of the $\xi$ variable for the node across Gibbs samples) for each node.} Each bar corresponds to one node (microbe). \revision{Within each plot the bars and intervals are colored depending on whether the node is found to be influential (dark) or not influential (light) based on the 95\% credible intervals. Each plot is split based on whether the nodes are truly influential (right) or not (left).}
}
\label{fig:nodes-sm3-R5}
\end{figure}

\begin{figure}[!ht]
\begin{subfigure}[t]{0.49\textwidth}
    \centering
    \includegraphics[scale=0.27]{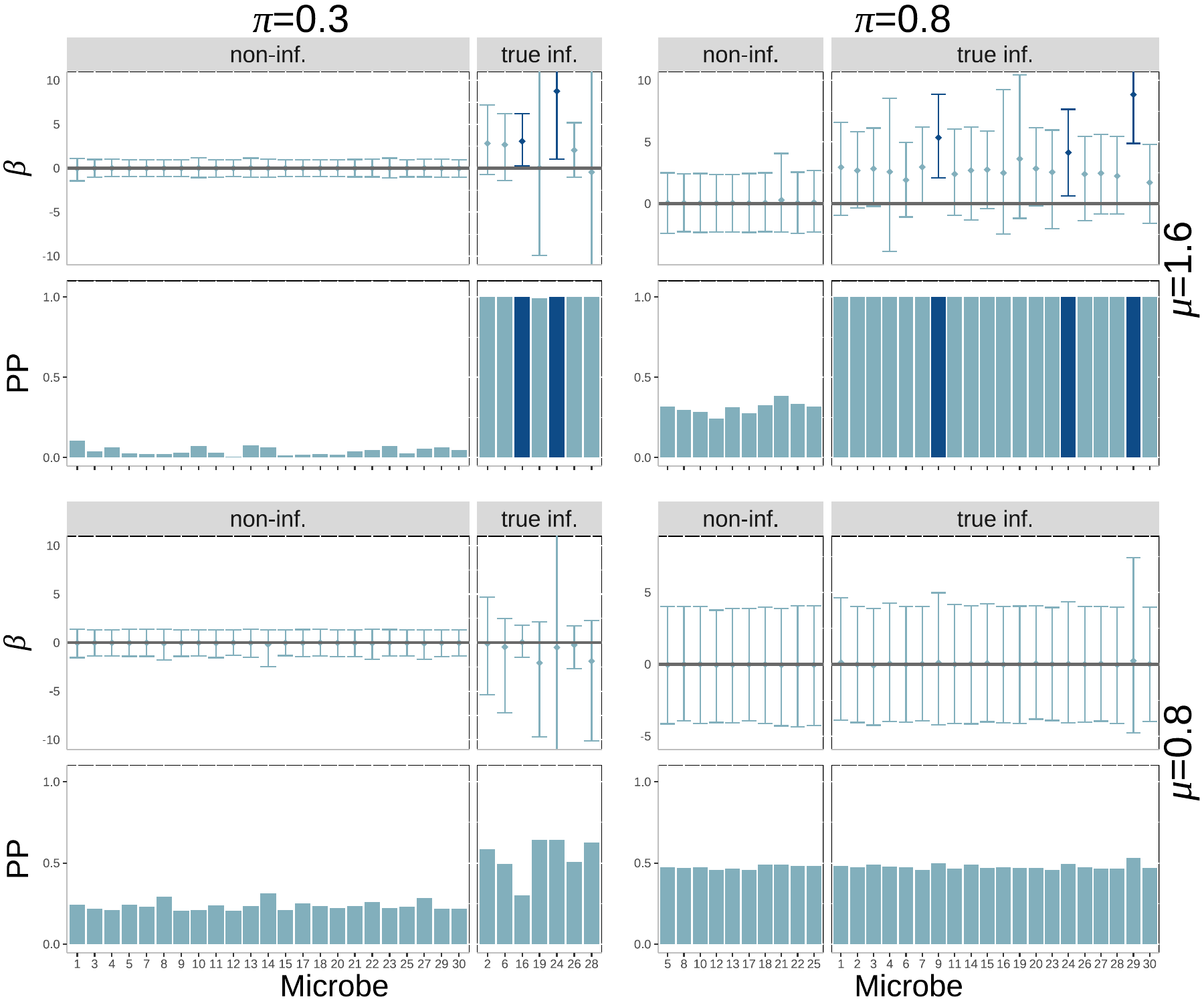}
\caption{$\mathbf{n=100}$, $\mathbf{R=7}$}
\end{subfigure}
\begin{subfigure}[t]{0.49\textwidth}
    \centering
    \includegraphics[scale=0.27]{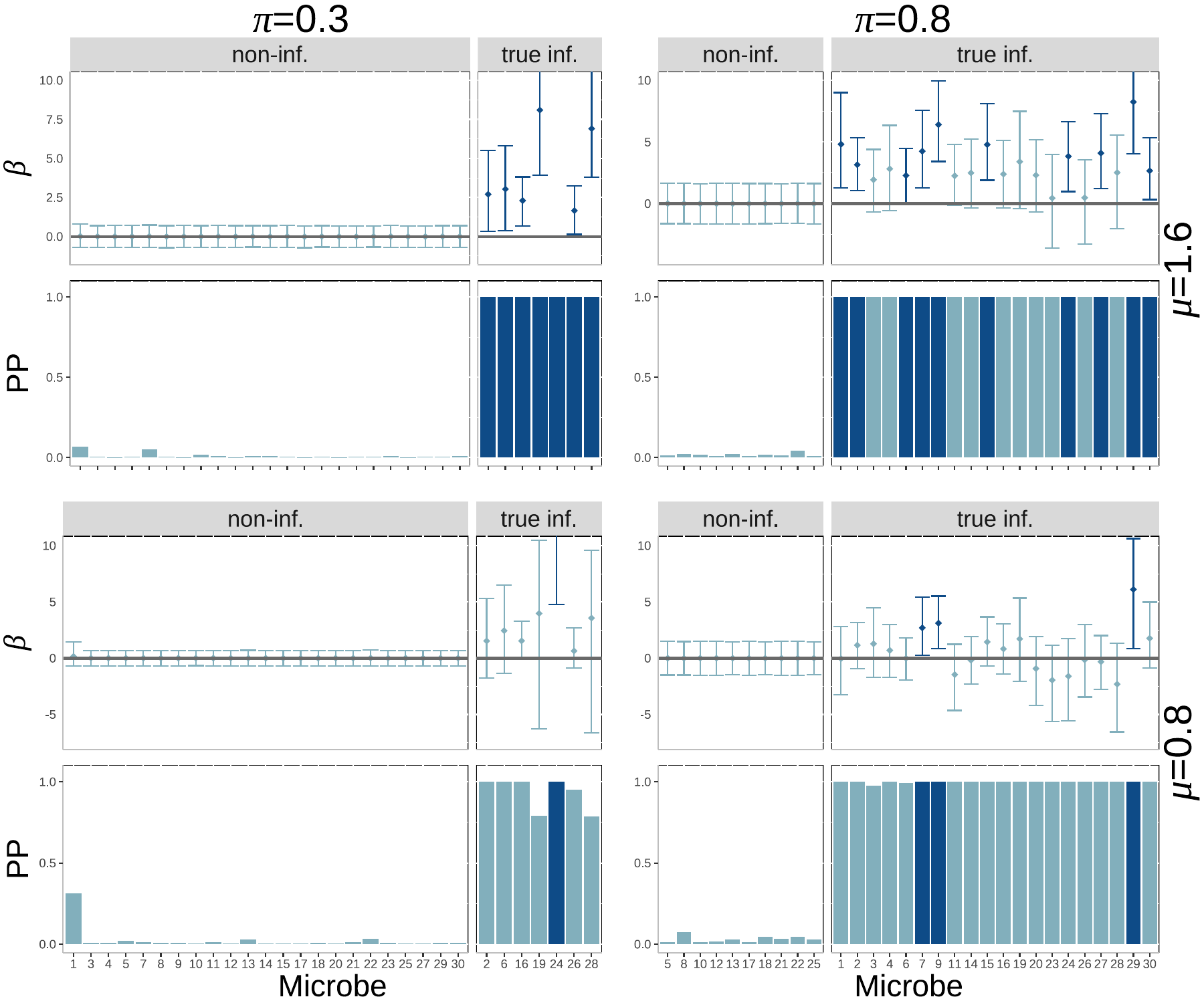}
    \caption{$\mathbf{n=500}$, $\mathbf{R=7}$}
\end{subfigure}
\caption[Posterior probability of influential nodes and coefficients for nodes (theoretical simulations $k=15$, $R=5,7,9$)]{{\bf Posterior probability of influential nodes and coefficients for nodes (theoretical simulations $k=15$, $R=7$).}
Different groups of four panels represent different sample sizes ($n=100,500$). \revision{Within each group, we have four panels corresponding to the two values of edge effect size ($\mu=0.8, 1.6$) and two values of probability of influential node ($\pi=0.3, 0.8$) which controls the sparsity of the regression coefficient matrix ($\mathbf B$). Within each of these panels we have two plots: 95\% credible intervals (top) and posterior probability of influence (bottom - calculated as the mean of the $\xi$ variable for the node across Gibbs samples) for each node.} Each bar corresponds to one node (microbe). \revision{Within each plot the bars and intervals are colored depending on whether the node is found to be influential (dark) or not influential (light) based on the 95\% credible intervals. Each plot is split based on whether the nodes are truly influential (right) or not (left).}
}
\label{fig:nodes-sm3-R7}
\end{figure}

\begin{figure}[!ht]
\begin{subfigure}[t]{0.49\textwidth}
    \centering
    \includegraphics[scale=0.27]{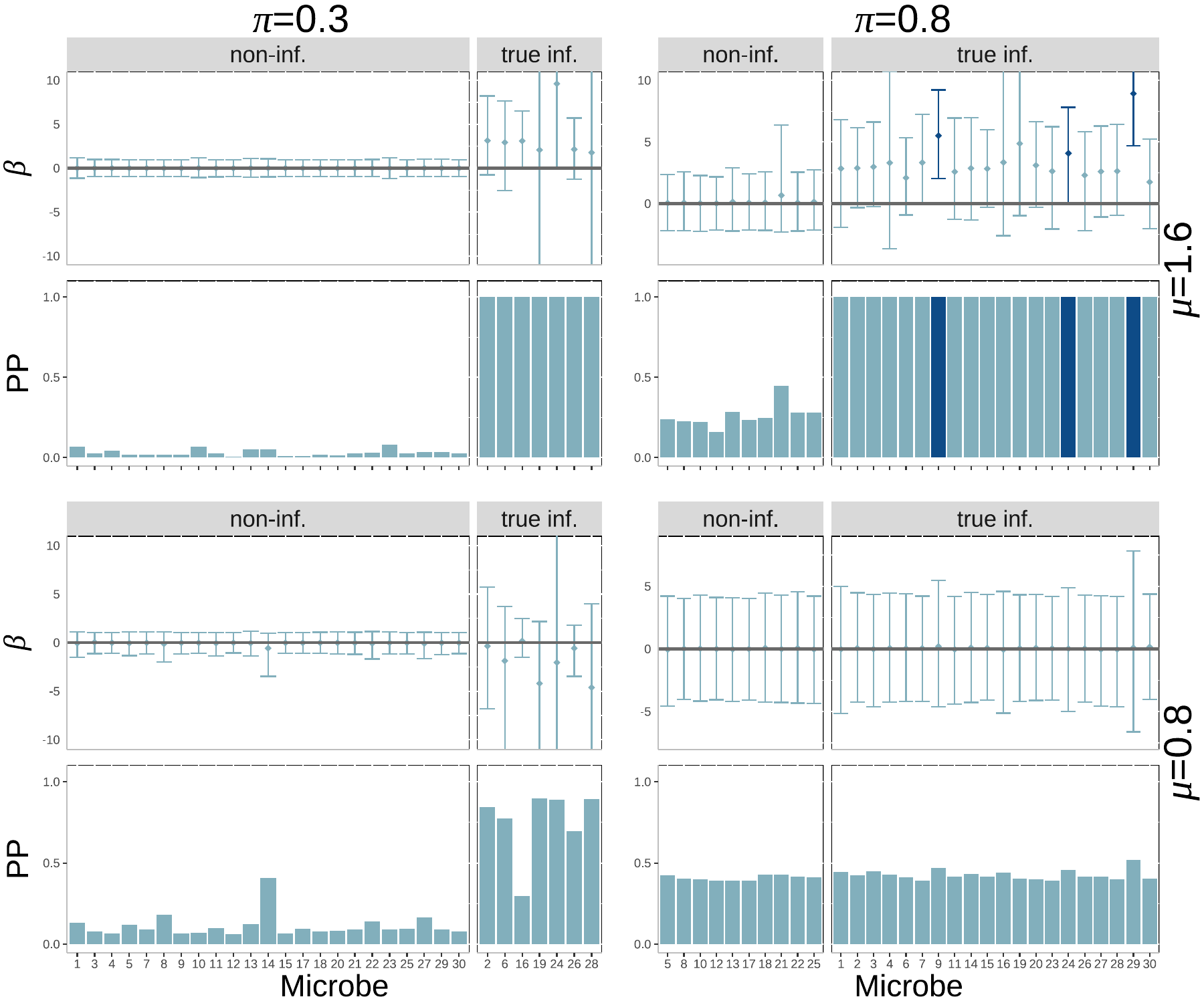}
    \caption{$\mathbf{n=100}$, $\mathbf{R=9}$}
\end{subfigure}
\begin{subfigure}[t]{0.49\textwidth}
    \centering
    \includegraphics[scale=0.27]{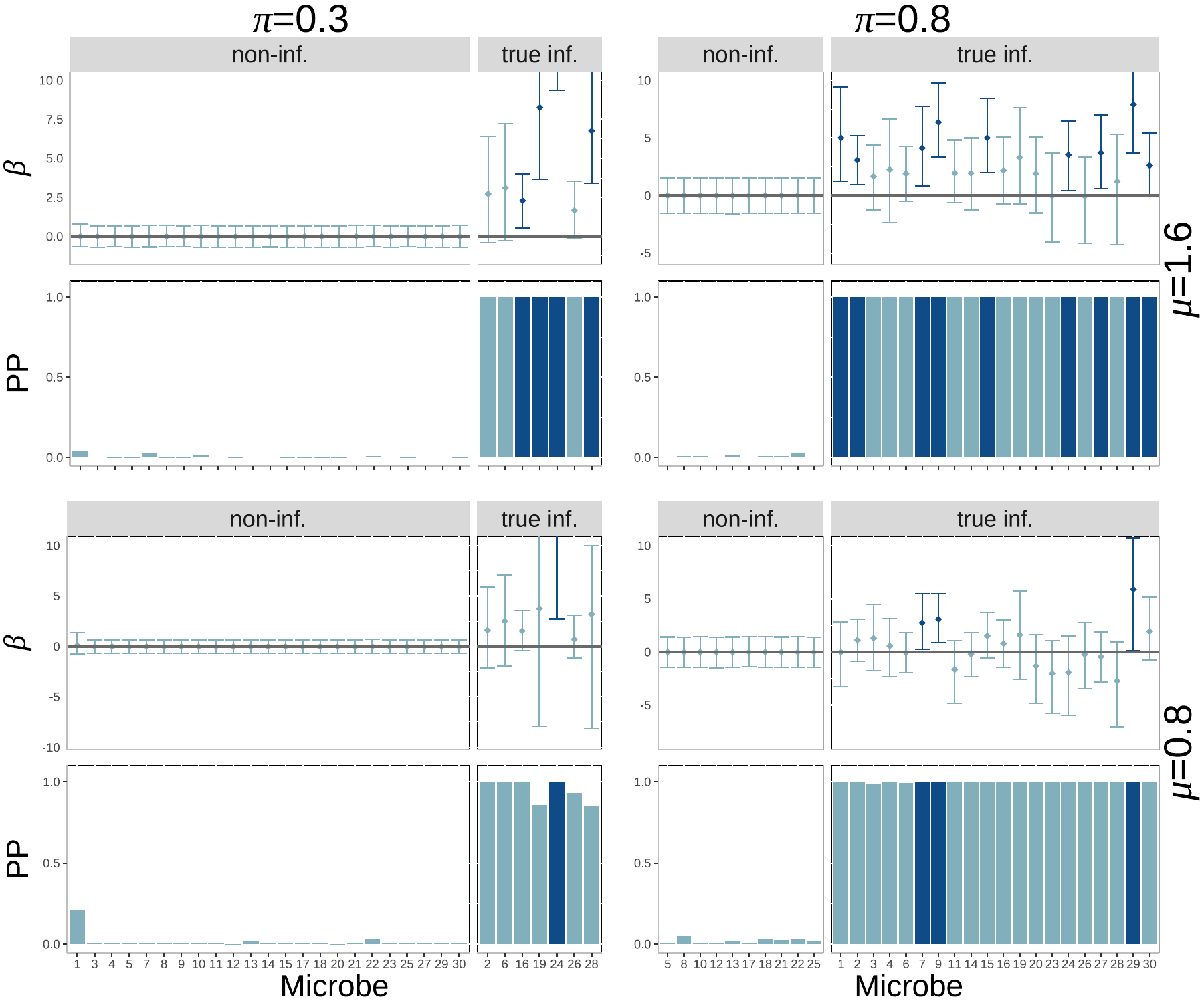}
    \caption{$\mathbf{n=500}$, $\mathbf{R=9}$}
\end{subfigure}
\caption[Posterior probability of influential nodes and coefficients for nodes (theoretical simulations $k=15$, $R=5,7,9$)]{{\bf Posterior probability of influential nodes and coefficients for nodes (theoretical simulations $k=15$, $R=9$).}
Different groups of four panels represent different sample sizes ($n=100,500$). \revision{Within each group, we have four panels corresponding to the two values of edge effect size ($\mu=0.8, 1.6$) and two values of probability of influential node ($\pi=0.3, 0.8$) which controls the sparsity of the regression coefficient matrix ($\mathbf B$). Within each of these panels we have two plots: 95\% credible intervals (top) and posterior probability of influence (bottom - calculated as the mean of the $\xi$ variable for the node across Gibbs samples) for each node.} Each bar corresponds to one node (microbe). \revision{Within each plot the bars and intervals are colored depending on whether the node is found to be influential (dark) or not influential (light) based on the 95\% credible intervals. Each plot is split based on whether the nodes are truly influential (right) or not (left).}
}
\label{fig:nodes-sm3-R9}
\end{figure}

\begin{figure}[!ht]
\centering
\includegraphics[scale=0.35]{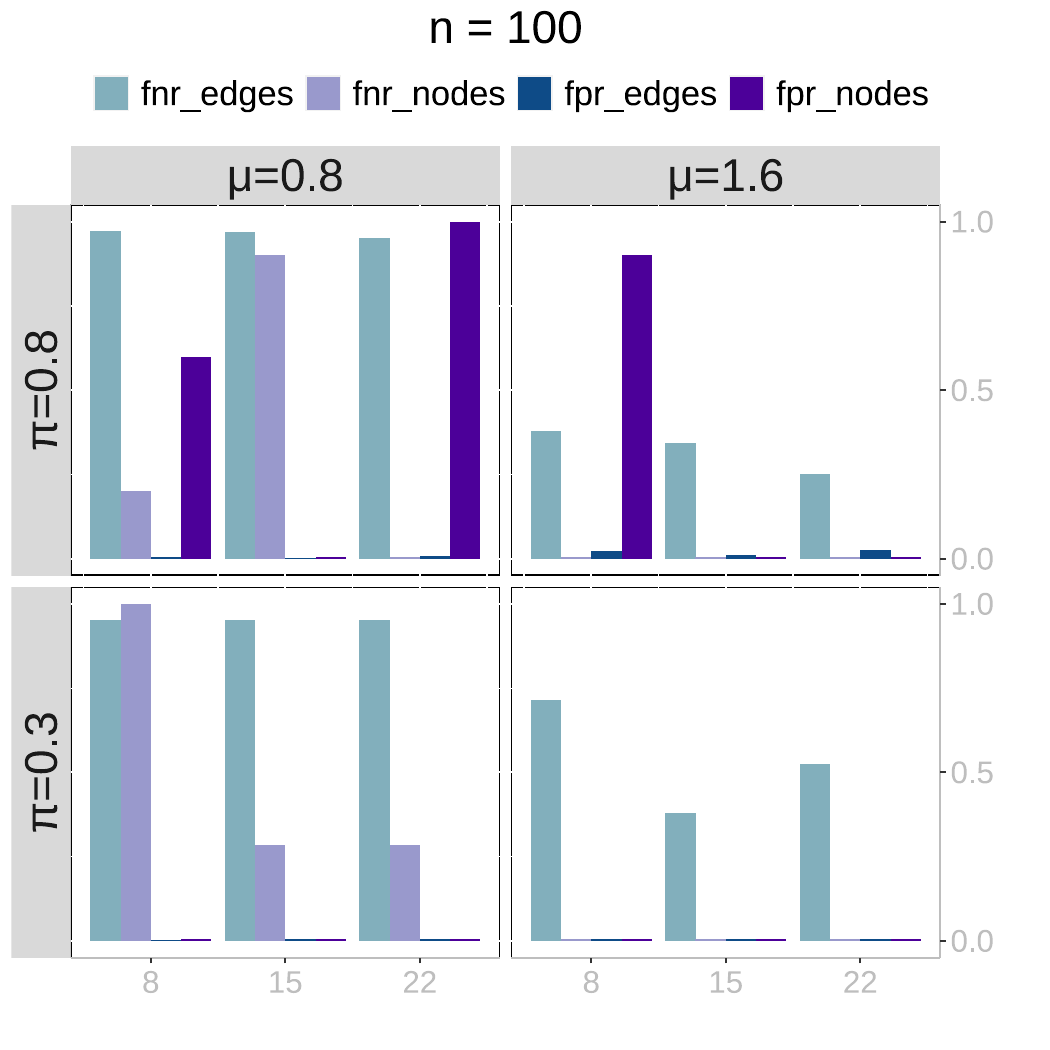}
\includegraphics[scale=0.35]{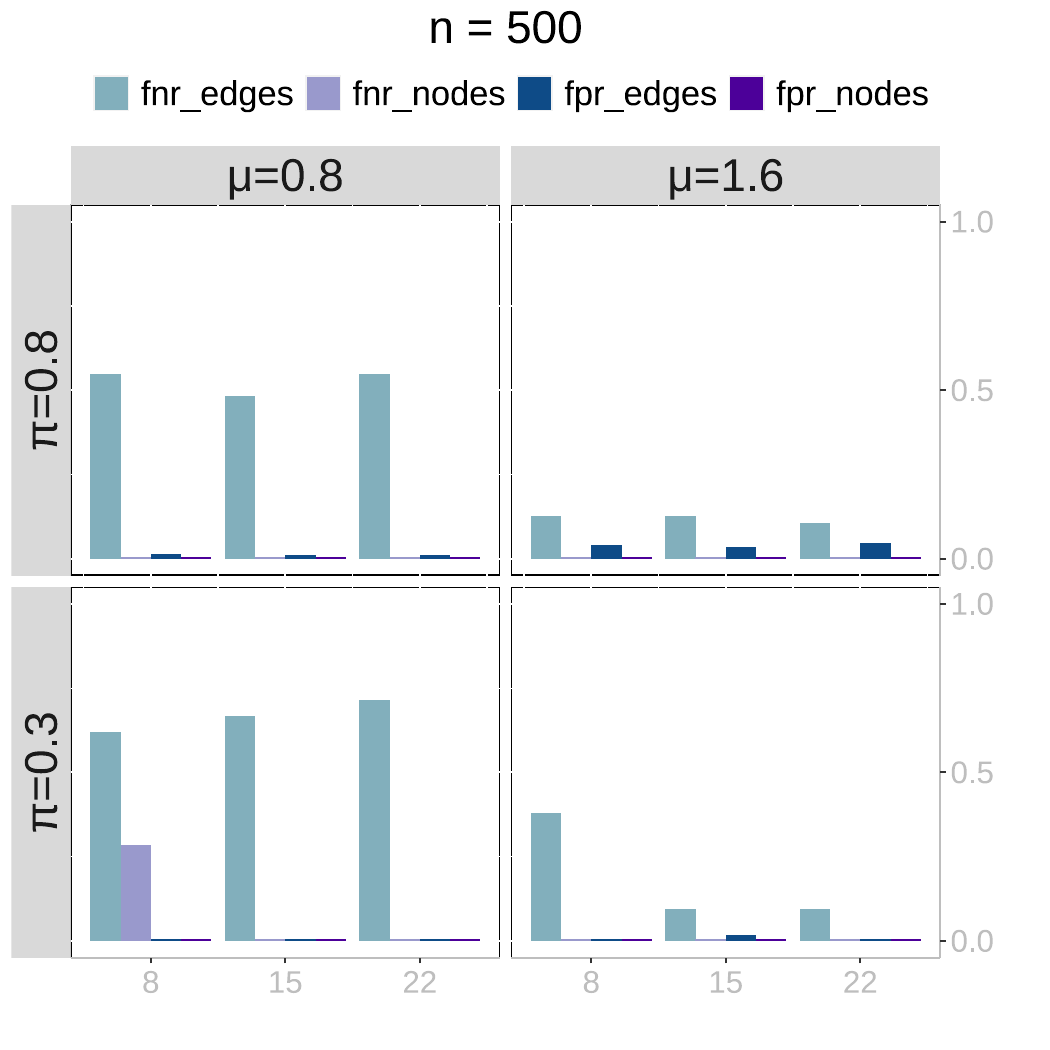}
\caption[False positive and false negative rates for influential edges and nodes (theoretical simulations)]{{\bf False positive and false negative rates for influential edges and nodes (theoretical simulations).} X axis corresponds to the number of sampled nodes (microbes) which relates to the sparsity of the adjacency matrices $\mathbf A_i$. Y axis corresponds to false positive or false negative rates for edges or nodes (depending on the color of the bar). Decisions to reject for edges are based on 95\% posterior credible intervals, and for nodes are based on whether the posterior probability of influence is greater than 0.5. Within each panel, we have four plots corresponding to the two values of edge effect size ($\mu=0.8, 1.6$) and two values of probability of influential node ($\pi=0.3, 0.8$) which controls the sparsity of the regression coefficient matrix ($\mathbf B$). Each bar corresponds to one rate: false positive rate and false negative rate for edges (dark and light blue) and false positive rate and false negative rate for nodes (dark and light purple).}
\label{fig:rates}
\end{figure}

\begin{figure}[!ht]
    \centering
    \begin{subfigure}[t]{\textwidth}
        \centering
        \includegraphics[scale=0.45]{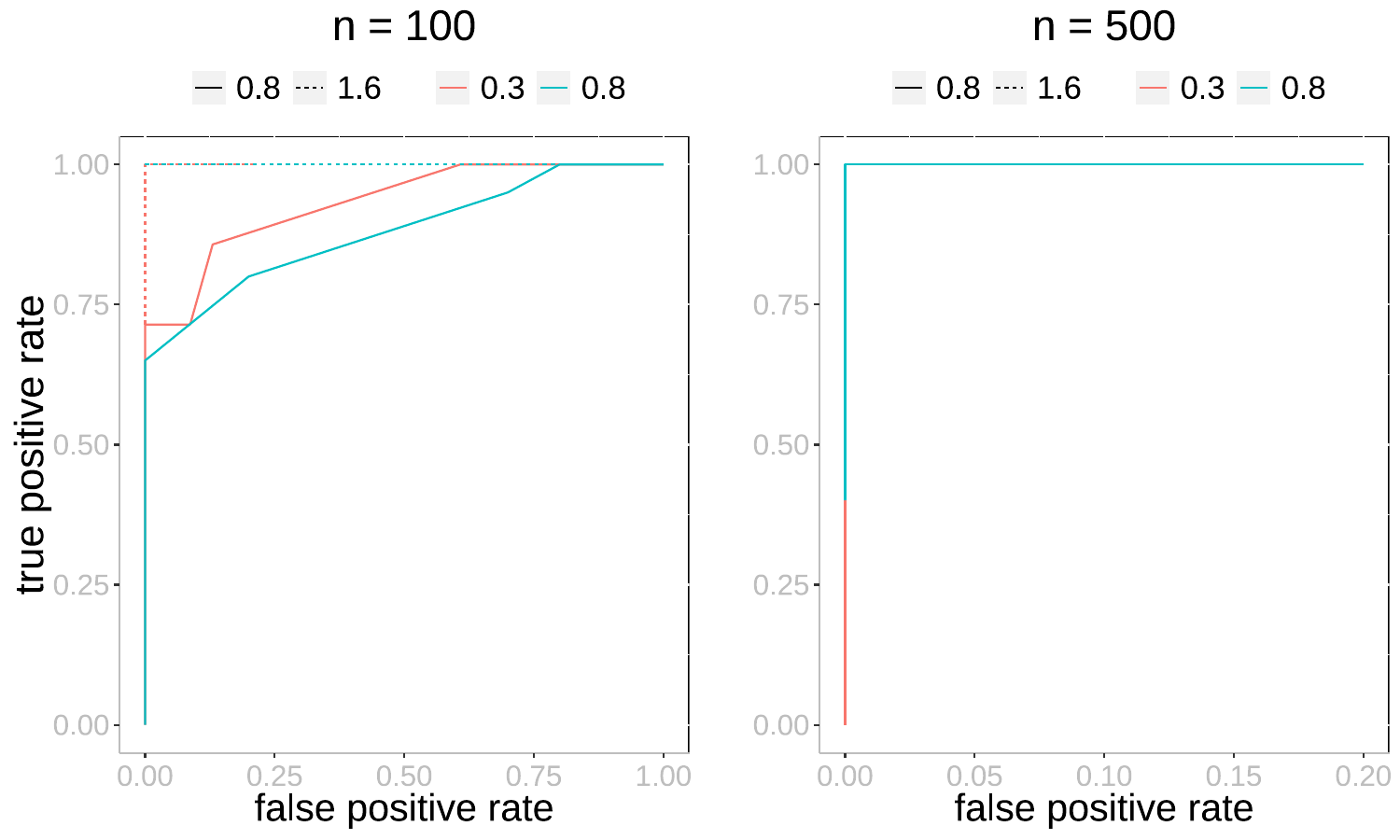}
        \caption{$\mathbf{k=8}$}
    \end{subfigure}
    \begin{subfigure}[t]{\textwidth}
        \centering
        \includegraphics[scale=0.45]{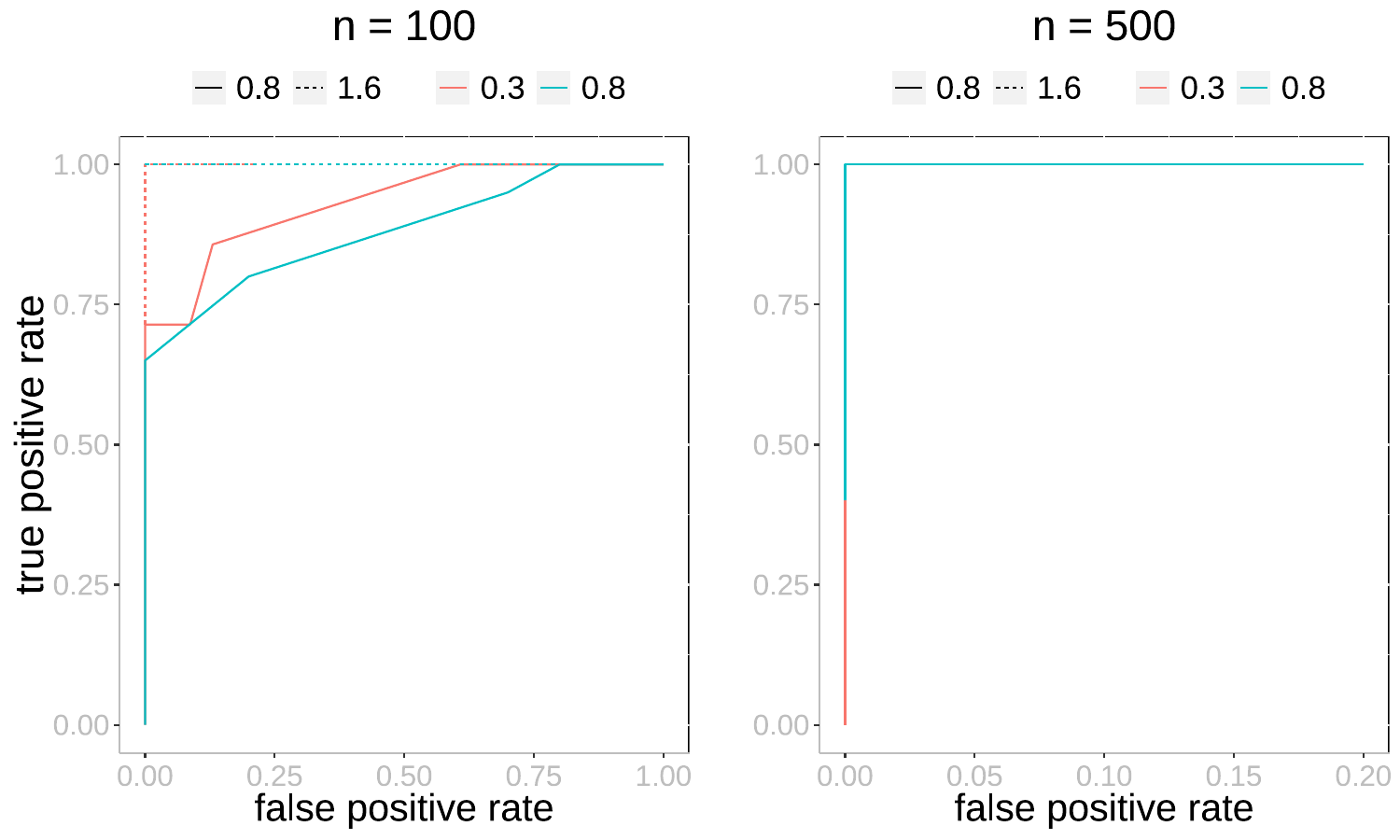}
        \caption{$\mathbf{k=22}$}
    \end{subfigure}
    \caption[ROC curve for node posterior probabilities (theoretical simulations)]{\revision{{\bf ROC curve for node posterior probabilities (theoretical simulations) for $R=7$.} X axis refers to false positive rate, Y axis refers to true positive rate. False positive and true positive rate changes as the arbitrary cutoff value for determining node influence changes. Top is $k=8$ microbes per sample, bottom is $k=22$. In each plot, the left pane gives the small sample size ($n=100$) case, while the right pane gives the large sample size ($n=500$) case. Dashed lines correspond to different values of the true mean for edge effects ($\mu = 0.8, 1.6$) and different colors correspond to different sparsity levels on the regression coefﬁcient matrix B ($\pi = 0.3, 0.8$).}
    }
    \label{fig:roc}
\end{figure}

\FloatBarrier
\subsubsection{Realistic case: Additive model}
\FloatBarrier

\begin{figure}[!ht]
    \centering
    \begin{subfigure}[b]{0.49\textwidth}
    \centering
    \includegraphics[scale=0.19]{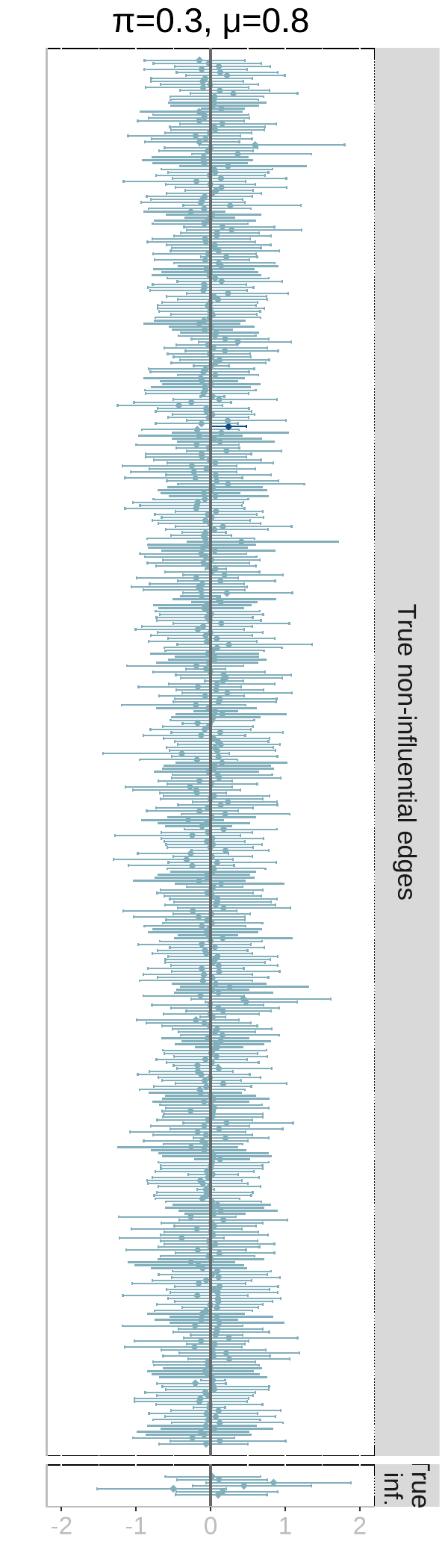}
    \includegraphics[scale=0.19]{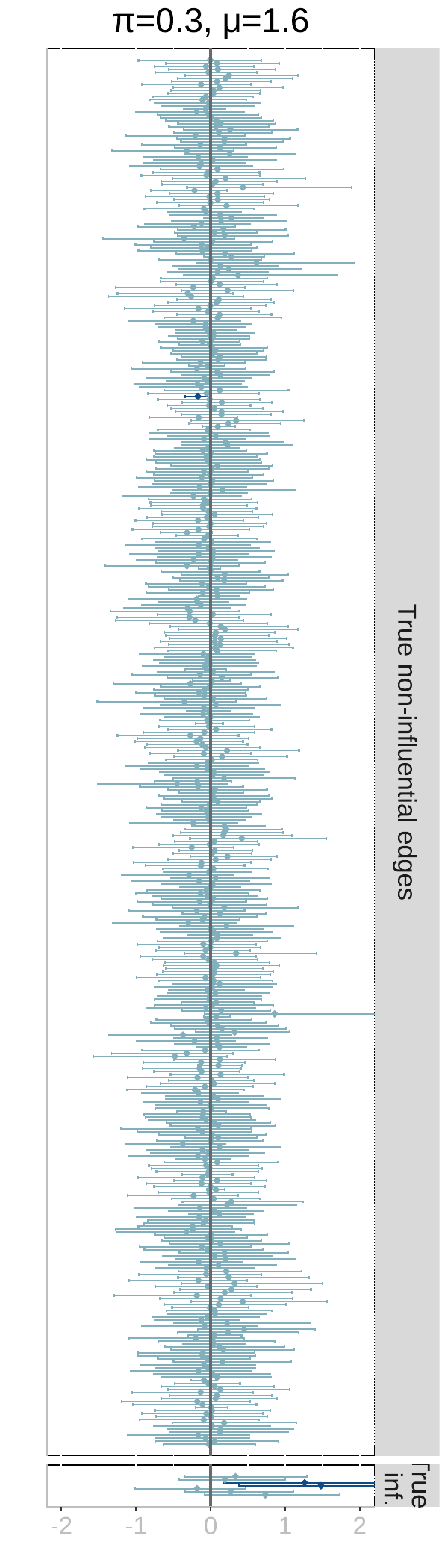}
    \includegraphics[scale=0.19]{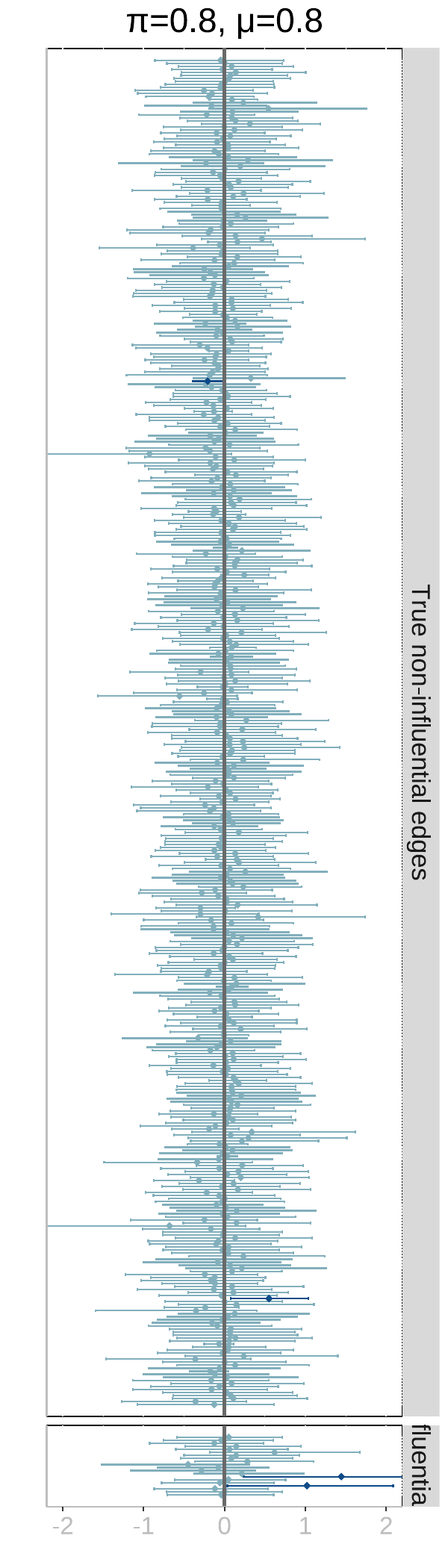}
    \includegraphics[scale=0.19]{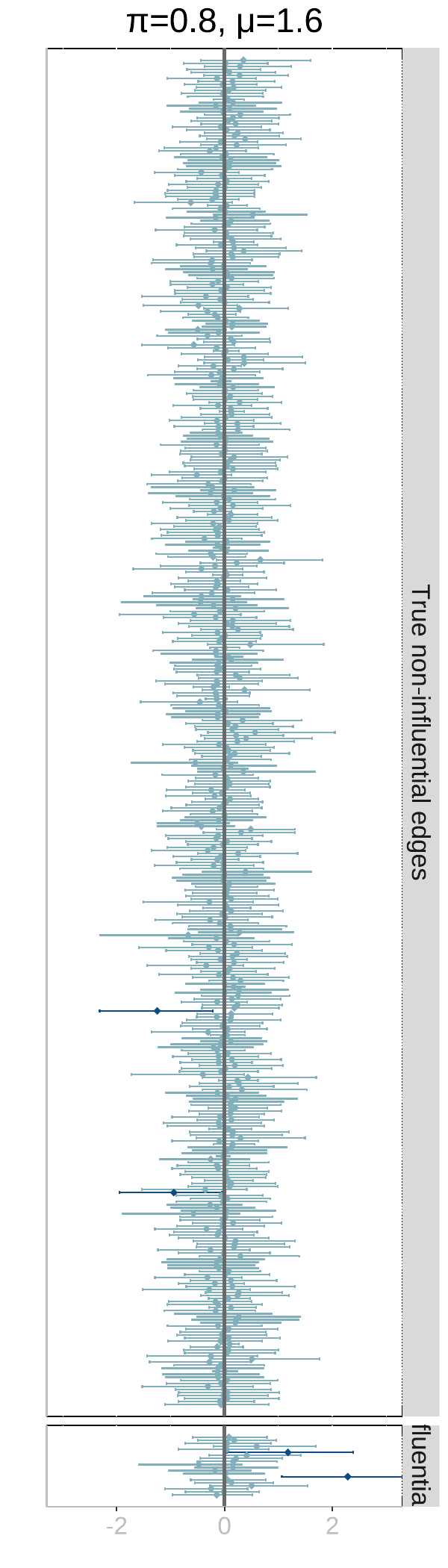}
    \caption{$\mathbf{n=500}$}
    \end{subfigure}
    \begin{subfigure}[b]{0.49\textwidth}
    \centering
    \includegraphics[scale=0.19]{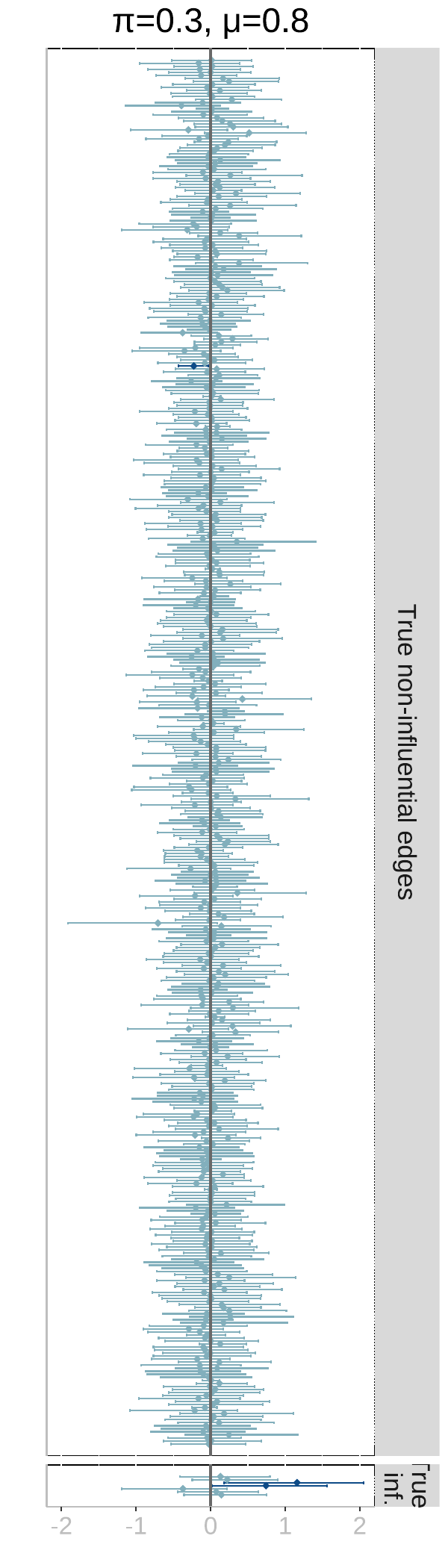}
    \includegraphics[scale=0.19]{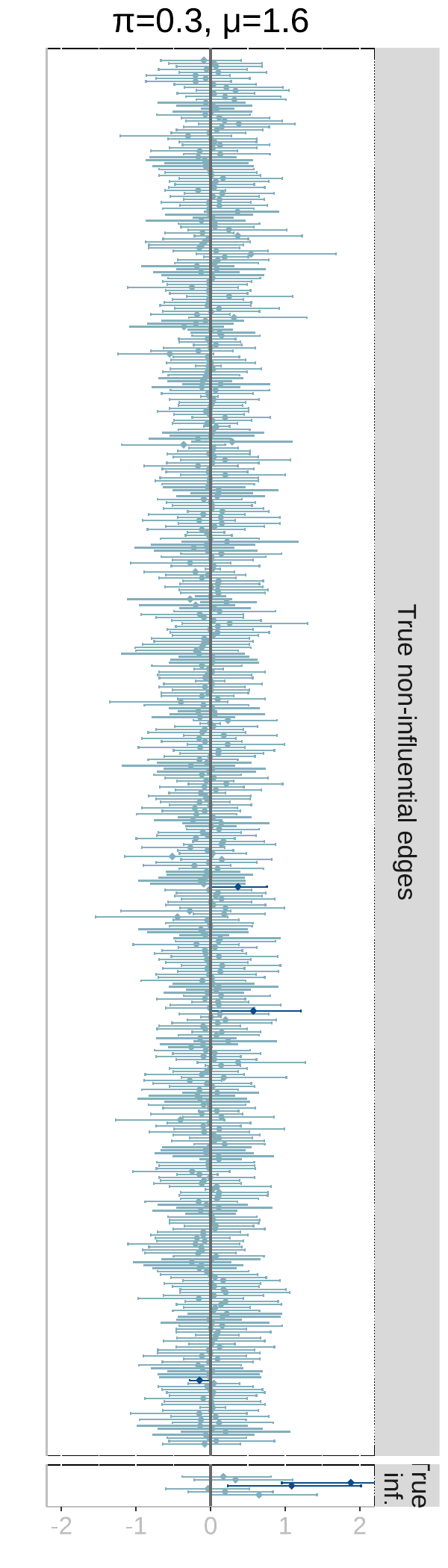}
    \includegraphics[scale=0.19]{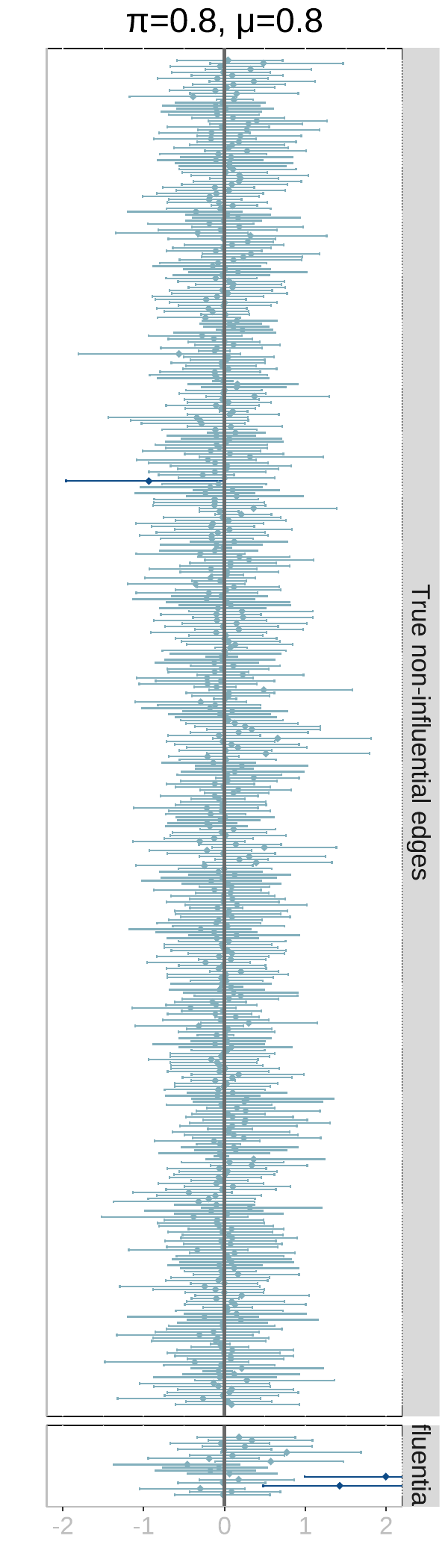}
    \includegraphics[scale=0.19]{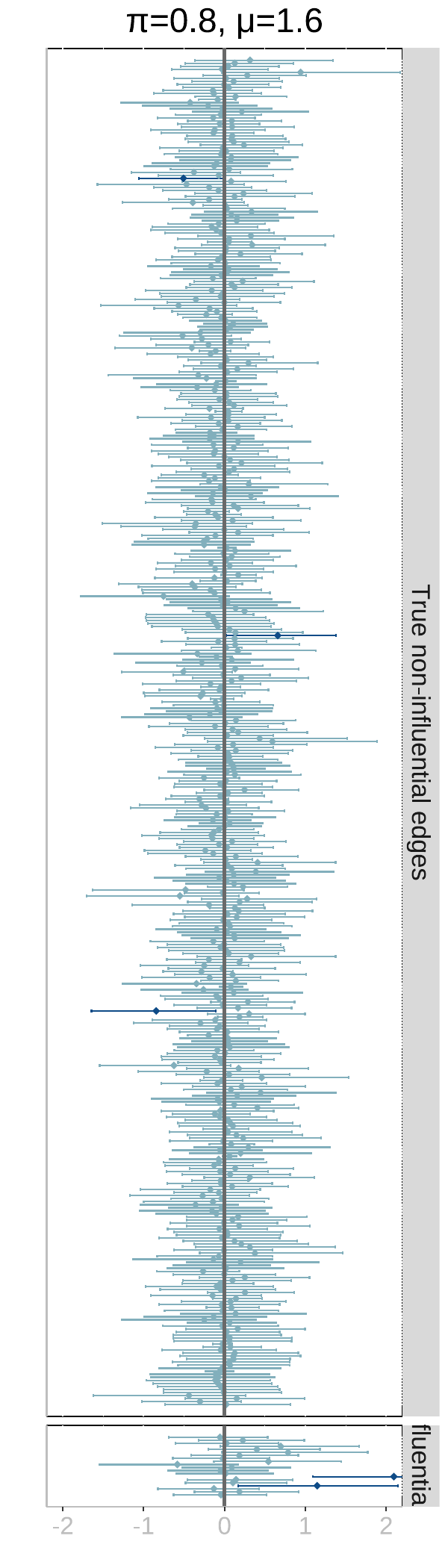}
    \caption{$\mathbf{n=1000}$}
    \end{subfigure}
    \caption[95 \% credible intervals for edge effects (additive model with random coefficients) with $k=8$ sampled nodes]{{\bf 95 \% credible intervals for edge effects (additive model with random coefficients) with $k=8$ sampled nodes.} Top (a): Sample size of $n=500$. Bottom (b): Sample size of $n=1000$. Each panel corresponds to a scenario of $\pi=0.3, 0.8$ (which controls the sparsity of the regression coefficient matrix $\mathbf B$) and $\mu=0.8,1.6$. We plot the 95 \% credible intervals for the regression coefficients per edge. In the additive model, all edges are non-influential. The color of the intervals depends on whether it intersects zero (light) and hence estimated to be non-influential or does not intersect zero (dark) and hence estimated to be influential by the model. Given that the additive model does not have interaction (edge) effects, these panels allow us to visualize false positives (dark intervals).}
    \label{fig:edges_add_rand}
\end{figure}

\begin{figure}[!ht]
    \centering
    \begin{subfigure}[b]{0.49\textwidth}
    \centering
    \includegraphics[scale=0.19]{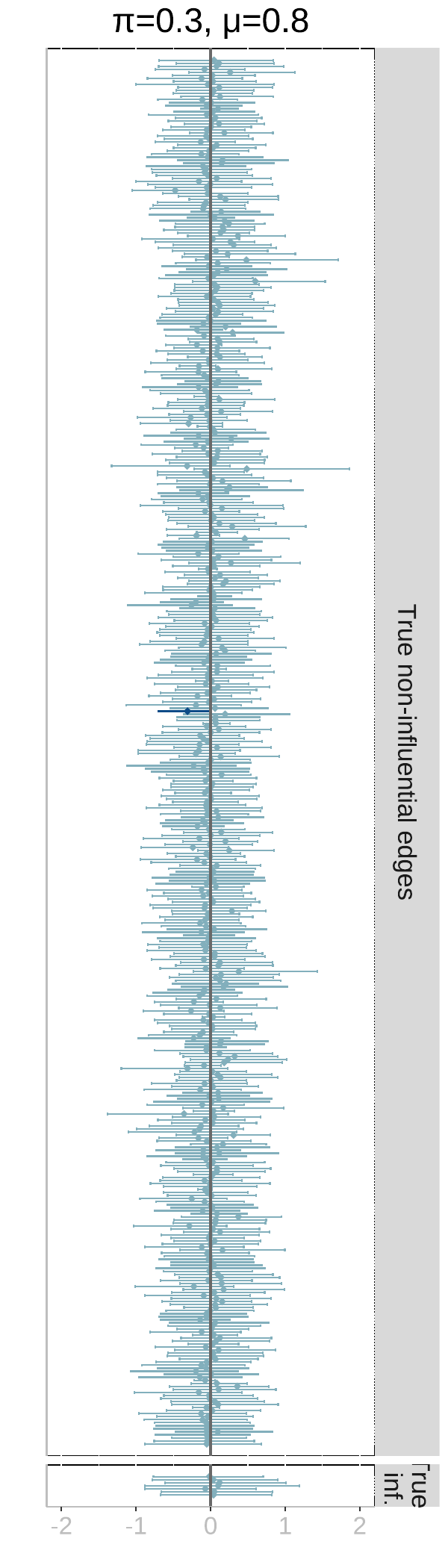}
    \includegraphics[scale=0.19]{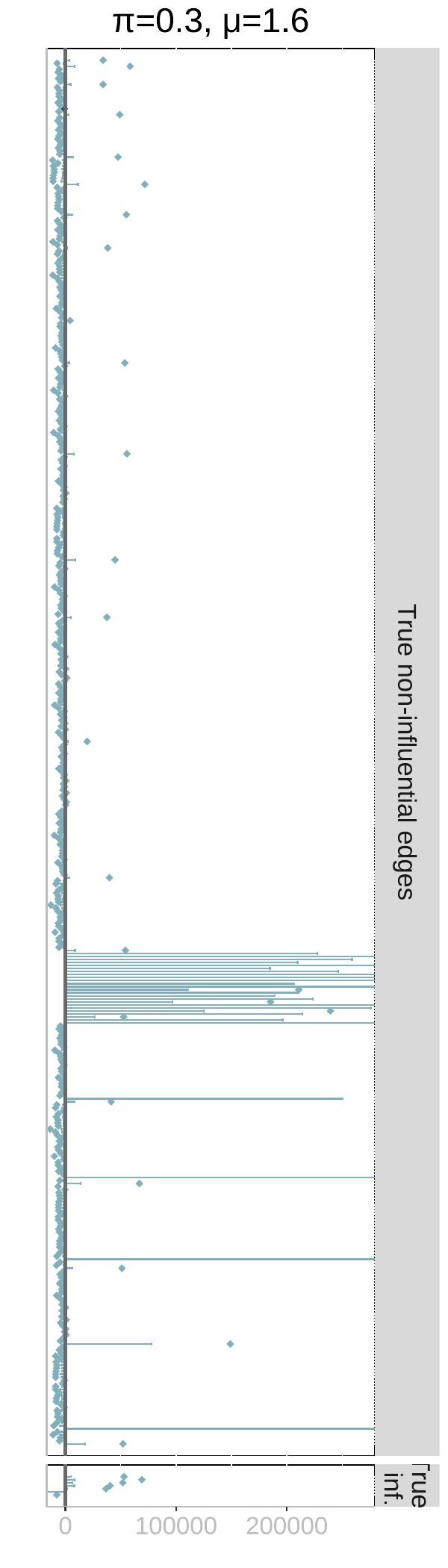}
    \includegraphics[scale=0.19]{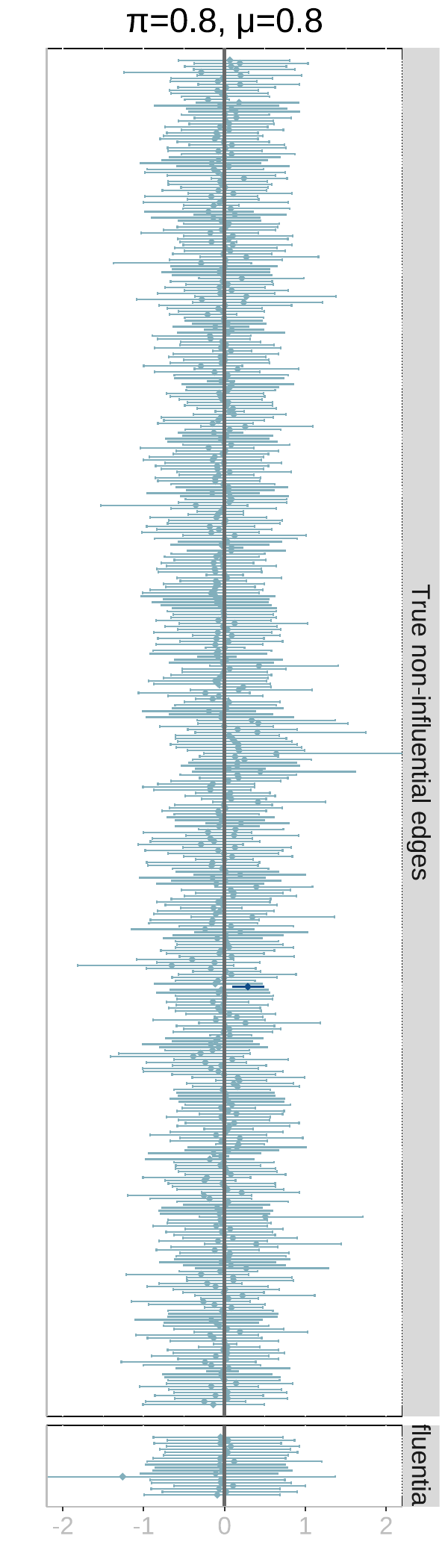}
    \includegraphics[scale=0.19]{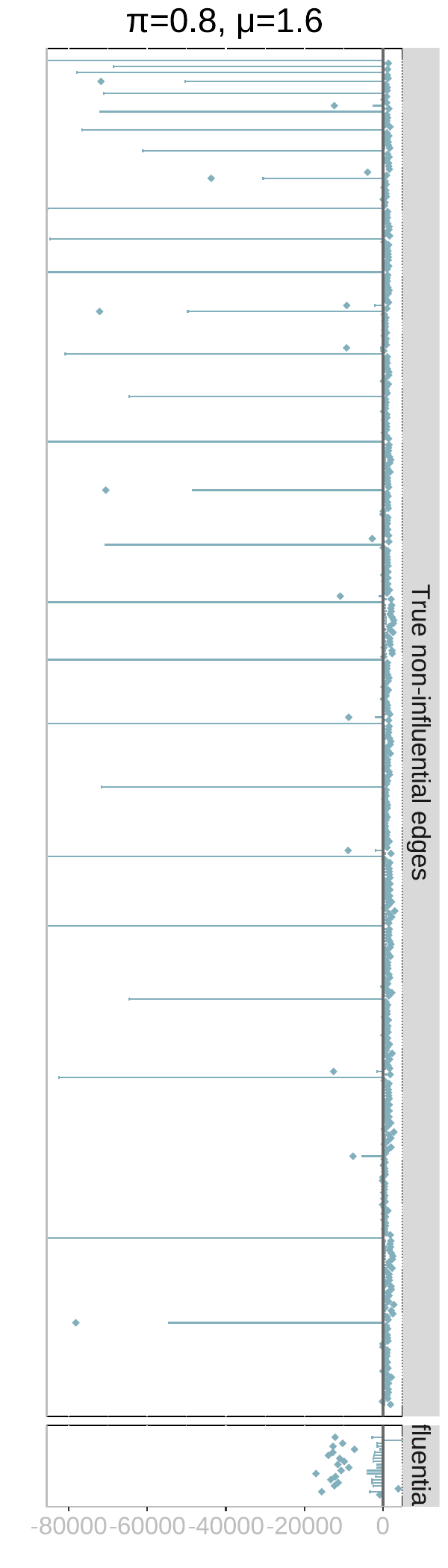}

    \caption{$\mathbf{n=500}$}
    \end{subfigure}
    \begin{subfigure}[b]{0.49\textwidth}
    \centering
    \includegraphics[scale=0.19]{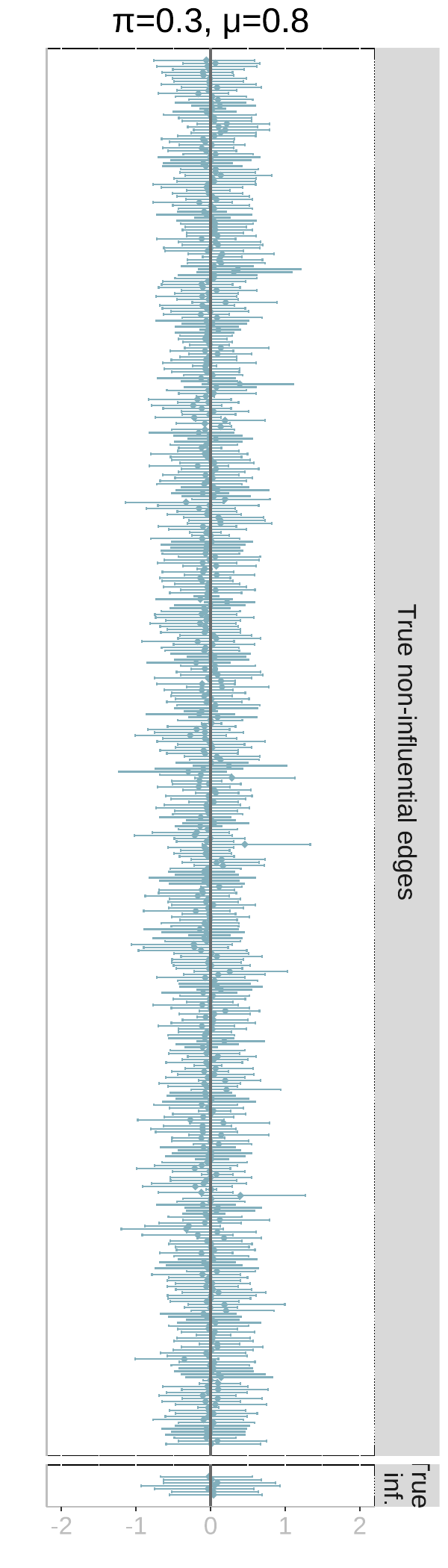}
    \includegraphics[scale=0.19]{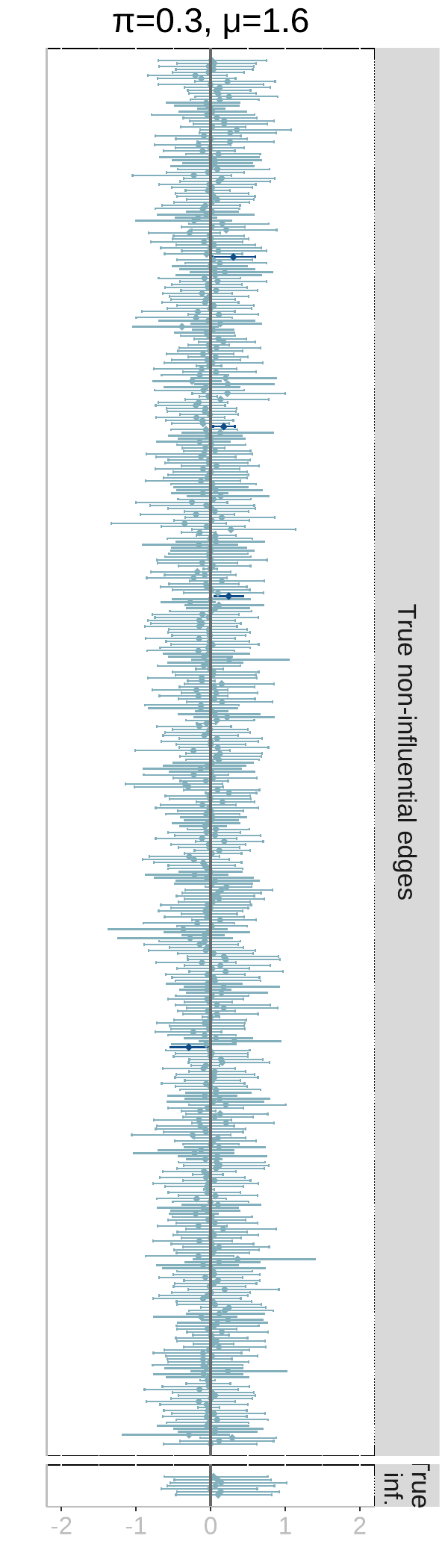}
    \includegraphics[scale=0.19]{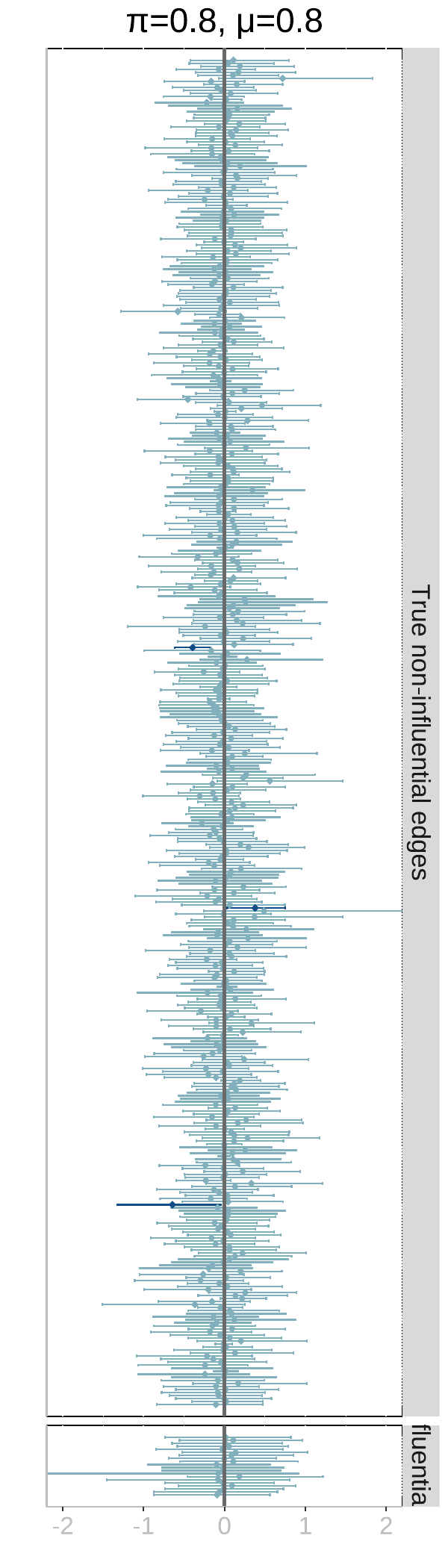}
    \includegraphics[scale=0.19]{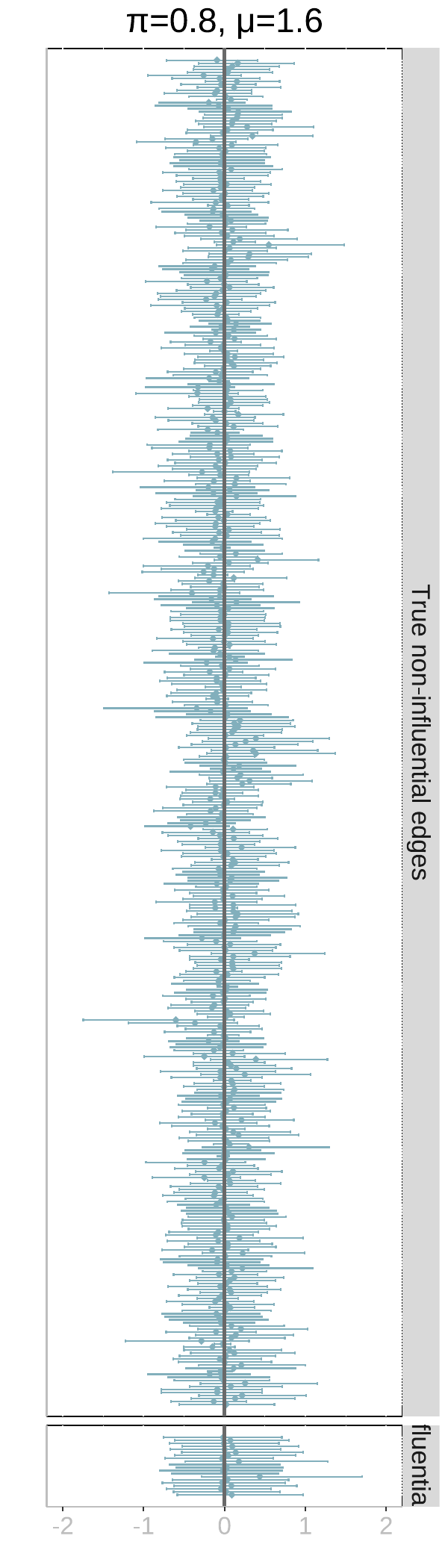} 
    \caption{$\mathbf{n=1000}$}
    \end{subfigure}
    \caption[95 \% credible intervals for edge effects (additive model with random coefficients) with $k=22$ sampled nodes]{{\bf 95 \% credible intervals for edge effects (additive model with random coefficients) with $k=22$ sampled nodes.} Top (a): Sample size of $n=500$. Bottom (b): Sample size of $n=1000$. Each panel corresponds to a scenario of $\pi=0.3, 0.8$ (which controls the sparsity of the regression coefficient matrix $\mathbf B$) and $\mu=0.8,1.6$. We plot the 95 \% credible intervals for the regression coefficients per edge. In the additive model, all edges are non-influential. The color of the intervals depends on whether it intersects zero (light) and hence estimated to be non-influential or does not intersect zero (dark) and hence estimated to be influential by the model. Given that the additive model does not have interaction (edge) effects, these panels allow us to visualize false positives (dark intervals).}
    \label{fig:edges_add_rand2}
\end{figure}

\begin{figure}[!ht]
\centering
\begin{subfigure}[b]{0.49\textwidth}
\centering
\includegraphics[scale=0.19]{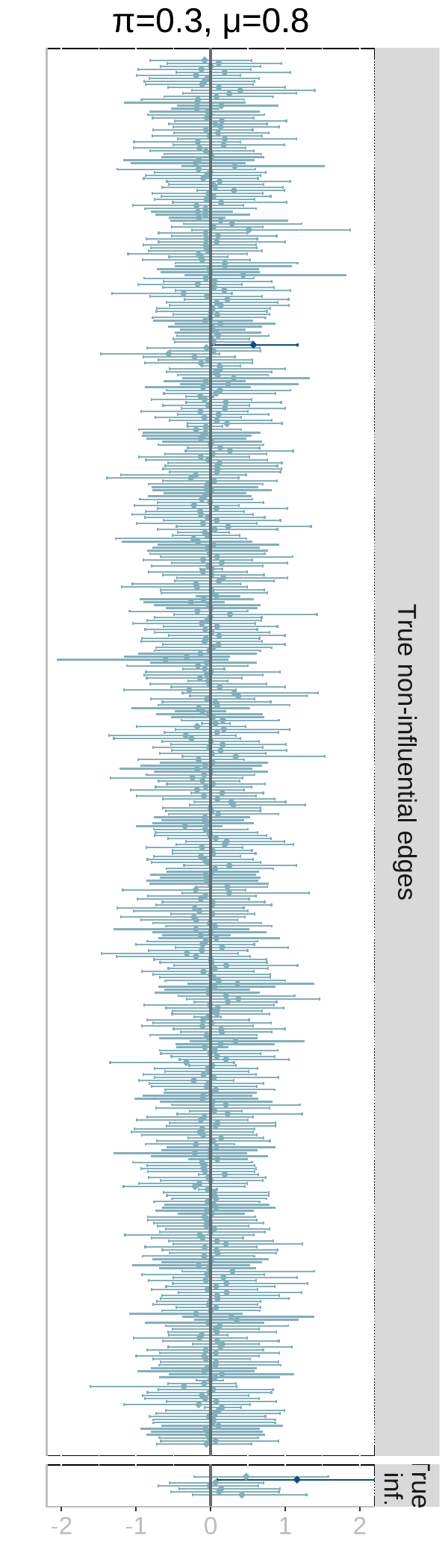}
\includegraphics[scale=0.19]{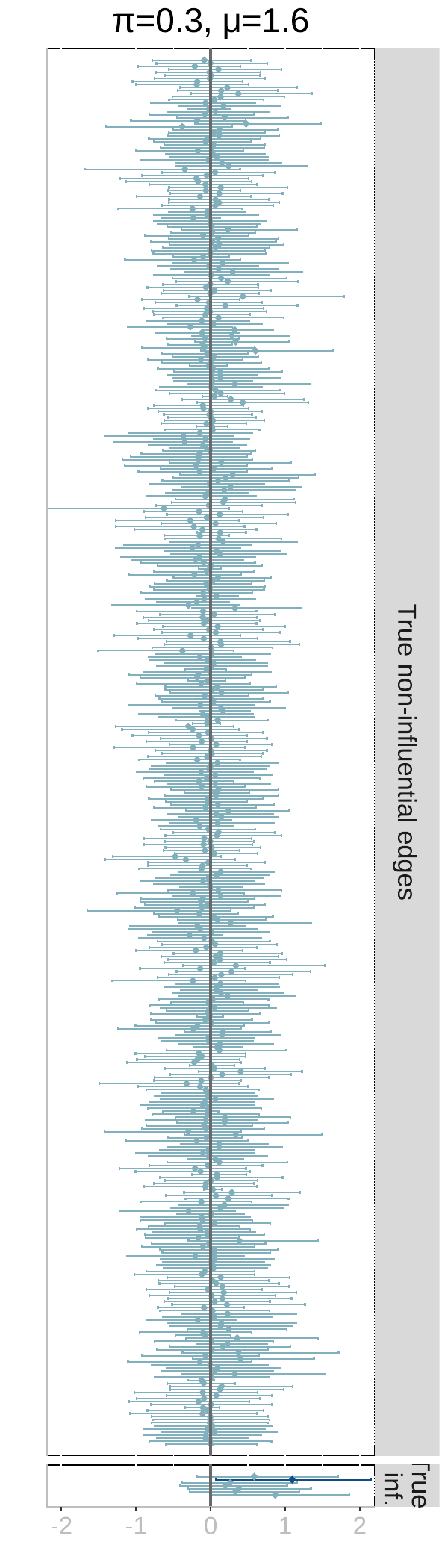}
\includegraphics[scale=0.19]{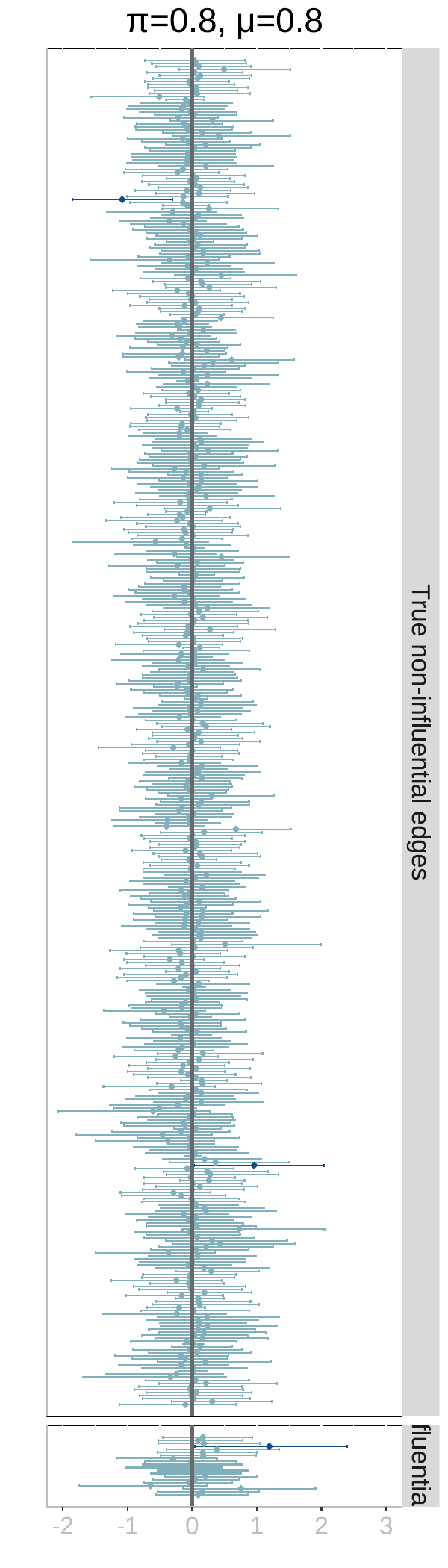}
\includegraphics[scale=0.19]{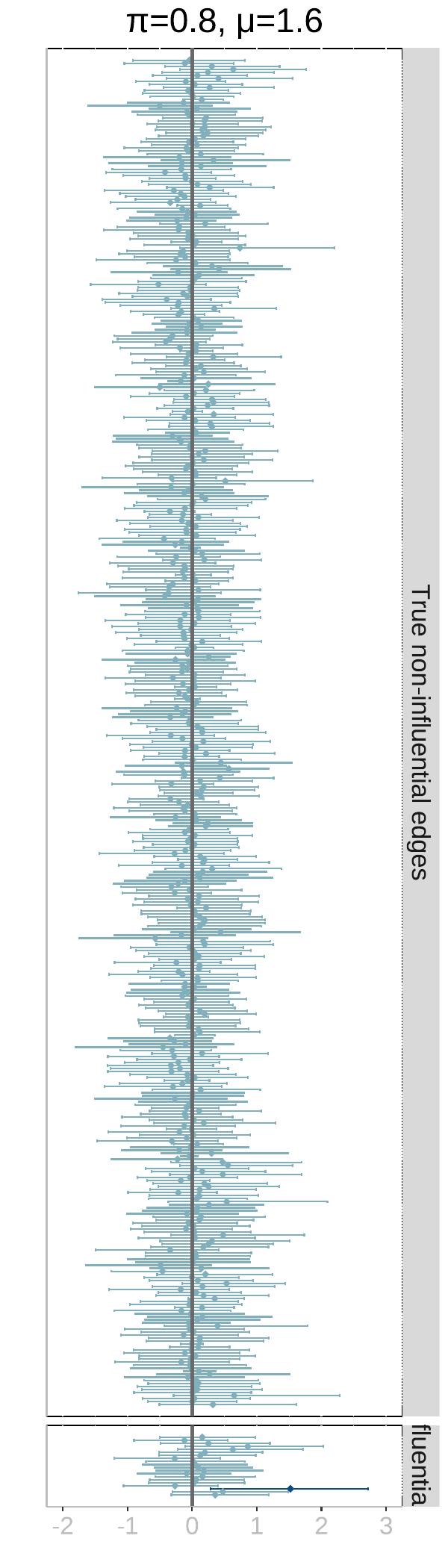}
\caption{$\mathbf{n=500}$}
\end{subfigure}
\begin{subfigure}[b]{0.49\textwidth}
\centering
\includegraphics[scale=0.19]{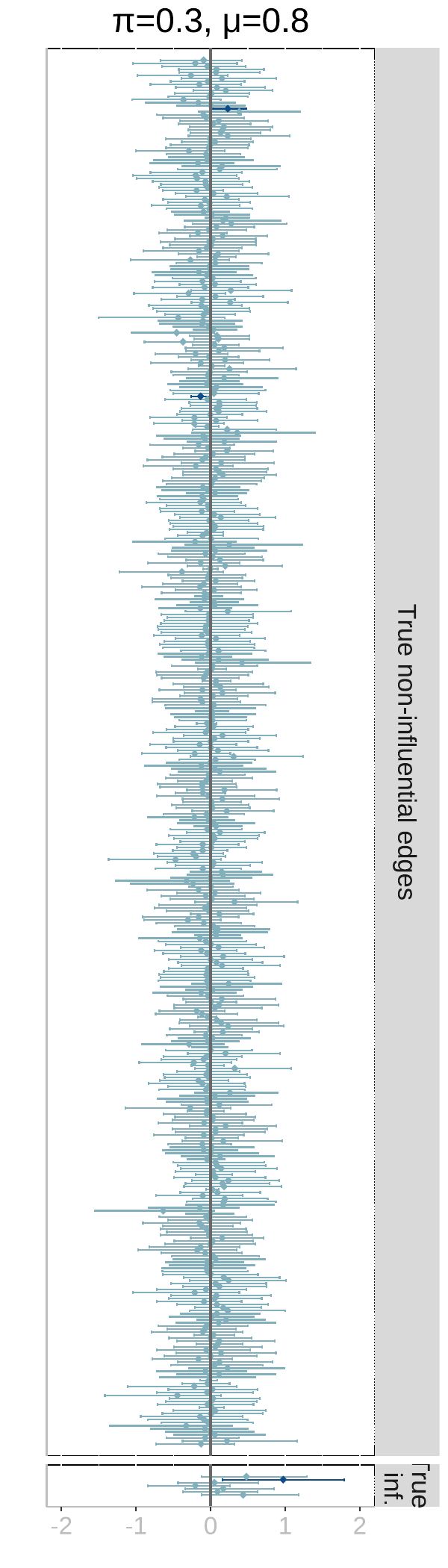}
\includegraphics[scale=0.19]{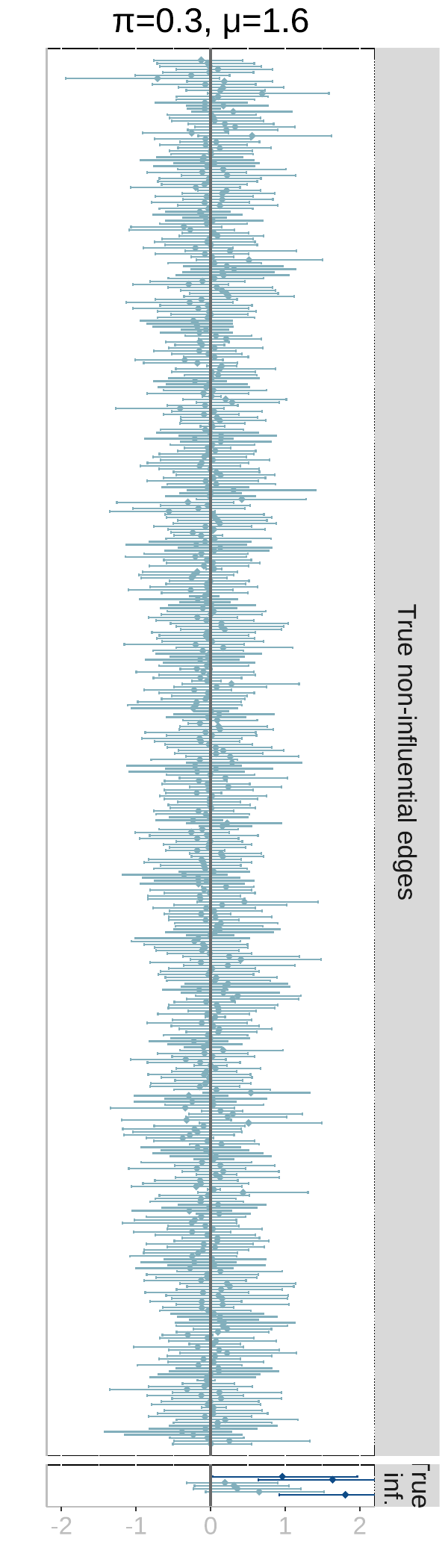}
\includegraphics[scale=0.19]{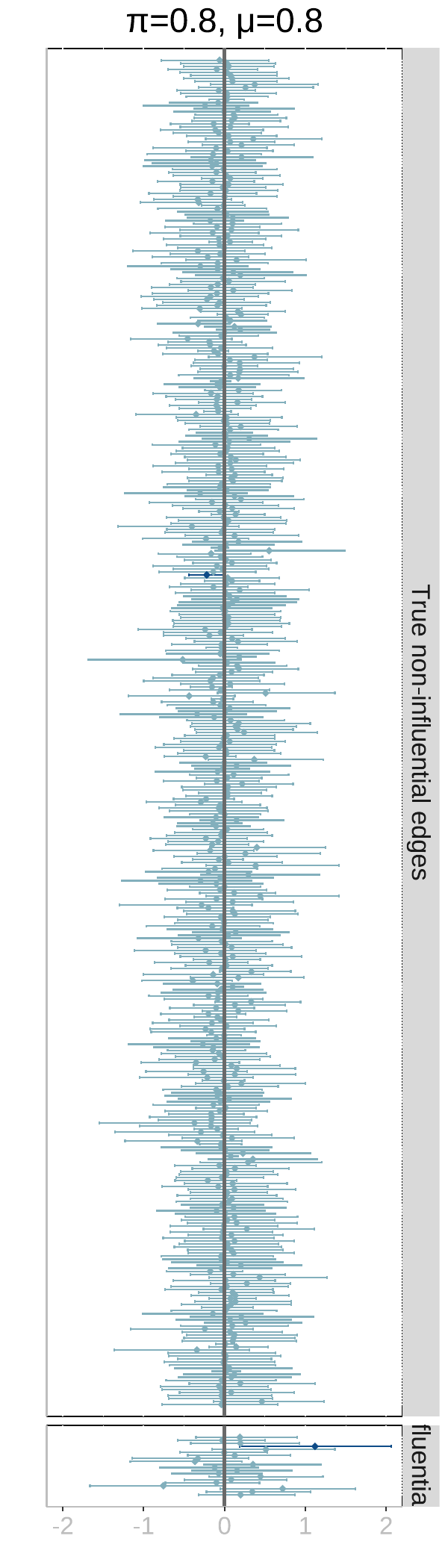}
\includegraphics[scale=0.19]{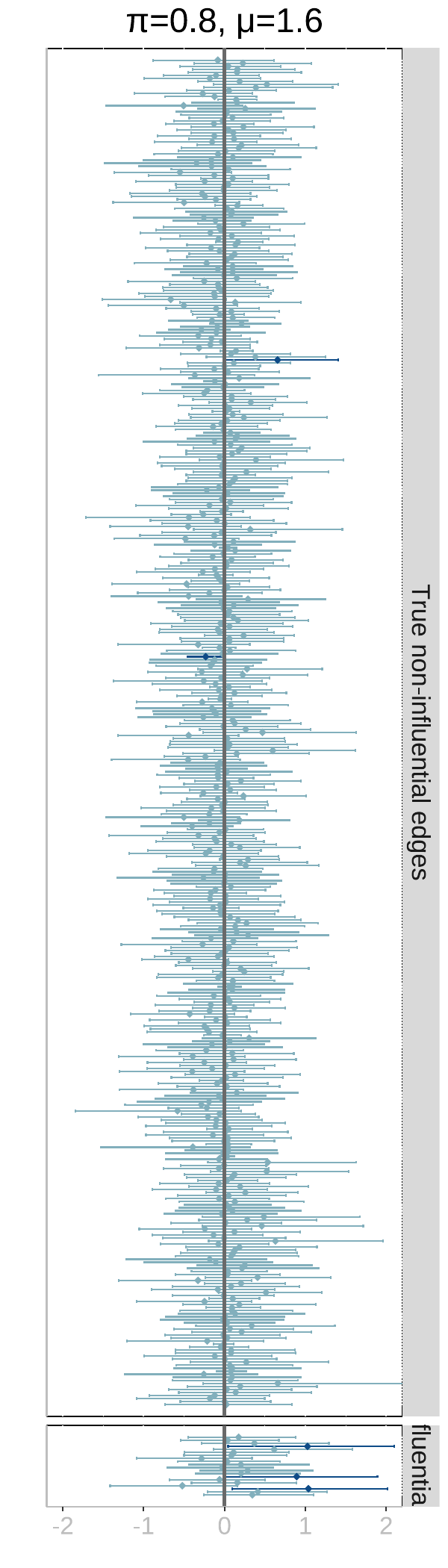} 
\caption{$\mathbf{n=1000}$}
\end{subfigure}
\caption[95 \% credible intervals for edge effects (additive model with phylogenetic coefficients) with $k=8$ sampled nodes]{{\bf 95 \% credible intervals for edge effects (additive model with phylogenetic coefficients) with $k=8$ sampled nodes.} Top (a): Sample size of $n=500$. Bottom (b): Sample size of $n=1000$. Each panel corresponds to a scenario of $\pi=0.3, 0.8$ (which controls the sparsity of the regression coefficient matrix $\mathbf B$) and $\mu=0.8,1.6$. We plot the 95 \% credible intervals  for the regression coefficients per edge. In the additive model, all edges are non-influential. The color of the intervals depends on whether it intersects zero (light) and hence estimated to be non-influential or does not intersect zero (dark) and hence estimated to be influential by the model. Given that the additive model does not have interaction (edge) effects, these panels allow us to visualize false positives (dark intervals). }
\label{fig:edges_add_phylo}
\end{figure}

\begin{figure}[!ht]
\centering
\begin{subfigure}[b]{0.49\textwidth}
\centering
\includegraphics[scale=0.19]{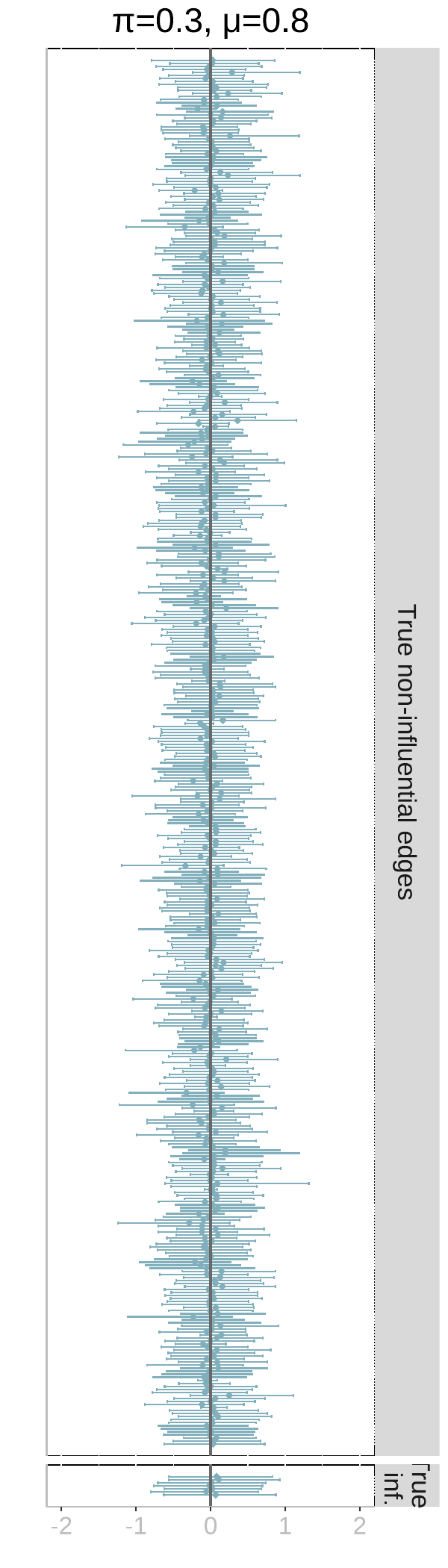}
\includegraphics[scale=0.19]{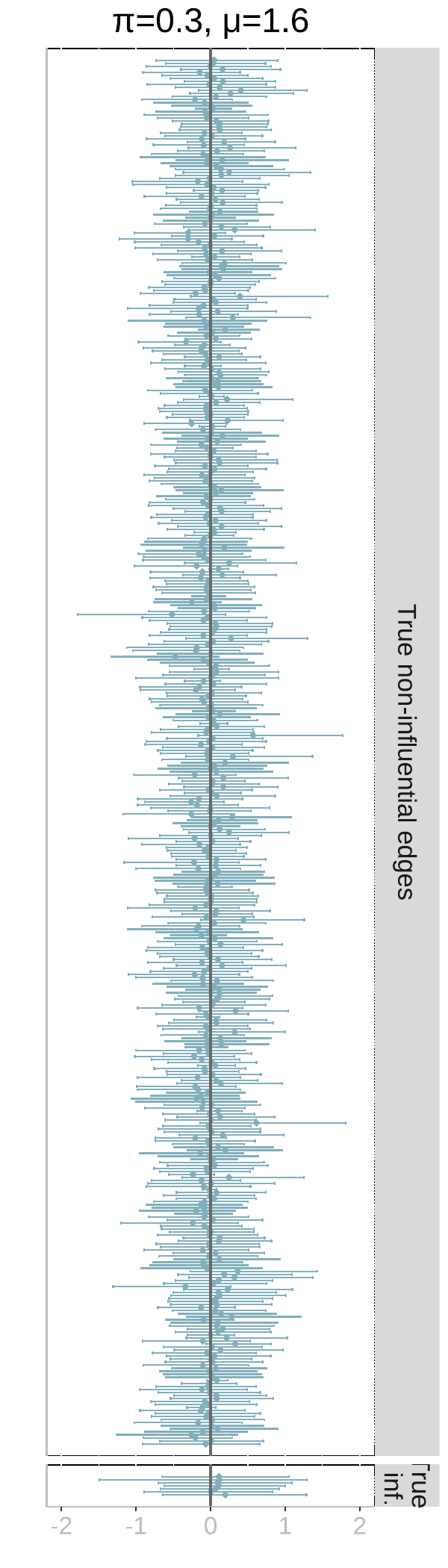}
\includegraphics[scale=0.19]{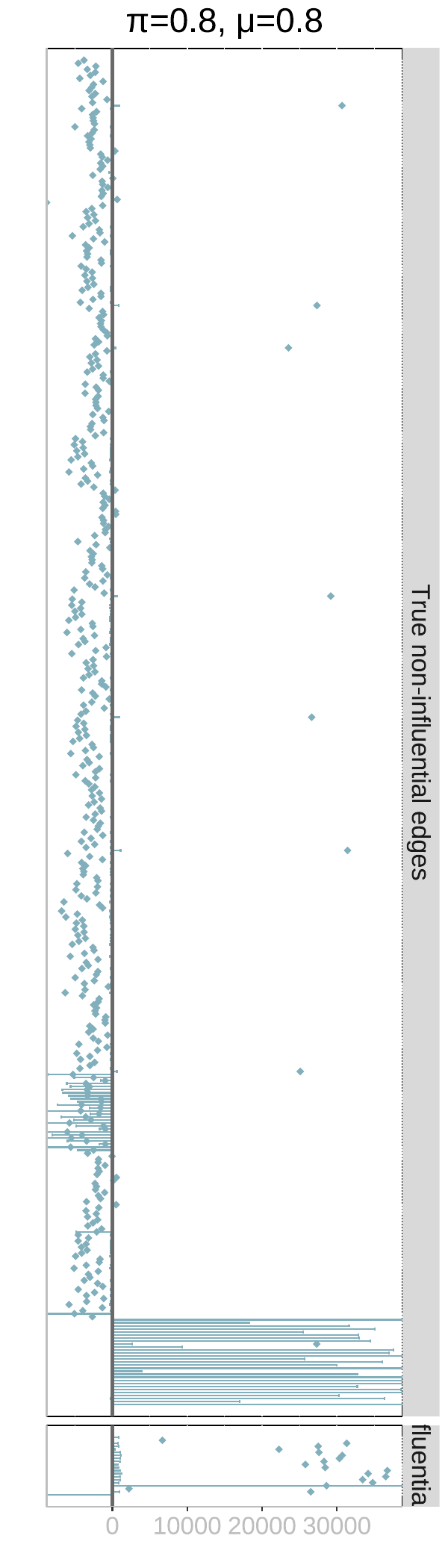}
\includegraphics[scale=0.19]{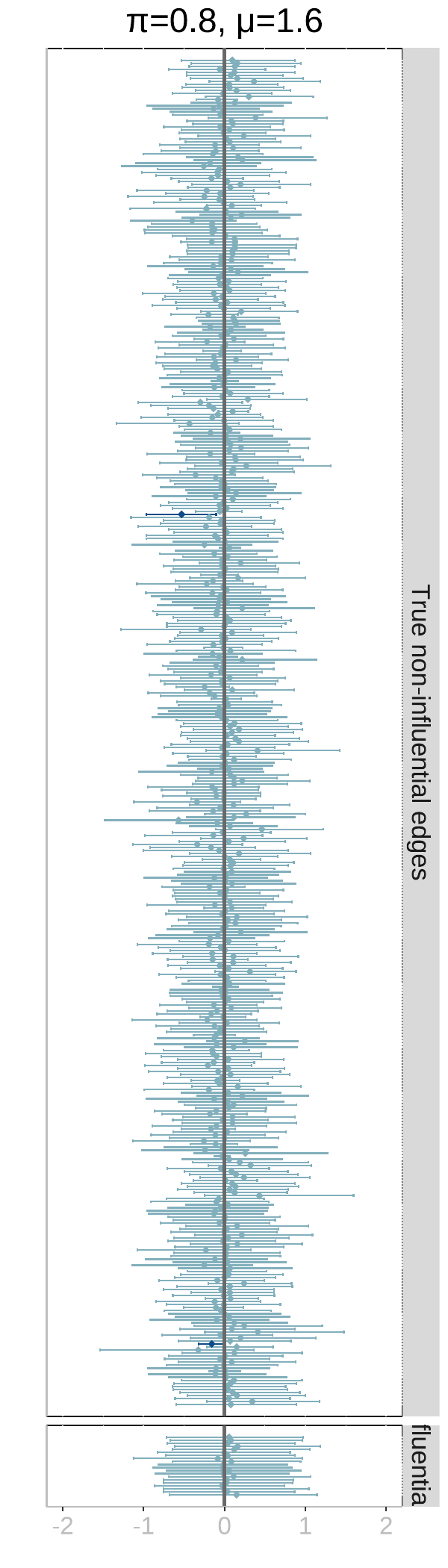}
\caption{$\mathbf{n=500}$}
\end{subfigure}
\begin{subfigure}[b]{0.49\textwidth}
\centering
\includegraphics[scale=0.19]{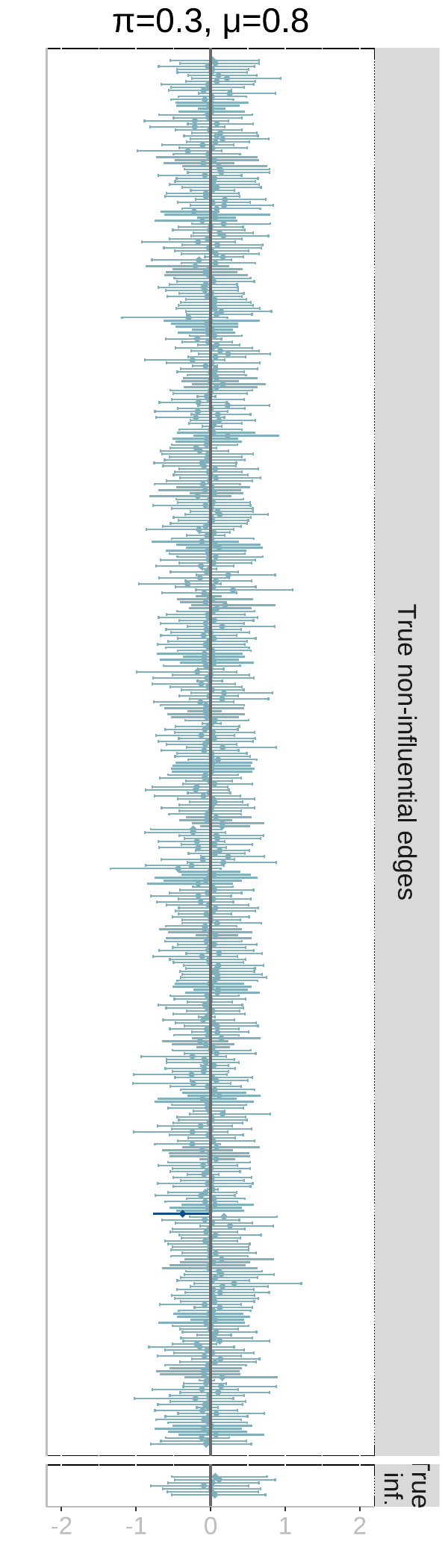}
\includegraphics[scale=0.19]{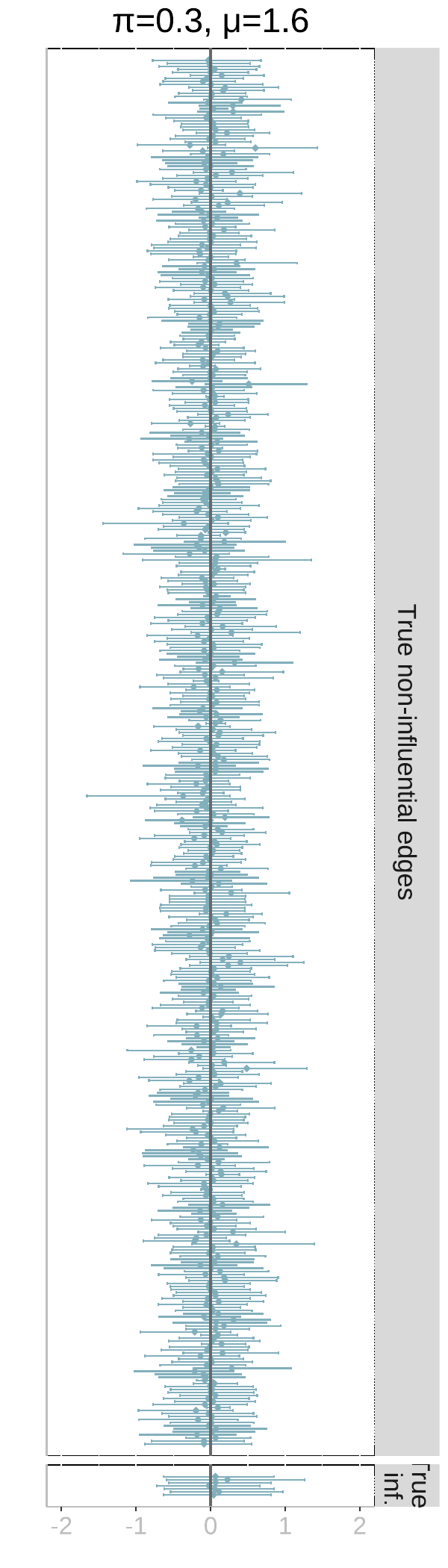}
\includegraphics[scale=0.19]{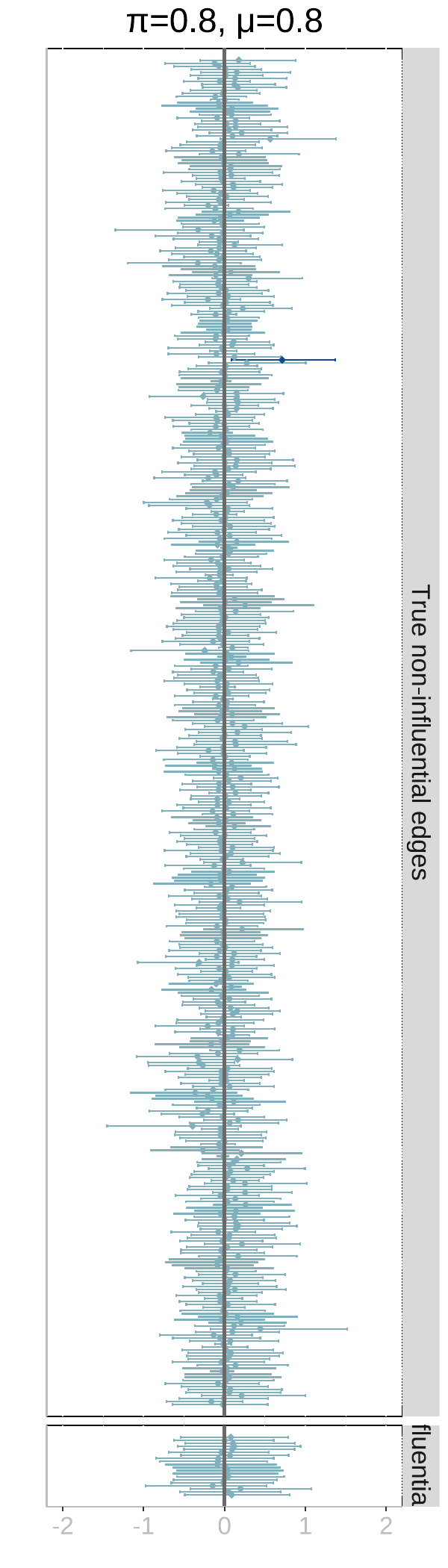}
\includegraphics[scale=0.19]{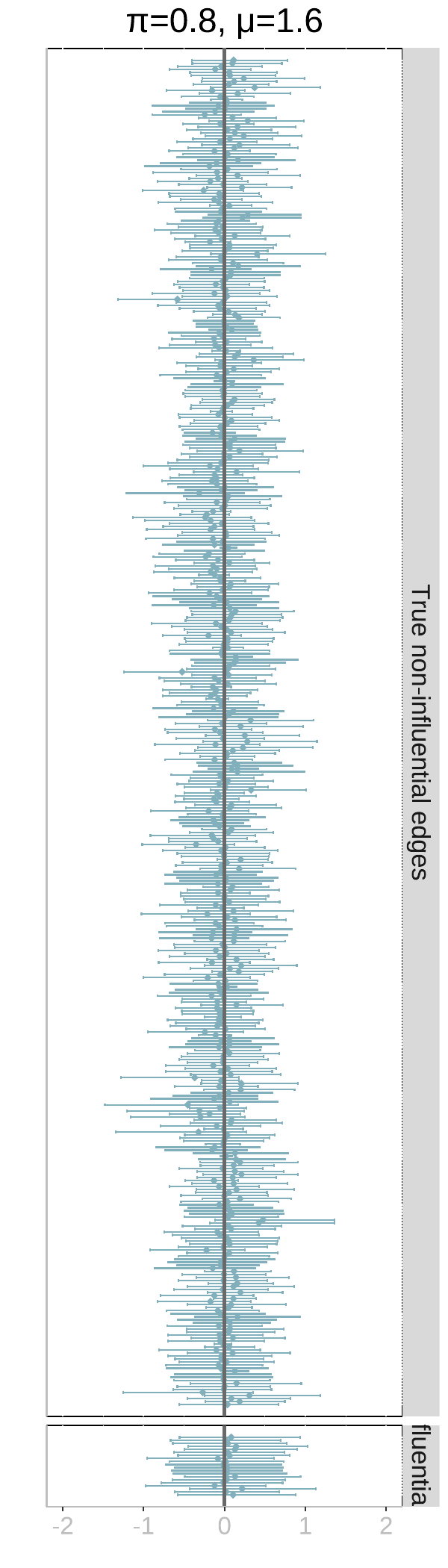} 
\caption{$\mathbf{n=1000}$}
\end{subfigure}
\caption[95 \% credible intervals for edge effects (additive model with phylogenetic coefficients) with $k=22$ sampled nodes]{{\bf 95 \% credible intervals  for edge effects (additive model with phylogenetic coefficients) with $k=22$ sampled nodes.} Top (a): Sample size of $n=500$. Bottom (b): Sample size of $n=1000$. Each panel corresponds to a scenario of $\pi=0.3, 0.8$ (which controls the sparsity of the regression coefficient matrix $\mathbf B$) and $\mu=0.8,1.6$. We plot the 95 \% credible intervals  for the regression coefficients per edge. In the additive model, all edges are non-influential. The color of the intervals depends on whether it intersects zero (light) and hence estimated to be non-influential or does not intersect zero (dark) and hence estimated to be influential by the model. Given that the additive model does not have interaction (edge) effects, these panels allow us to visualize false positives (dark intervals).}
\label{fig:edges_add_phylo2}
\end{figure}

\begin{figure}[!ht]
    \centering
    \begin{subfigure}[t]{0.49\textwidth}
        \includegraphics[scale=0.27]{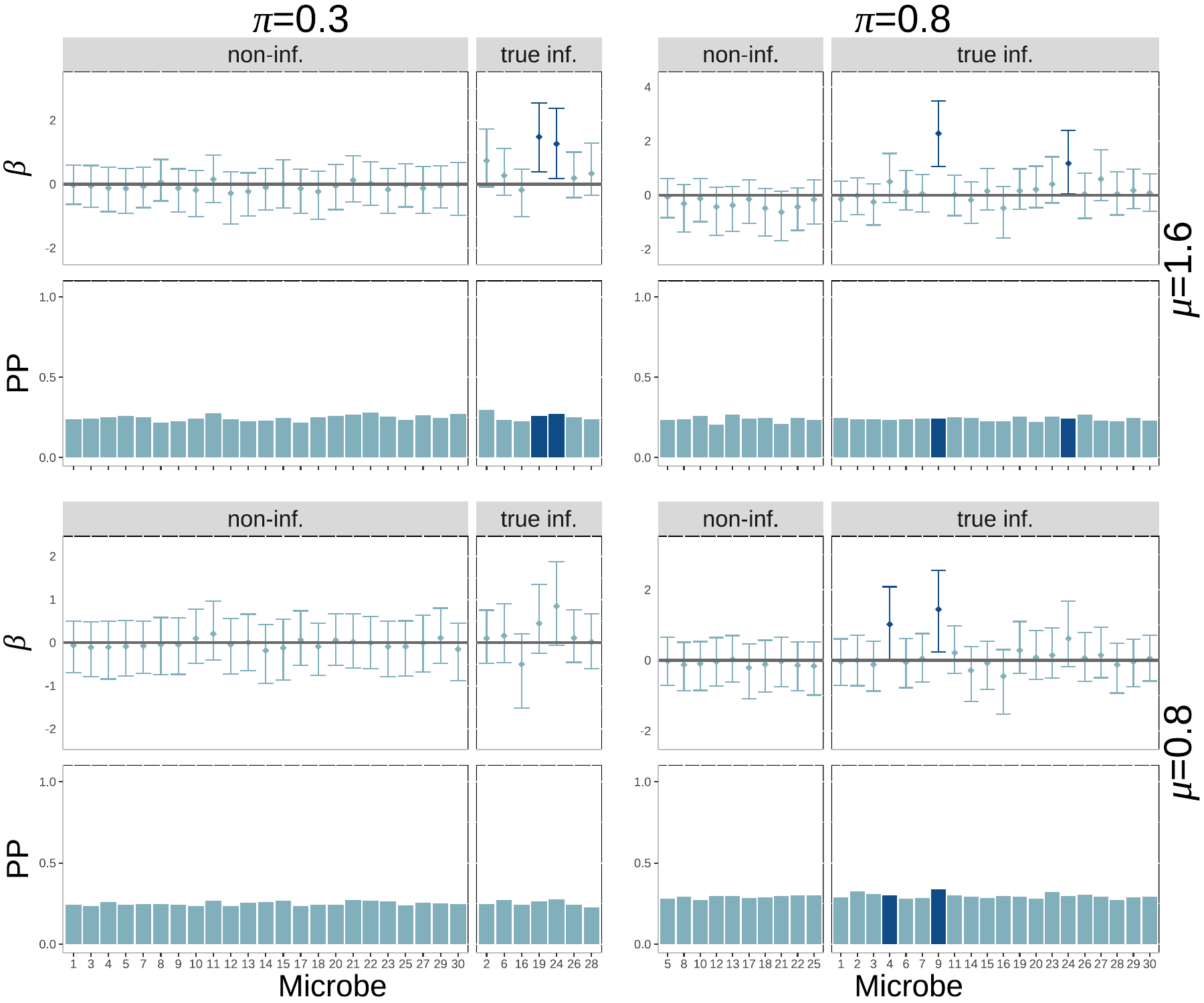}
        \caption{$\mathbf{n=500}$, $\mathbf{k=8}$}
    \end{subfigure}
    \begin{subfigure}[t]{0.49\textwidth}
        \includegraphics[scale=0.27]{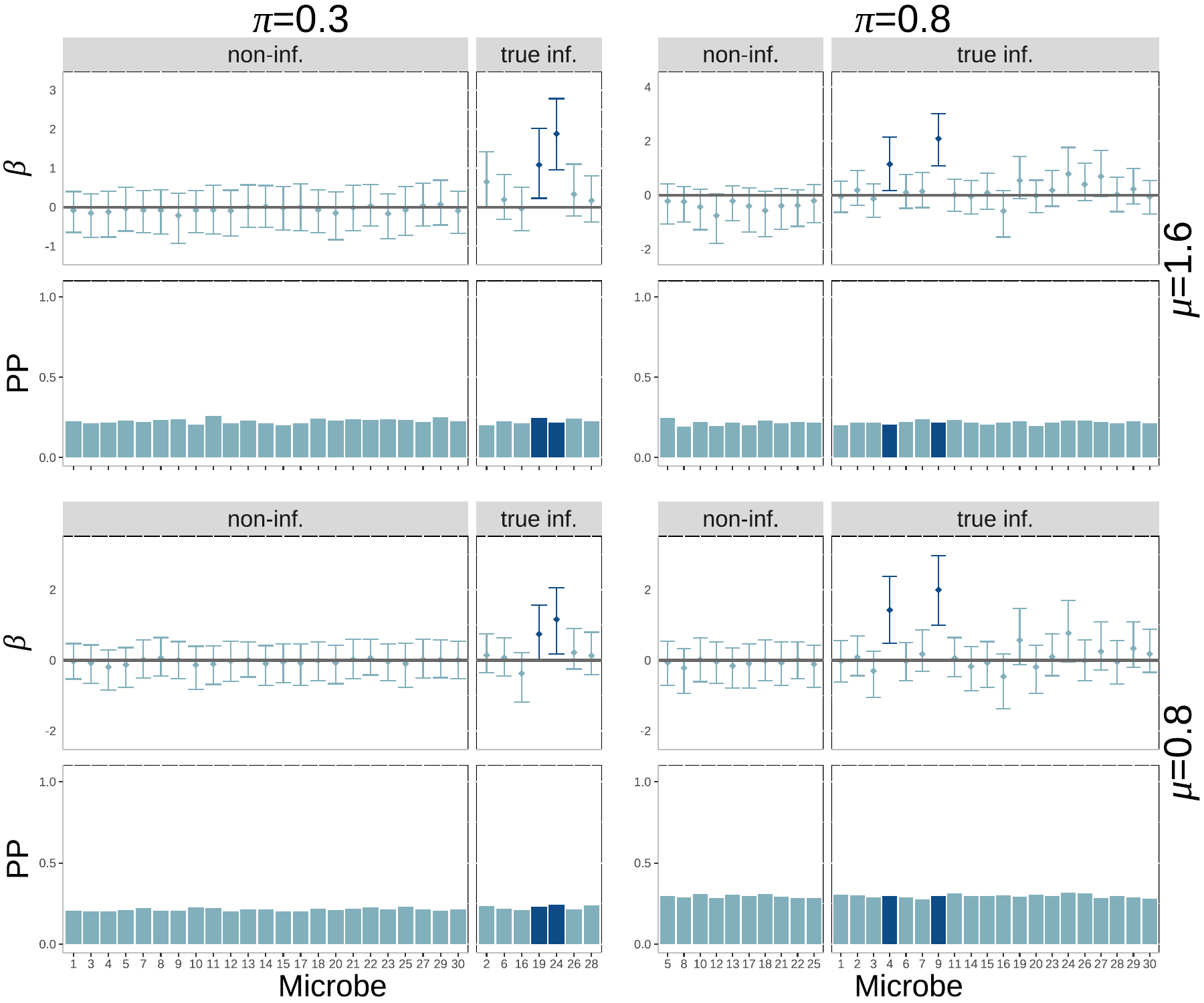} 
        \caption{$\mathbf{n=1000}$, $\mathbf{k=8}$}
    \end{subfigure}\\
    \begin{subfigure}[t]{0.49\textwidth}
        \includegraphics[scale=0.27]{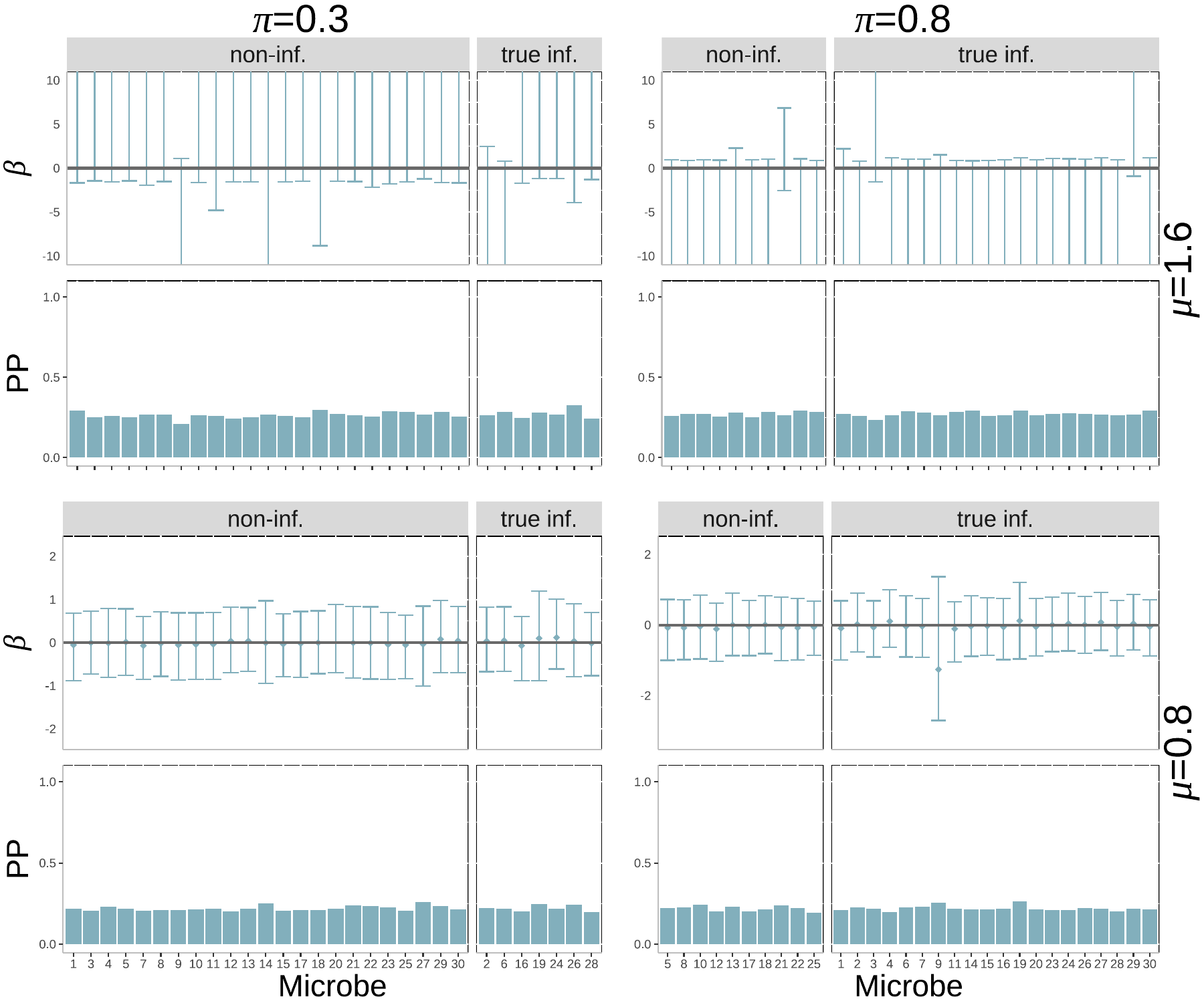}
        \caption{$\mathbf{n=500}$, $\mathbf{k=22}$}
    \end{subfigure}
    \begin{subfigure}[t]{0.49\textwidth}
        \includegraphics[scale=0.27]{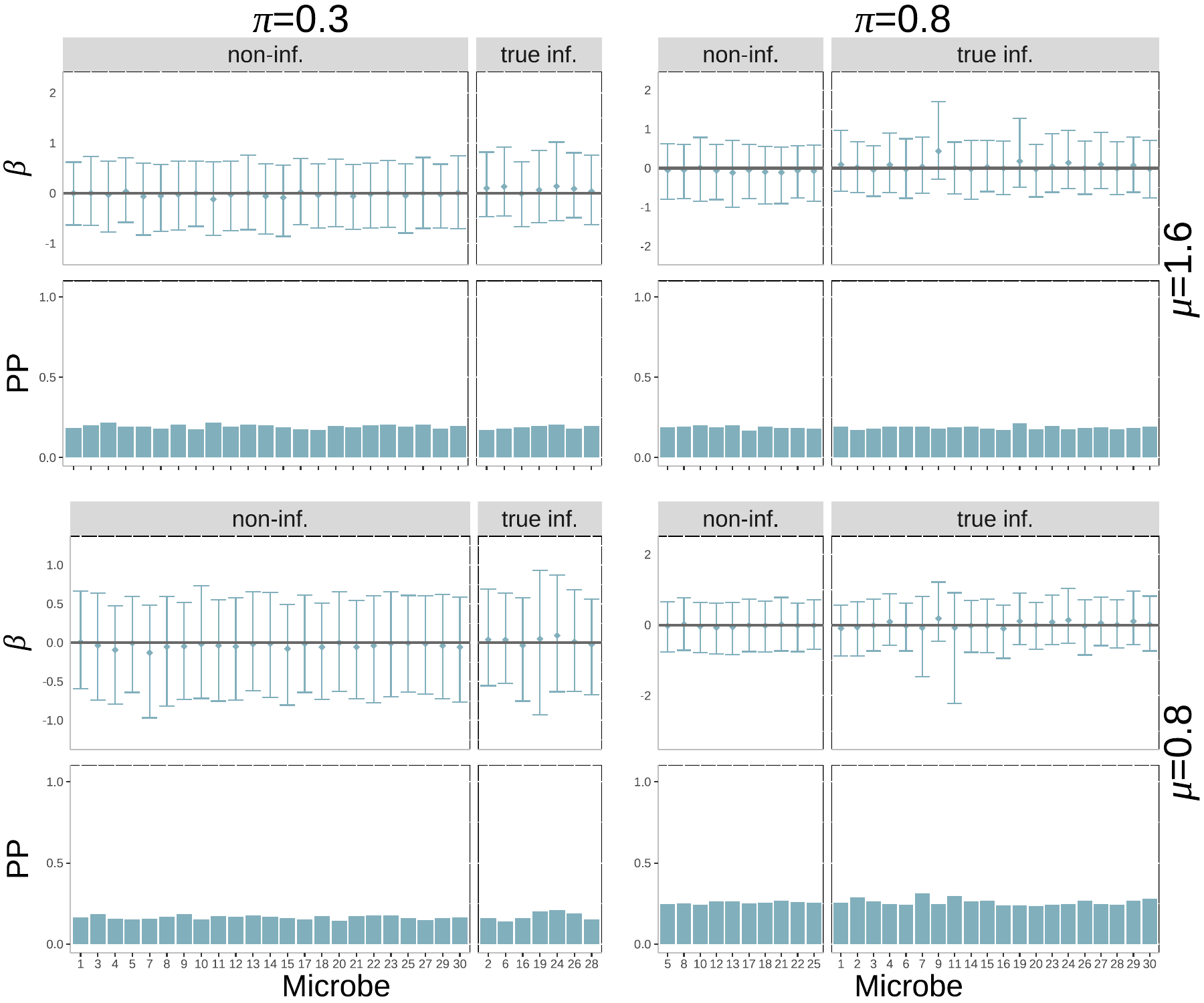}
        \caption{$\mathbf{n=1000}$, $\mathbf{k=22}$}
    \end{subfigure}
    \caption[Posterior probability of influential nodes and coefficients for nodes (additive model with random coefficients)]{{\bf Posterior probability of influential nodes and coefficients for nodes (additive model with random coefficients).}
    Different \revision{groups of four} panels represent different number of sampled microbes ($k=8,22$) which controls the sparsity of the adjacency matrix and different sample sizes ($n=500,1000$). \revision{Within each group, we have four panels corresponding to the two values of edge effect size ($\mu=0.8, 1.6$) and two values of probability of influential node ($\pi=0.3, 0.8$) which controls the sparsity of the regression coefficient matrix ($\mathbf B$). Within each of these panels we have two plots: 95\% credible intervals (top) and posterior probability of influence (bottom - calculated as the mean of the $\xi$ variable for the node across Gibbs samples) for each node.} Each bar corresponds to one node (microbe). \revision{Within each plot the bars and intervals are colored depending on whether the node is found to be influential (dark) or not influential (light) based on the 95\% credible intervals. Each plot is split based on whether the nodes are truly influential (right) or not (left).}
    } 
    \label{fig:nodes_add_rand}
\end{figure}

\begin{figure}[!ht]
    \centering
    \begin{subfigure}[t]{0.49\textwidth}
        \centering
        \includegraphics[scale=0.27]{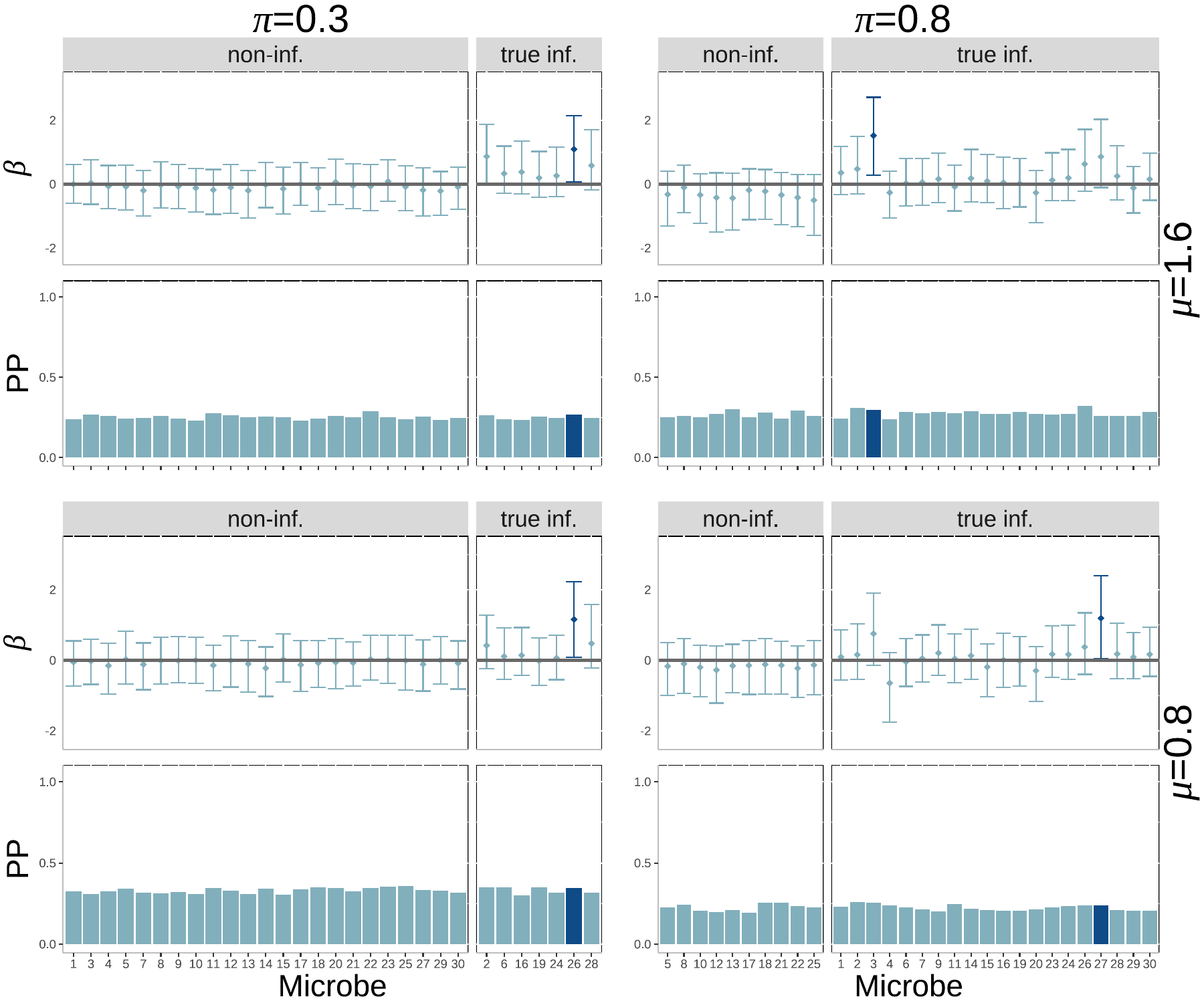}
        \caption{$\mathbf{n=500}$, $\mathbf{k=8}$}
    \end{subfigure}
    \begin{subfigure}[t]{0.49\textwidth}
        \includegraphics[scale=0.27]{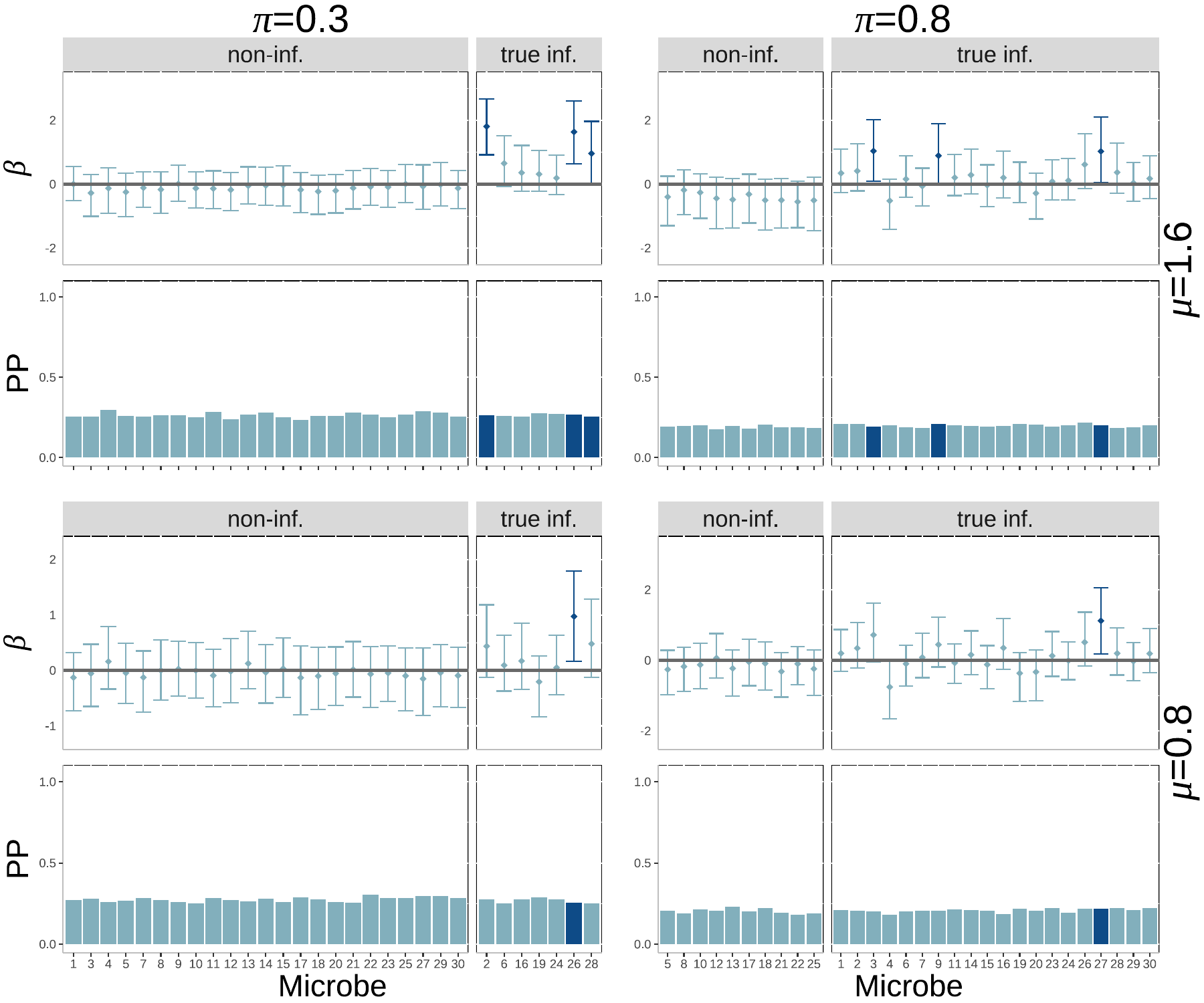}
        \caption{$\mathbf{n=1000}$, $\mathbf{k=8}$}
    \end{subfigure}\\
    \begin{subfigure}[t]{0.49\textwidth}
        \includegraphics[scale=0.27]{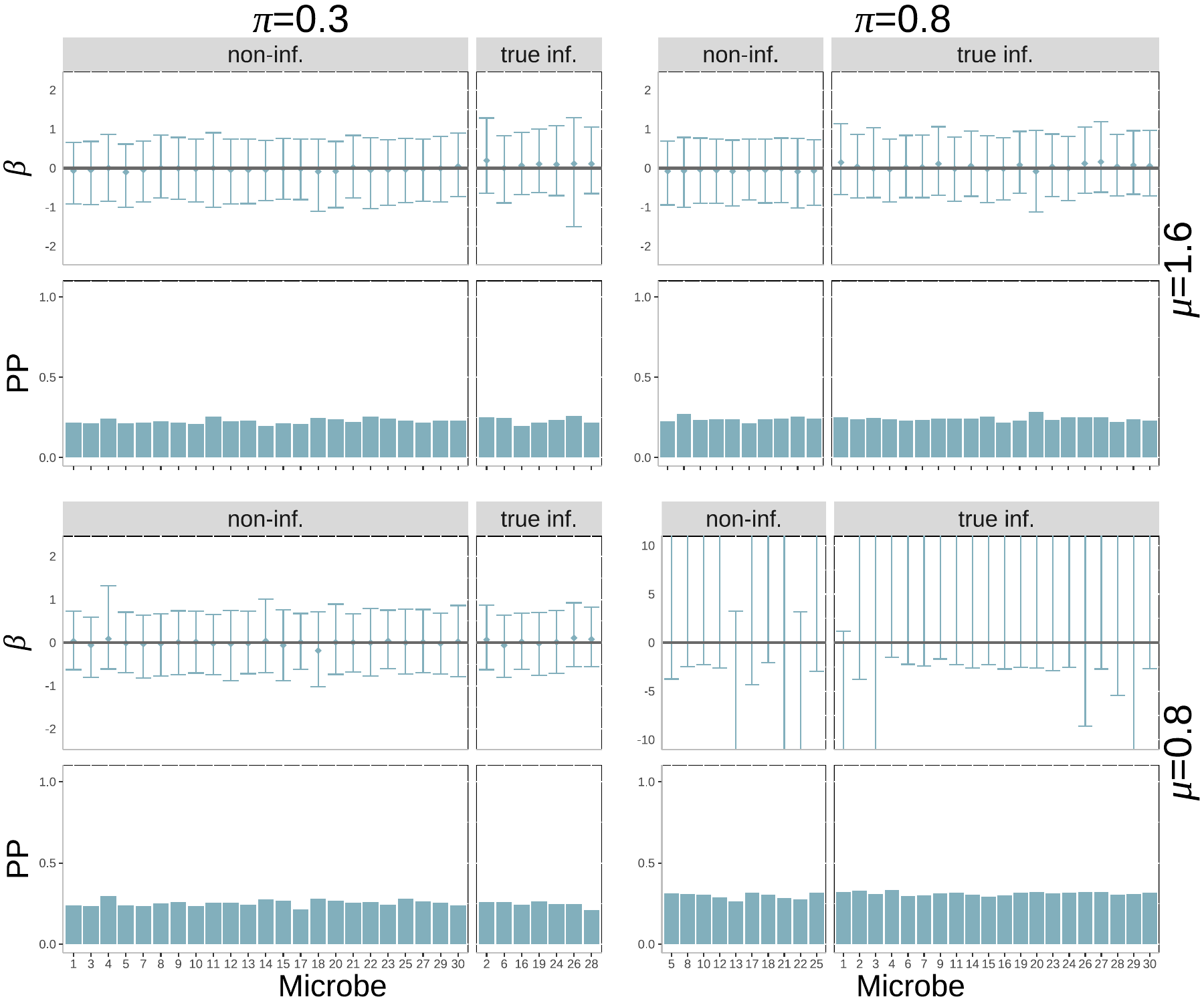}
        \caption{$\mathbf{n=500}$, $\mathbf{k=22}$}
    \end{subfigure}
    \begin{subfigure}[t]{0.49\textwidth}
        \includegraphics[scale=0.27]{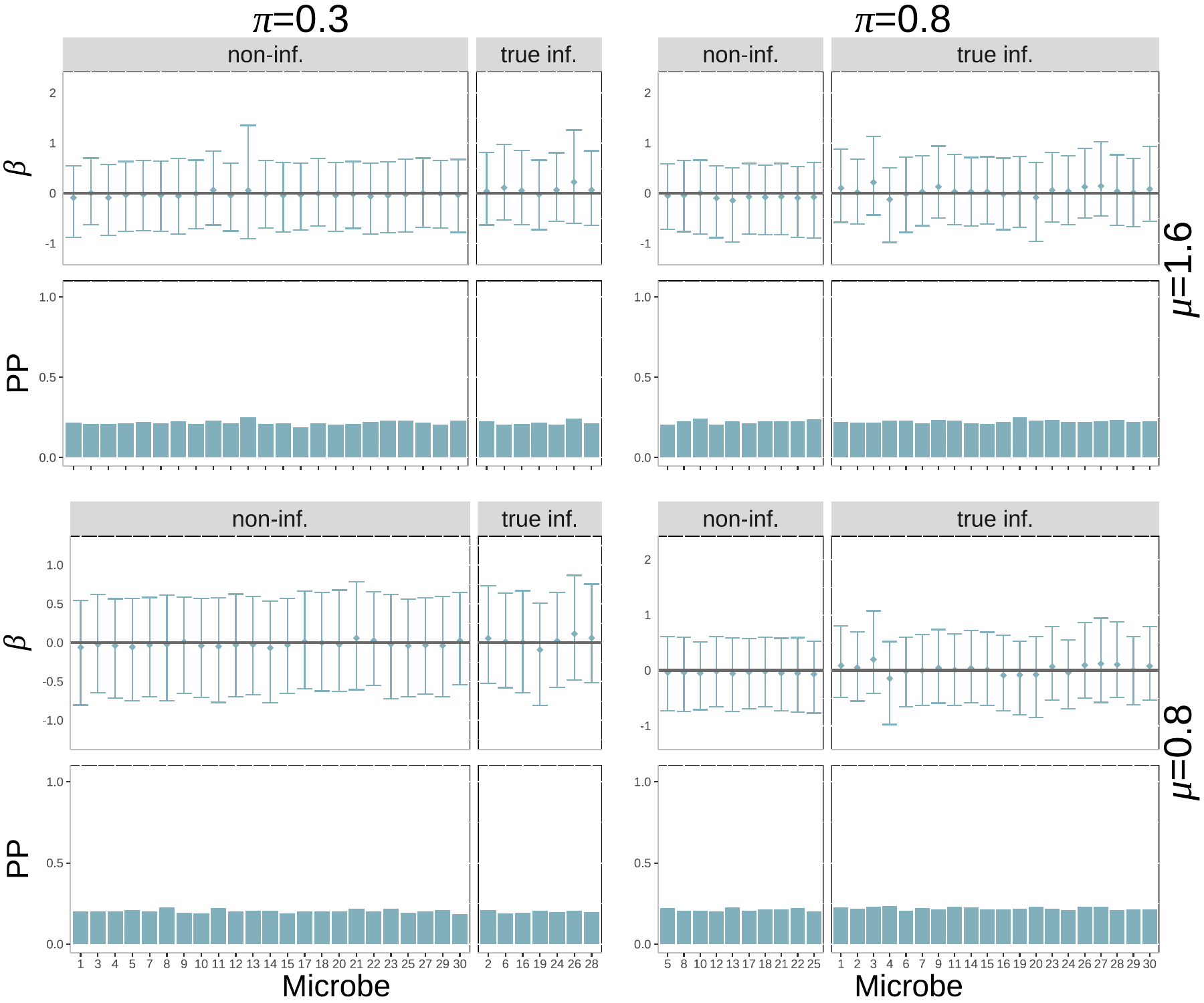}
        \caption{$\mathbf{n=1000}$, $\mathbf{k=22}$}
    \end{subfigure}
    \caption[Posterior probability of influential nodes and coefficients for nodes (additive model with phylogenetic coefficients)]{{\bf Posterior probability of influential nodes and coefficients for nodes (additive model with phylogenetic coefficients).}
    Different groups of four panels represent different number of sampled microbes ($k=8,22$) which controls the sparsity of the adjacency matrix and different sample sizes ($n=500,1000$). \revision{Within each group, we have four panels corresponding to the two values of edge effect size ($\mu=0.8, 1.6$) and two values of probability of influential node ($\pi=0.3, 0.8$) which controls the sparsity of the regression coefficient matrix ($\mathbf B$). Within each of these panels we have two plots: 95\% credible intervals (top) and posterior probability of influence (bottom - calculated as the mean of the $\xi$ variable for the node across Gibbs samples) for each node.} Each bar corresponds to one node (microbe). \revision{Within each plot the bars and intervals are colored depending on whether the node is found to be influential (dark) or not influential (light) based on the 95\% credible intervals. Each plot is split based on whether the nodes are truly influential (right) or not (left).}
    } 
    \label{fig:nodes_add_phylo}
    \end{figure}

\FloatBarrier
\subsubsection{Realistic case: Interaction model}
\FloatBarrier

\begin{figure}[!ht]
\centering
\begin{subfigure}[b]{0.49\textwidth}
\centering
\includegraphics[scale=0.19]{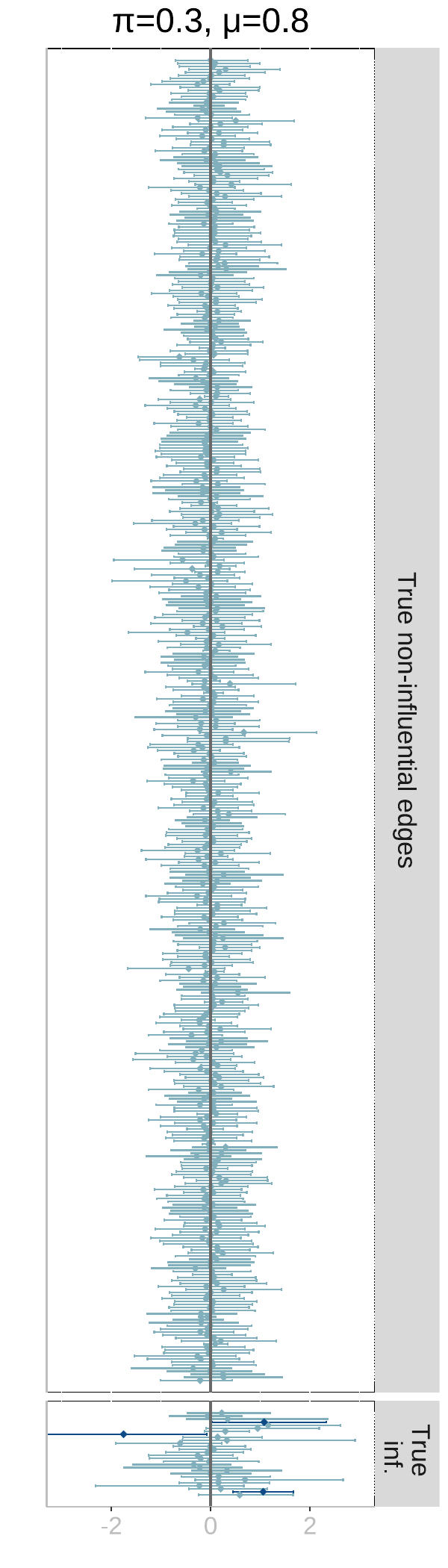}
\includegraphics[scale=0.19]{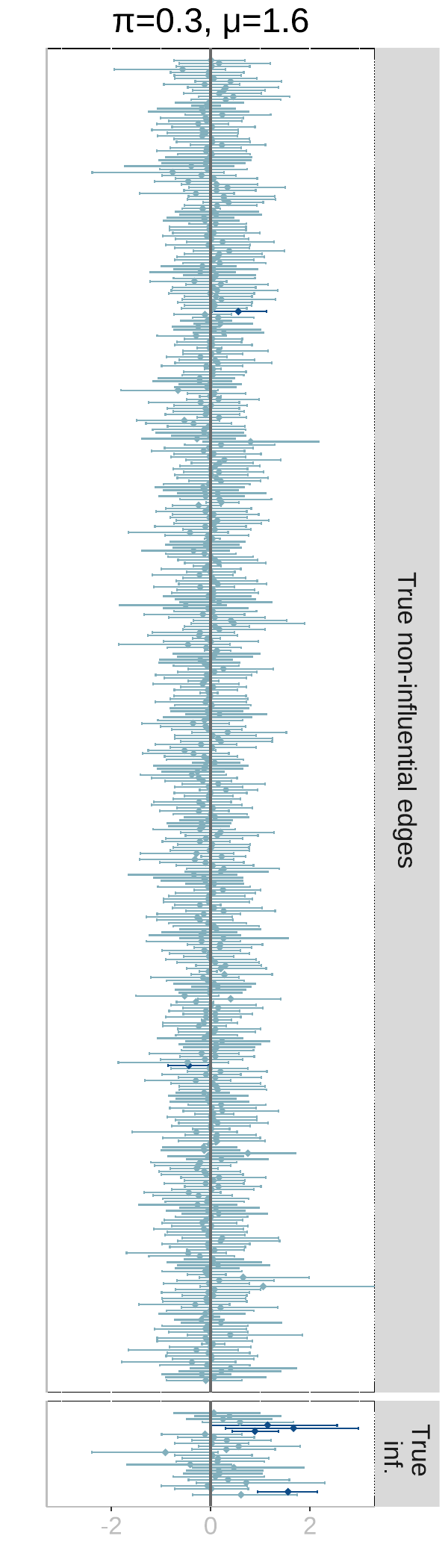}
\includegraphics[scale=0.19]{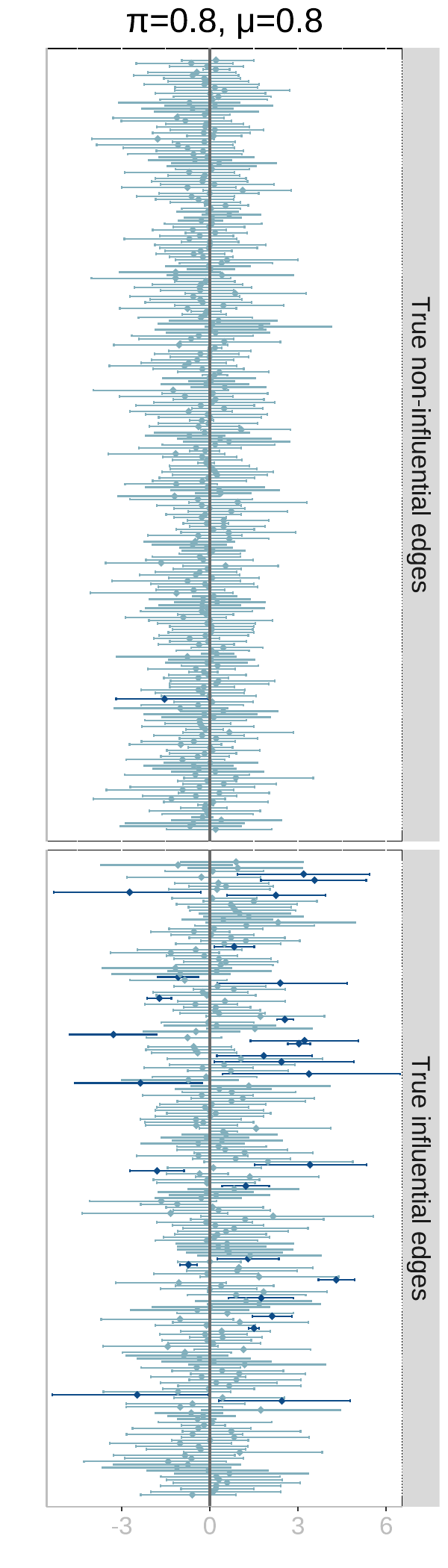}
\includegraphics[scale=0.19]{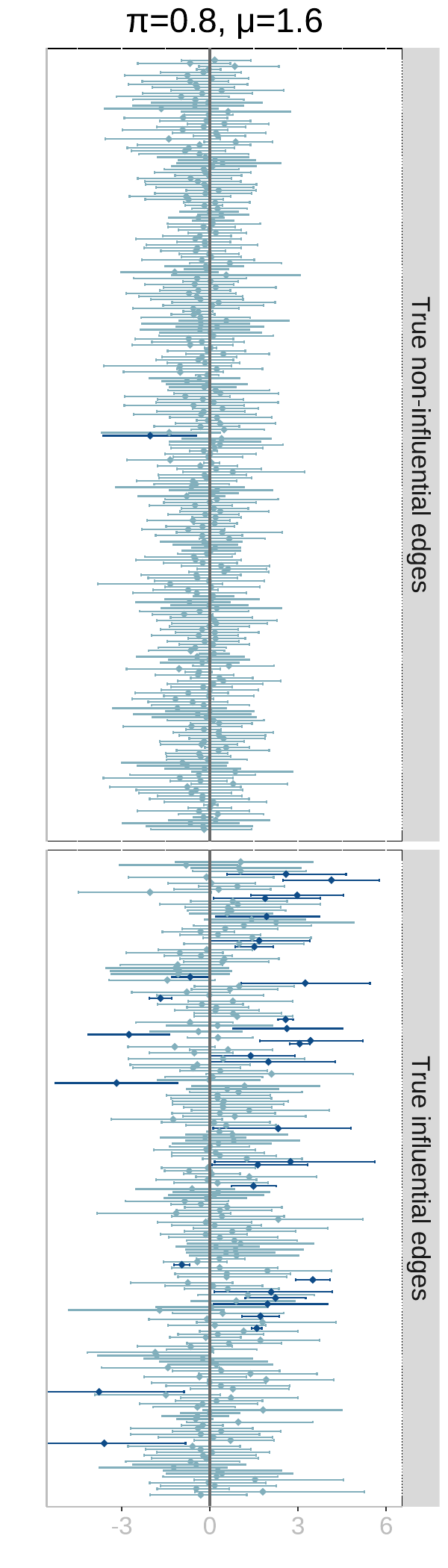}
\caption{$\mathbf{n=500}$}
\end{subfigure}
\begin{subfigure}[b]{0.49\textwidth}
\centering
\includegraphics[scale=0.19]{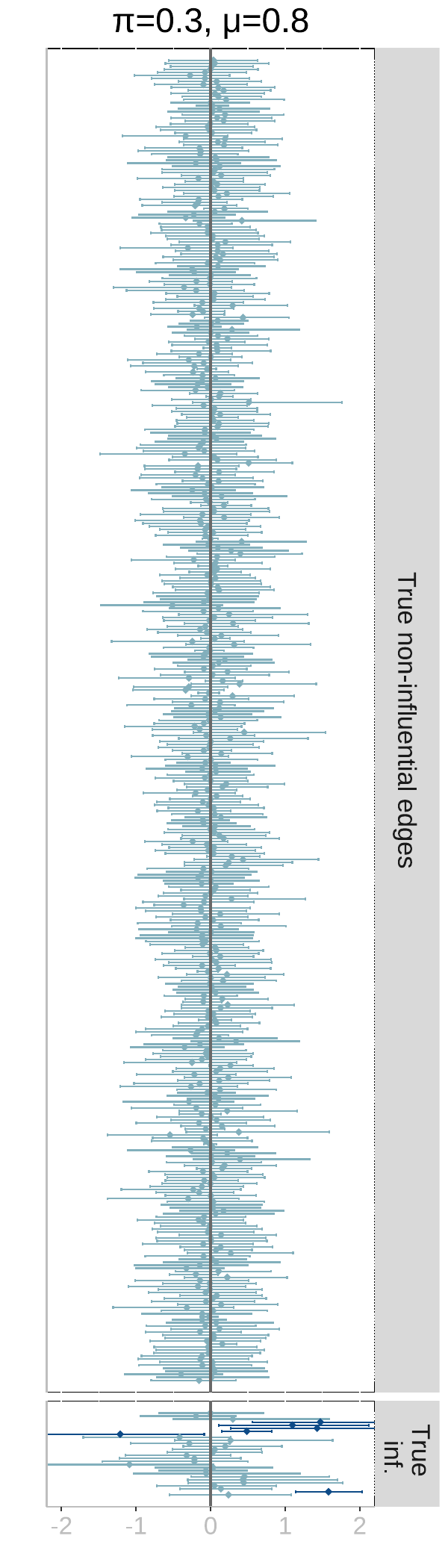}
\includegraphics[scale=0.19]{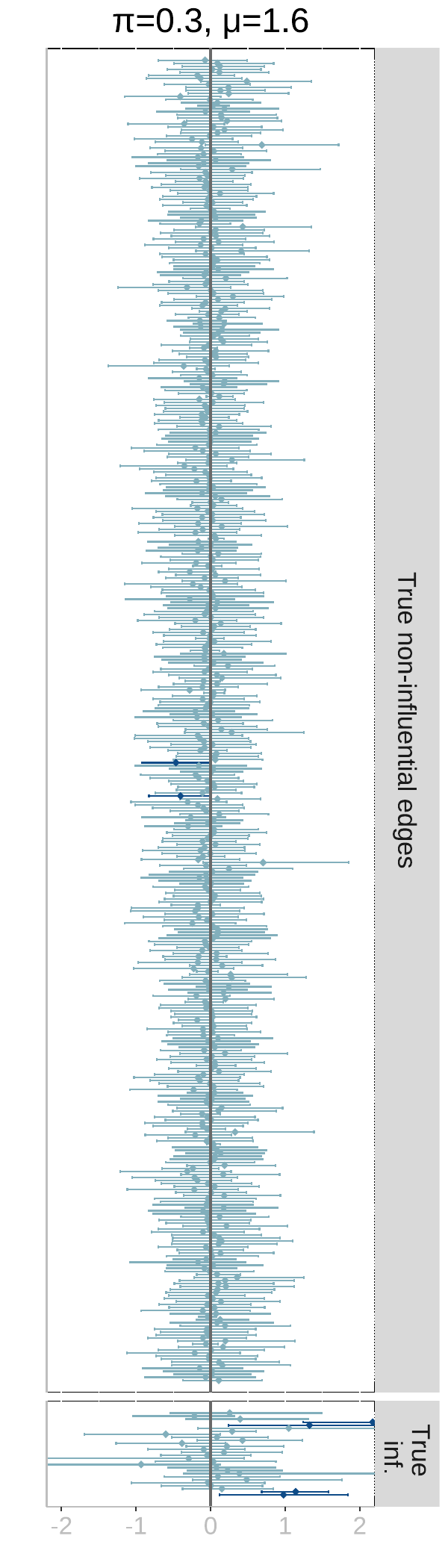}
\includegraphics[scale=0.19]{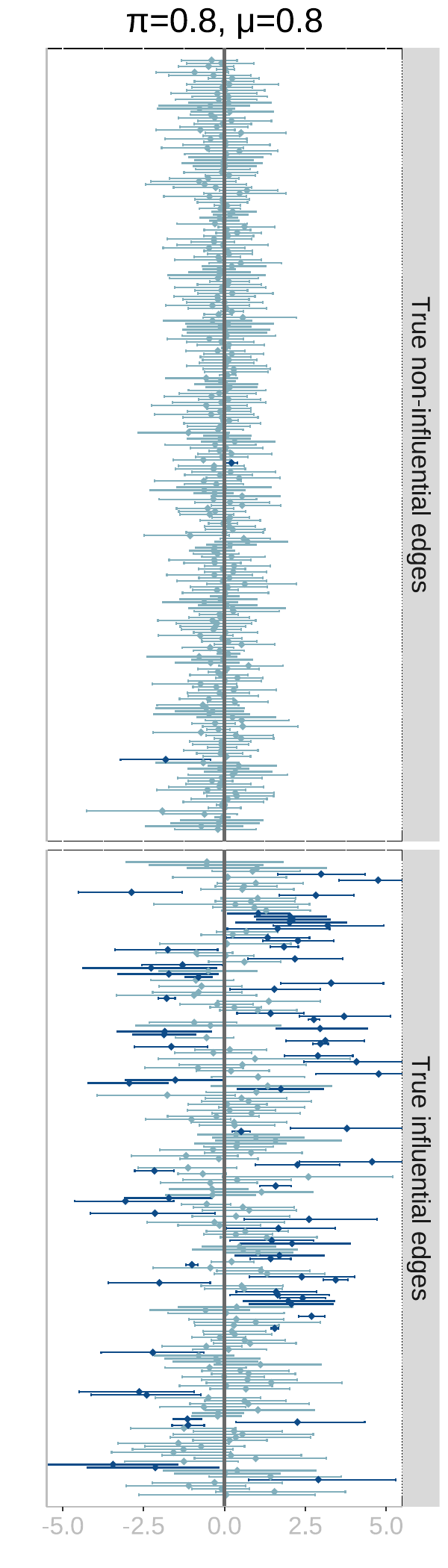}
\includegraphics[scale=0.19]{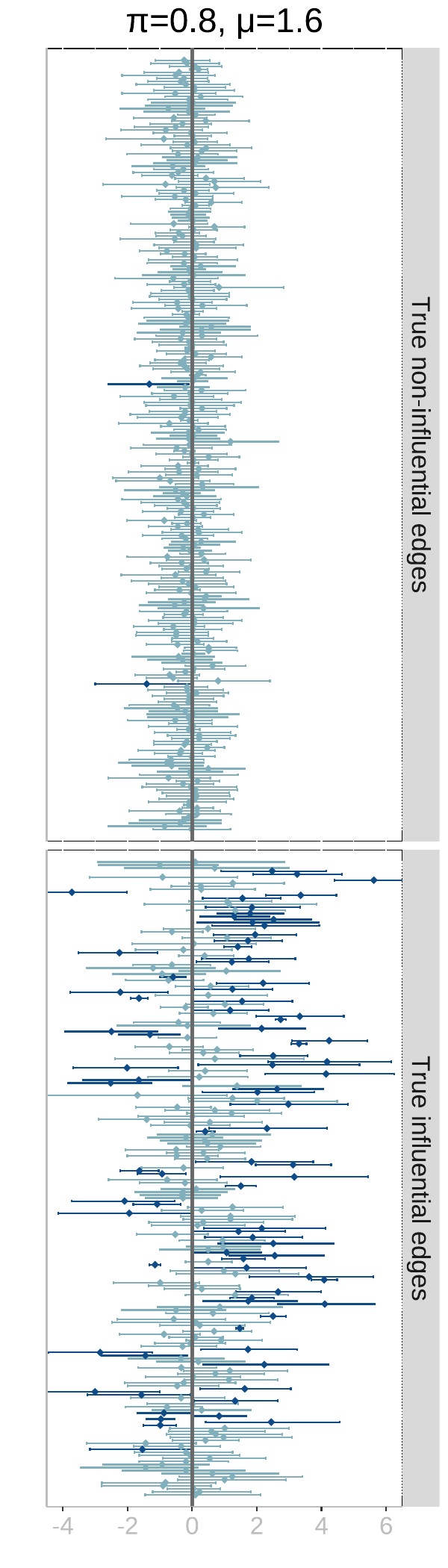} 
\caption{$\mathbf{n=1000}$}
\end{subfigure}
\caption[95 \% credible intervals for edge effects (interaction model with random coefficients) with $k=8$ sampled nodes]{{\bf 95 \% credible intervals for edge effects (interaction model with random coefficients) with $k=8$ sampled nodes.} Top (a): Sample size of $n=500$. Bottom (b): Sample size of $n=1000$. Each panel corresponds to a scenario of $\pi=0.3, 0.8$ (which controls the sparsity of the regression coefficient matrix $\mathbf B$) and $\mu=0.8,1.6$. We plot the 95 \% credible intervals for the regression coefficients per edge ordered depending on whether they are truly non-influential edges (top of each panel) or truly influential edges (bottom of each panel). The color of the intervals depends on whether it intersects zero (light) and hence estimated to be non-influential or does not intersect zero (dark) and hence estimated to be influential by the model. These panels allow us to visualize false positives (dark intervals on the top panel) or false negatives (light intervals on the bottom panel).}
\label{fig:edges_int_rand}
\end{figure}

\begin{figure}[!ht]
\centering
\begin{subfigure}[b]{0.49\textwidth}
\centering
\includegraphics[scale=0.19]{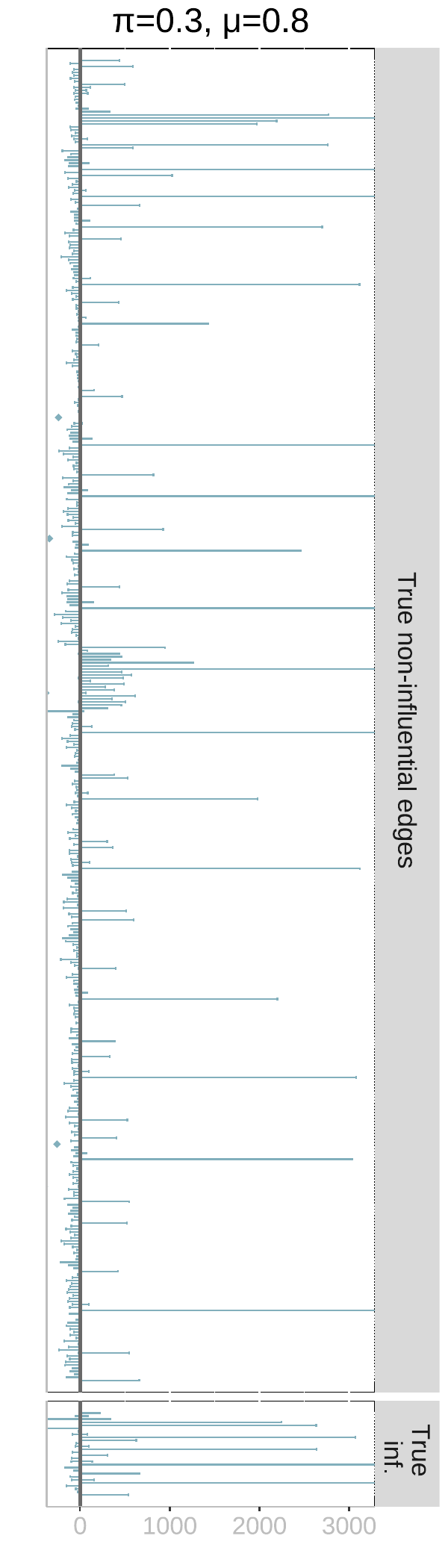}
\includegraphics[scale=0.19]{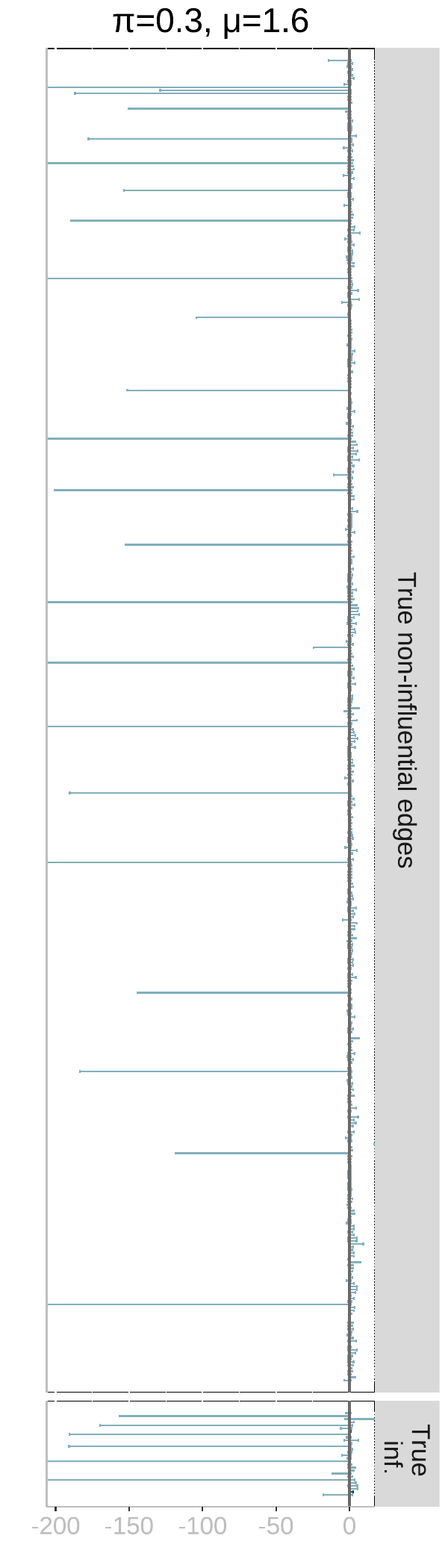}
\includegraphics[scale=0.19]{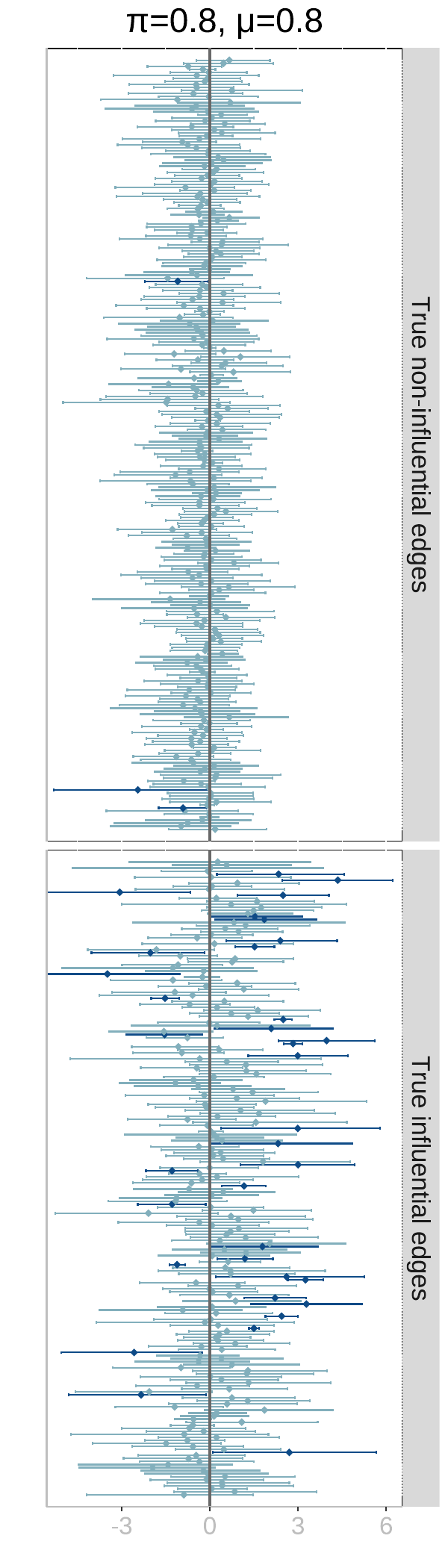}
\includegraphics[scale=0.19]{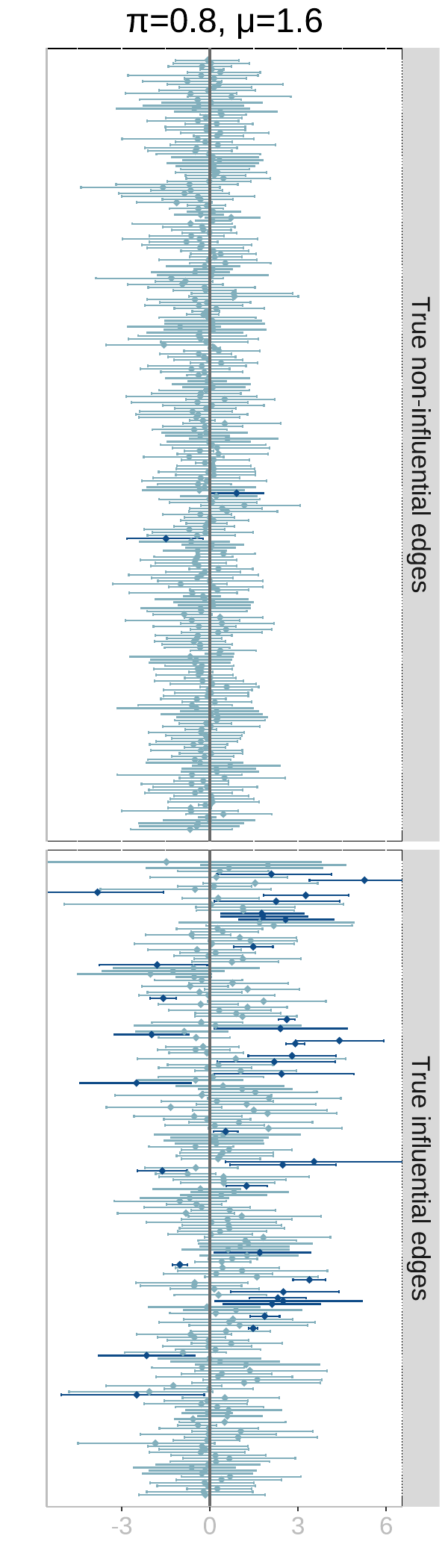}
\caption{$\mathbf{n=500}$}
\end{subfigure}
\begin{subfigure}[b]{0.49\textwidth}
\centering
\includegraphics[scale=0.19]{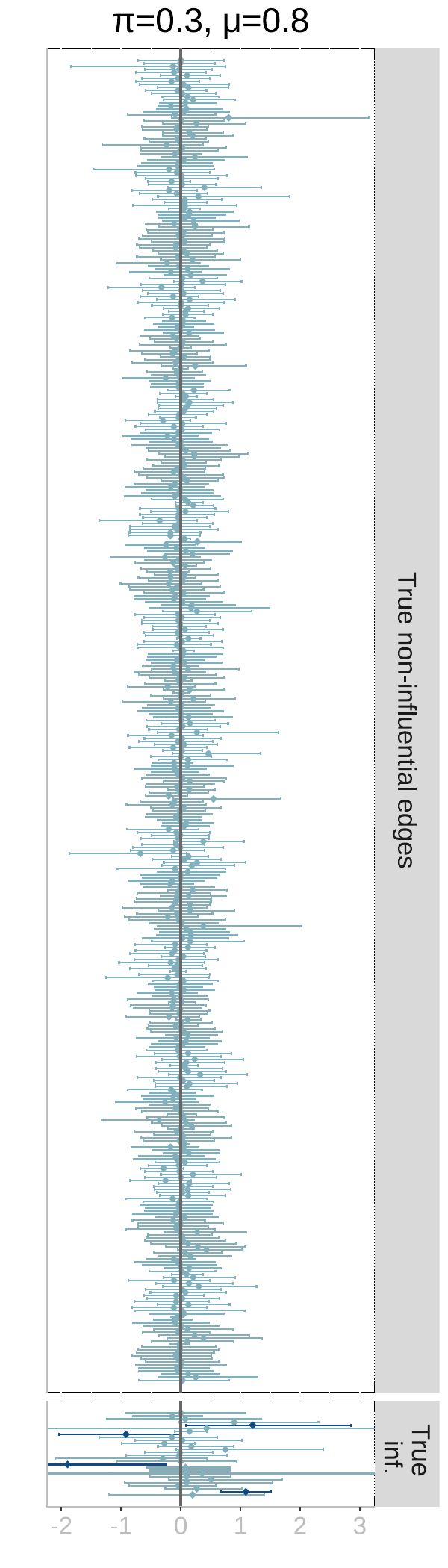}
\includegraphics[scale=0.19]{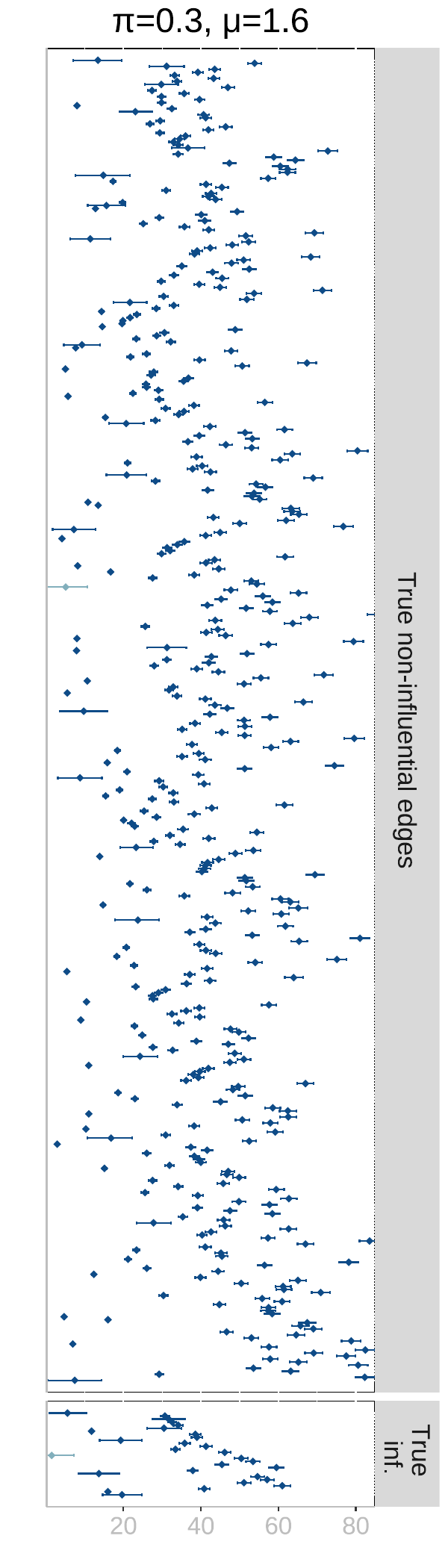}
\includegraphics[scale=0.19]{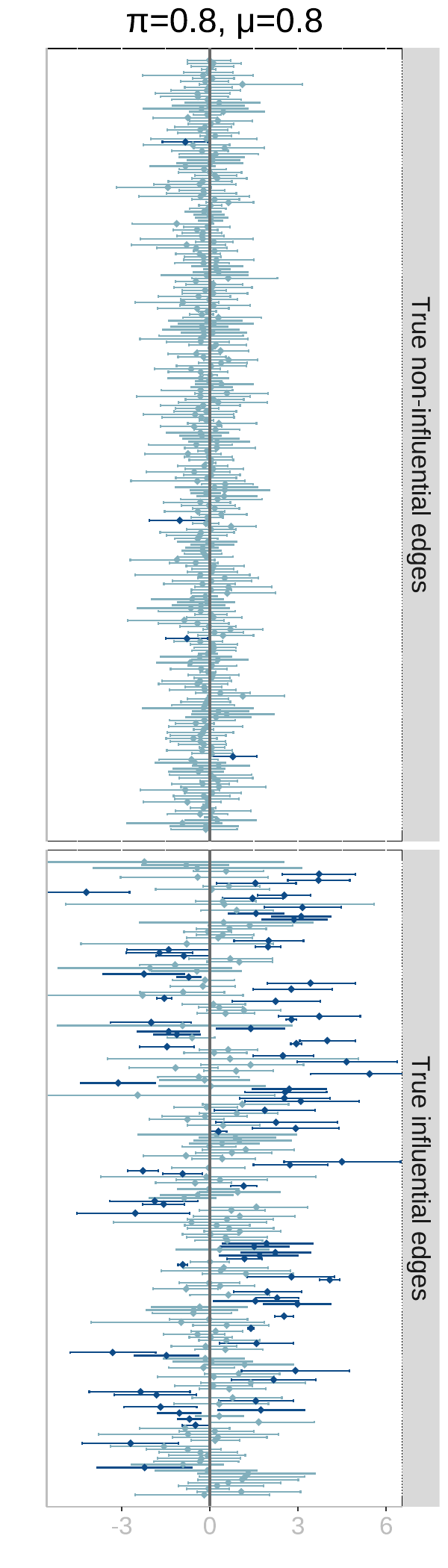}
\includegraphics[scale=0.19]{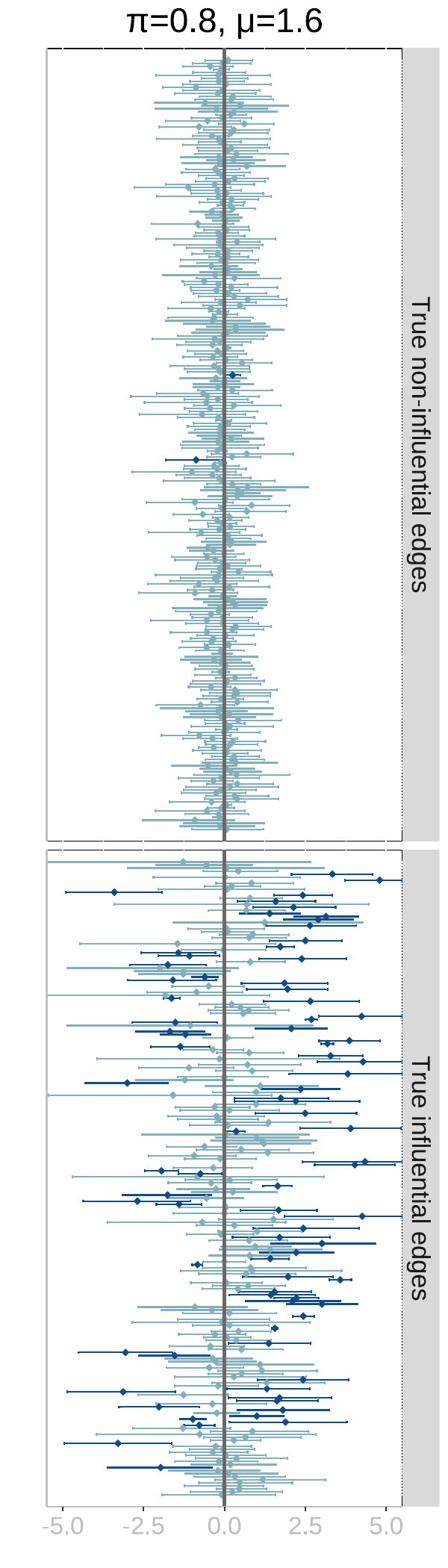}
\caption{$\mathbf{n=1000}$}
\end{subfigure}
\caption[95 \% credible intervals for edge effects (interaction model with random coefficients) with $k=22$ sampled nodes]{{\bf 95 \% credible intervals for edge effects (interaction model with random coefficients) with $k=22$ sampled nodes.} Top (a): Sample size of $n=500$. Bottom (b): Sample size of $n=1000$. Each panel corresponds to a scenario of $\pi=0.3, 0.8$ (which controls the sparsity of the regression coefficient matrix $\mathbf B$) and $\mu=0.8,1.6$. We plot the 95 \% credible intervals for the regression coefficients per edge ordered depending on whether they are truly non-influential edges (top of each panel) or truly influential edges (bottom of each panel). The color of the intervals depends on whether it intersects zero (light) and hence estimated to be non-influential or does not intersect zero (dark) and hence estimated to be influential by the model. These panels allow us to visualize false positives (dark intervals on the top panel) or false negatives (light intervals on the bottom panel).}
\label{fig:edges_int_rand2}
\end{figure}

\begin{figure}[!ht]
\centering
\begin{subfigure}[b]{0.49\textwidth}
\centering
\includegraphics[scale=0.19]{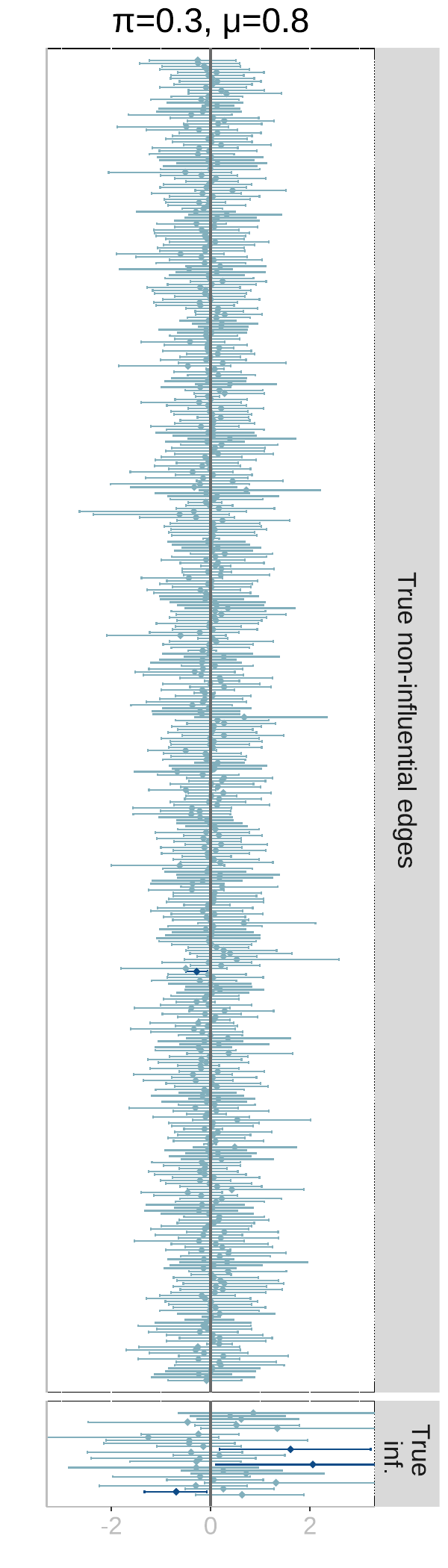}
\includegraphics[scale=0.19]{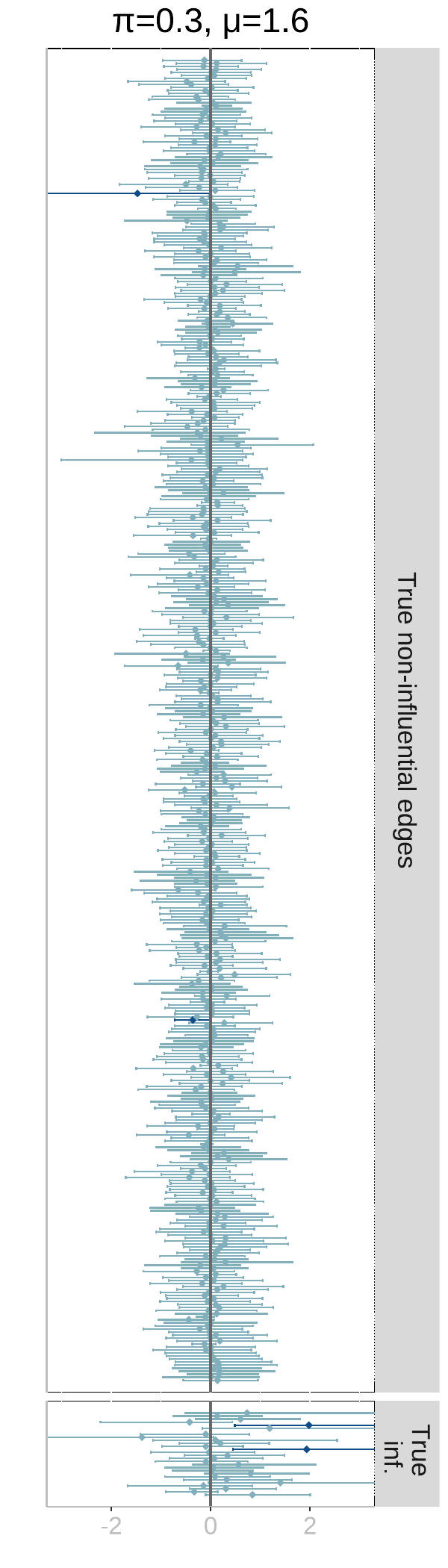}
\includegraphics[scale=0.19]{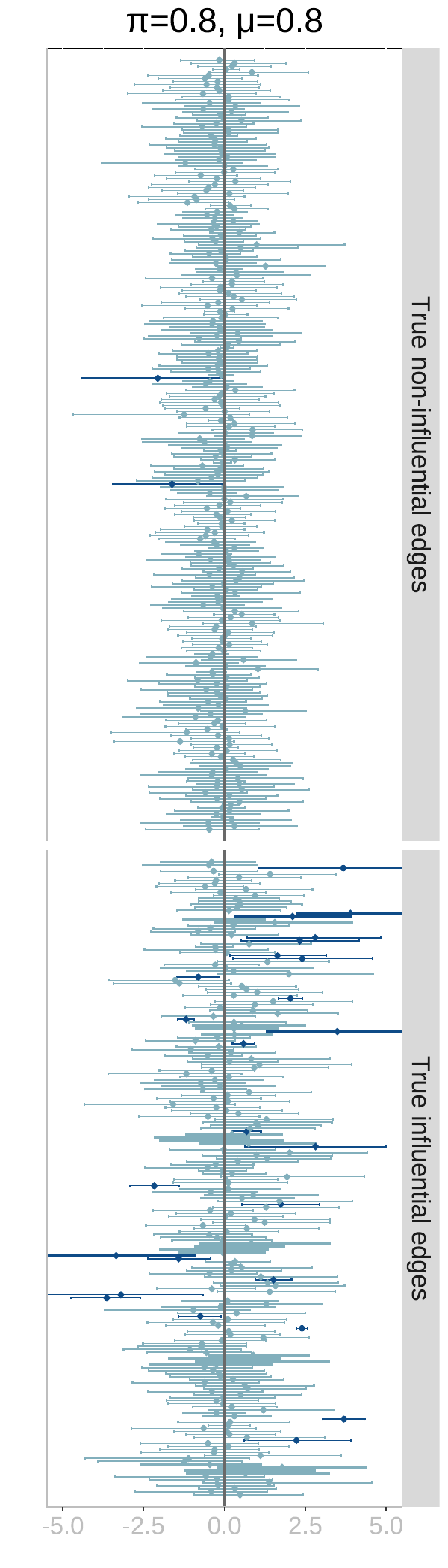}
\includegraphics[scale=0.19]{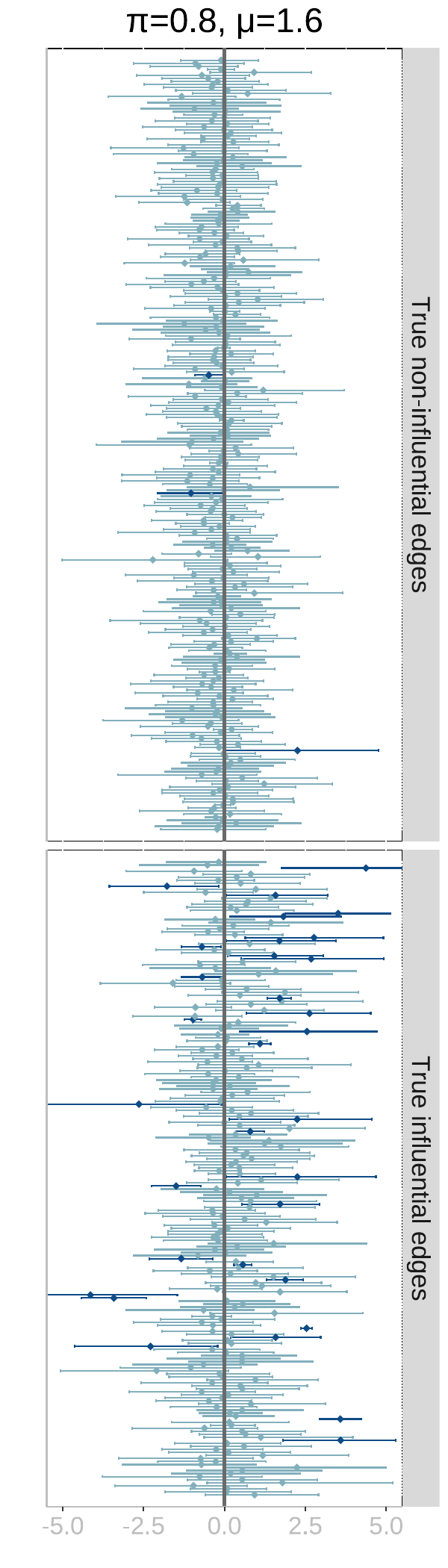}
\caption{$\mathbf{n=500}$}
\end{subfigure}
\begin{subfigure}[b]{0.49\textwidth}
\centering
\includegraphics[scale=0.19]{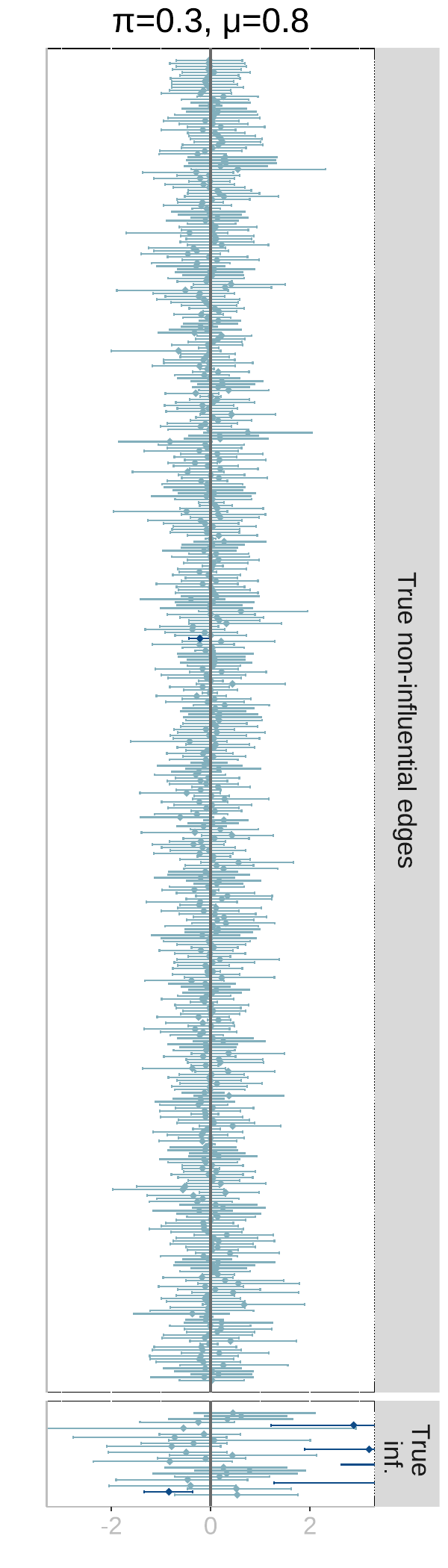}
\includegraphics[scale=0.19]{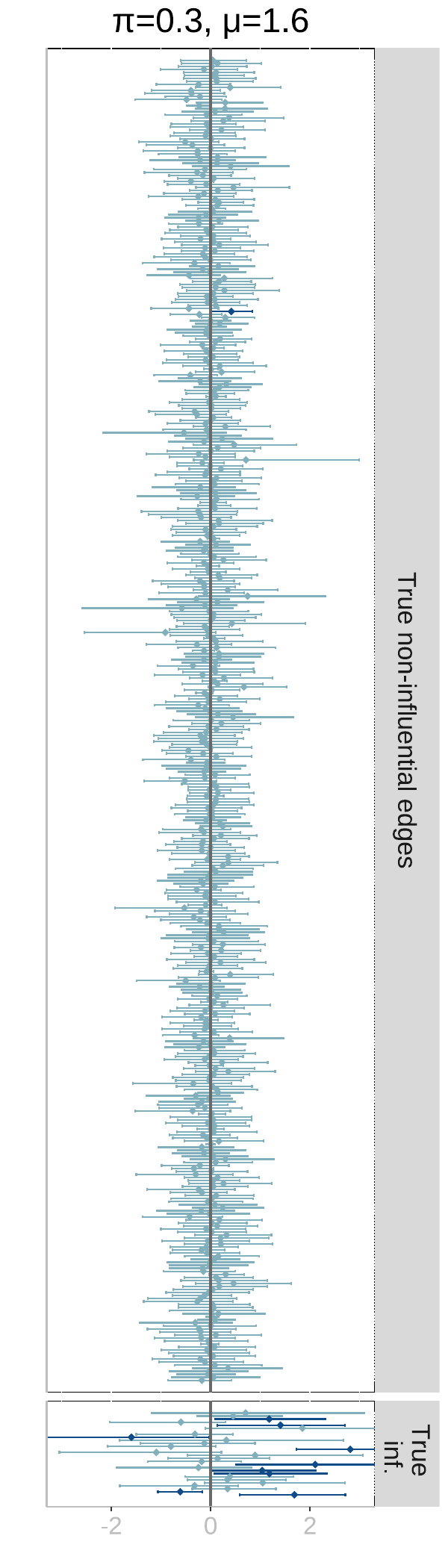}
\includegraphics[scale=0.19]{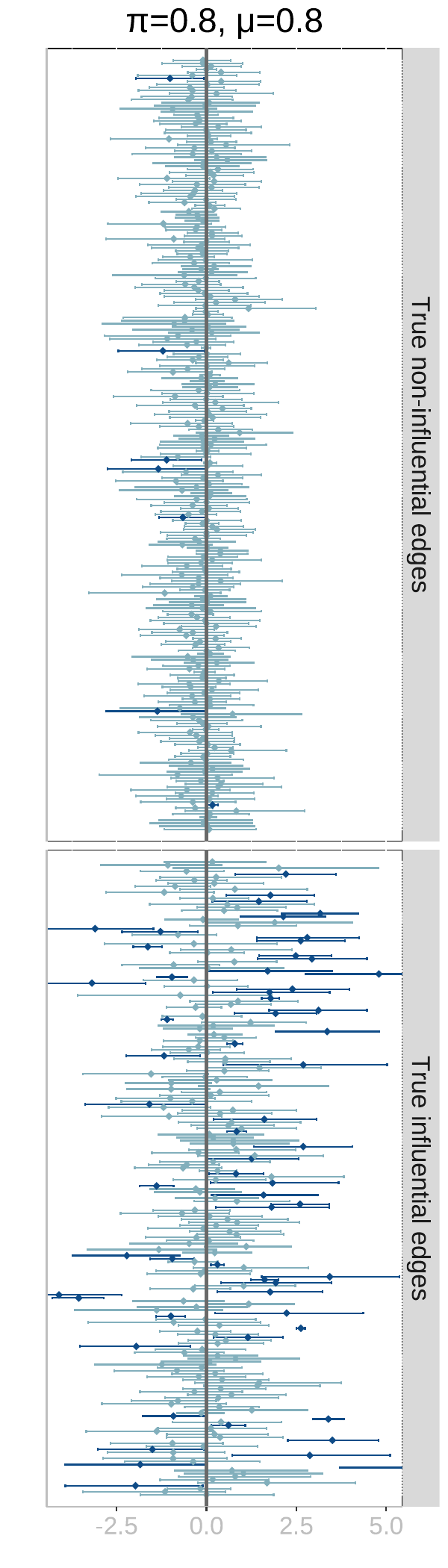}
\includegraphics[scale=0.19]{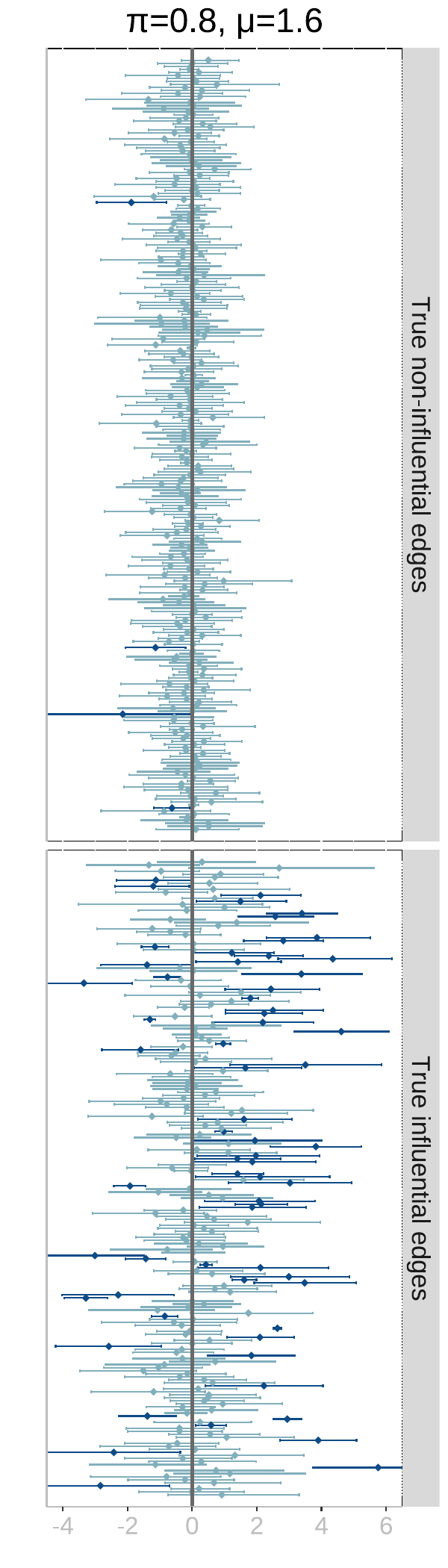}
\caption{$\mathbf{n=1000}$}
\end{subfigure}
\caption[95 \% credible intervals for edge effects (interaction model with phylogenetic coefficients) with $k=8$ sampled nodes]{{\bf 95 \% credible intervals for edge effects (interaction model with phylogenetic coefficients) with $k=8$ sampled nodes.} Top (a): Sample size of $n=500$. Bottom (b): Sample size of $n=1000$. Each panel corresponds to a scenario of $\pi=0.3, 0.8$ (which controls the sparsity of the regression coefficient matrix $\mathbf B$) and $\mu=0.8,1.6$. We plot the 95 \% credible intervals for the regression coefficients per edge ordered depending on whether they are truly non-influential edges (top of each panel) or truly influential edges (bottom of each panel). The color of the intervals depends on whether it intersects zero (light) and hence estimated to be non-influential or does not intersect zero (dark) and hence estimated to be influential by the model. These panels allow us to visualize false positives (dark intervals on the top panel) or false negatives (light intervals on the bottom panel).}
\label{fig:edges_int_phylo}
\end{figure}

\begin{figure}[!ht]
\centering
\begin{subfigure}[b]{0.49\textwidth}
\centering
\includegraphics[scale=0.19]{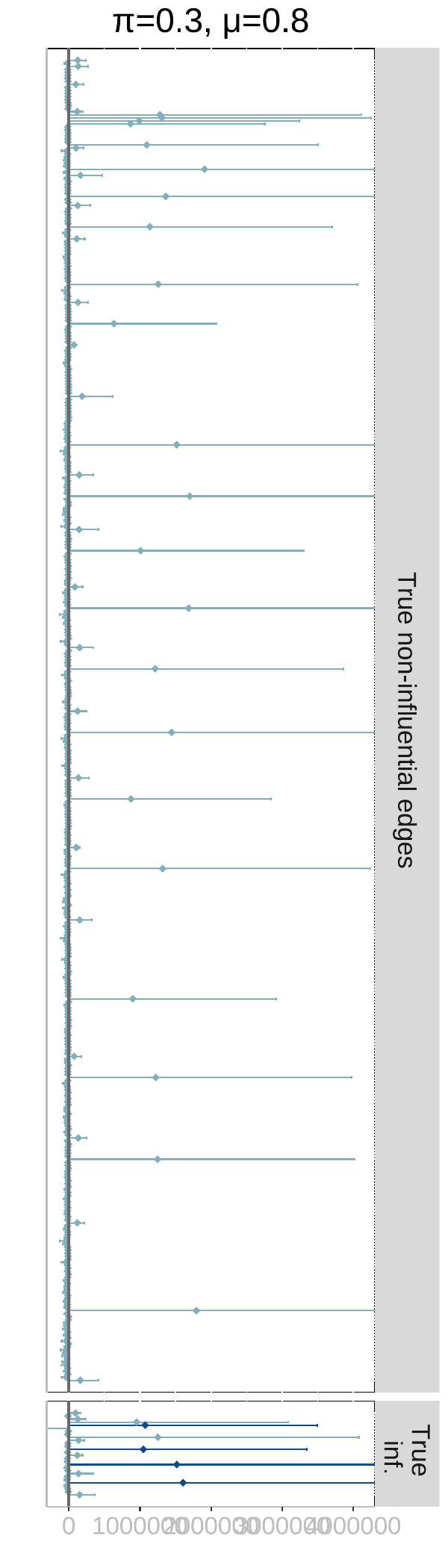}
\includegraphics[scale=0.19]{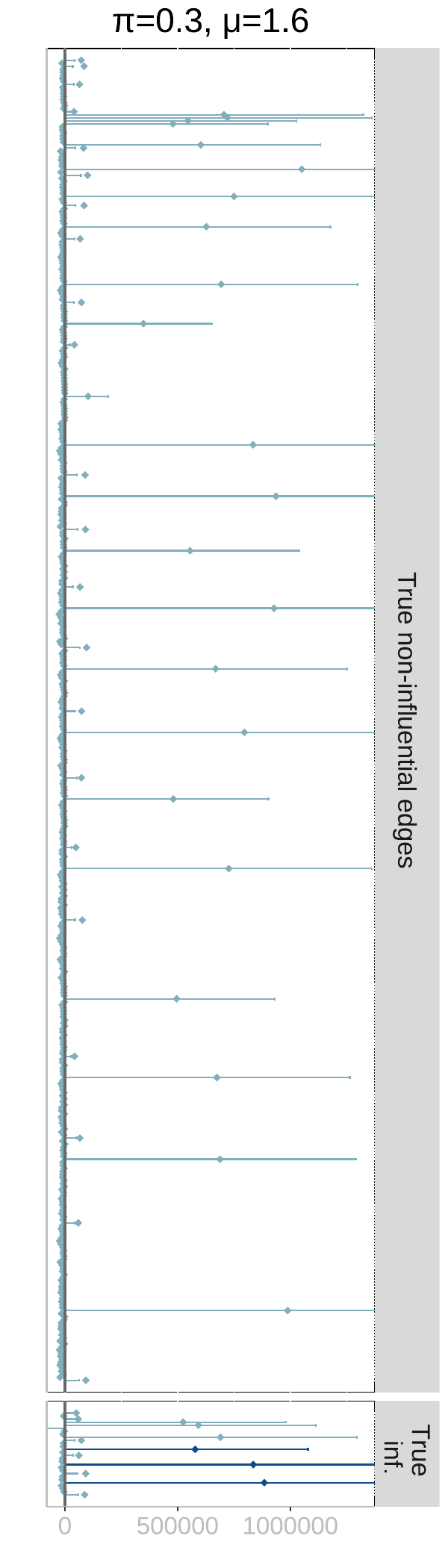}
\includegraphics[scale=0.19]{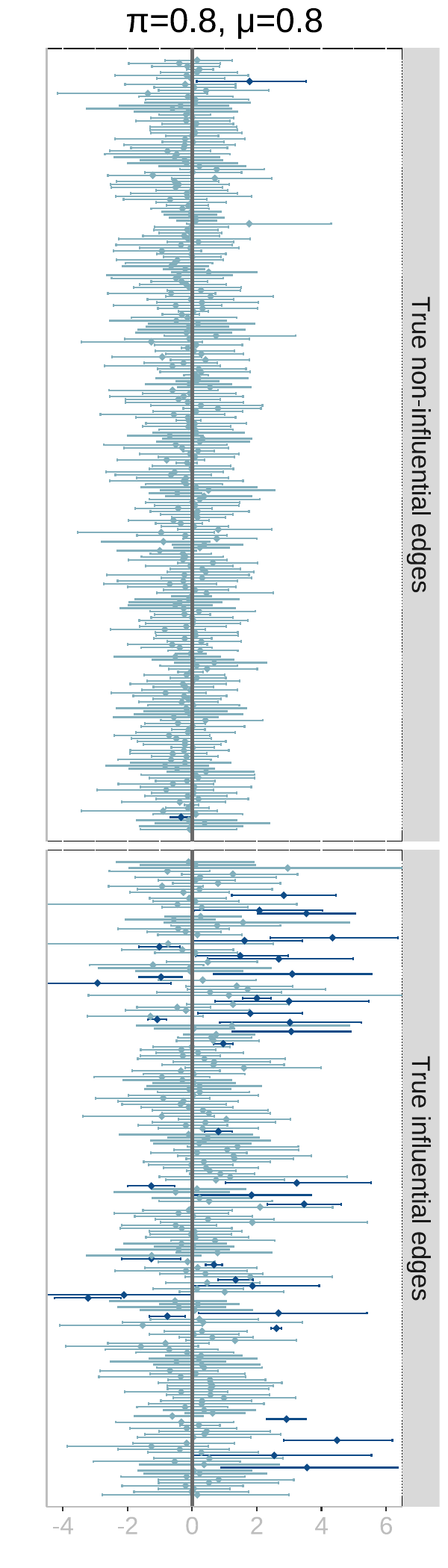}
\includegraphics[scale=0.19]{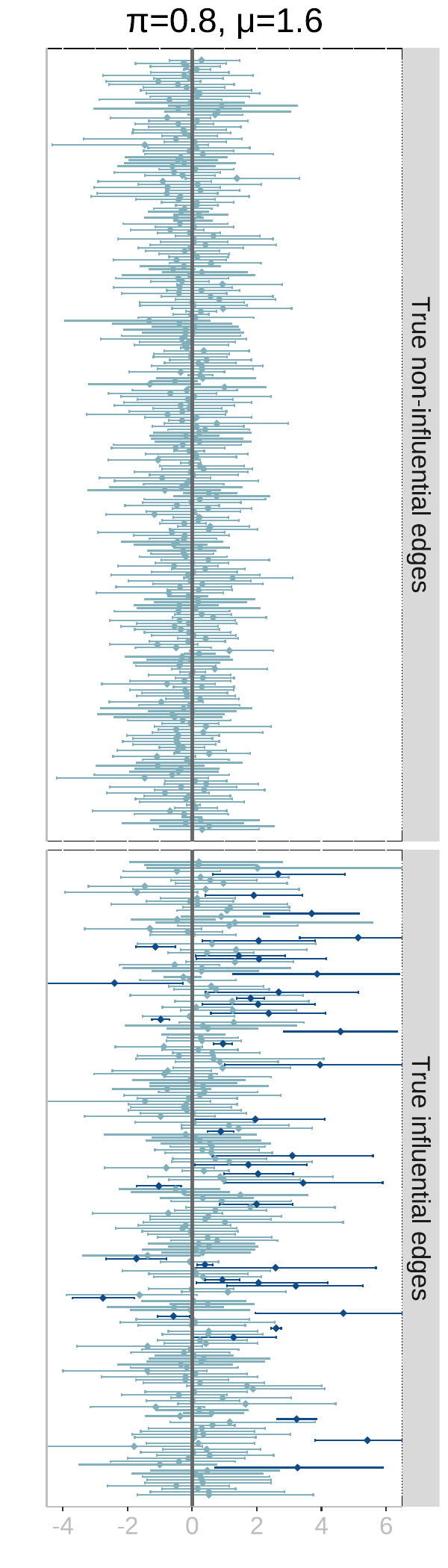}
\caption{$\mathbf{n=500}$}
\end{subfigure}
\begin{subfigure}[b]{0.49\textwidth}
\centering
\includegraphics[scale=0.19]{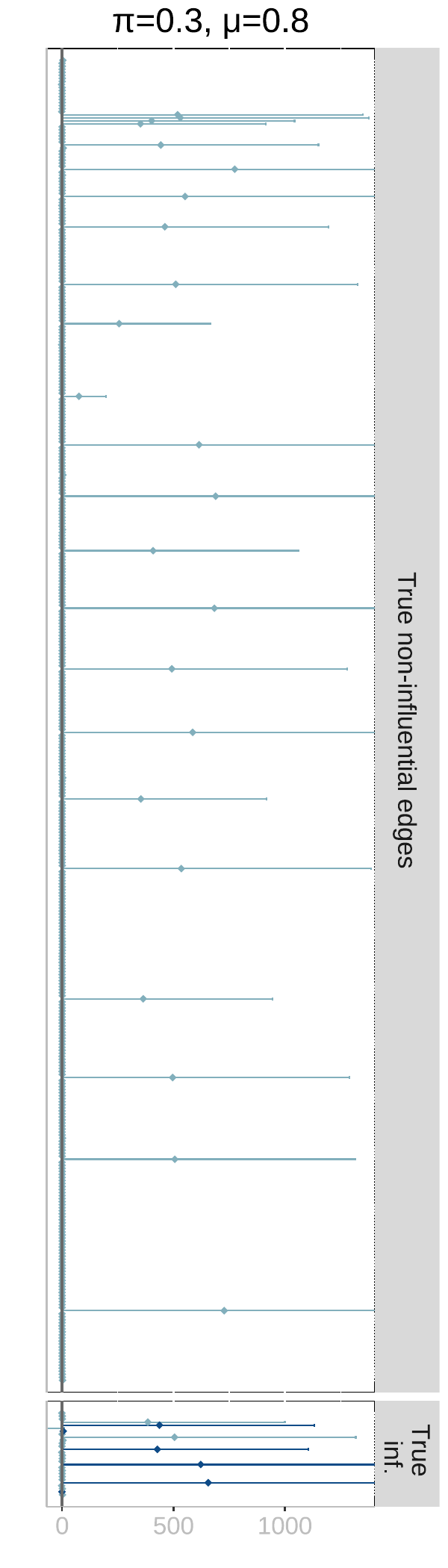}
\includegraphics[scale=0.19]{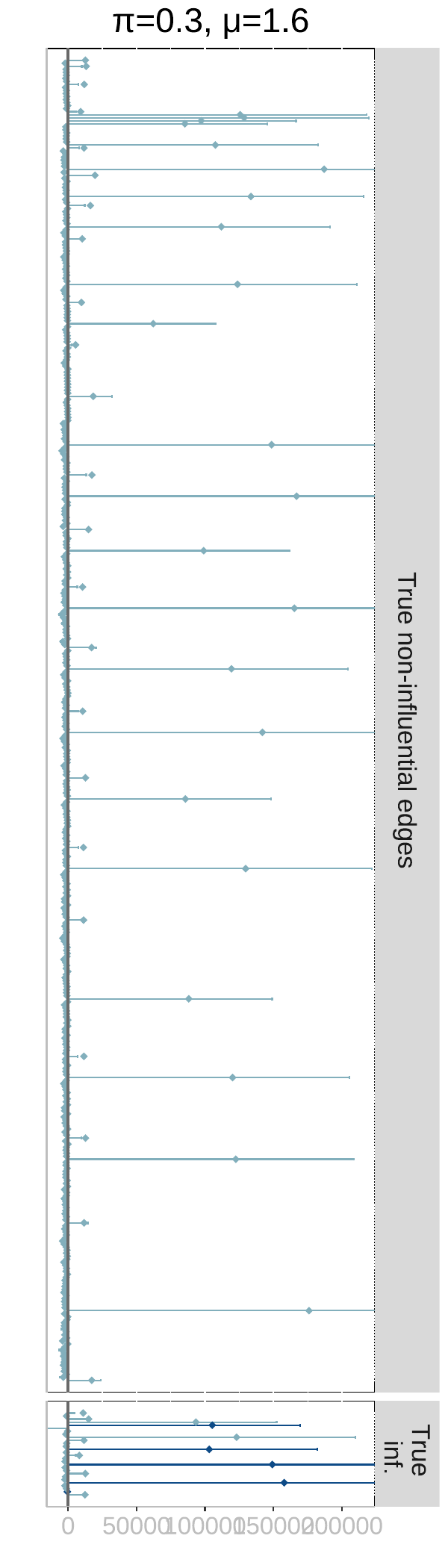}
\includegraphics[scale=0.19]{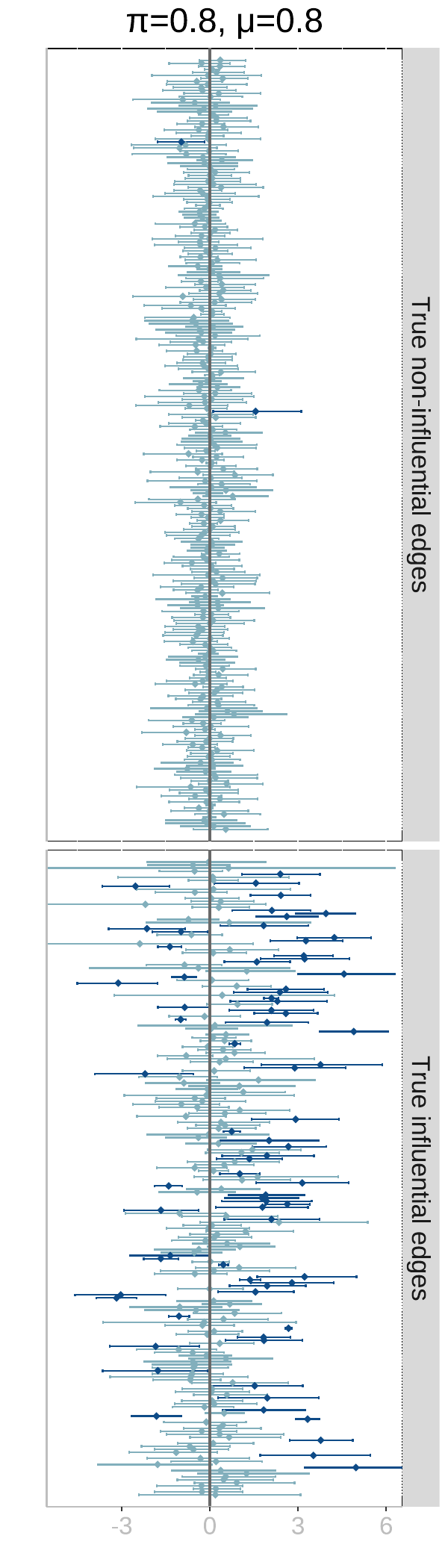}
\includegraphics[scale=0.19]{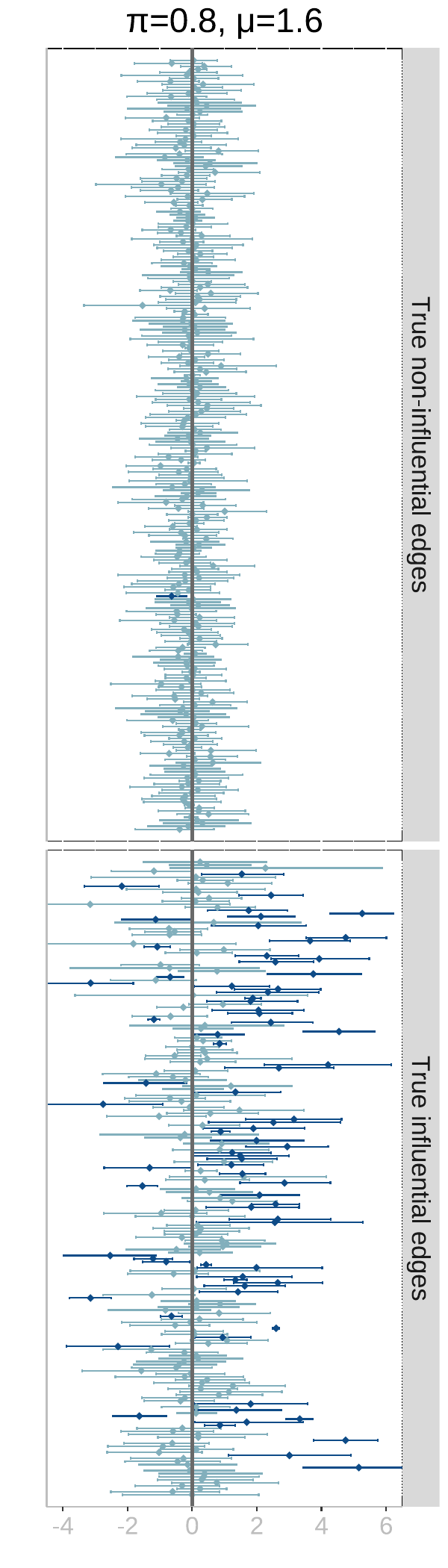}
\caption{$\mathbf{n=1000}$}
\end{subfigure}
\caption[95 \% credible intervals for edge effects (interaction model with phylogenetic coefficients) with $k=22$ sampled nodes]{{\bf 95 \% credible intervals for edge effects (interaction model with phylogenetic coefficients) with $k=22$ sampled nodes.} Top (a): Sample size of $n=500$. Bottom (b): Sample size of $n=1000$. Each panel corresponds to a scenario of $\pi=0.3, 0.8$ (which controls the sparsity of the regression coefficient matrix $\mathbf B$) and $\mu=0.8,1.6$. We plot the 95 \% credible intervals for the regression coefficients per edge ordered depending on whether they are truly non-influential edges (top of each panel) or truly influential edges (bottom of each panel). The color of the intervals depends on whether it intersects zero (light) and hence estimated to be non-influential or does not intersect zero (dark) and hence estimated to be influential by the model. These panels allow us to visualize false positives (dark intervals on the top panel) or false negatives (light intervals on the bottom panel).}
\label{fig:edges_int_phylo2}
\end{figure}

\begin{figure}[!ht]
    \centering
    \begin{subfigure}[t]{0.49\textwidth}
        \includegraphics[scale=0.25]{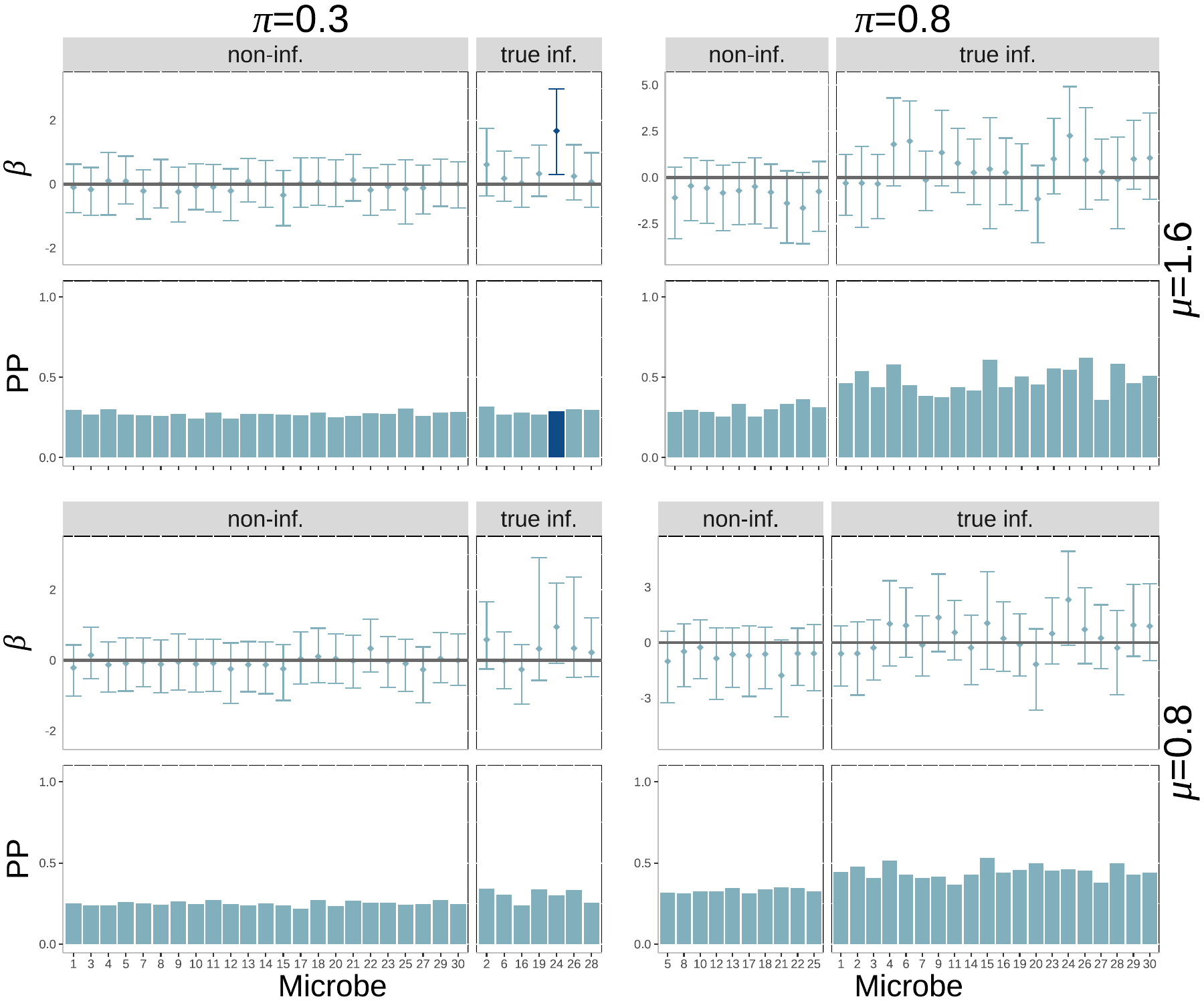}
        \caption{$\mathbf{n=500}$,$\mathbf{k=8}$}
    \end{subfigure}
    \begin{subfigure}[t]{0.49\textwidth}
        \includegraphics[scale=0.25]{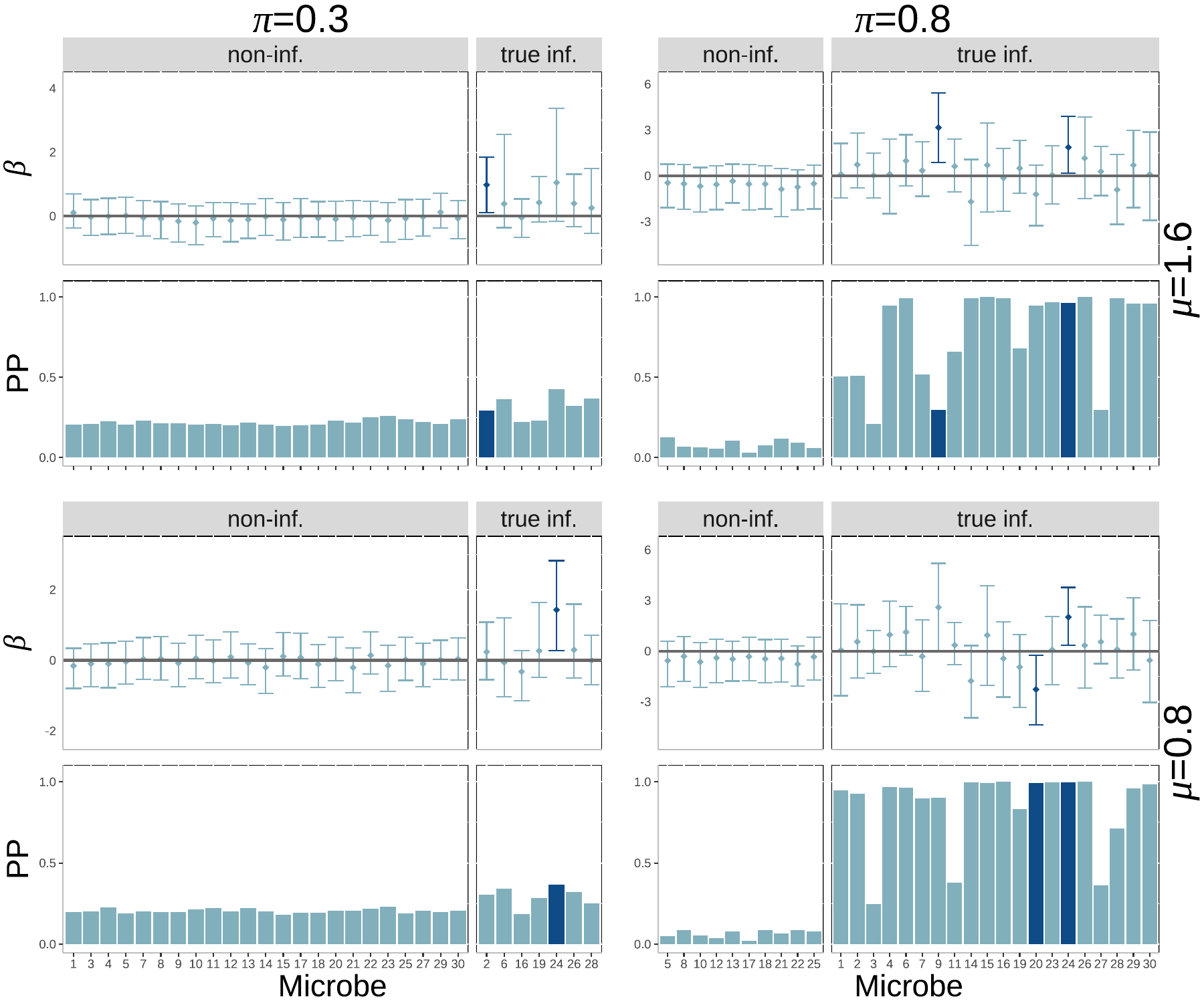}
        \caption{$\mathbf{n=1000}$,$\mathbf{k=8}$}
    \end{subfigure}\\
    \begin{subfigure}[t]{0.49\textwidth}
        \includegraphics[scale=0.25]{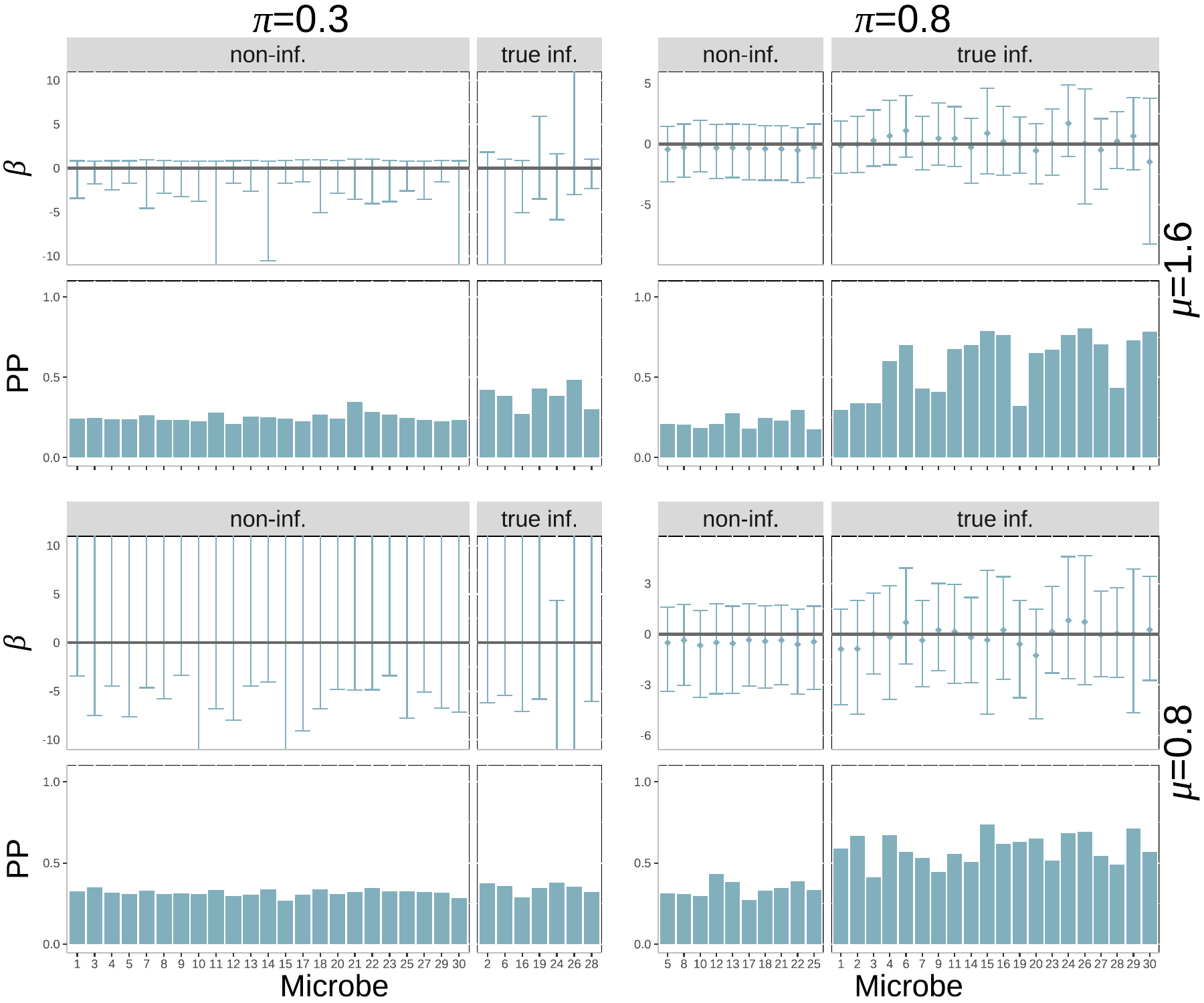}
        \caption{$\mathbf{n=500}$,$\mathbf{k=22}$}
    \end{subfigure}
    \begin{subfigure}[t]{0.49\textwidth}
        \includegraphics[scale=0.25]{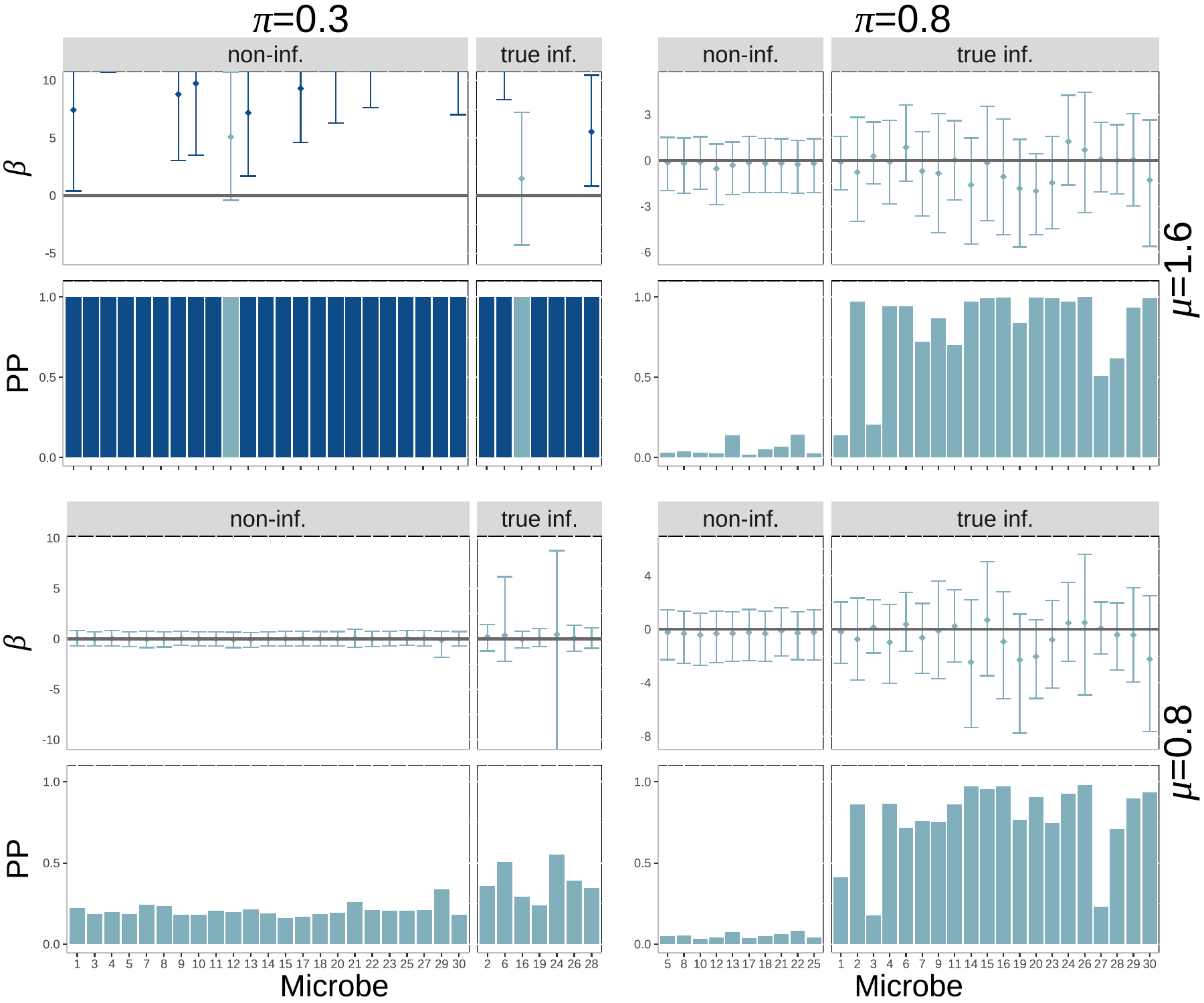}
        \caption{$\mathbf{n=1000}$,$\mathbf{k=22}$}
    \end{subfigure}
    \caption[Posterior probability of influential nodes and coefficients for nodes (interaction model with random coefficients)]{{\bf Posterior probability of influential nodes and coefficients for nodes (interaction model with random coefficients).}
    Different \revision{groups of four} panels represent different number of sampled microbes ($k=8,22$) which controls the sparsity of the adjacency matrix and different sample sizes ($n=500,1000$). \revision{Within each group, we have four panels corresponding to the two values of edge effect size ($\mu=0.8, 1.6$) and two values of probability of influential node ($\pi=0.3, 0.8$) which controls the sparsity of the regression coefficient matrix ($\mathbf B$). Within each of these panels we have two plots: 95\% credible intervals (top) and posterior probability of influence (bottom - calculated as the mean of the $\xi$ variable for the node across Gibbs samples) for each node.} Each bar corresponds to one node (microbe). \revision{Within each plot the bars and intervals are colored depending on whether the node is found to be influential (dark) or not influential (light) based on the 95\% credible intervals. Each plot is split based on whether the nodes are truly influential (right) or not (left).}
    } 
    \label{fig:nodes_int_rand}
\end{figure}

\begin{figure}[!ht]
    \centering
    \begin{subfigure}[t]{0.49\textwidth}
        \includegraphics[scale=0.27]{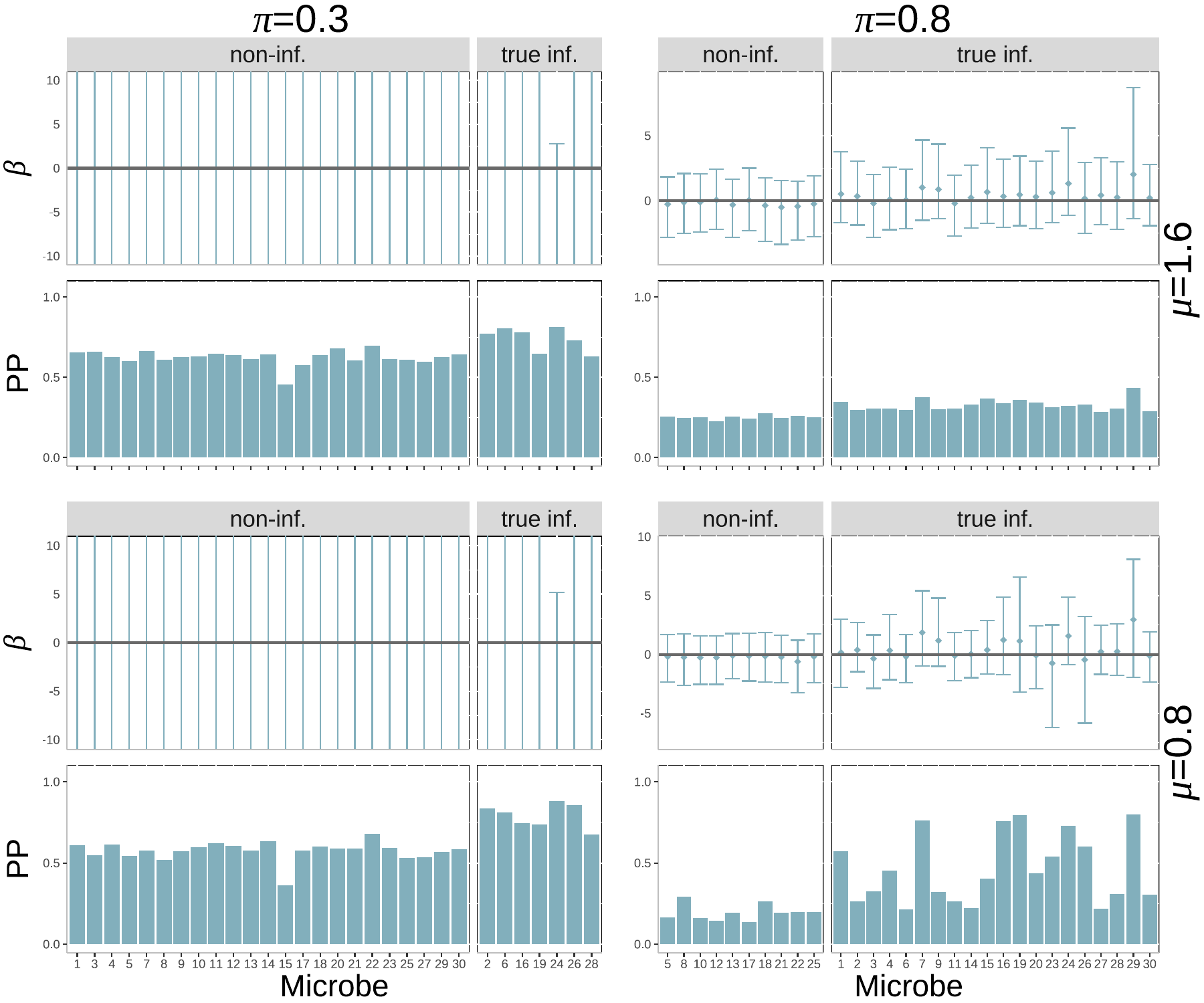}
        \caption{$\mathbf{n=500}$}
    \end{subfigure}
    \begin{subfigure}[t]{0.49\textwidth}
        \includegraphics[scale=0.27]{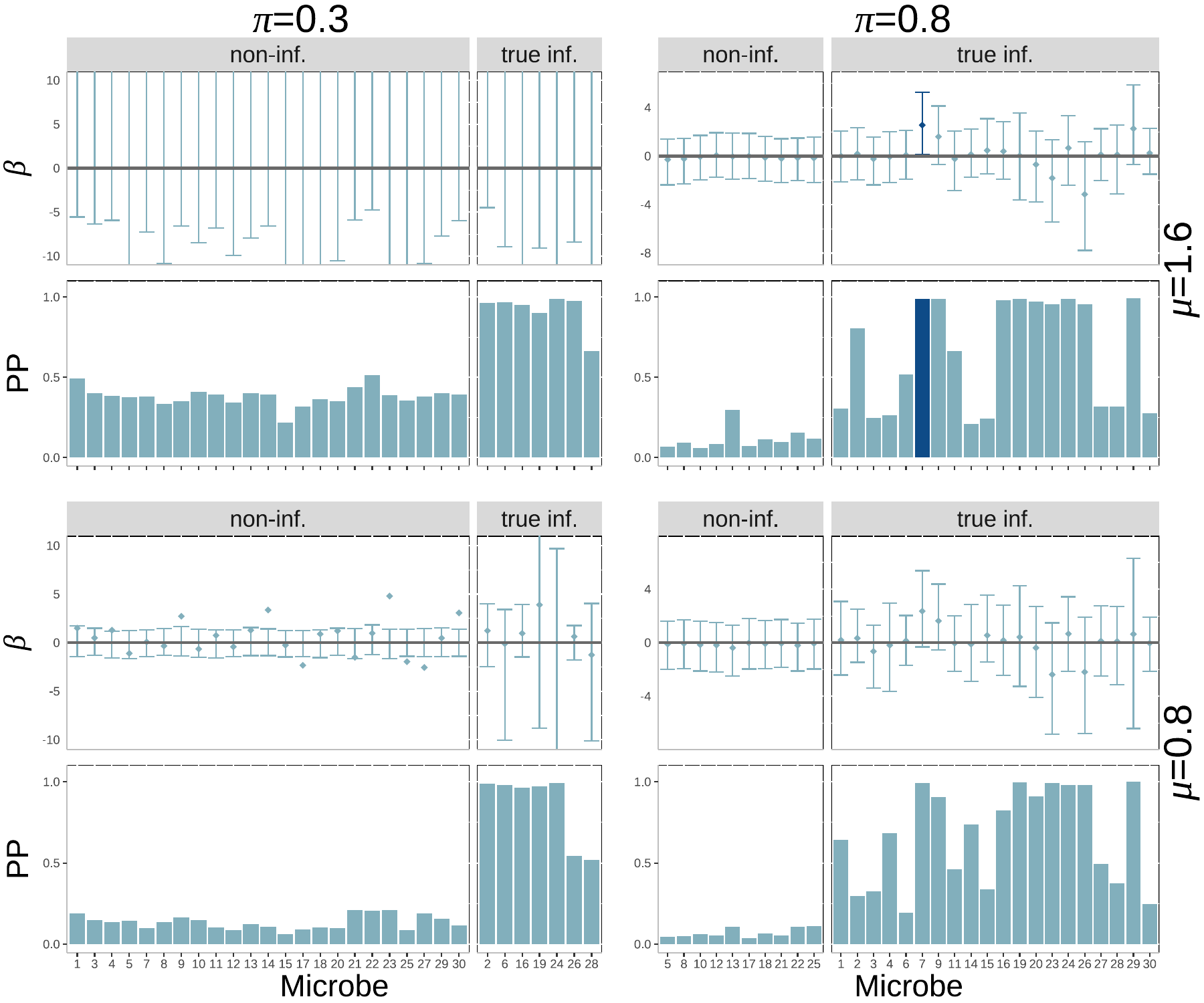}
        \caption{$\mathbf{n=1000}$}
    \end{subfigure}
    \caption[Posterior probability of influential nodes and coefficients for nodes (interaction model with phylogenetic coefficients)]{{\bf Posterior probability of influential nodes and coefficients for nodes (interaction model with phylogenetic coefficients).}
    Different \revision{groups of four} panels represent different number of sampled microbes ($k=22$) which controls the sparsity of the adjacency matrix and different sample sizes ($n=500,1000$). \revision{Within each group, we have four panels corresponding to the two values of edge effect size ($\mu=0.8, 1.6$) and two values of probability of influential node ($\pi=0.3, 0.8$) which controls the sparsity of the regression coefficient matrix ($\mathbf B$). Within each of these panels we have two plots: 95\% credible intervals (top) and posterior probability of influence (bottom - calculated as the mean of the $\xi$ variable for the node across Gibbs samples) for each node.} Each bar corresponds to one node (microbe). \revision{Within each plot the bars and intervals are colored depending on whether the node is found to be influential (dark) or not influential (light) based on the 95\% credible intervals. Each plot is split based on whether the nodes are truly influential (right) or not (left).}
    } 
    \label{fig:nodes_int_phylo_adx}
\end{figure}

%
%
%


\FloatBarrier
\subsubsection{Realistic case: Functional redundancy}
\FloatBarrier

\begin{figure}[!ht]
\centering
\begin{subfigure}[t]{0.49\textwidth}
    \includegraphics[scale=0.27]{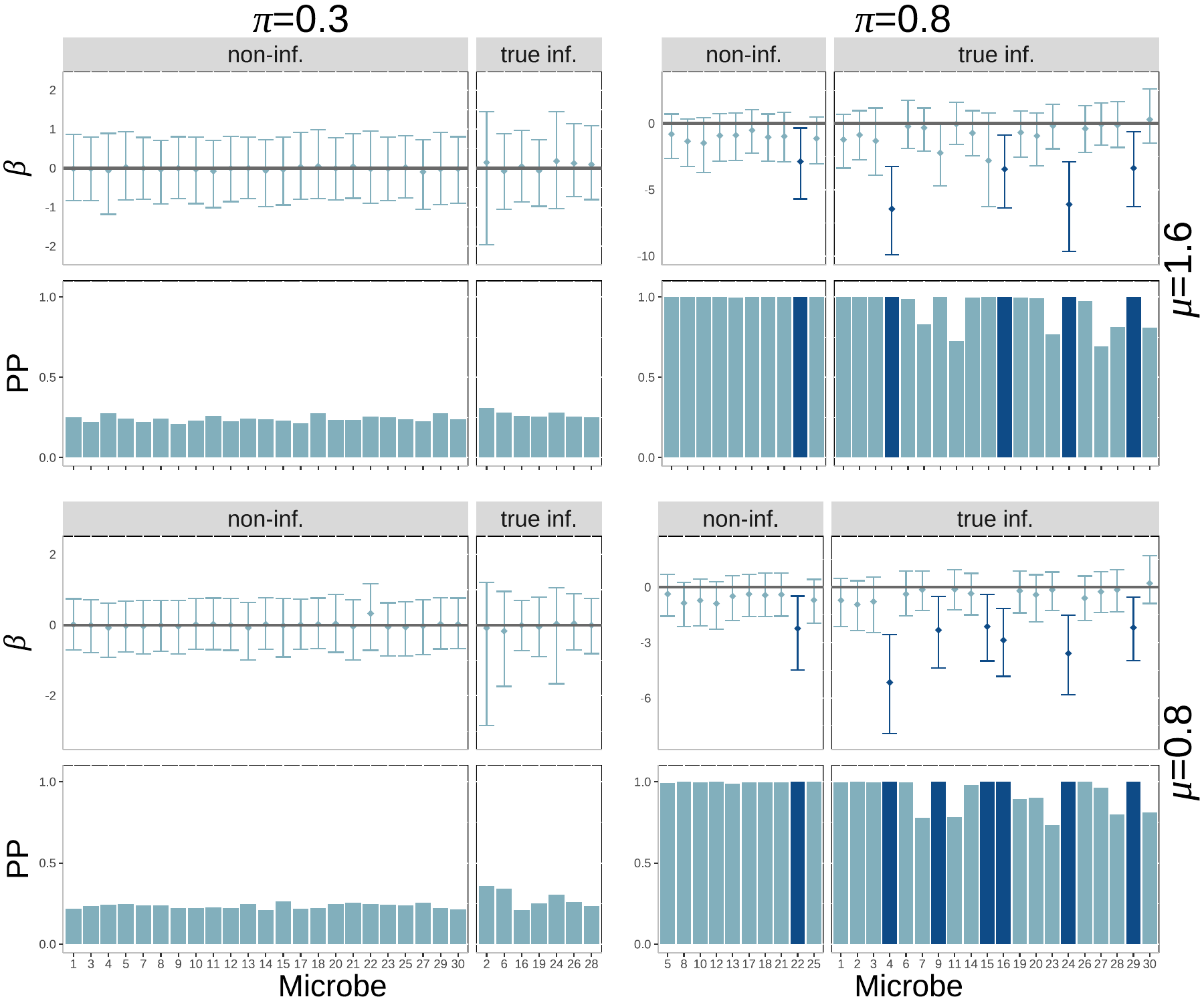}
    \caption{$\mathbf{n=500}$}
\end{subfigure}
\begin{subfigure}[t]{0.49\textwidth}
    \includegraphics[scale=0.27]{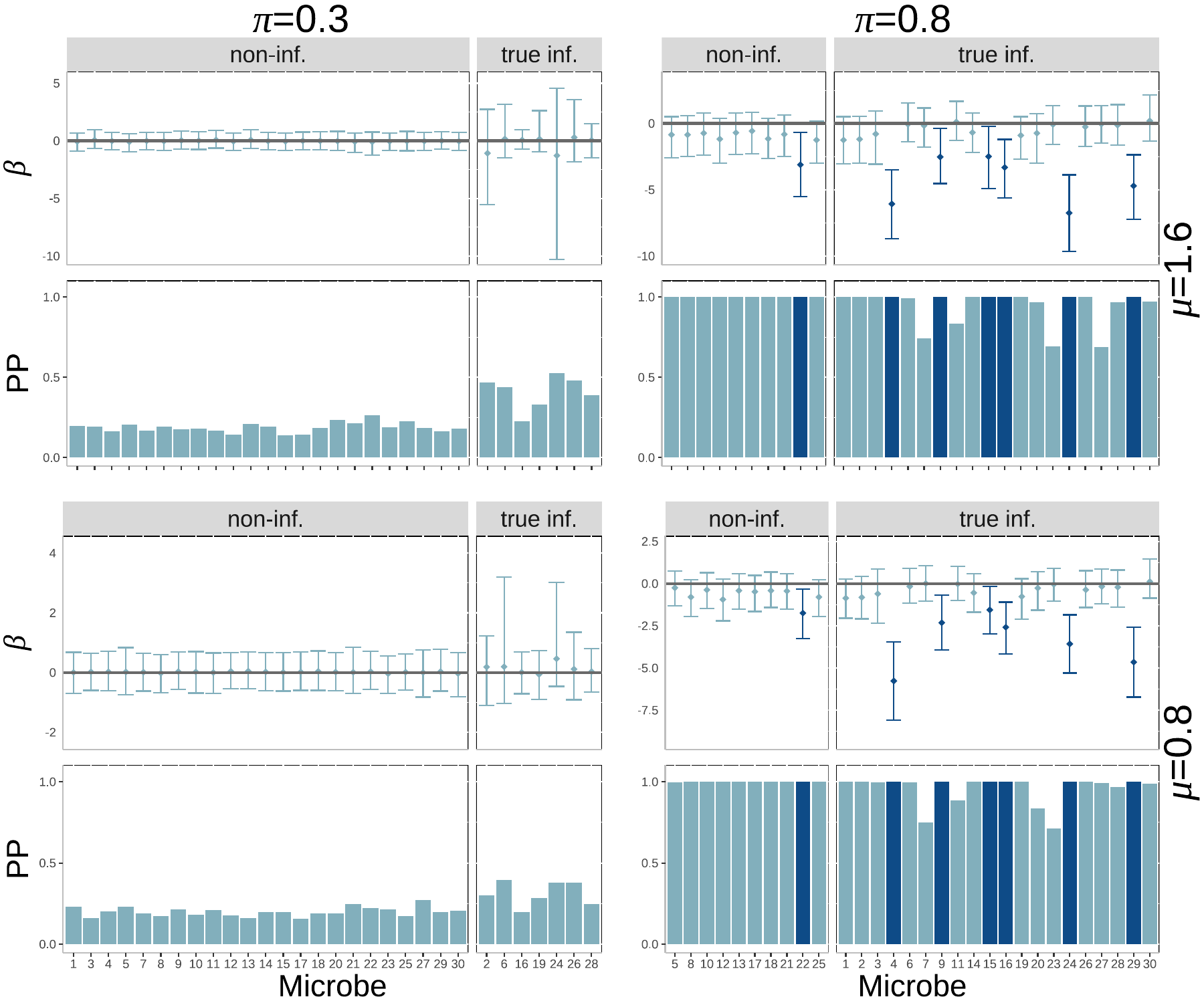}
    \caption{$\mathbf{n=1000}$}
\end{subfigure}
\caption[Posterior probability of influential nodes and coefficients for nodes (functional redundancy model with random coefficients )]{{\bf Posterior probability of influential nodes and coefficients for nodes (functional redundancy model with random coefficients).}
Different \revision{groups of four panels (quadrants) represent different sample sizes ($n=100,500$) and different number of sampled microbes ($k=22$). Within each group, we have four panels corresponding to the two values of edge effect size ($\mu=0.8, 1.6$) and two values of probability of influential node ($\pi=0.3, 0.8$) which controls the sparsity of the regression coefficient matrix ($\mathbf B$). Within each of these panels we have two plots: 95\% credible intervals (top) and posterior probability of influence (bottom - calculated as the mean of the $\xi$ variable for the node across Gibbs samples) for each node. Each bar corresponds to one node (microbe). Within each plot the bars and intervals are colored depending on whether the node is found to be influential (dark) or not influential (light) based on the 95\% credible intervals. Each plot is split based on whether the nodes are truly influential (right) or not (left).}
}
\label{fig:nodes_red_rand2}
\end{figure}

\begin{figure}[!ht]
\centering
\begin{subfigure}[b]{0.49\textwidth}
\centering
\includegraphics[scale=0.19]{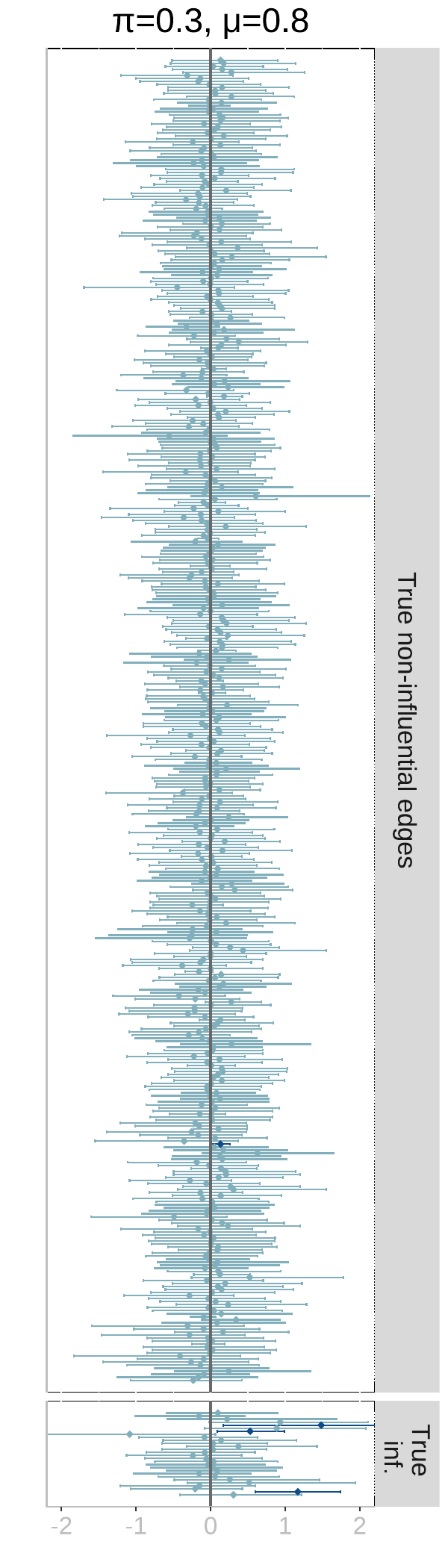}
\includegraphics[scale=0.19]{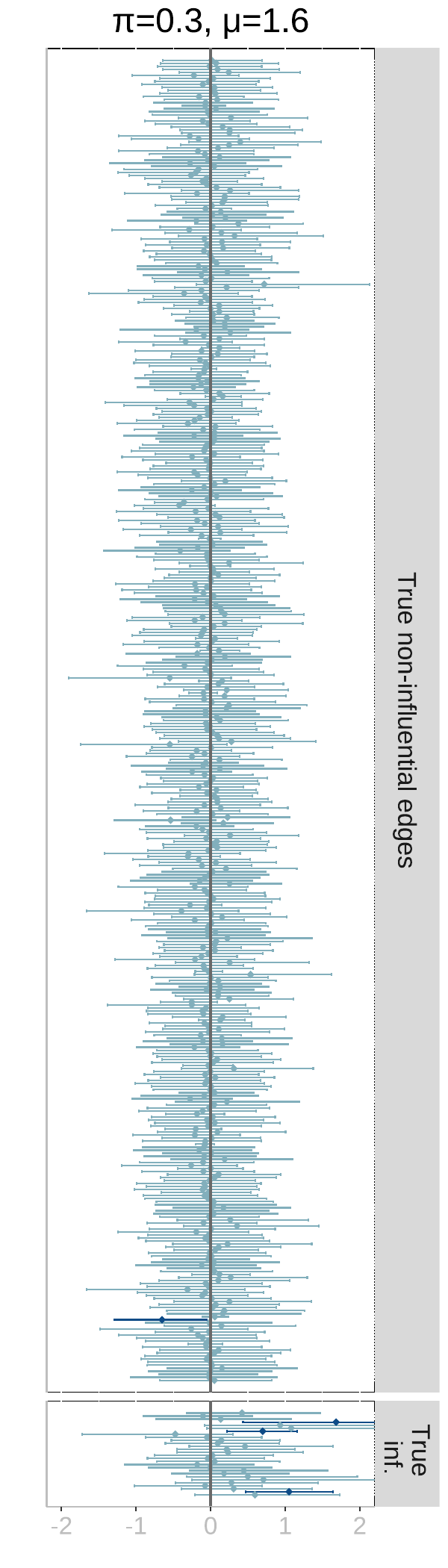}
\includegraphics[scale=0.19]{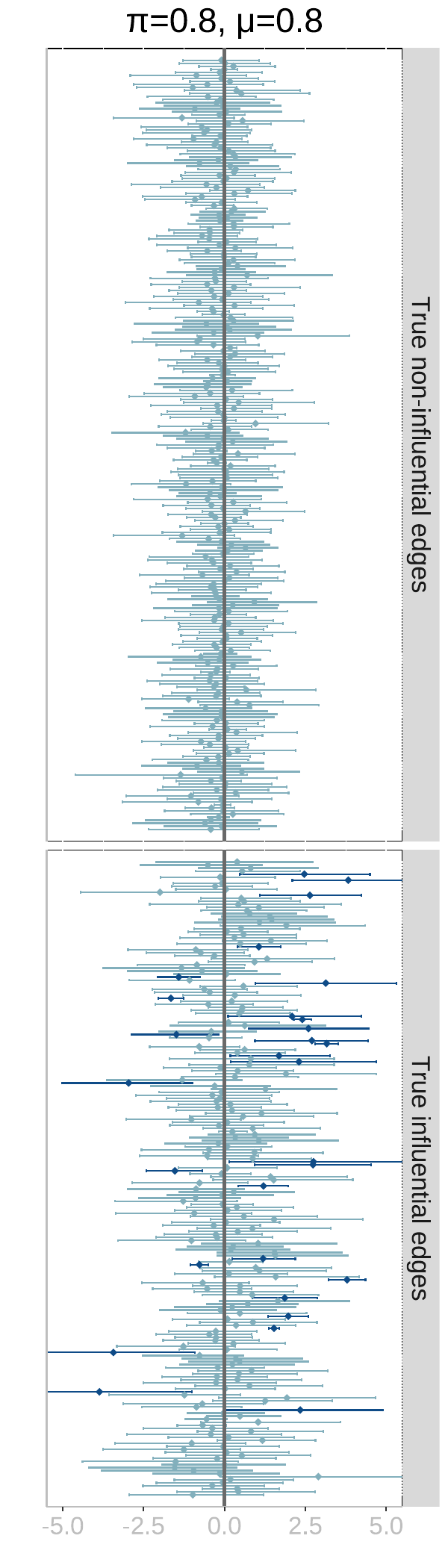}
\includegraphics[scale=0.19]{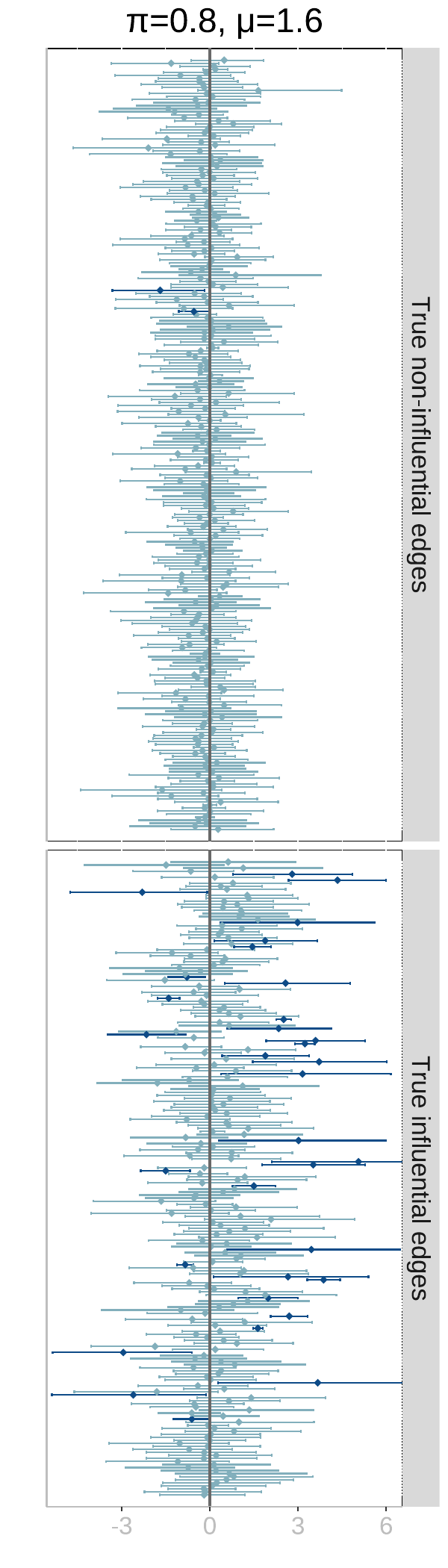}
\caption{$\mathbf{n=500}$}
\end{subfigure}
\begin{subfigure}[b]{0.49\textwidth}
\centering
\includegraphics[scale=0.19]{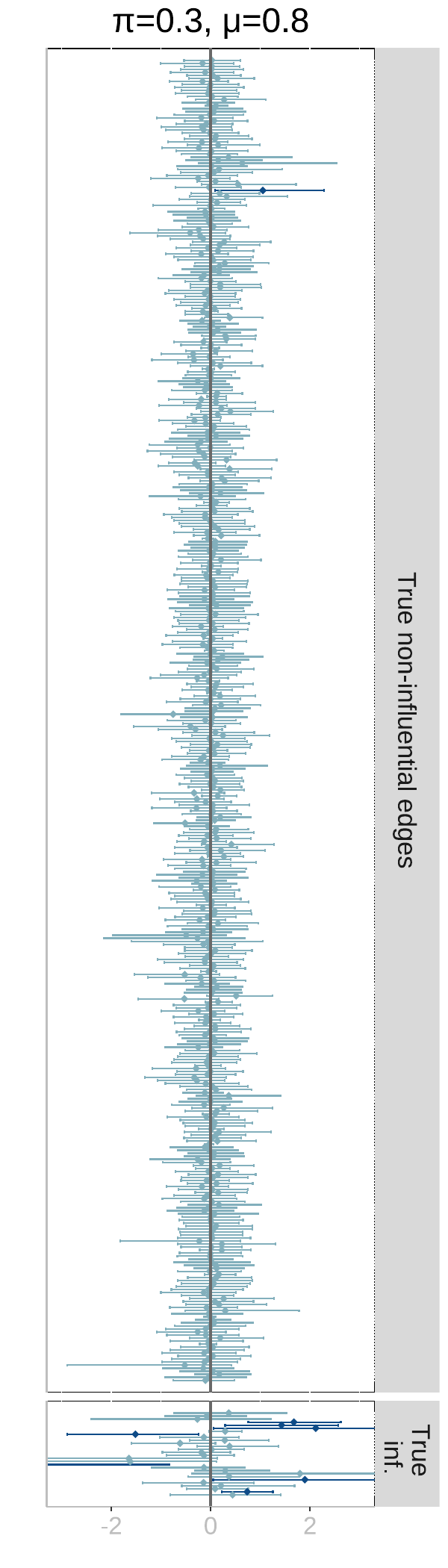}
\includegraphics[scale=0.19]{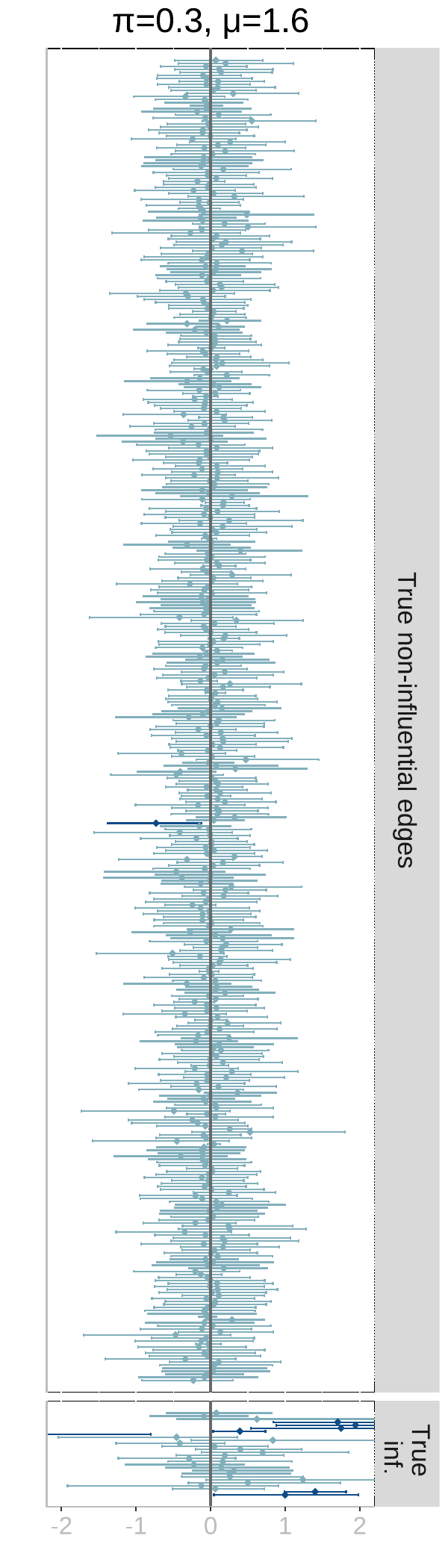}
\includegraphics[scale=0.19]{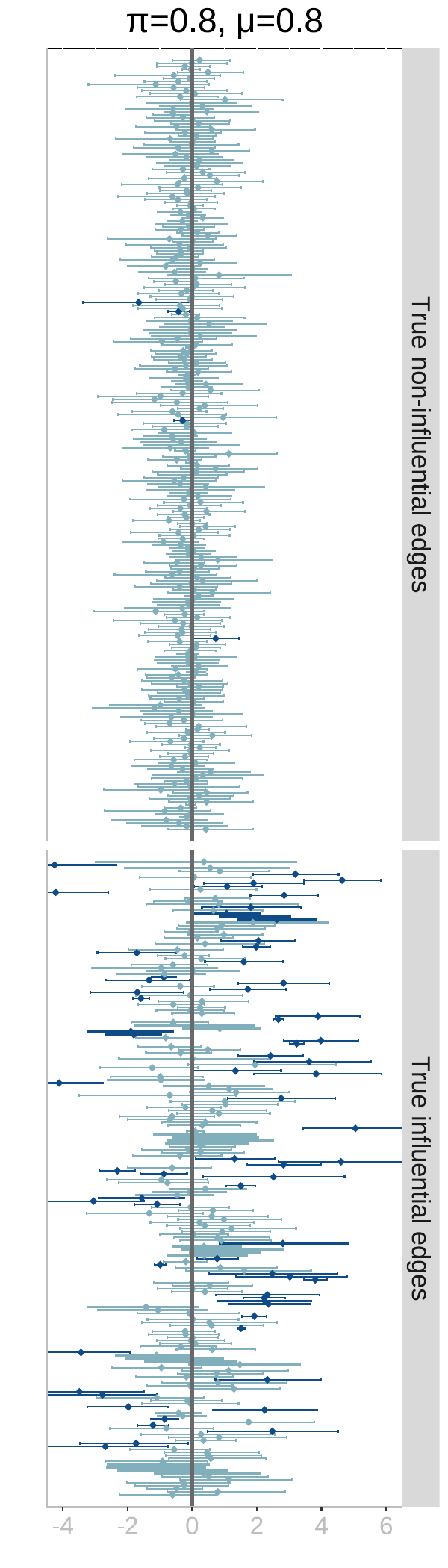}
\includegraphics[scale=0.19]{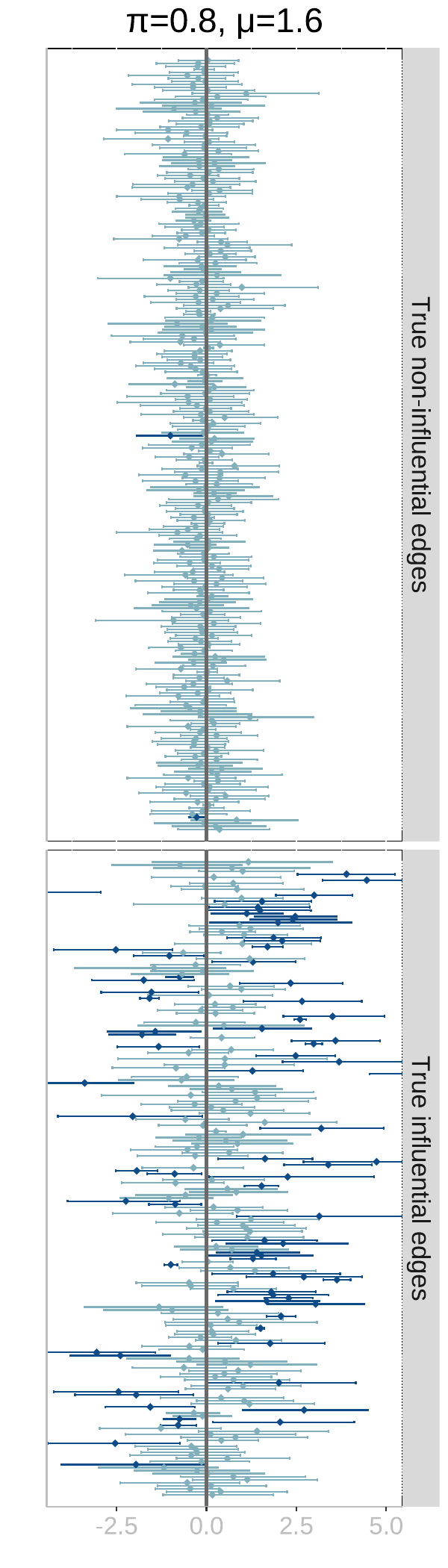}
\caption{$\mathbf{n=1000}$}
\end{subfigure}
\caption[95 \% credible intervals for edge effects (functional redundancy model with random coefficients) with $k=8$ sampled nodes]{{\bf 95 \% credible intervals for edge effects (functional redundancy model with random coefficients) with $k=8$ sampled nodes.} Top (a): Sample size of $n=500$. Bottom (b): Sample size of $n=1000$. Each panel corresponds to a scenario of $\pi=0.3, 0.8$ (which controls the sparsity of the regression coefficient matrix $\mathbf B$) and $\mu=0.8,1.6$. We plot the 95 \% credible intervals for the regression coefficients per edge ordered depending on whether they are truly non-influential edges (top of each panel) or truly influential edges (bottom of each panel). The color of the intervals depends on whether it intersects zero (light) and hence estimated to be non-influential or does not intersect zero (dark) and hence estimated to be influential by the model. These panels allow us to visualize false positives (dark intervals on the top panel) or false negatives (light intervals on the bottom panel).}
\label{fig:edges_red_rand}
\end{figure}

\begin{figure}[!ht]
\centering
\begin{subfigure}[b]{0.49\textwidth}
\centering
\includegraphics[scale=0.19]{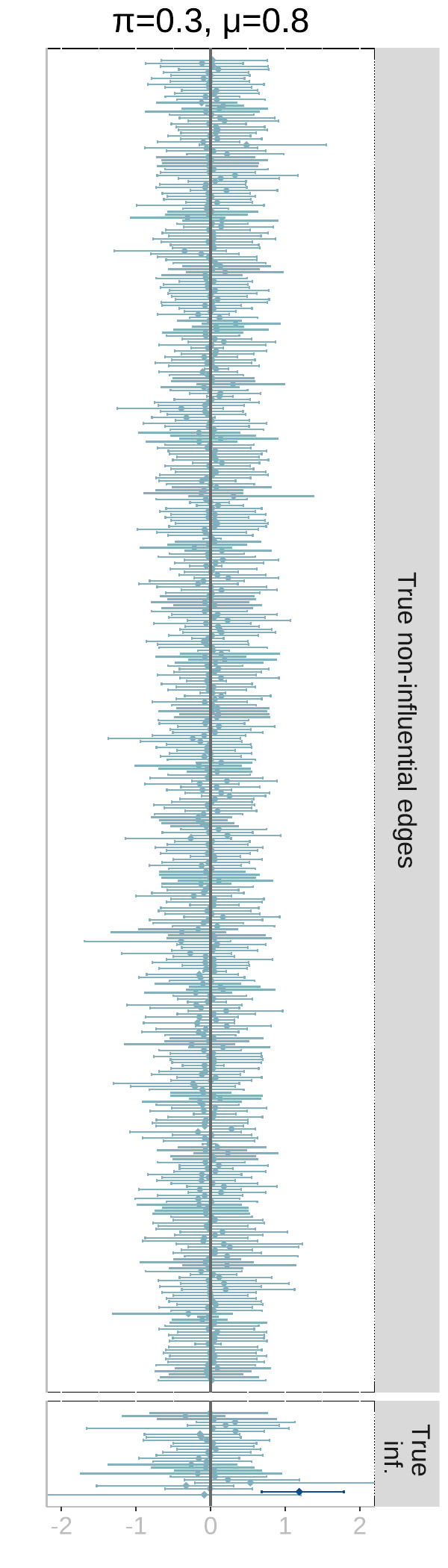}
\includegraphics[scale=0.19]{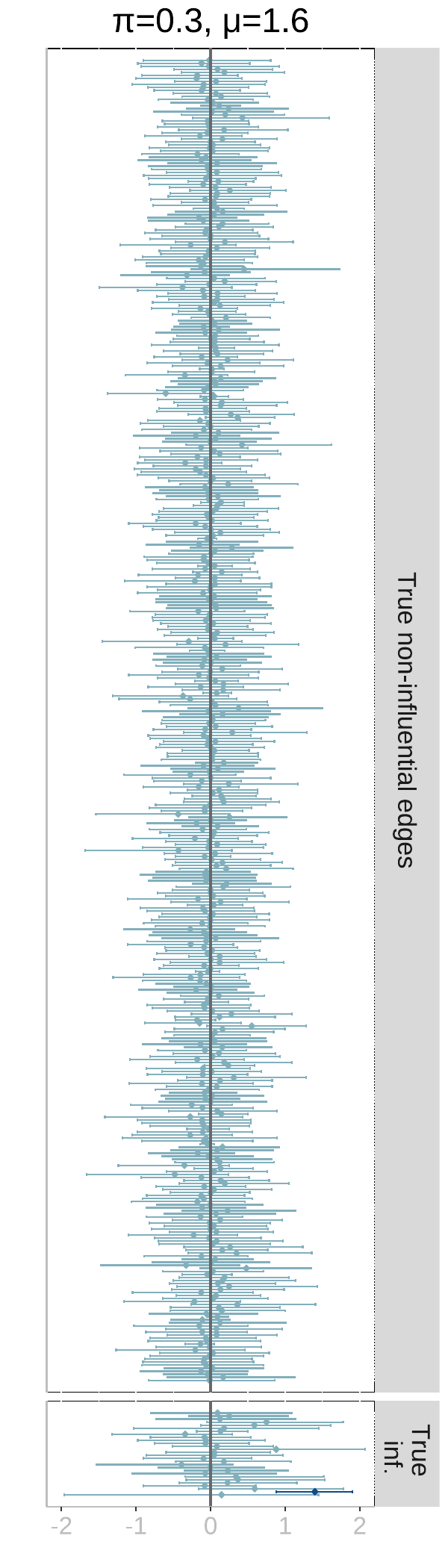}
\includegraphics[scale=0.19]{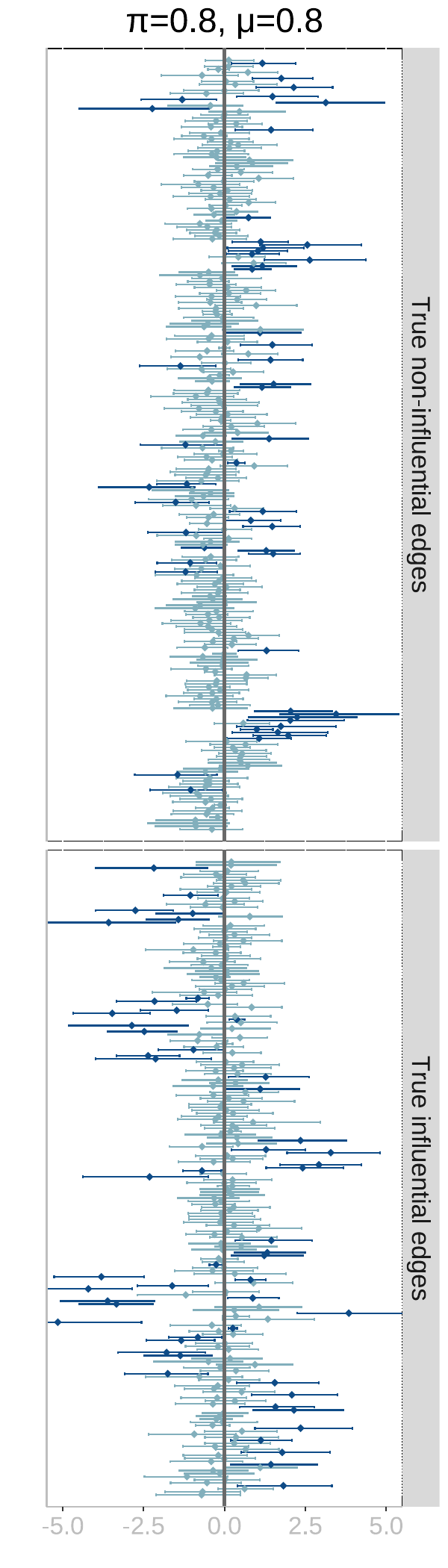}
\includegraphics[scale=0.19]{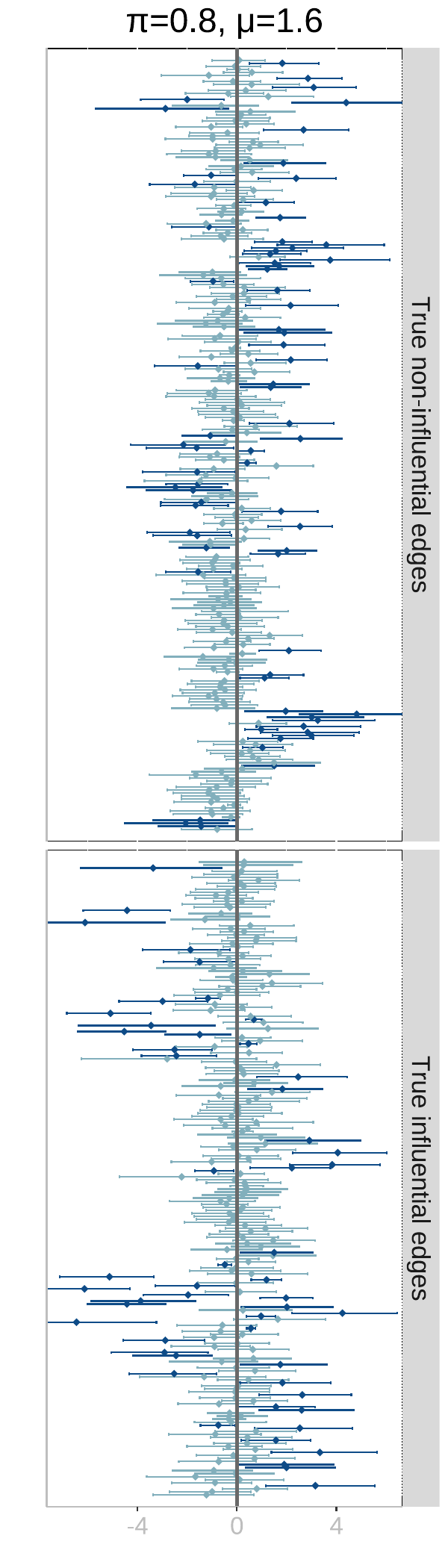}
\caption{$\mathbf{n=500}$}
\end{subfigure}
\begin{subfigure}[b]{0.49\textwidth}
\centering
\includegraphics[scale=0.19]{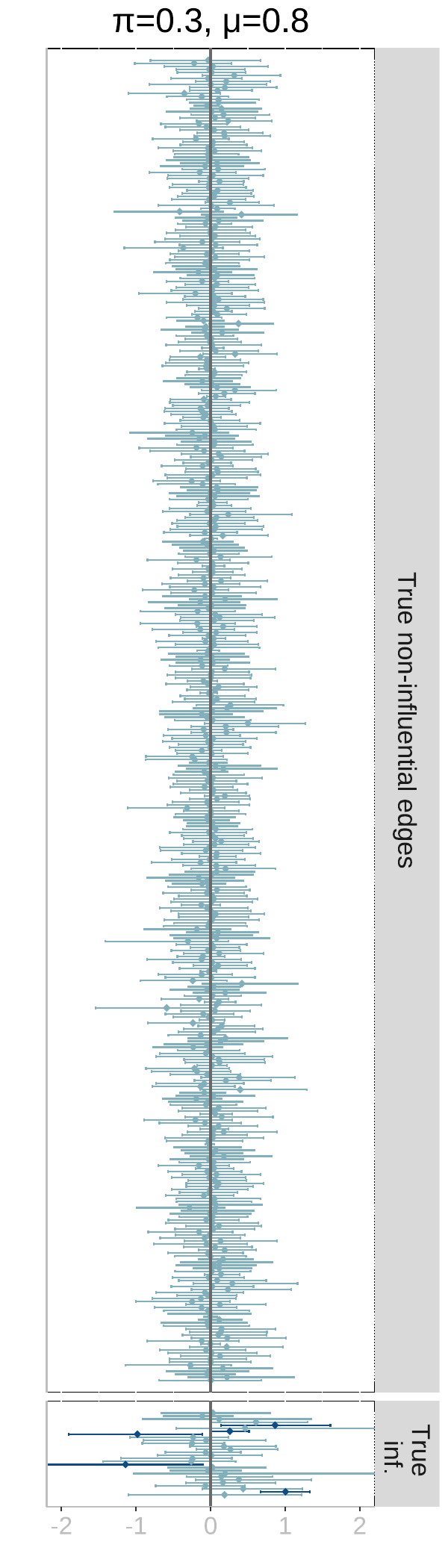}
\includegraphics[scale=0.19]{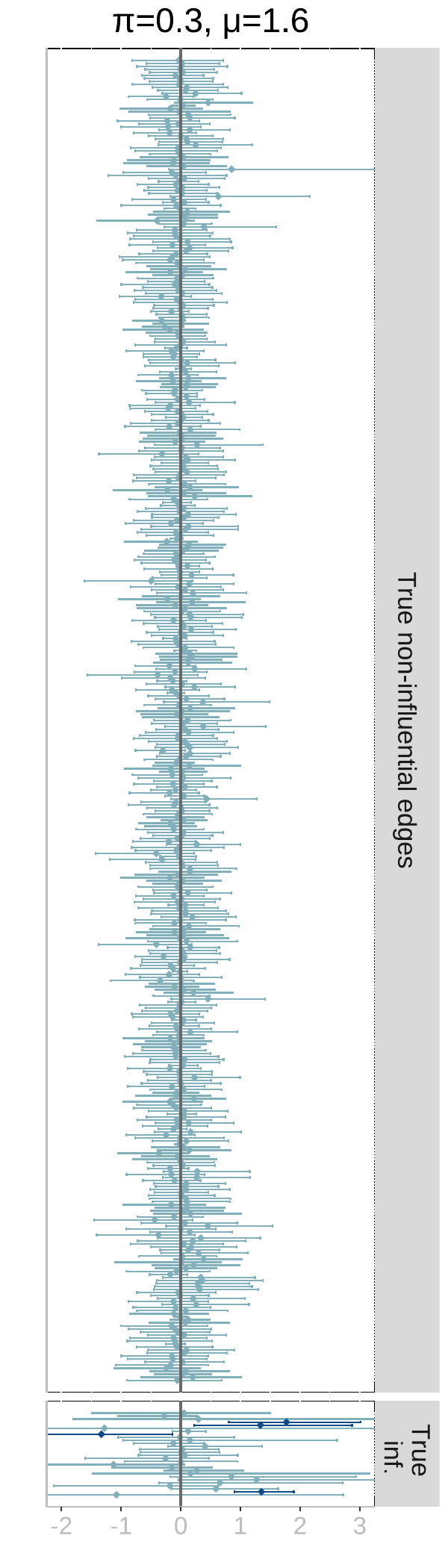}
\includegraphics[scale=0.19]{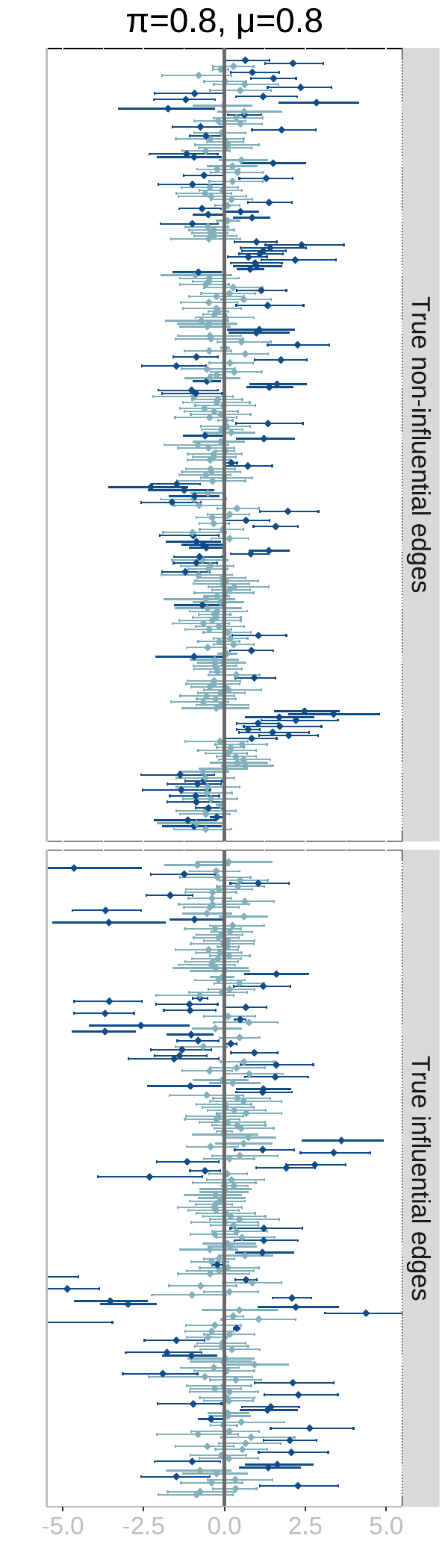}
\includegraphics[scale=0.19]{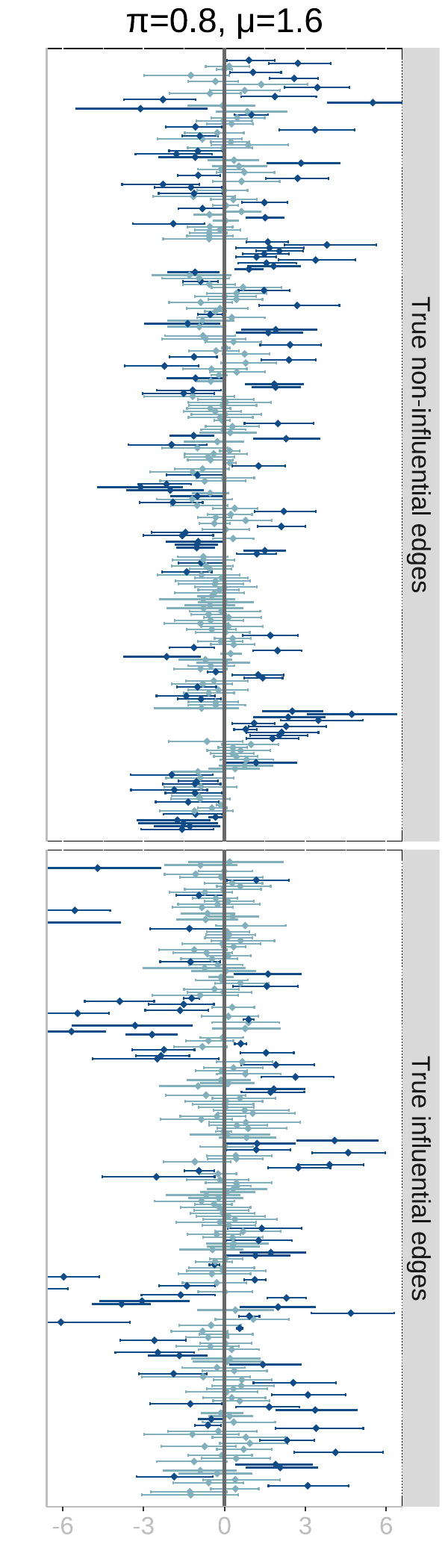}
\caption{$\mathbf{n=500}$}
\end{subfigure}
\caption[95 \% credible intervals for edge effects (functional redundancy model with random coefficients) with $k=22$ sampled nodes]{{\bf 95 \% credible intervals for edge effects (functional redundancy model with random coefficients) with $k=22$ sampled nodes.} Top (a): Sample size of $n=500$. Bottom (b): Sample size of $n=1000$. Each panel corresponds to a scenario of $\pi=0.3, 0.8$ (which controls the sparsity of the regression coefficient matrix $\mathbf B$) and $\mu=0.8,1.6$. We plot the 95 \% credible intervals for the regression coefficients per edge ordered depending on whether they are truly non-influential edges (top of each panel) or truly influential edges (bottom of each panel). The color of the intervals depends on whether it intersects zero (light) and hence estimated to be non-influential or does not intersect zero (dark) and hence estimated to be influential by the model. These panels allow us to visualize false positives (dark intervals on the top panel) or false negatives (light intervals on the bottom panel).}
\label{fig:edges_red_rand2}
\end{figure}

\begin{figure}[!ht]
\centering
\begin{subfigure}[b]{0.49\textwidth}
\centering
\includegraphics[scale=0.19]{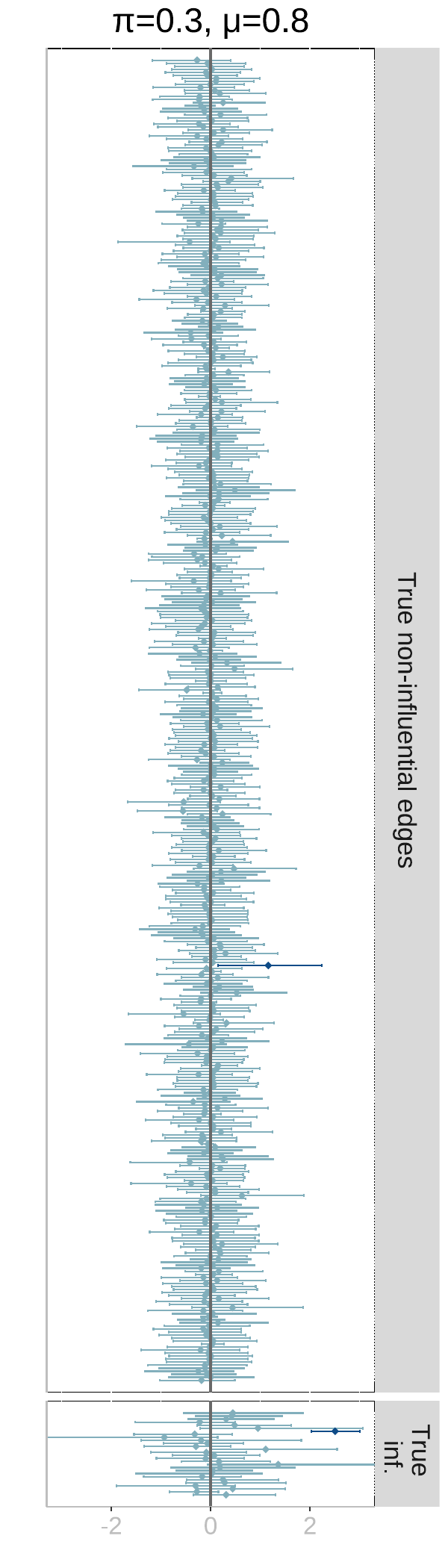}
\includegraphics[scale=0.19]{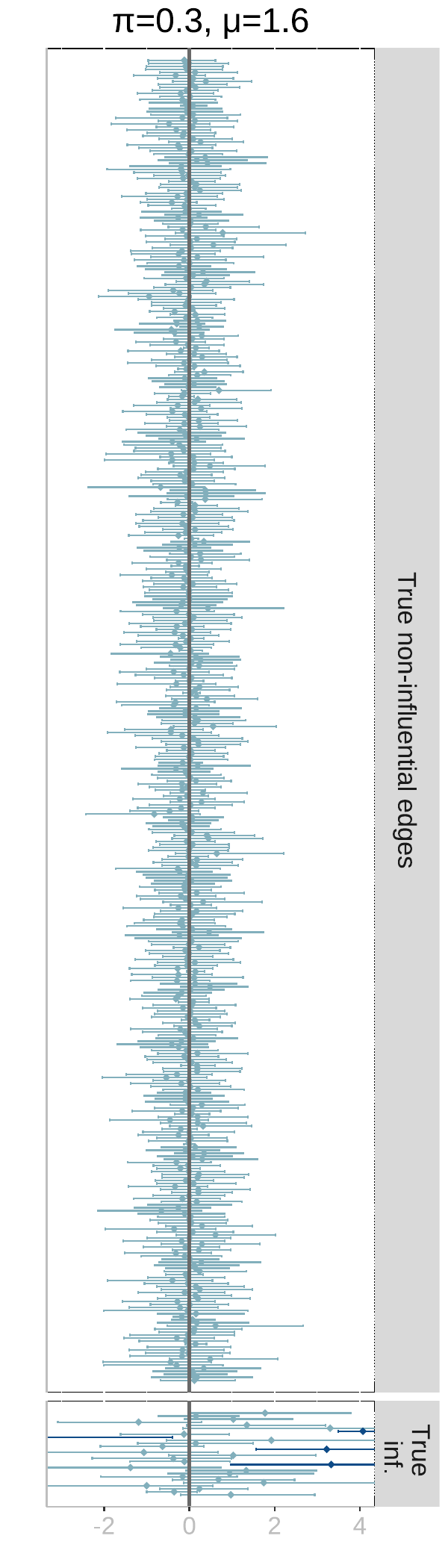}
\includegraphics[scale=0.19]{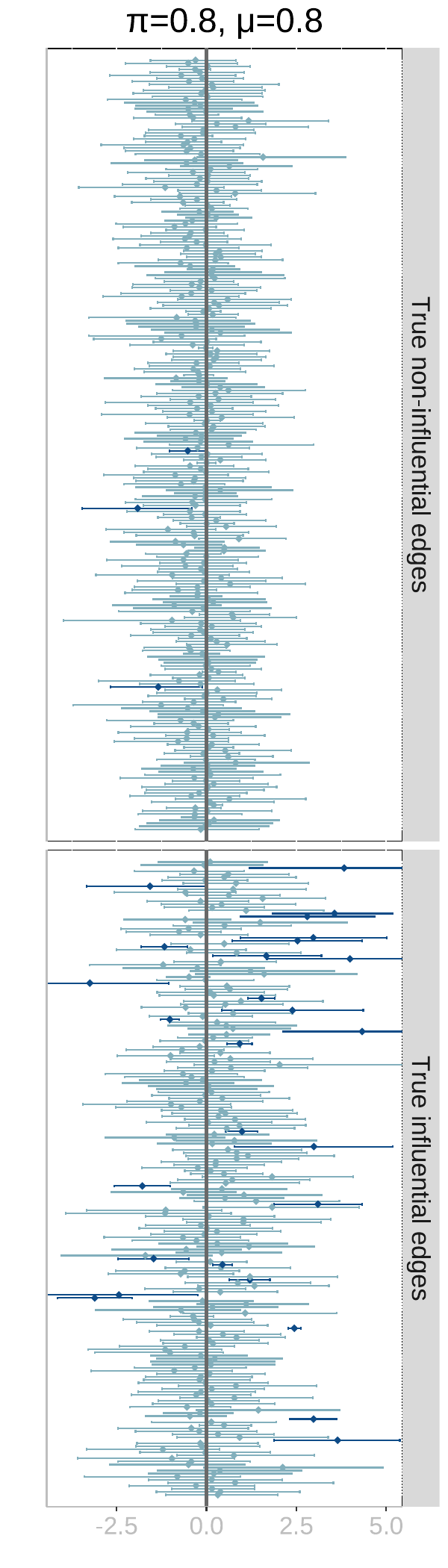}
\includegraphics[scale=0.19]{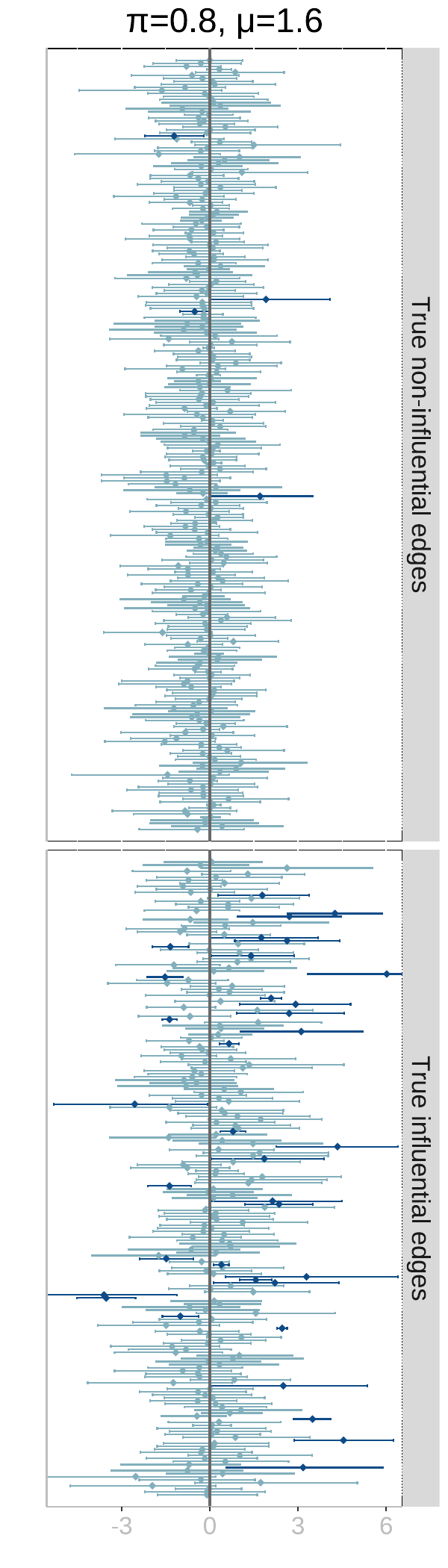}
\caption{$\mathbf{n=500}$}
\end{subfigure}
\begin{subfigure}[b]{0.49\textwidth}
\centering
\includegraphics[scale=0.19]{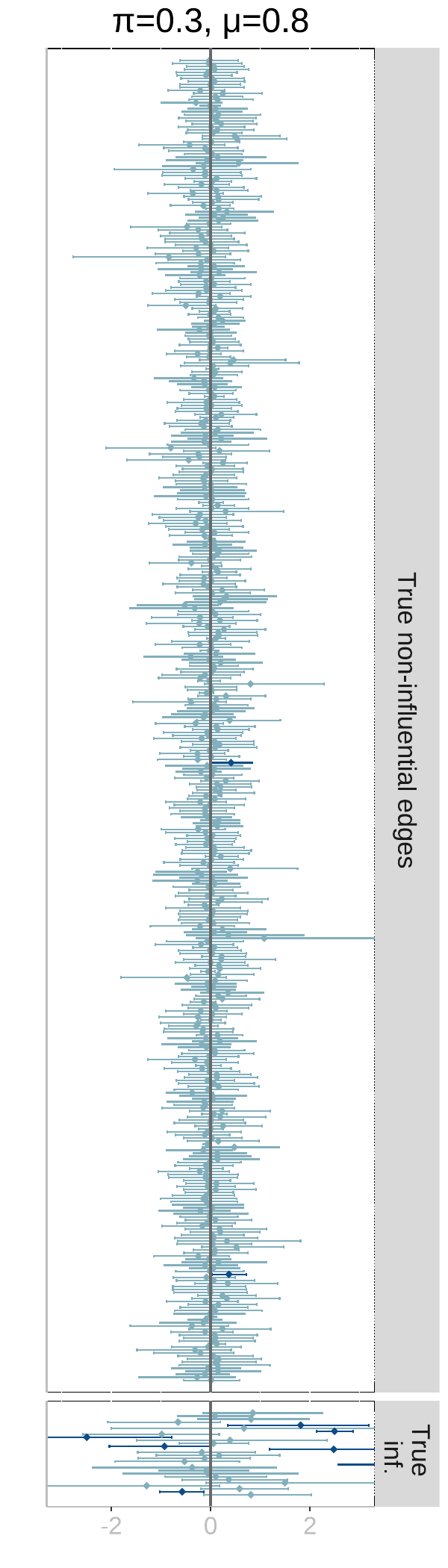}
\includegraphics[scale=0.19]{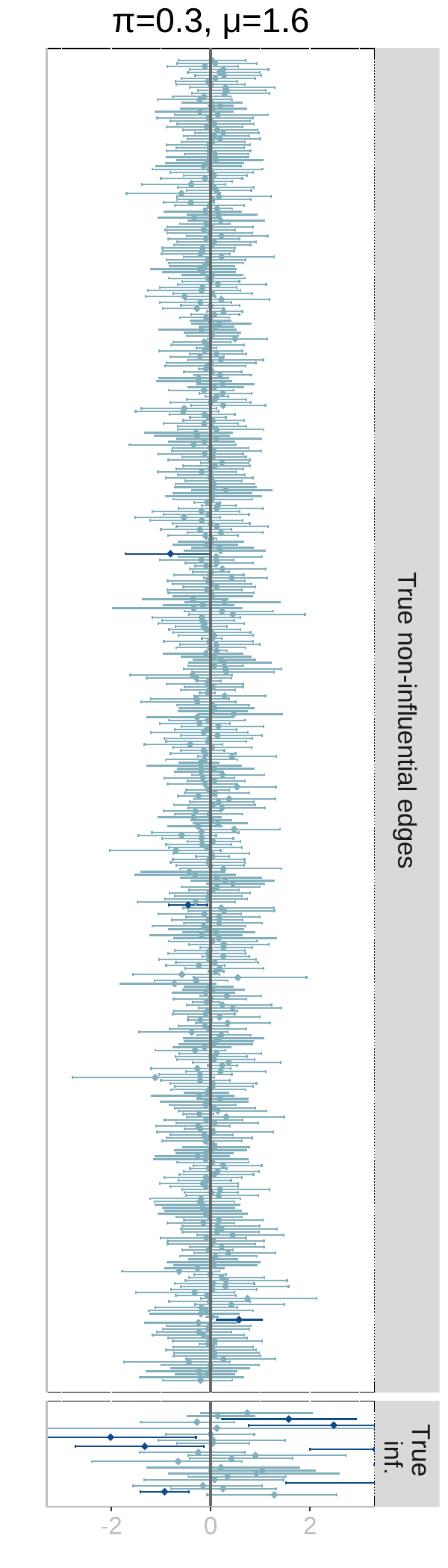}
\includegraphics[scale=0.19]{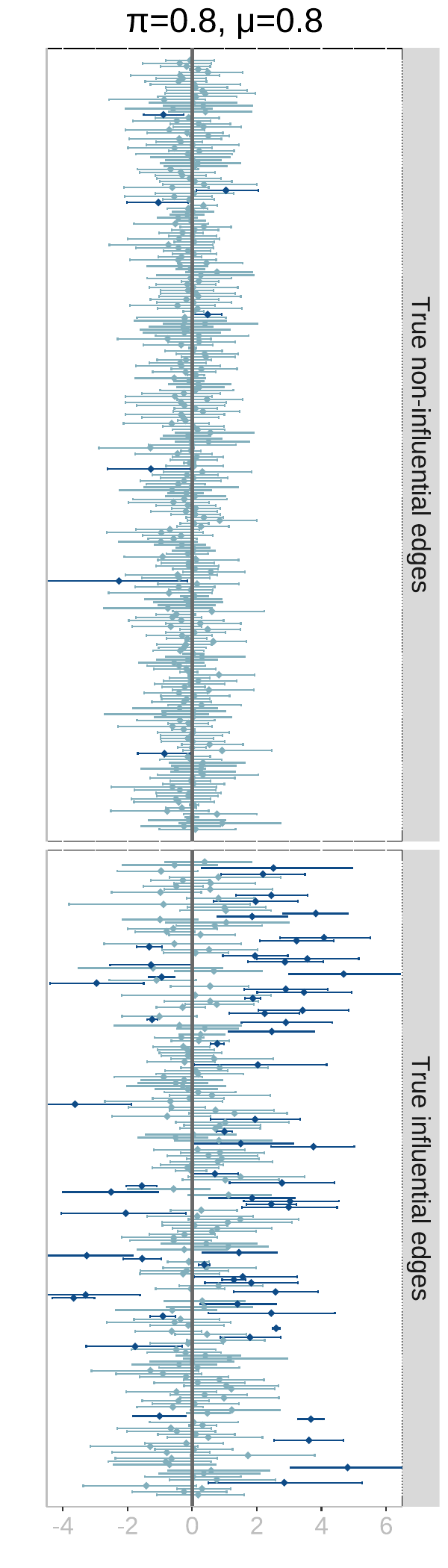}
\includegraphics[scale=0.19]{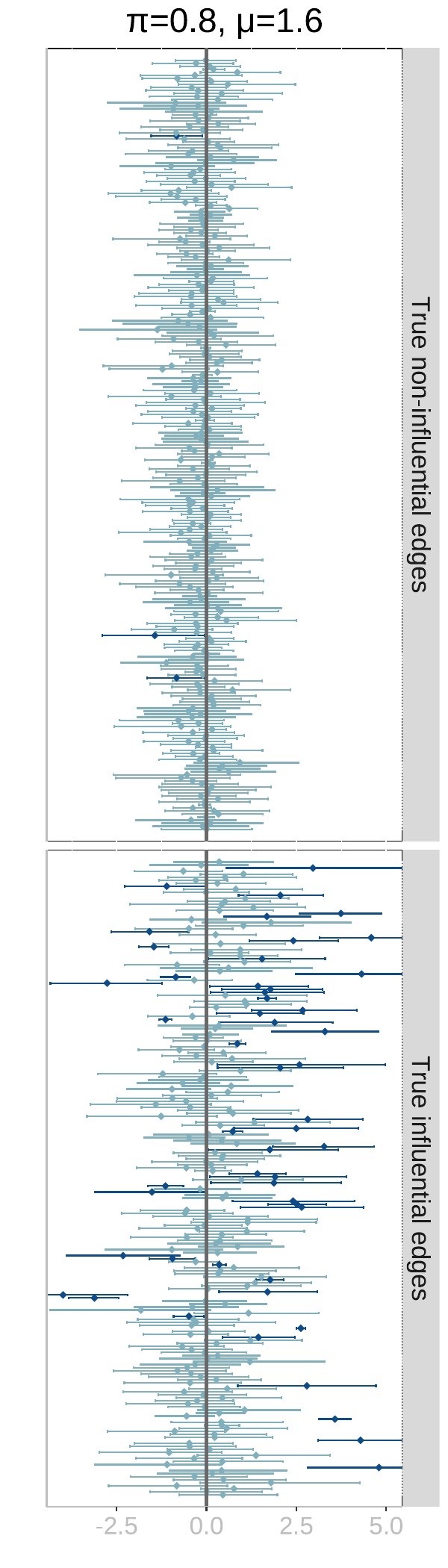}
\caption{$\mathbf{n=1000}$}
\end{subfigure}
\caption[95 \% credible intervals for edge effects (functional redundancy model with phylogenetic coefficients) with $k=8$ sampled nodes]{{\bf 95 \% credible intervals for edge effects (functional redundancy model with phylogenetic coefficients) with $k=8$ sampled nodes.} Top (a): Sample size of $n=500$. Bottom (b): Sample size of $n=1000$. Each panel corresponds to a scenario of $\pi=0.3, 0.8$ (which controls the sparsity of the regression coefficient matrix $\mathbf B$) and $\mu=0.8,1.6$. We plot the 95 \% credible intervals for the regression coefficients per edge ordered depending on whether they are truly non-influential edges (top of each panel) or truly influential edges (bottom of each panel). The color of the intervals depends on whether it intersects zero (light) and hence estimated to be non-influential or does not intersect zero (dark) and hence estimated to be influential by the model. These panels allow us to visualize false positives (dark intervals on the top panel) or false negatives (light intervals on the bottom panel).}
\label{fig:edges_red_phylo}
\end{figure}

\begin{figure}[!ht]
    \centering
    \begin{subfigure}[b]{0.49\textwidth}
    \centering
    \includegraphics[scale=0.19]{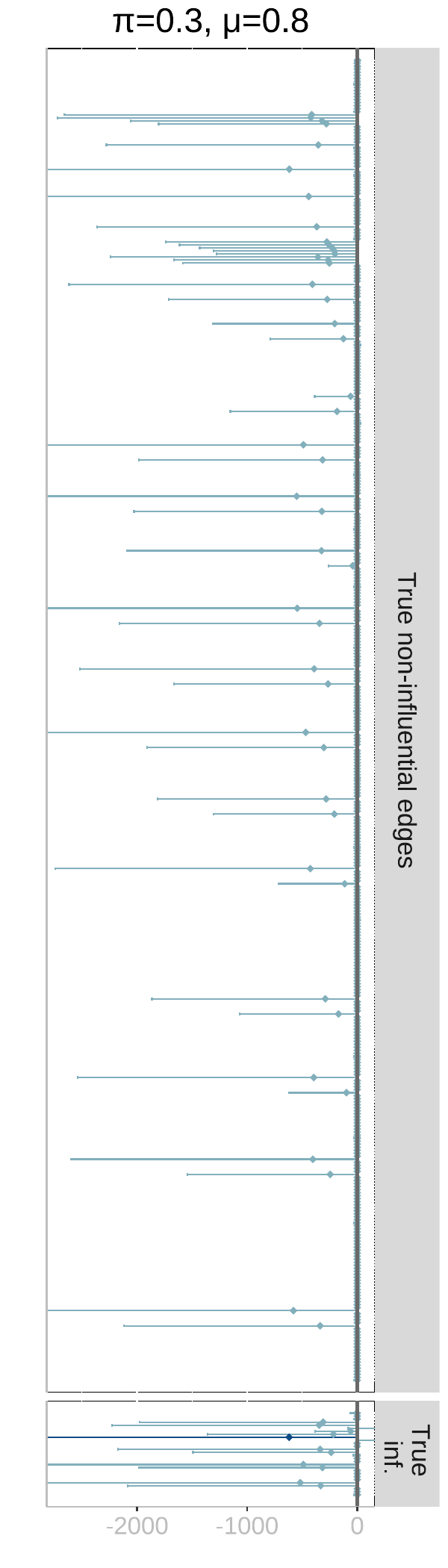}
    \includegraphics[scale=0.19]{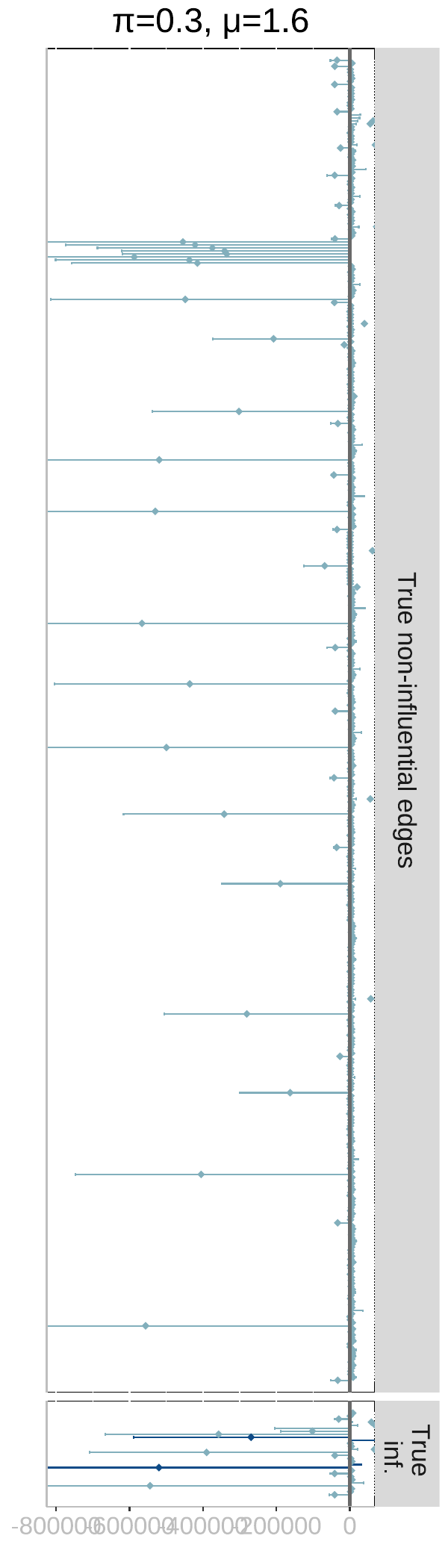}
    \includegraphics[scale=0.19]{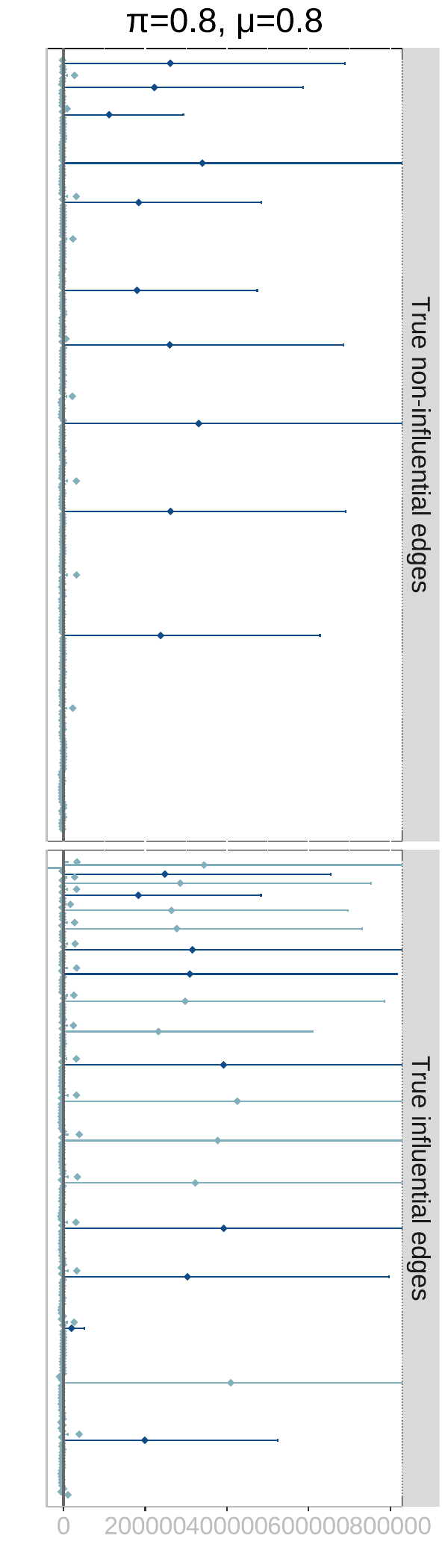}
    \includegraphics[scale=0.19]{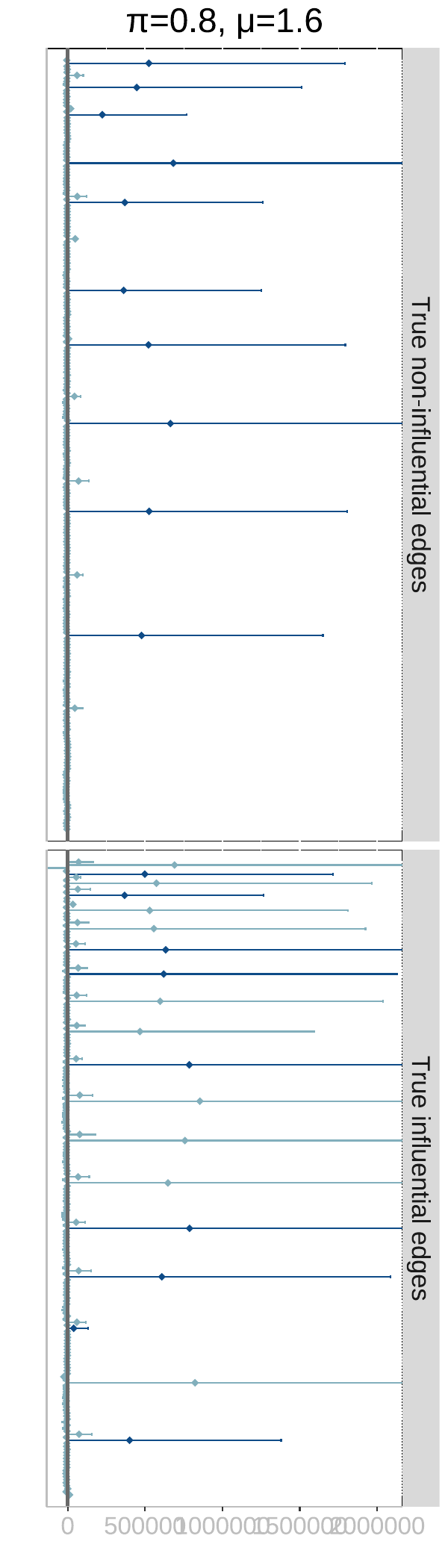}
    \caption{$\mathbf{n=500}$}
    \end{subfigure}
    \begin{subfigure}[b]{0.49\textwidth}
    \centering
    \includegraphics[scale=0.19]{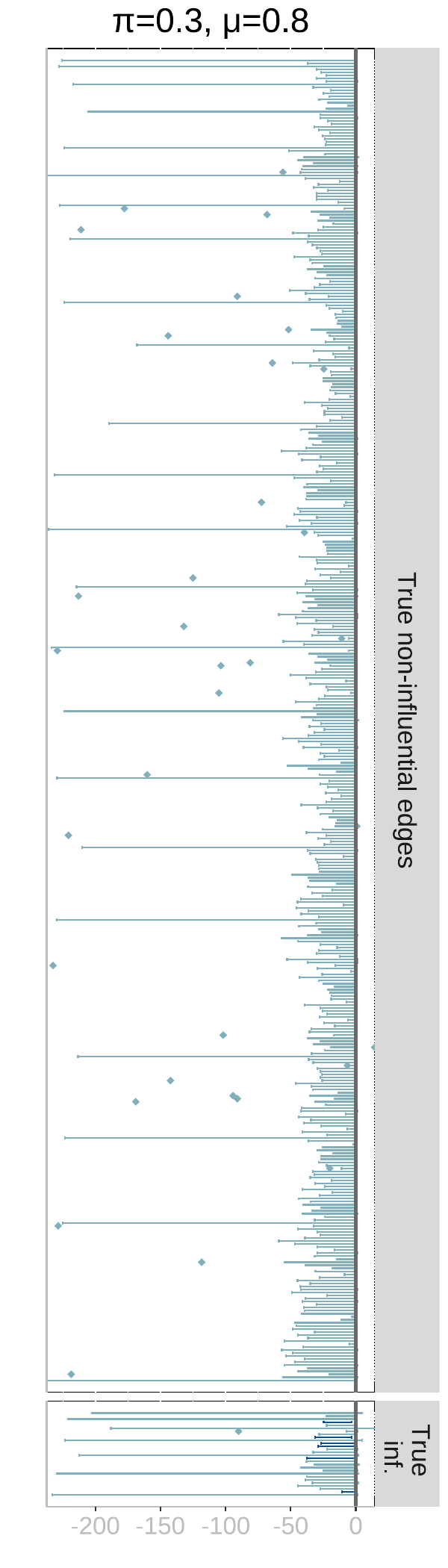}
    \includegraphics[scale=0.19]{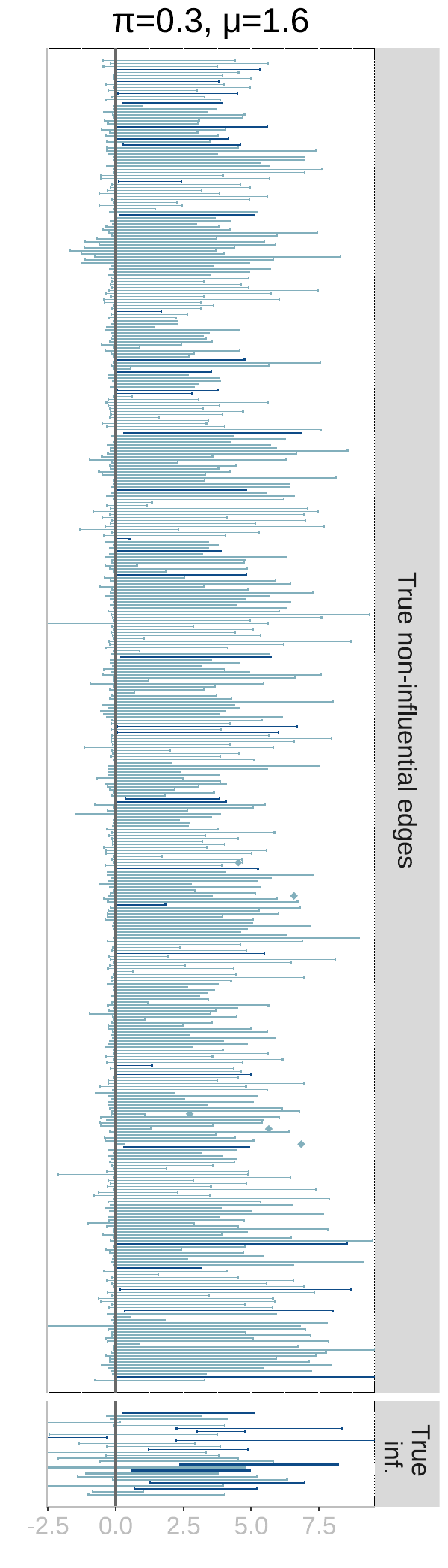}
    \includegraphics[scale=0.19]{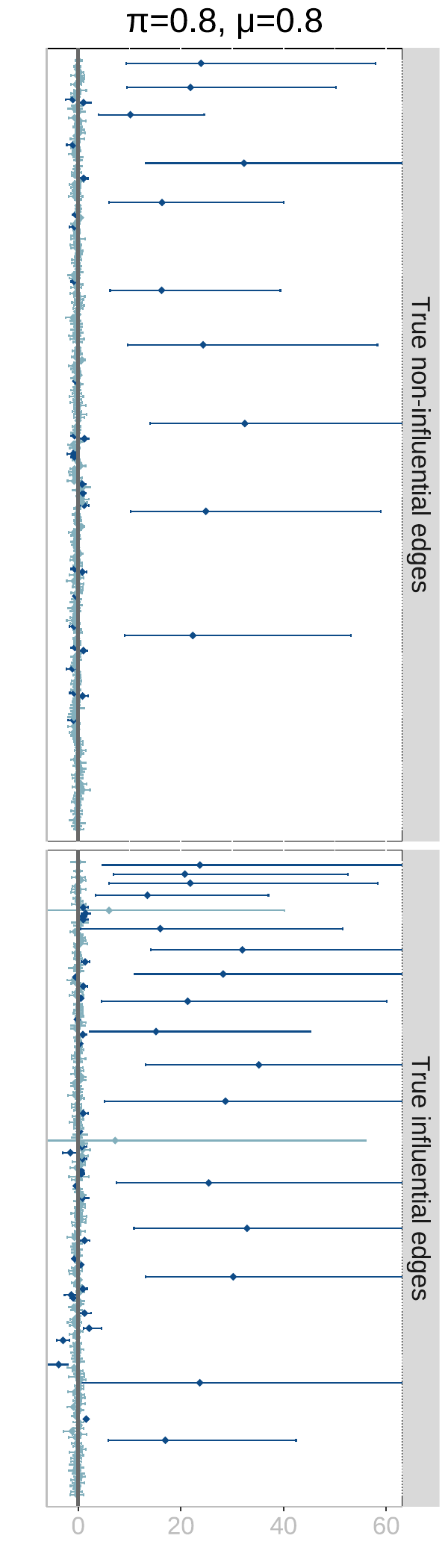}
    \includegraphics[scale=0.19]{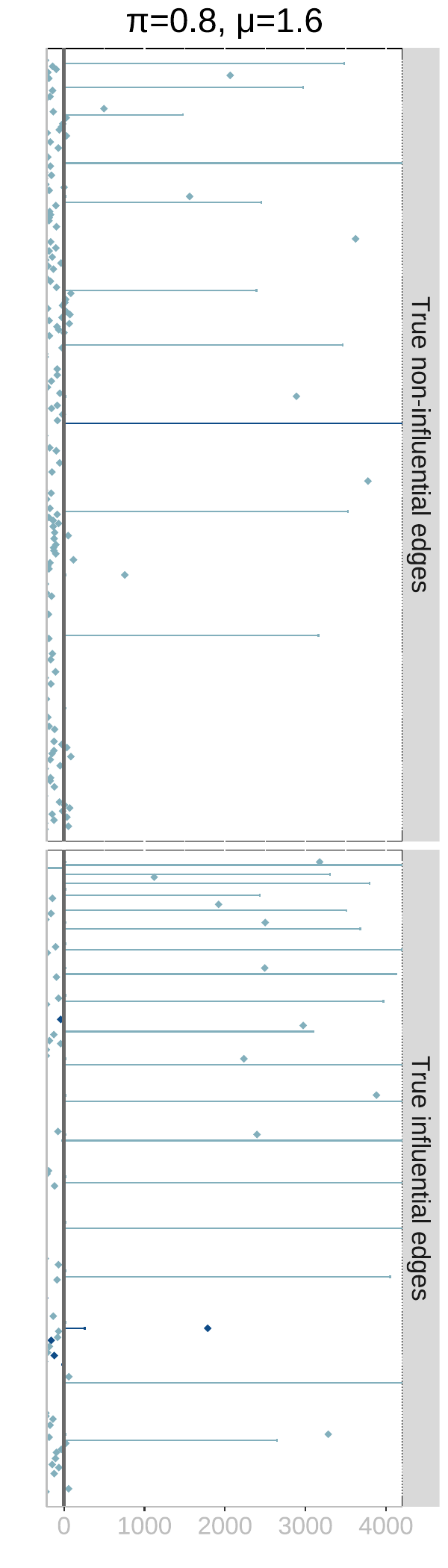}
    \caption{$\mathbf{n=1000}$}
    \end{subfigure}
    \caption[95 \% credible intervals for edge effects (functional redundancy model with phylogenetic coefficients) with $k=22$ sampled nodes]{{\bf 95 \% credible intervals for edge effects (functional redundancy model with phylogenetic coefficients) with $k=22$ sampled nodes.} Top (a): Sample size of $n=500$. Bottom (b): Sample size of $n=1000$. Each panel corresponds to a scenario of $\pi=0.3, 0.8$ (which controls the sparsity of the regression coefficient matrix $\mathbf B$) and $\mu=0.8,1.6$. We plot the 95 \% credible intervals for the regression coefficients per edge ordered depending on whether they are truly non-influential edges (top of each panel) or truly influential edges (bottom of each panel). The color of the intervals depends on whether it intersects zero (light) and hence estimated to be non-influential or does not intersect zero (dark) and hence estimated to be influential by the model. These panels allow us to visualize false positives (dark intervals on the top panel) or false negatives (light intervals on the bottom panel).}
    \label{fig:edges_red_phylo2}
\end{figure}

\begin{figure}[!ht]
    \centering
    \begin{subfigure}[t]{0.49\textwidth}
        \centering
        \includegraphics[scale=0.27]{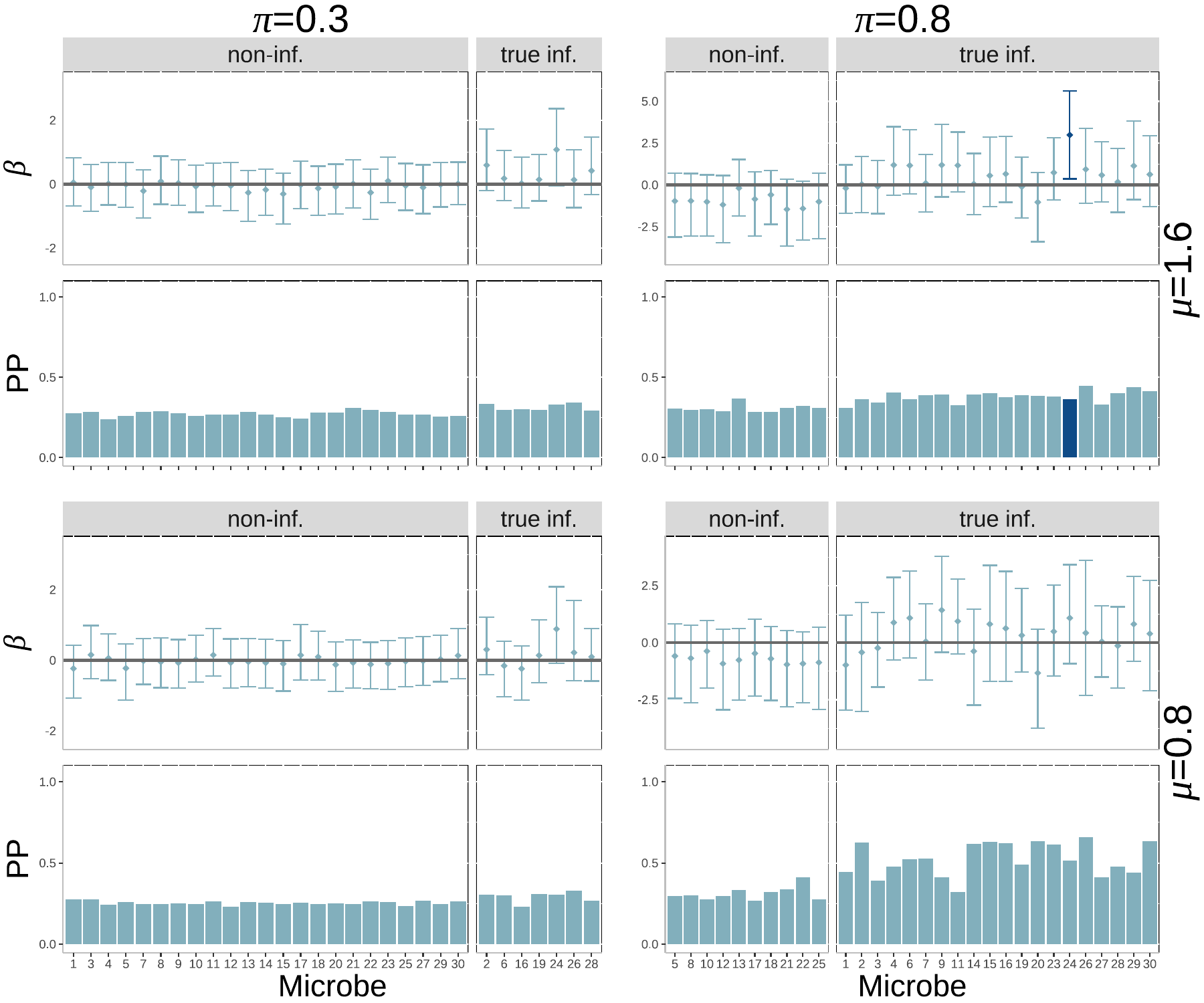}
        \caption{$\mathbf{n=500}$}
    \end{subfigure}
    \begin{subfigure}[t]{0.49\textwidth}
        \centering
        \includegraphics[scale=0.27]{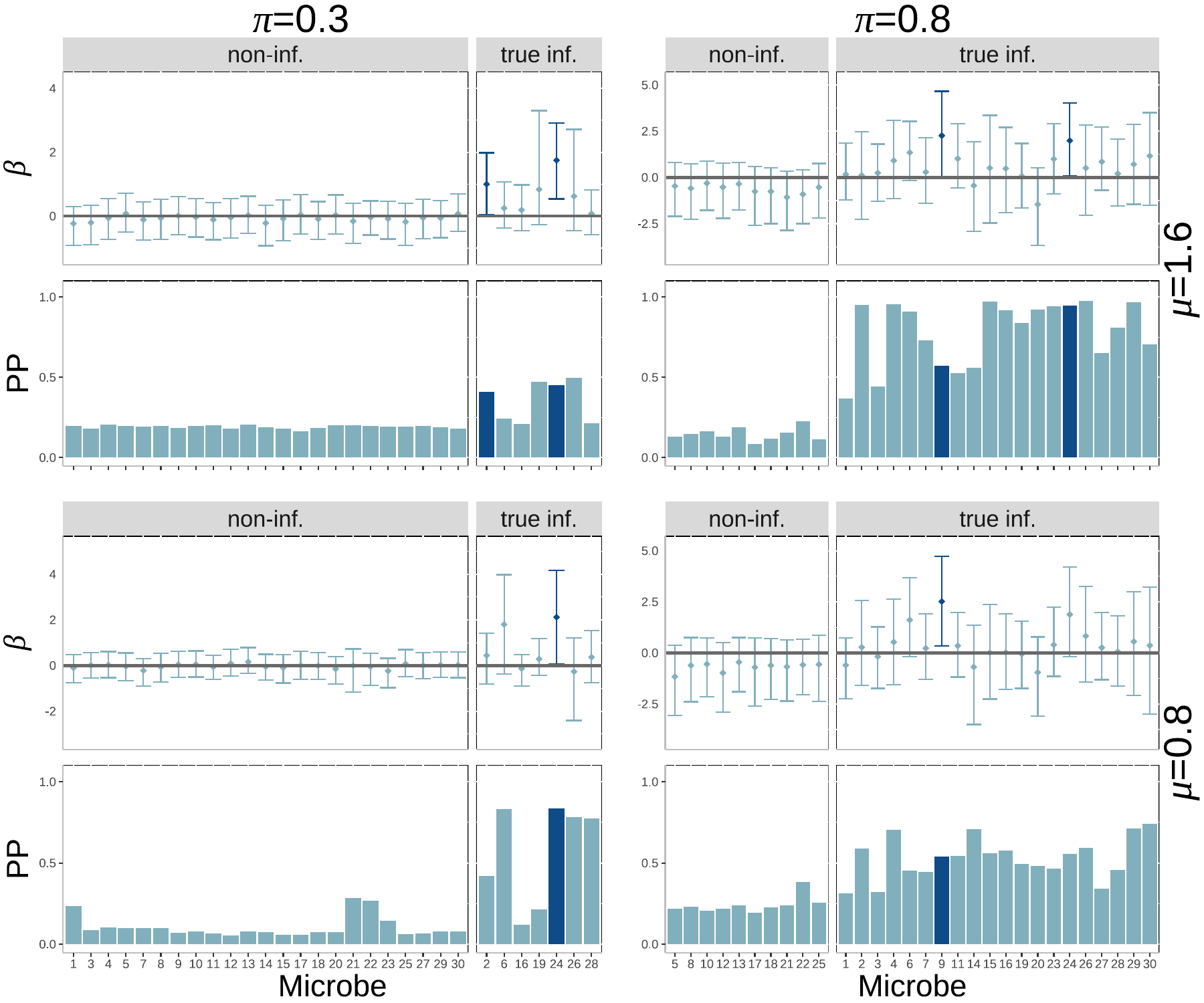}
        \caption{$\mathbf{n=1000}$}
    \end{subfigure}
    \caption[Posterior probability of influential nodes and coefficients for nodes (functional redundancy model with random coefficients)]{{\bf Posterior probability of influential nodes and coefficients for nodes (functional redundancy model with random coefficients) with $k=8$ sampled nodes.}
    Different \revision{groups of four} panels represent different sample sizes ($n=500,1000$). \revision{Within each group, we have four panels corresponding to the two values of edge effect size ($\mu=0.8, 1.6$) and two values of probability of influential node ($\pi=0.3, 0.8$) which controls the sparsity of the regression coefficient matrix ($\mathbf B$). Within each of these panels we have two plots: 95\% credible intervals (top) and posterior probability of influence (bottom - calculated as the mean of the $\xi$ variable for the node across Gibbs samples) for each node.} Each bar corresponds to one node (microbe). \revision{Within each plot the bars and intervals are colored depending on whether the node is found to be influential (dark) or not influential (light) based on the 95\% credible intervals. Each plot is split based on whether the nodes are truly influential (right) or not (left).}
    } 
    \label{fig:nodes_red_rand_adx_k8}
\end{figure}

\begin{figure}[!ht]
    \centering
    \begin{subfigure}[t]{0.49\textwidth}
        \centering
        \includegraphics[scale=0.27]{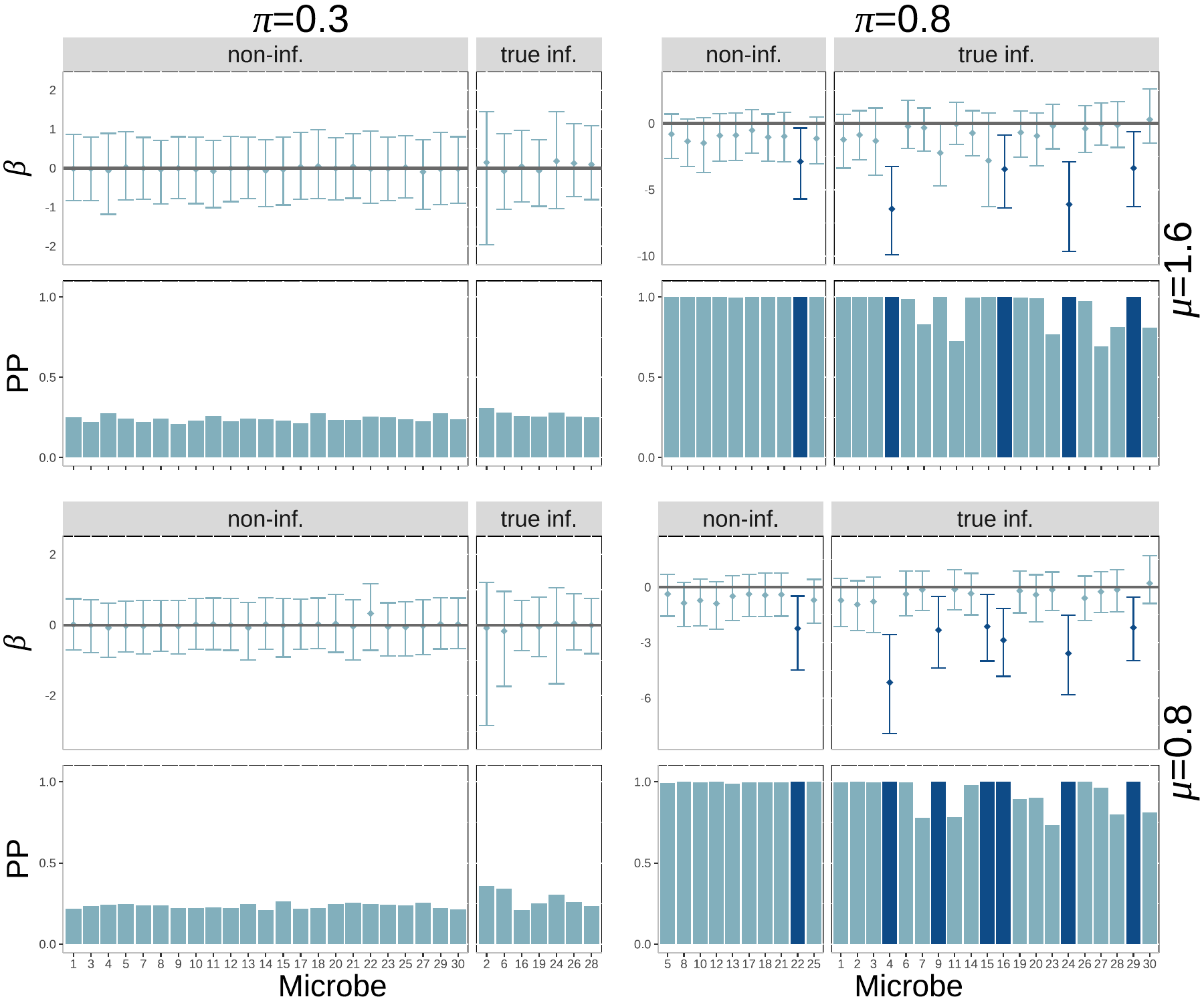}
        \caption{$\mathbf{n=500}$}
    \end{subfigure}
    \begin{subfigure}[t]{0.49\textwidth}
        \centering
        \includegraphics[scale=0.27]{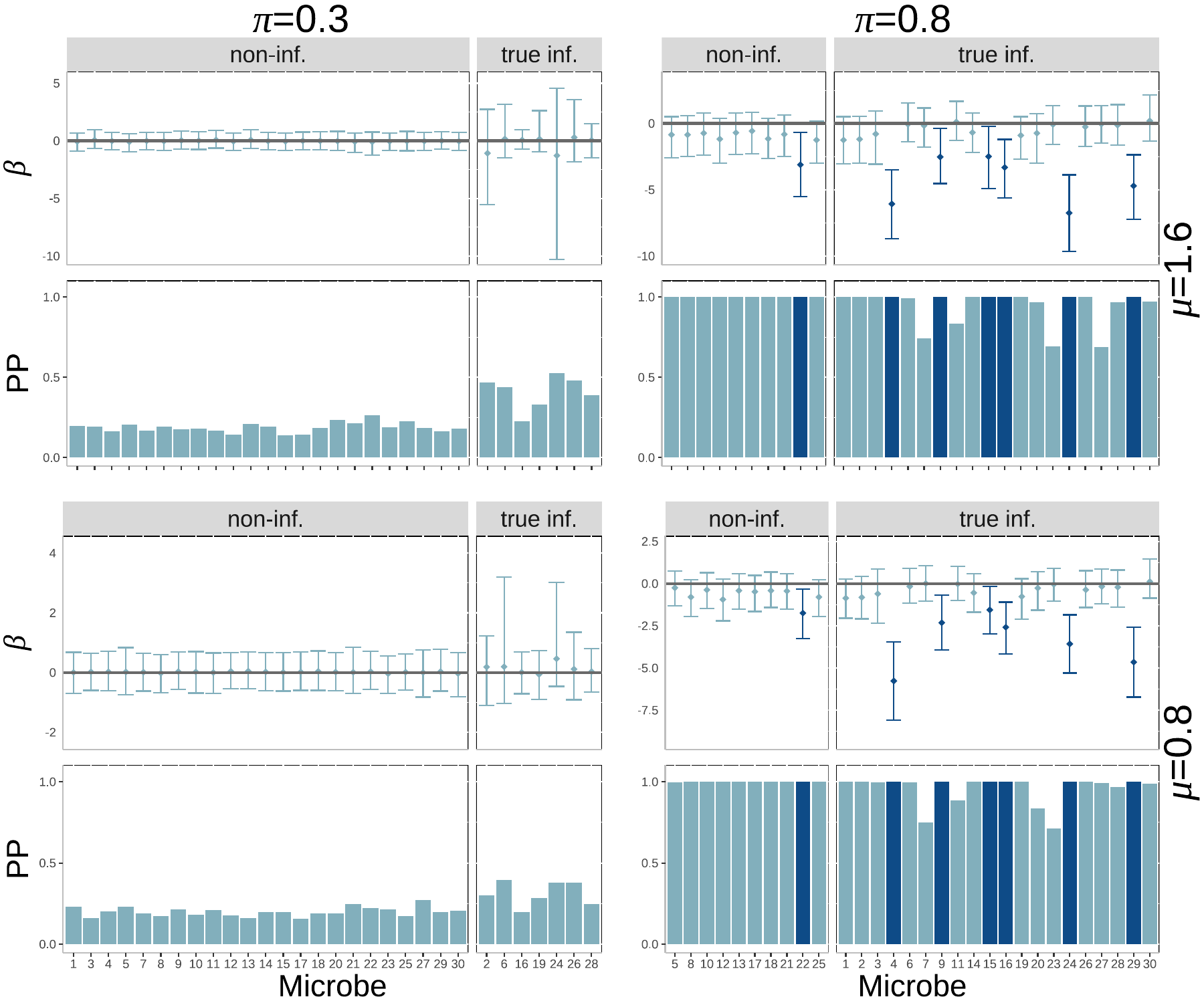}
        \caption{$\mathbf{n=1000}$}
    \end{subfigure}
    \caption[Posterior probability of influential nodes and coefficients for nodes (functional redundancy model with random coefficients)]{{\bf Posterior probability of influential nodes and coefficients for nodes (functional redundancy model with random coefficients) with $k=22$ sampled nodes.}
    Different \revision{groups of four} panels represent different sample sizes ($n=500,1000$). \revision{Within each group, we have four panels corresponding to the two values of edge effect size ($\mu=0.8, 1.6$) and two values of probability of influential node ($\pi=0.3, 0.8$) which controls the sparsity of the regression coefficient matrix ($\mathbf B$). Within each of these panels we have two plots: 95\% credible intervals (top) and posterior probability of influence (bottom - calculated as the mean of the $\xi$ variable for the node across Gibbs samples) for each node.} Each bar corresponds to one node (microbe). \revision{Within each plot the bars and intervals are colored depending on whether the node is found to be influential (dark) or not influential (light) based on the 95\% credible intervals. Each plot is split based on whether the nodes are truly influential (right) or not (left).}
    } 
    \label{fig:nodes_red_rand_adx_k22}
\end{figure}

\begin{figure}[!ht]
    \centering
    \begin{subfigure}[t]{0.49\textwidth}
        \centering
        \includegraphics[scale=0.27]{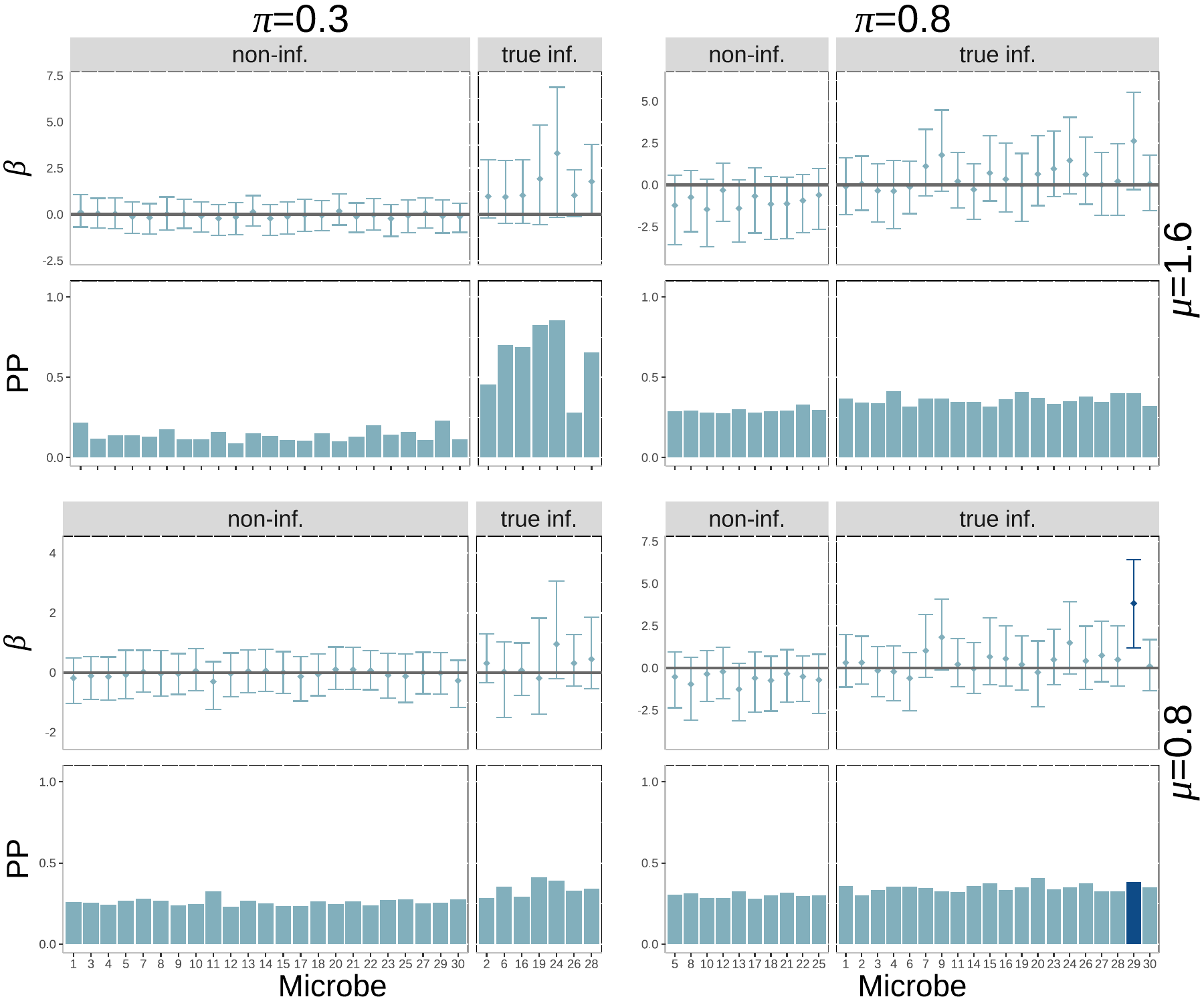}
        \caption{$\mathbf{n=500}$}
    \end{subfigure}
    \begin{subfigure}[t]{0.49\textwidth}
        \centering
        \includegraphics[scale=0.27]{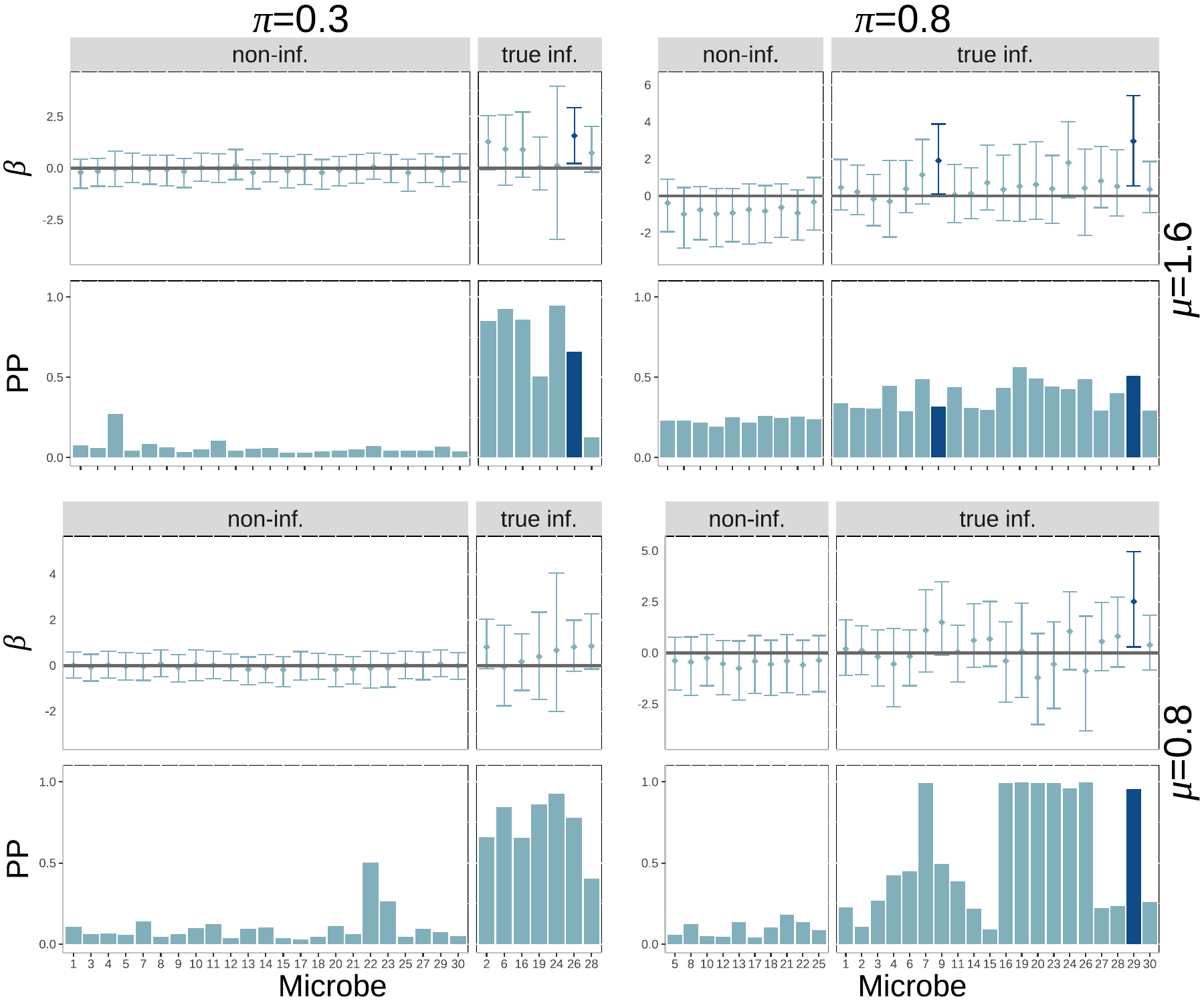}
        \caption{$\mathbf{n=1000}$}
    \end{subfigure}
    \caption[Posterior probability of influential nodes and coefficients for nodes (functional redundancy model with phylogenetic coefficients)]{{\bf Posterior probability of influential nodes and coefficients for nodes (functional redundancy model with phylogenetic coefficients) with $k=8$ sampled nodes.}
    Different \revision{groups of four} panels represent different sample sizes ($n=500,1000$). \revision{Within each group, we have four panels corresponding to the two values of edge effect size ($\mu=0.8, 1.6$) and two values of probability of influential node ($\pi=0.3, 0.8$) which controls the sparsity of the regression coefficient matrix ($\mathbf B$). Within each of these panels we have two plots: 95\% credible intervals (top) and posterior probability of influence (bottom - calculated as the mean of the $\xi$ variable for the node across Gibbs samples) for each node.} Each bar corresponds to one node (microbe). \revision{Within each plot the bars and intervals are colored depending on whether the node is found to be influential (dark) or not influential (light) based on the 95\% credible intervals. Each plot is split based on whether the nodes are truly influential (right) or not (left).}
    } 
    \label{fig:nodes_red_phylo_k8}
\end{figure}

\begin{figure}[!ht]
    \begin{subfigure}[t]{0.49\textwidth}
        \centering
        \includegraphics[scale=0.27]{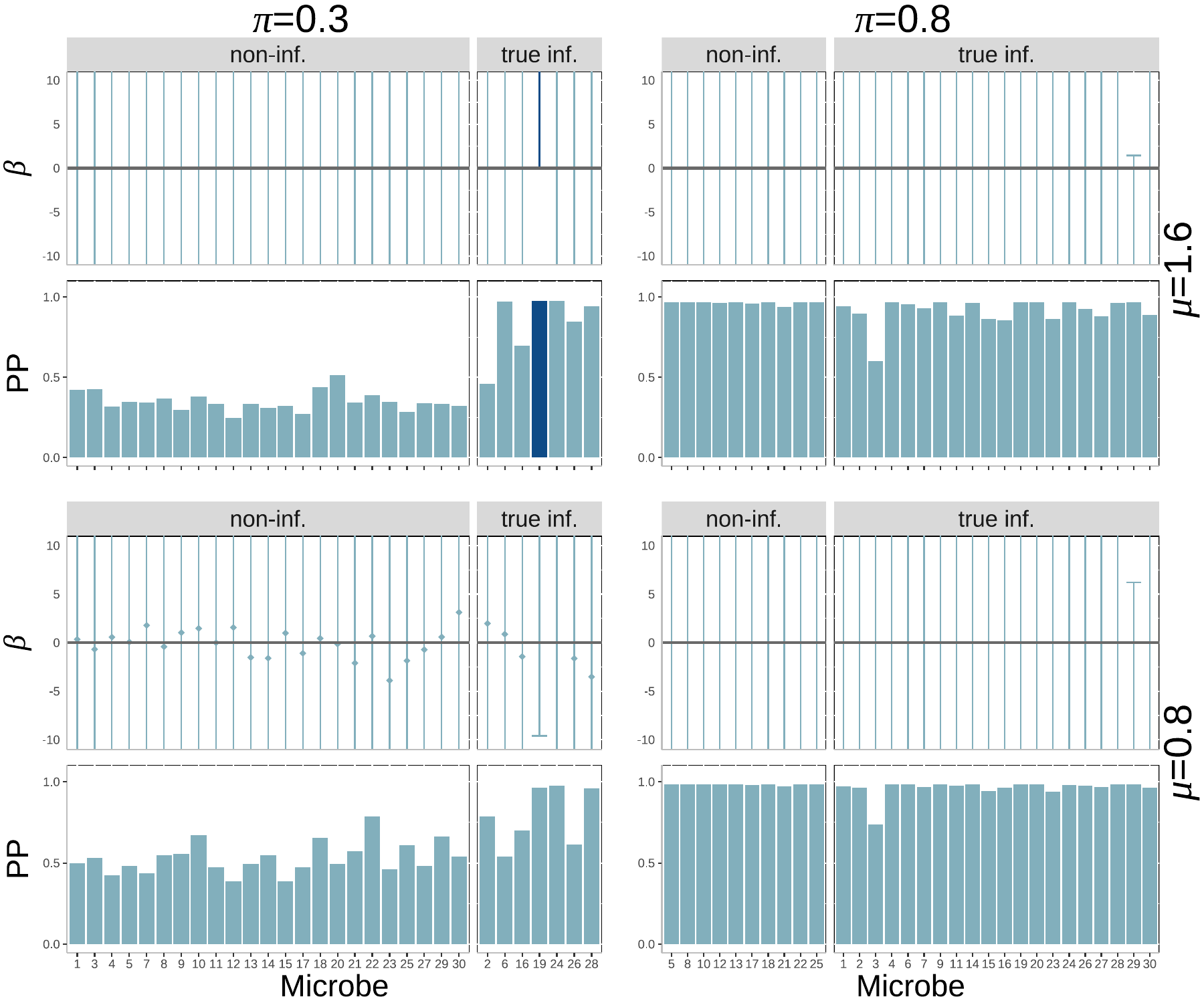}
        \caption{$\mathbf{n=500}$}
    \end{subfigure}
    \begin{subfigure}[t]{0.49\textwidth}
        \centering
        \includegraphics[scale=0.27]{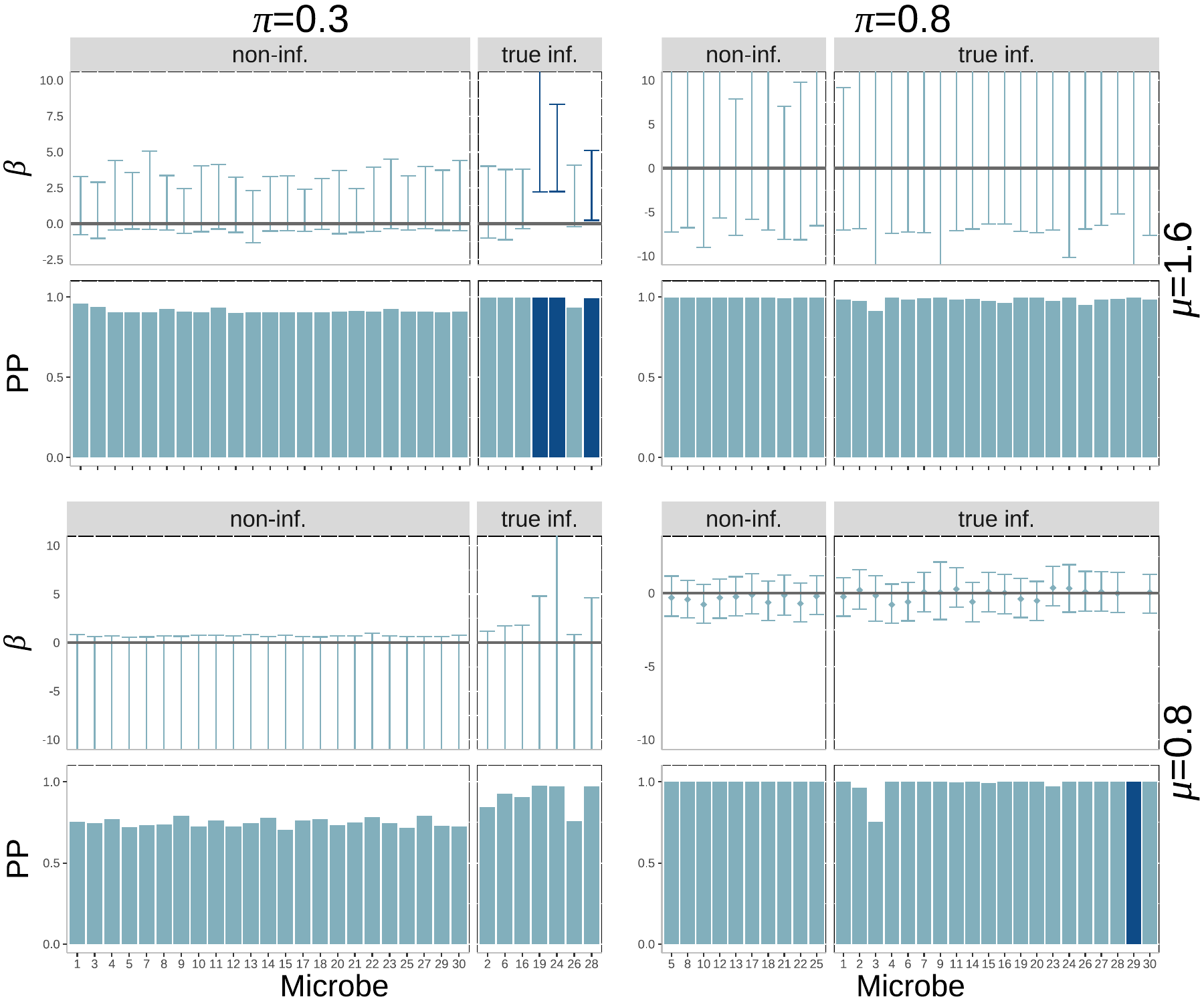}
        \caption{$\mathbf{n=1000}$}
    \end{subfigure}
    \caption[Posterior probability of influential nodes and coefficients for nodes (functional redundancy model with phylogenetic coefficients)]{{\bf Posterior probability of influential nodes and coefficients for nodes (functional redundancy model with phylogenetic coefficients) with $k=22$ sampled nodes.}
    Different \revision{groups of four} panels represent different sample sizes ($n=500,1000$). \revision{Within each group, we have four panels corresponding to the two values of edge effect size ($\mu=0.8, 1.6$) and two values of probability of influential node ($\pi=0.3, 0.8$) which controls the sparsity of the regression coefficient matrix ($\mathbf B$). Within each of these panels we have two plots: 95\% credible intervals (top) and posterior probability of influence (bottom - calculated as the mean of the $\xi$ variable for the node across Gibbs samples) for each node.} Each bar corresponds to one node (microbe). \revision{Within each plot the bars and intervals are colored depending on whether the node is found to be influential (dark) or not influential (light) based on the 95\% credible intervals. Each plot is split based on whether the nodes are truly influential (right) or not (left).}
    } 
    \label{fig:nodes_red_phylo_k22}
\end{figure}

\FloatBarrier
\subsubsection{The Effect of Data Augmentation}
\FloatBarrier

\begin{figure}[!ht]
    \centering
    \includegraphics[scale=0.40]{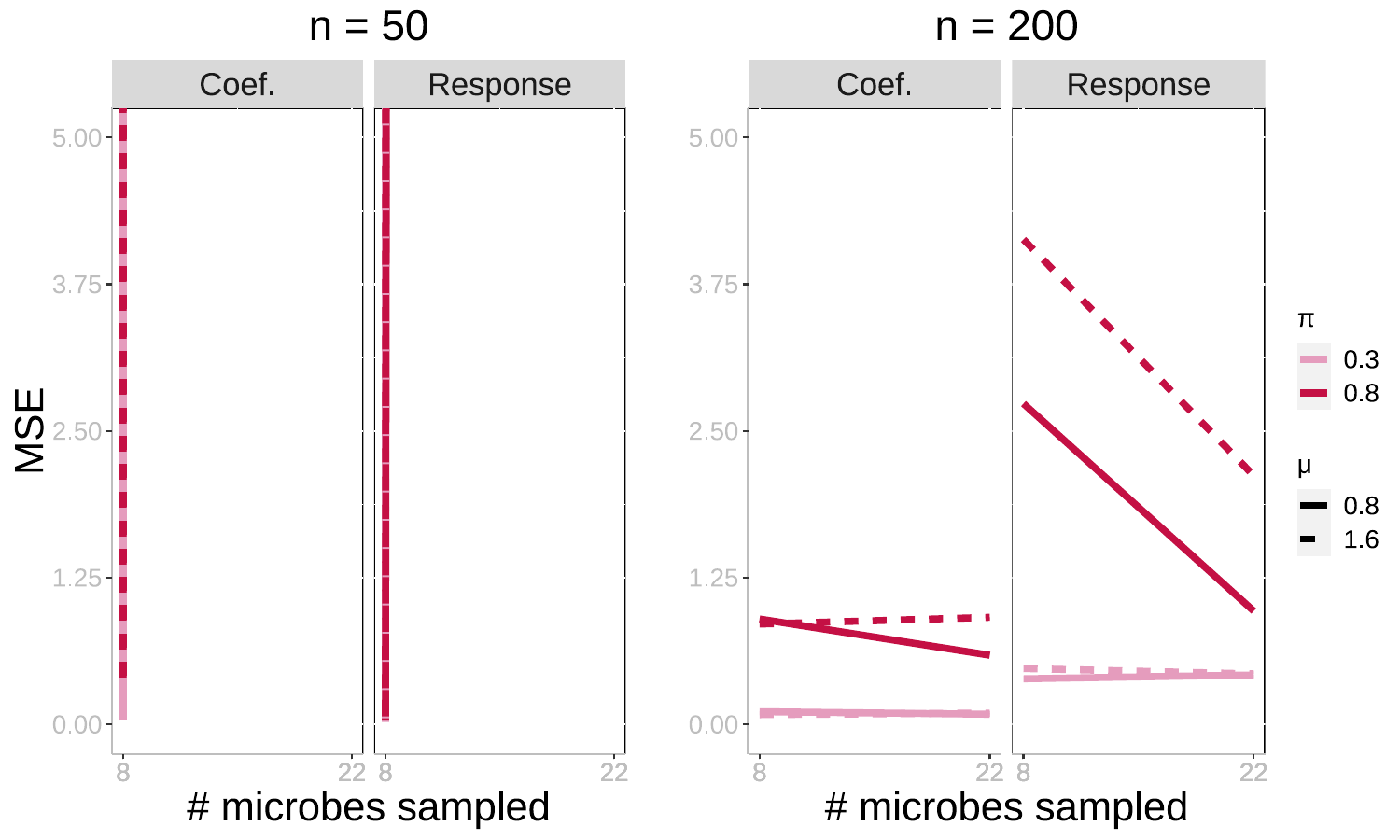}
    \caption[Mean Square Error for regression coefficients and response (functional redundancy model with random coefficients)]{{\bf Mean Square Error for regression coefficients and response \revision{(functional redundancy model with random coefficients) for $R=5$, $\nu=12$}.} \revision{X axis corresponds to the number of sampled nodes (microbes) which relates to the sparsity of the adjacency matrix $\mathbf X$. Dashed \revision{or solid} lines correspond to different values of the true mean for edge effects ($\mu=0.8, 1.6$) and different colors correspond to different sparsity levels on the regression coefficient matrix $\mathbf B$ ($\pi=0.3,0.8$). The plot on the left is for the original, unaugmented data only ($n=50$), while the plot on the right includes augmented data \revision{($n=200$)}. Lines which are missing indicate that MSE values for both numbers of microbes were above the maximum cutoff for the plot.}}
    \label{fig:mse_red_rand_aug}
\end{figure}

\begin{figure}[!ht]
    \centering
    \begin{subfigure}[t]{0.49\textwidth}
        \centering
        \includegraphics[scale=0.27]{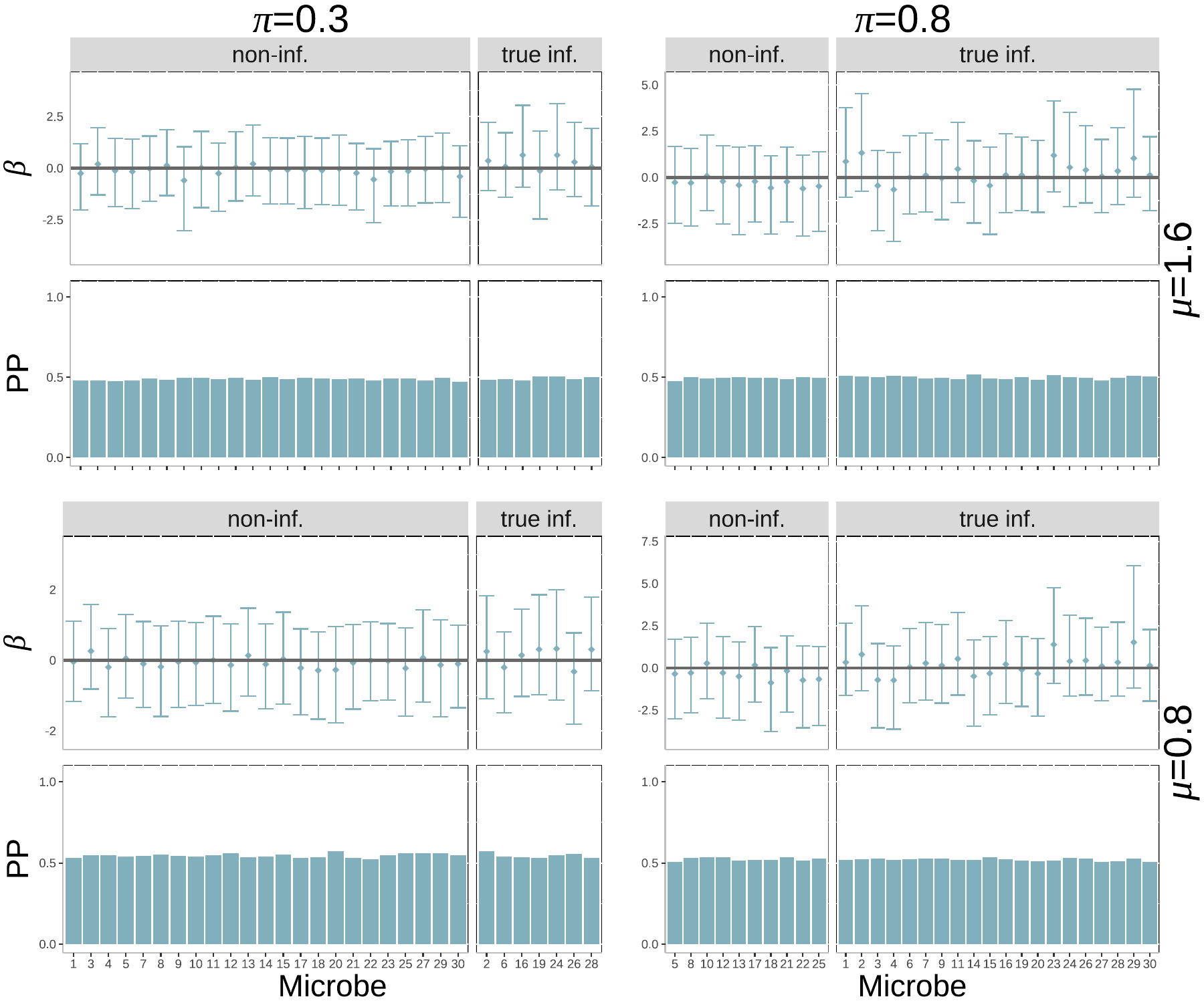}
        \caption{$\mathbf{n=50}$}
    \end{subfigure}
    \begin{subfigure}[t]{0.49\textwidth}
        \centering
        \includegraphics[scale=0.27]{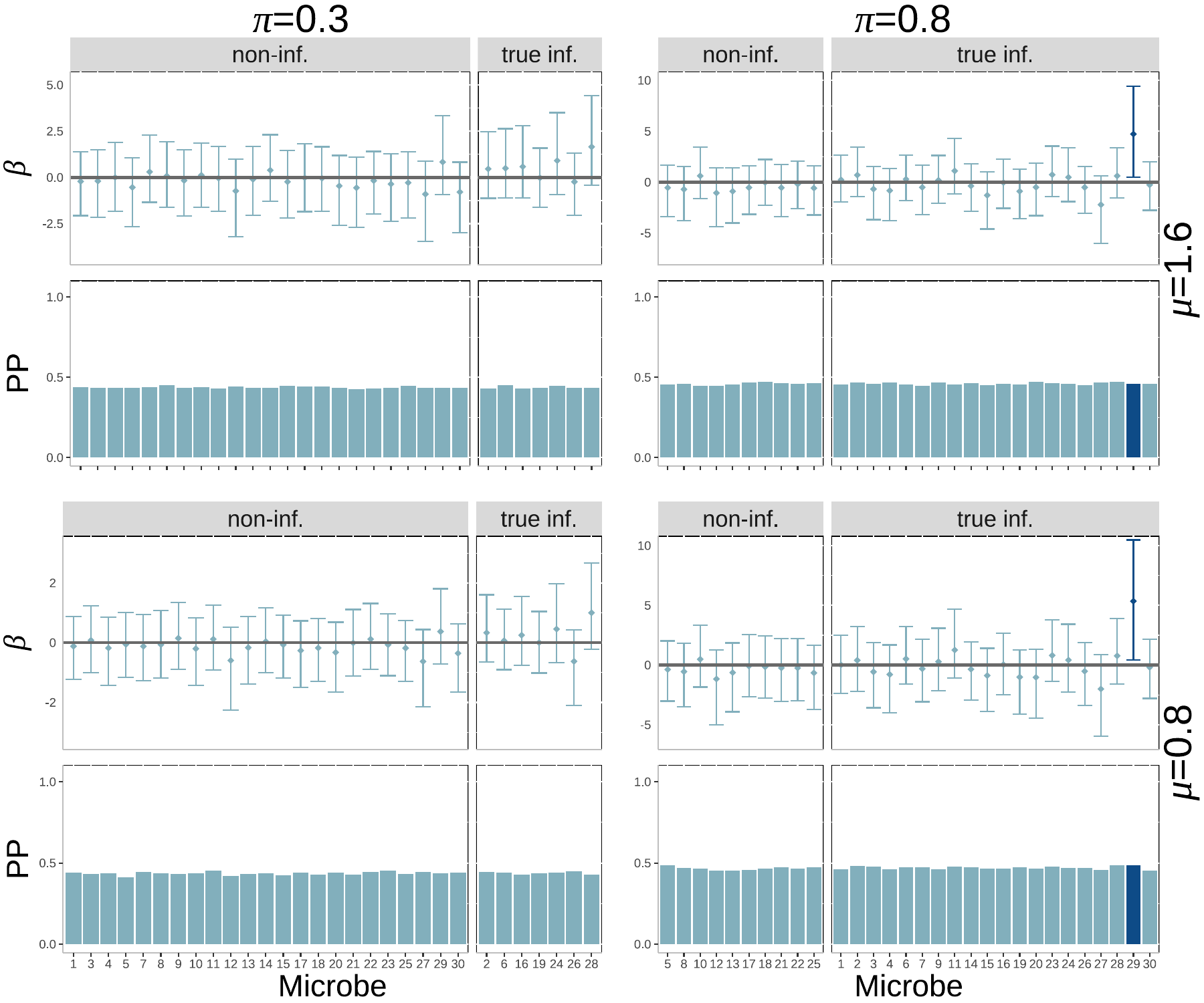}
        \caption{$\mathbf{n=200}$}
    \end{subfigure}
    \caption[Posterior probability of influential nodes and coefficients for nodes (functional redundancy model with phylogenetic coefficients)]{{\bf Posterior probability of influential nodes and coefficients for nodes (functional redundancy model with phylogenetic coefficients) with $k=8$ sampled microbes.}
    Different \revision{groups of four} panels represent different sample sizes ($n=50,200$). \revision{Within each group, we have four panels corresponding to the two values of edge effect size ($\mu=0.8, 1.6$) and two values of probability of influential node ($\pi=0.3, 0.8$) which controls the sparsity of the regression coefficient matrix ($\mathbf B$). Within each of these panels we have two plots: 95\% credible intervals (top) and posterior probability of influence (bottom - calculated as the mean of the $\xi$ variable for the node across Gibbs samples) for each node.} Each bar corresponds to one node (microbe). \revision{Within each plot the bars and intervals are colored depending on whether the node is found to be influential (dark) or not influential (light) based on the 95\% credible intervals. Each plot is split based on whether the nodes are truly influential (right) or not (left).}} 
    \label{fig:nodes_red_rand_aug_adx_k8}
\end{figure}

\begin{figure}[!ht]
    \centering
    \begin{subfigure}[t]{0.49\textwidth}
        \centering
        \includegraphics[scale=0.27]{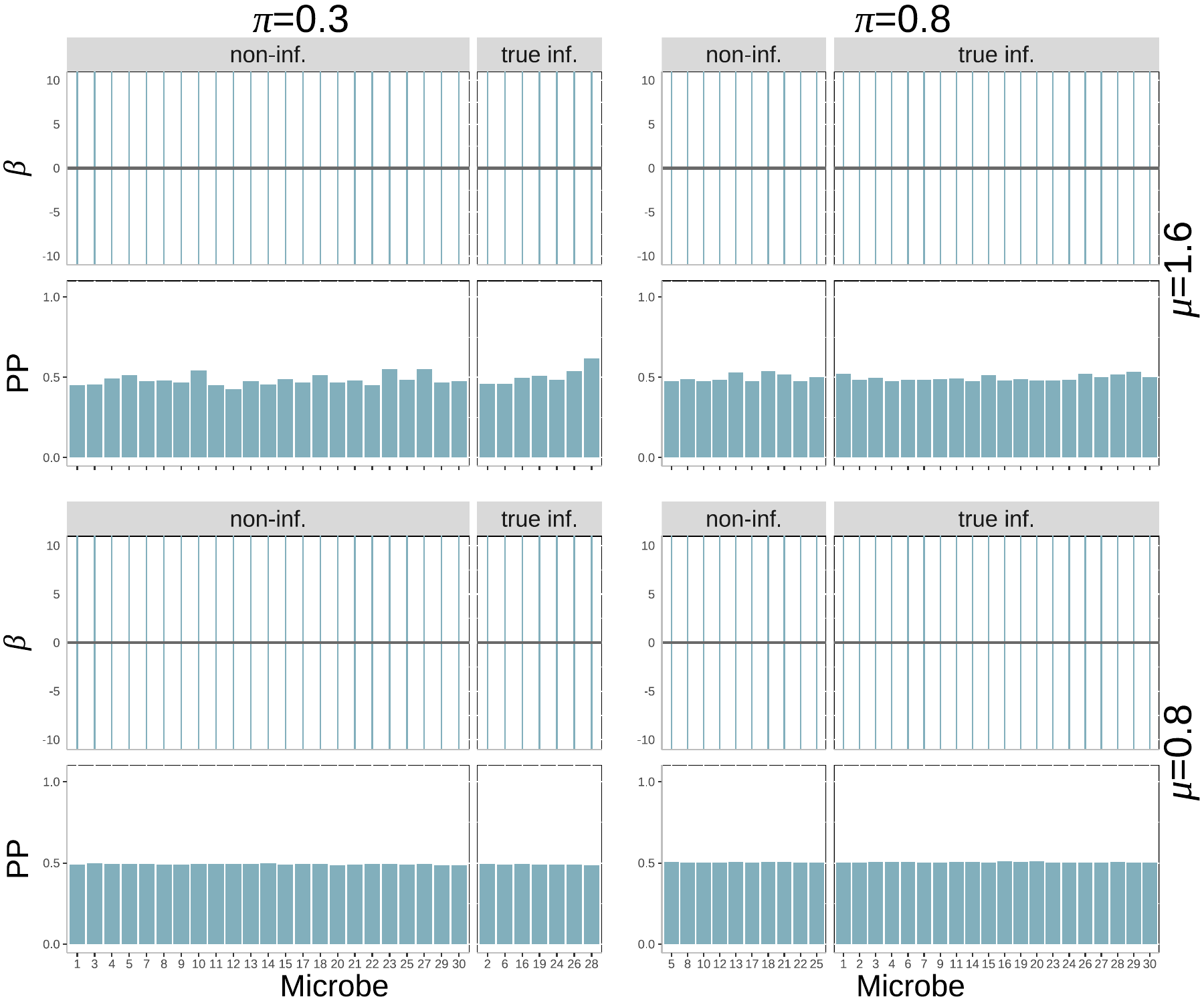}
        \caption{$\mathbf{n=50}$}
    \end{subfigure}
    \begin{subfigure}[t]{0.49\textwidth}
        \centering
        \includegraphics[scale=0.27]{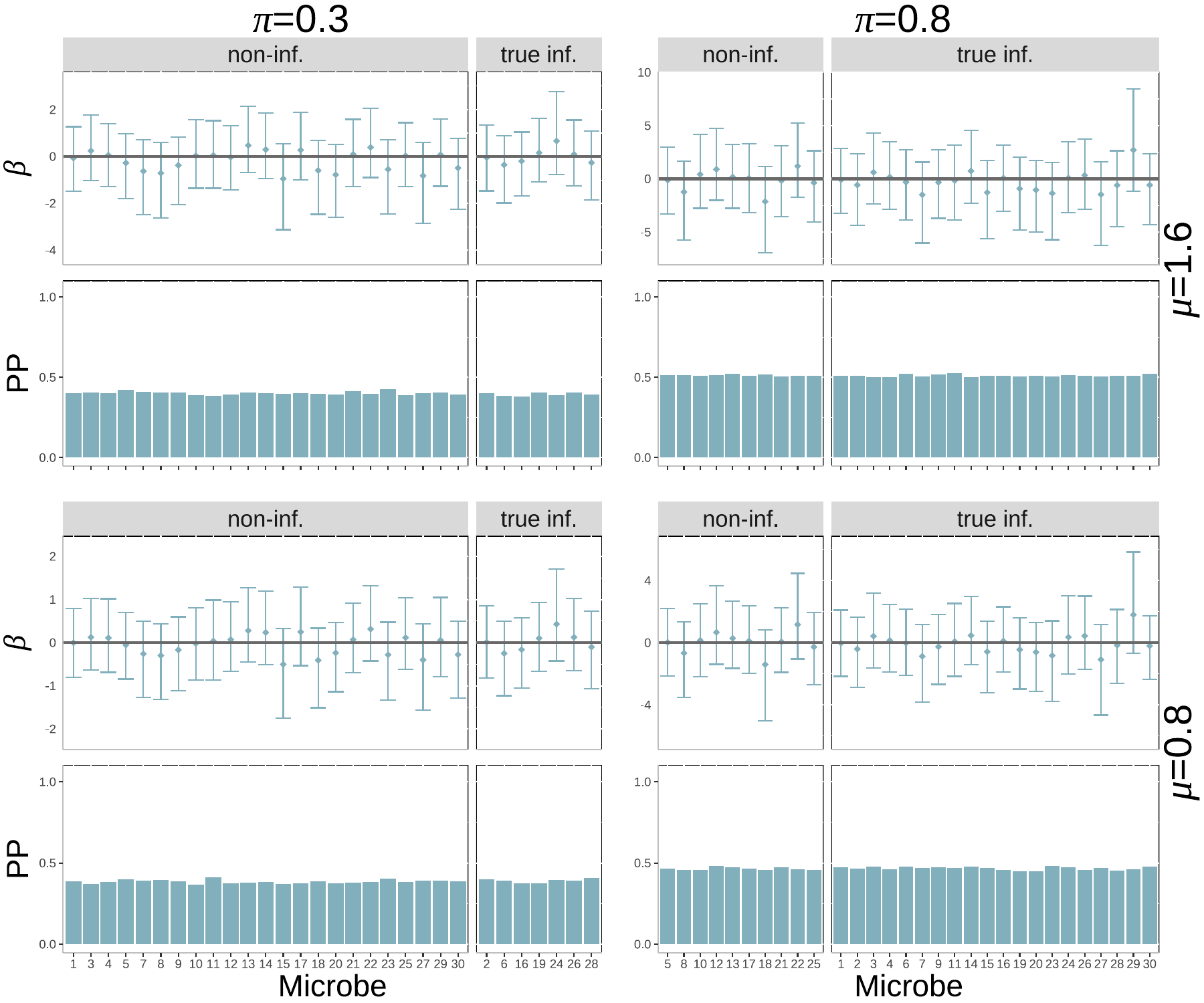}
        \caption{$\mathbf{n=200}$}
    \end{subfigure}
    \caption[Posterior probability of influential nodes and coefficients for nodes (functional redundancy model with phylogenetic coefficients)]{{\bf Posterior probability of influential nodes and coefficients for nodes (functional redundancy model with phylogenetic coefficients) with $k=22$ sampled microbes.}
    Different \revision{groups of four} panels represent different sample sizes ($n=50,200$). \revision{Within each group, we have four panels corresponding to the two values of edge effect size ($\mu=0.8, 1.6$) and two values of probability of influential node ($\pi=0.3, 0.8$) which controls the sparsity of the regression coefficient matrix ($\mathbf B$). Within each of these panels we have two plots: 95\% credible intervals (top) and posterior probability of influence (bottom - calculated as the mean of the $\xi$ variable for the node across Gibbs samples) for each node.} Each bar corresponds to one node (microbe). \revision{Within each plot the bars and intervals are colored depending on whether the node is found to be influential (dark) or not influential (light) based on the 95\% credible intervals. Each plot is split based on whether the nodes are truly influential (right) or not (left).}} 
    \label{fig:nodes_red_rand_aug_adx_k22}
\end{figure}

\FloatBarrier
\subsubsection{False positive and negative rates}
\FloatBarrier

\begin{figure}[!ht]
\centering
\includegraphics[scale=0.34]{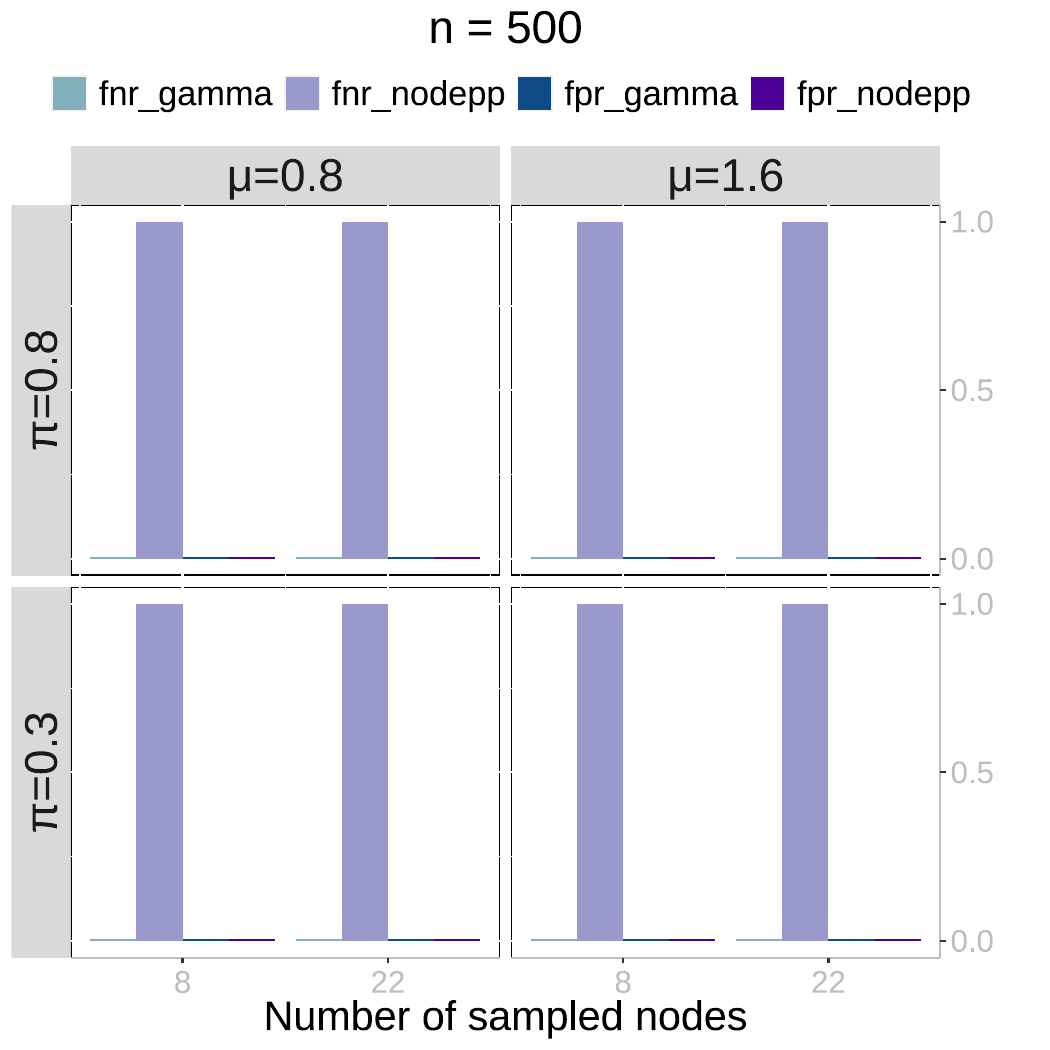}
\includegraphics[scale=0.34]{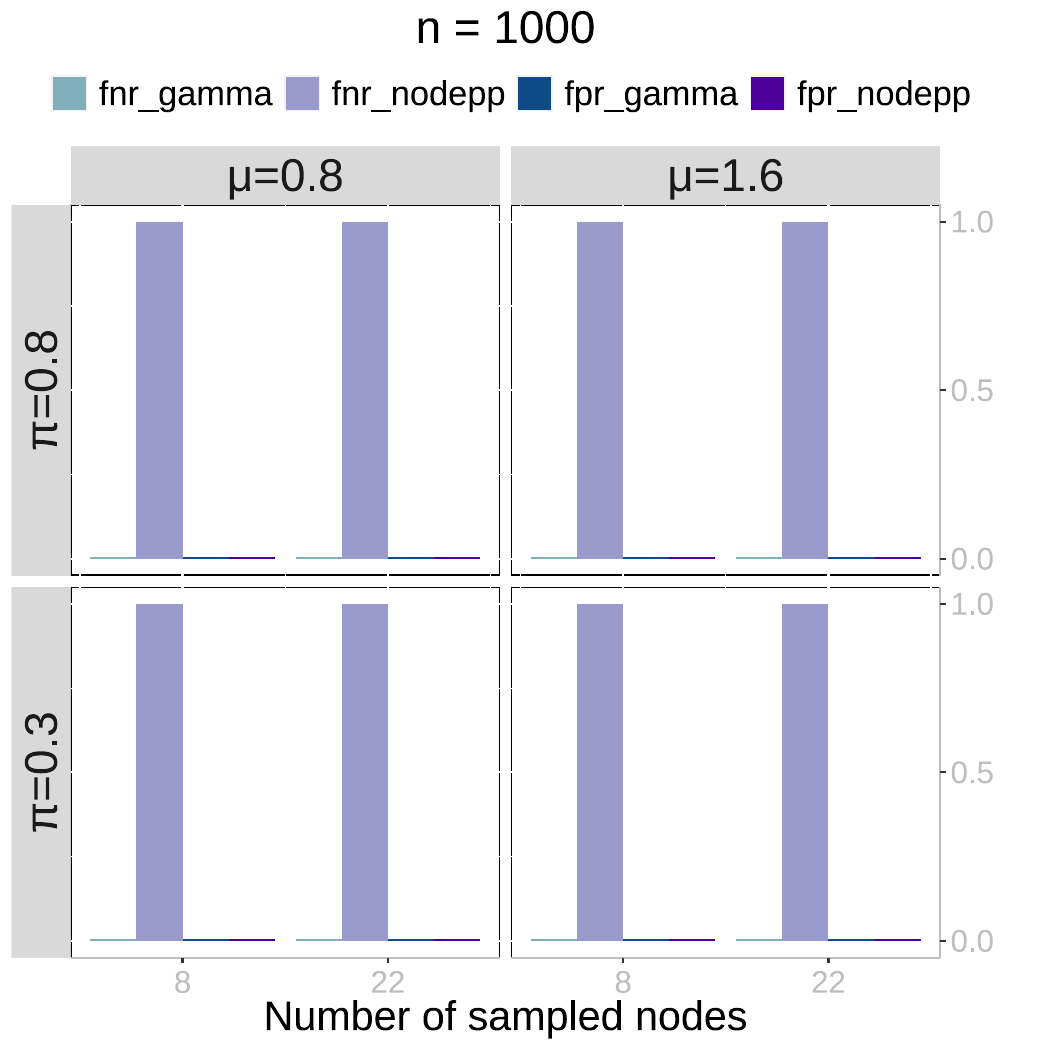}\\
\includegraphics[scale=0.34]{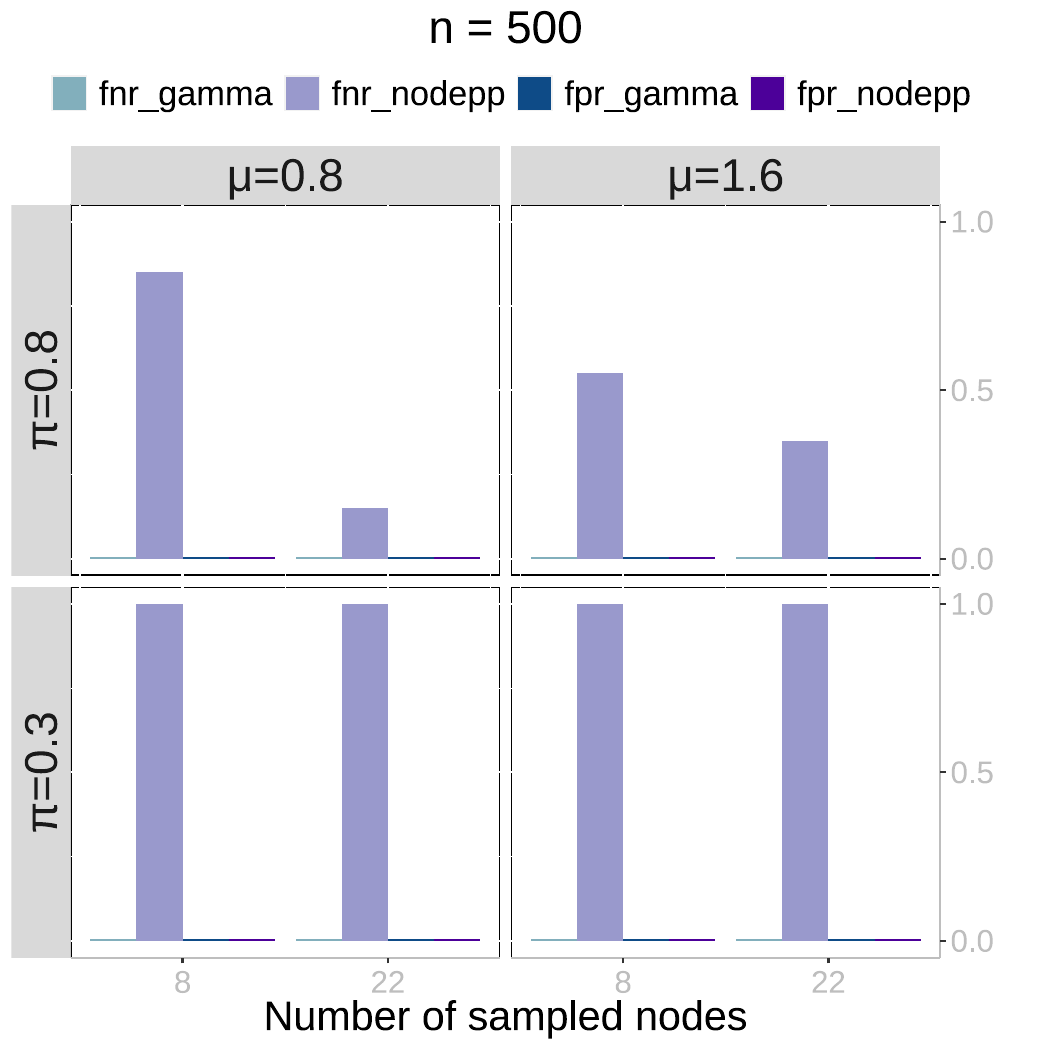}
\includegraphics[scale=0.34]{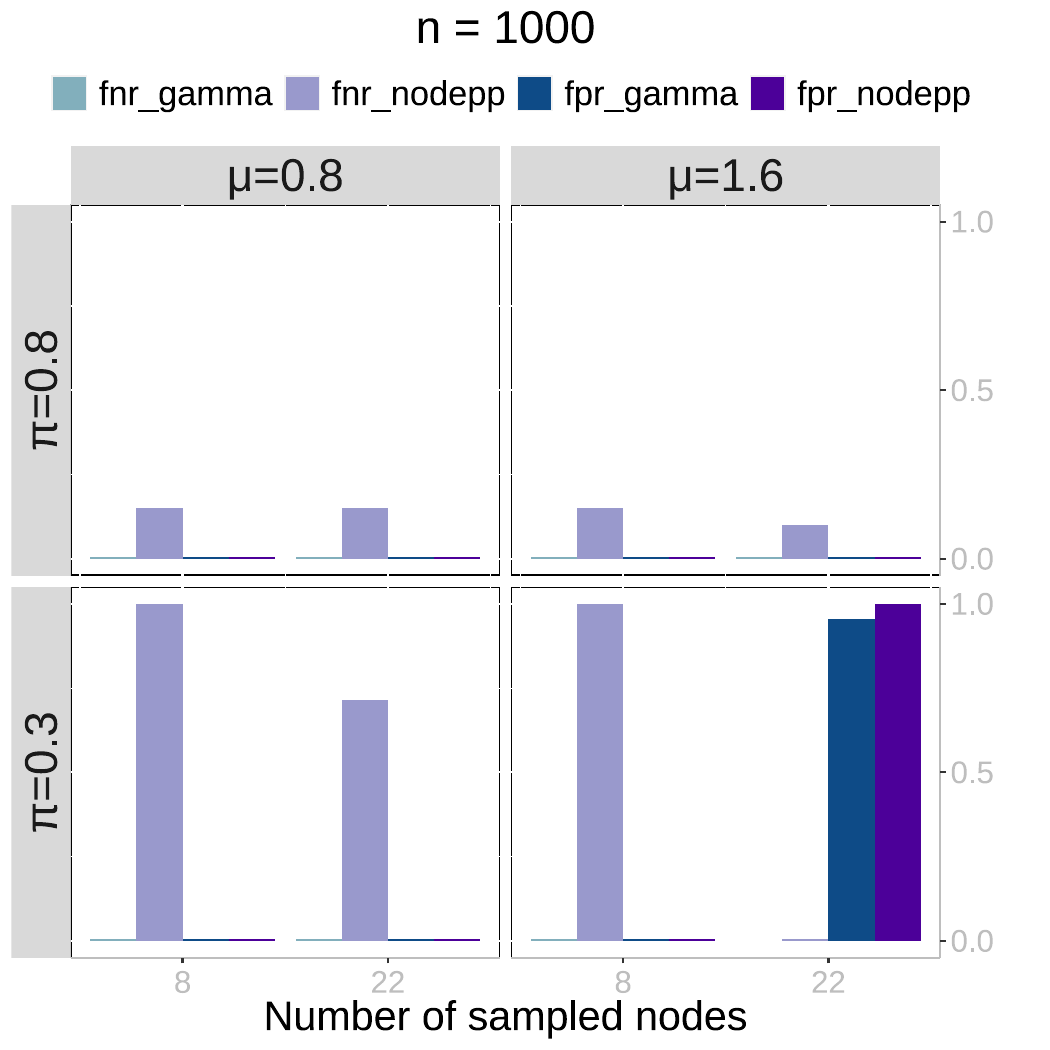}\\
\includegraphics[scale=0.34]{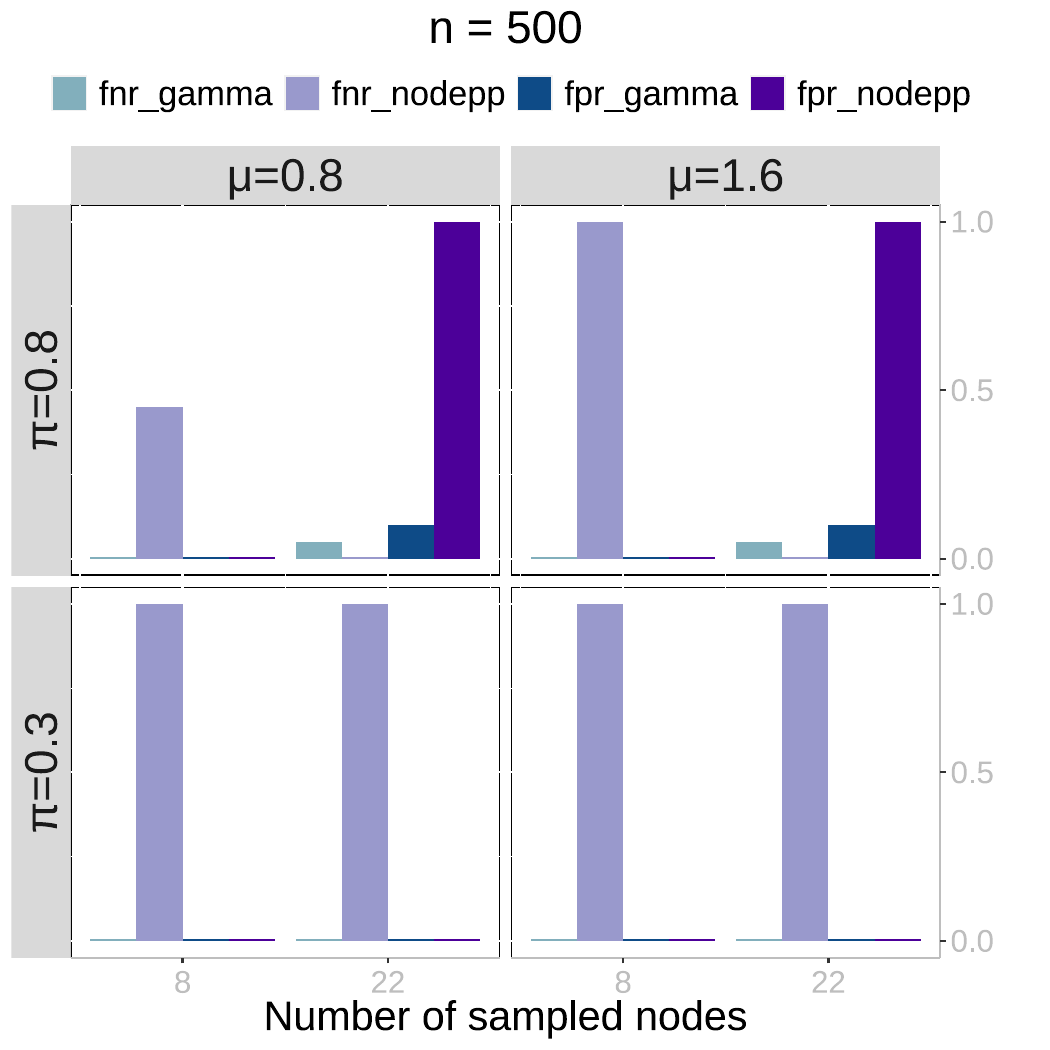}
\includegraphics[scale=0.34]{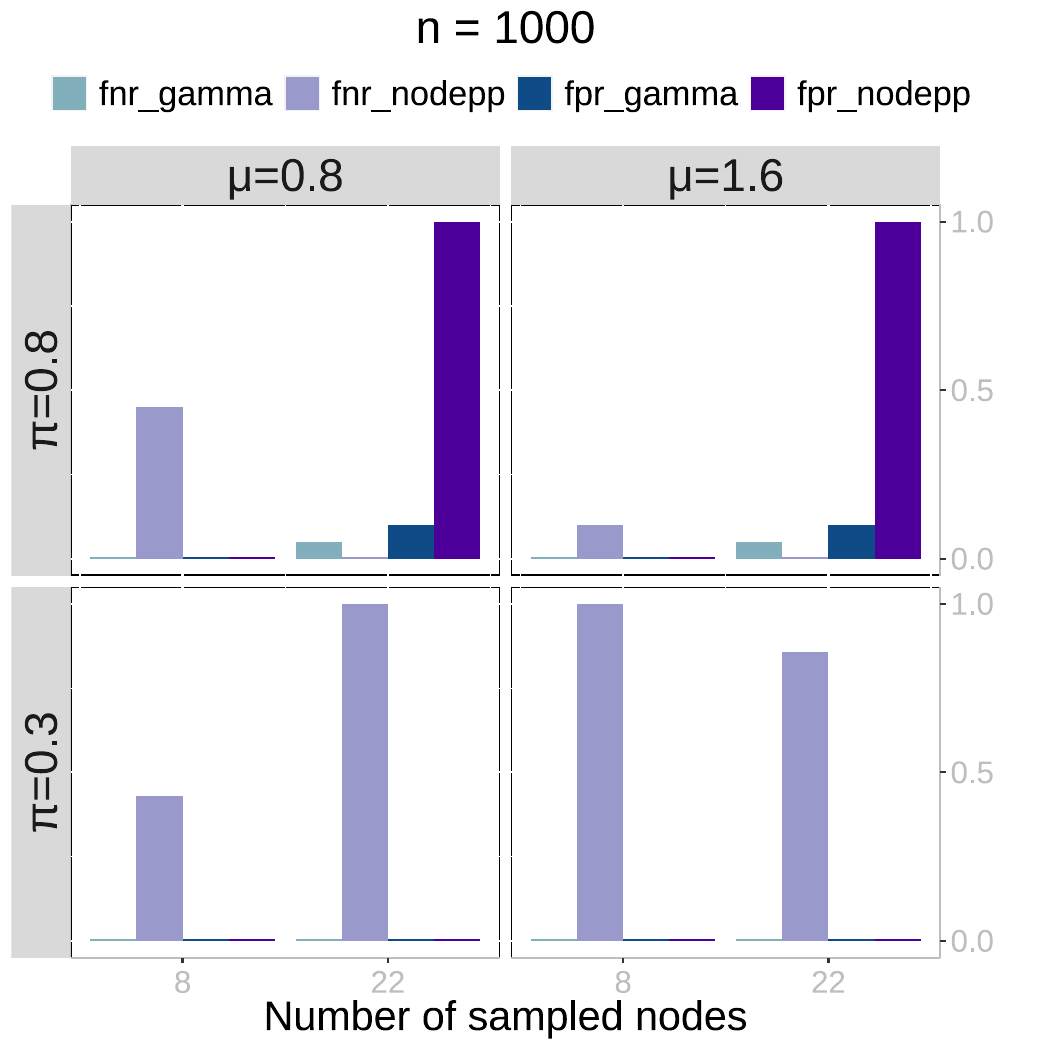}
\caption[False positive and false negative rates for influential edges and nodes for additive (top), interaction (middle) and functional redundancy (bottom) models with random coefficients]{{\bf False positive and false negative rates for influential edges and nodes for additive (top), interaction (middle) and functional redundancy (bottom) models with random coefficients.} X axis corresponds to the number of sampled nodes (microbes) which relates to the sparsity of the adjacency matrix $\mathbf X$. Decisions to reject for edges are based on 95\% posterior credible intervals, and for nodes are based on whether the posterior probability of influence is greater than 0.5. Within each panel, we have four plots corresponding to the two values of edge effect size ($\mu=0.8, 1.6$) and two values of probability of influential edge ($\pi=0.3, 0.8$) which controls the sparsity of the regression coefficient matrix ($\mathbf B$). Each bar corresponds to one rate: false positive rate and false negative rate for edges (dark and light blue) and false positive rate and false negative rate for nodes (dark and light purple).}
\label{fig:rates-rand}
\end{figure}

\begin{figure}[!ht]
\centering
\includegraphics[scale=0.34]{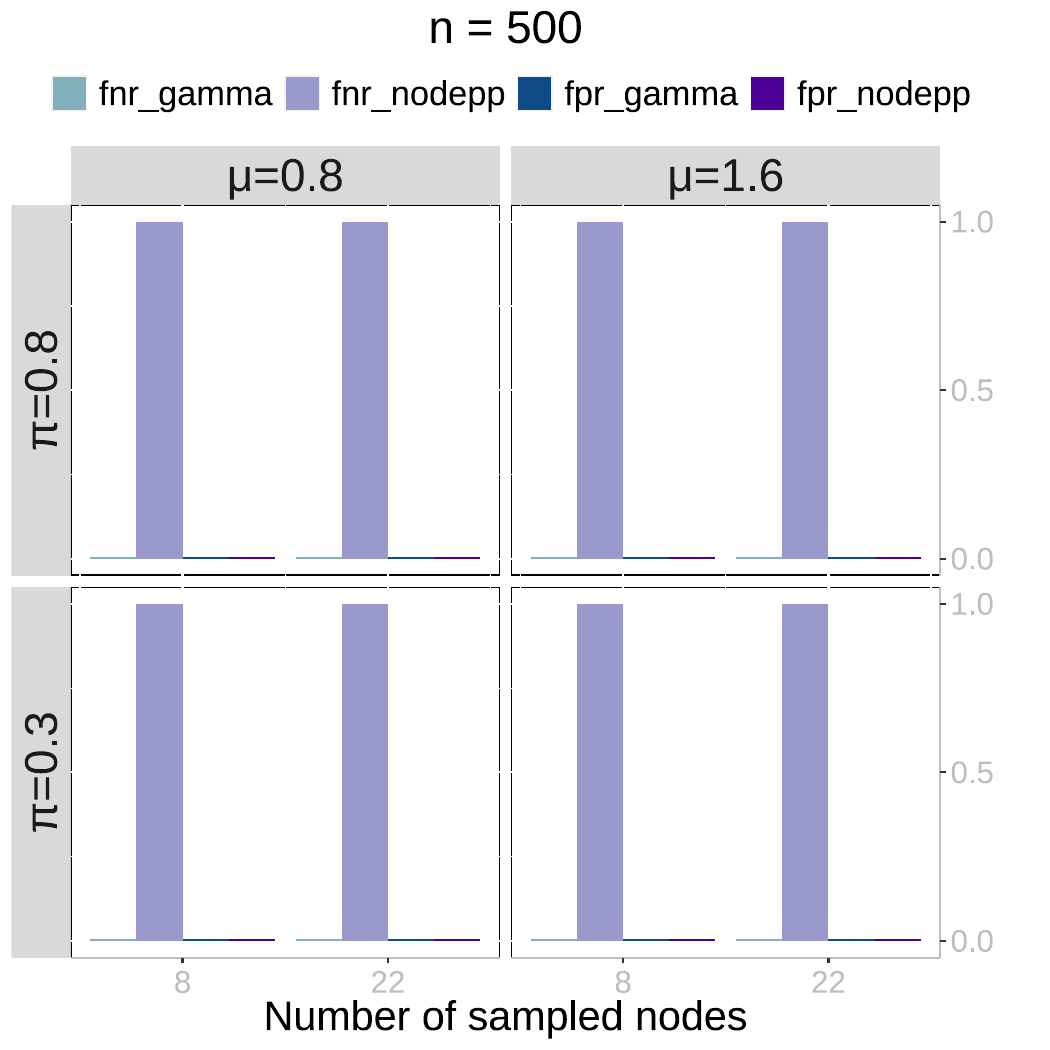}
\includegraphics[scale=0.34]{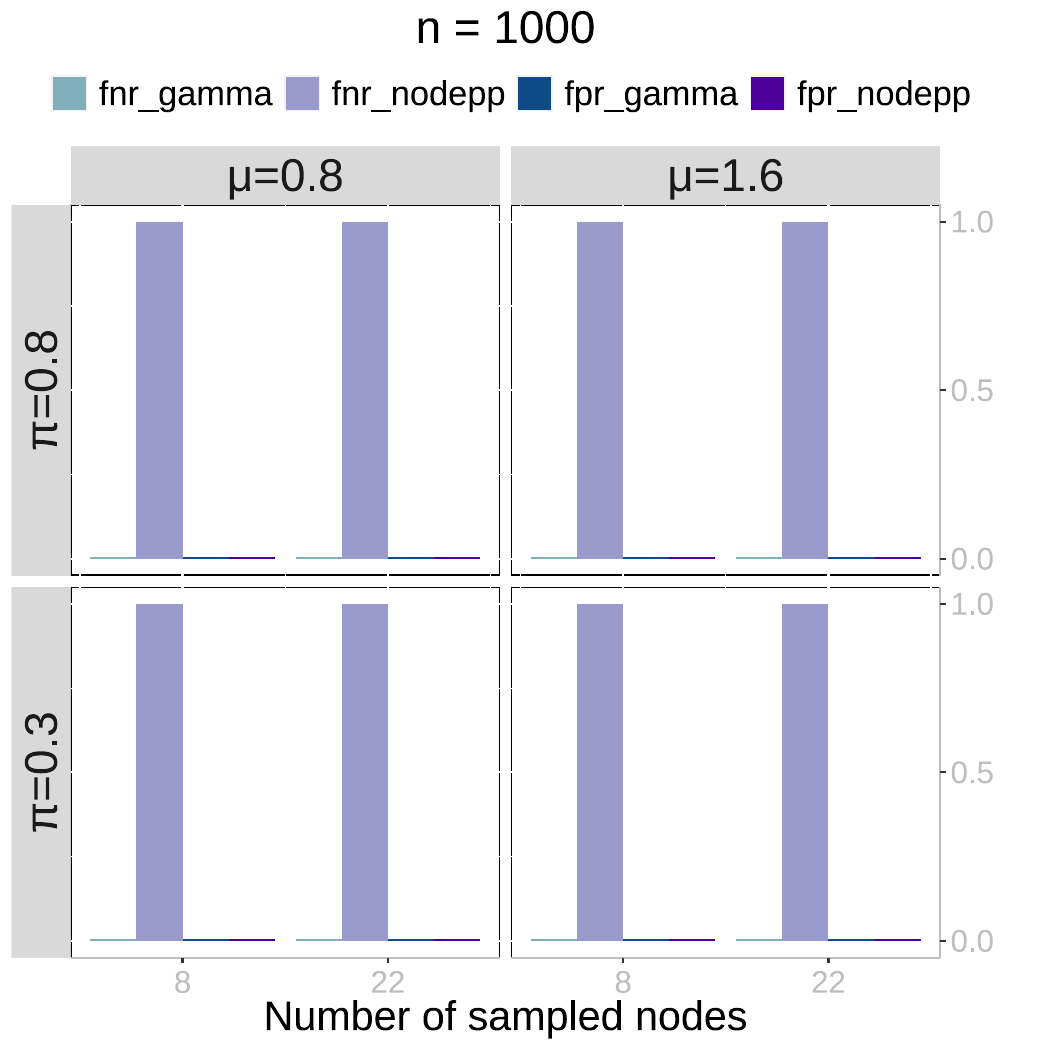}\\
\includegraphics[scale=0.34]{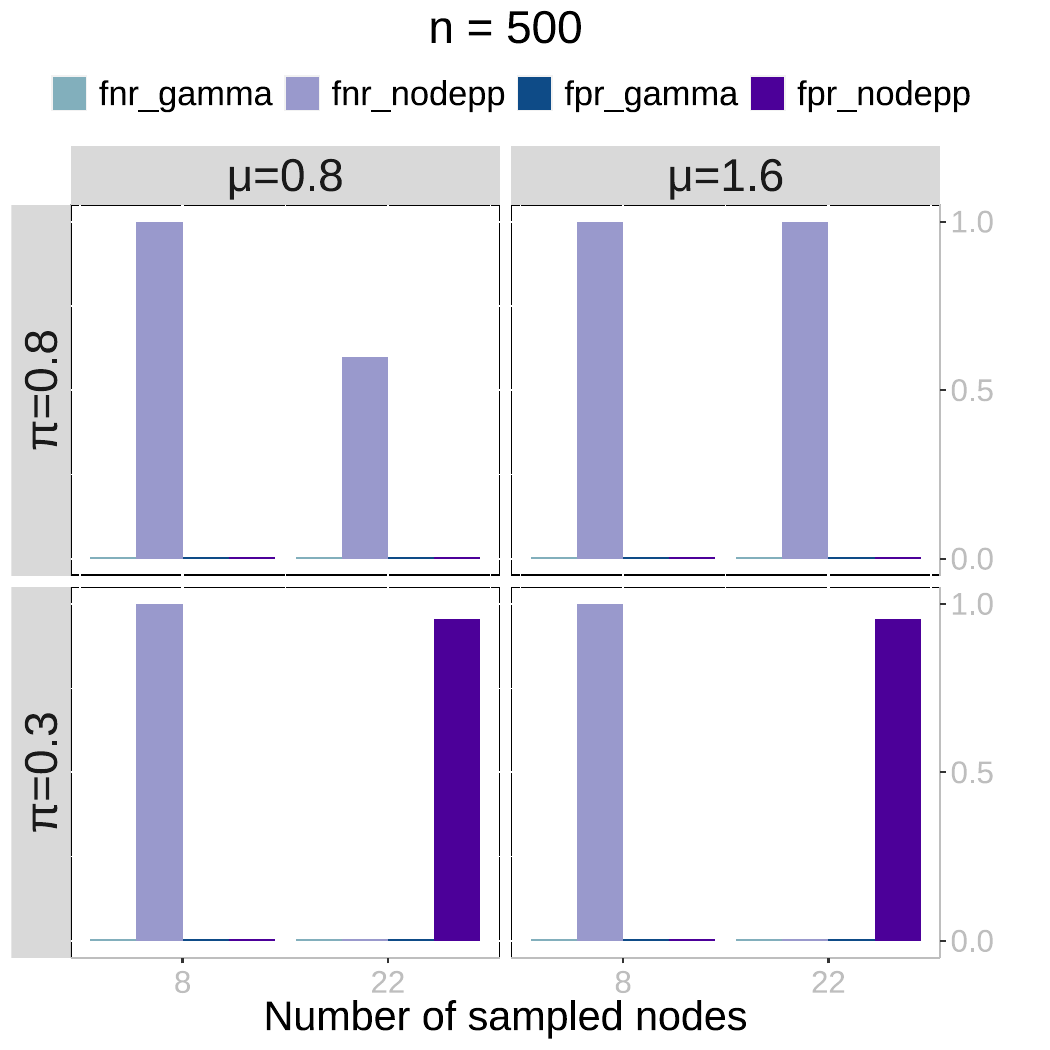}
\includegraphics[scale=0.34]{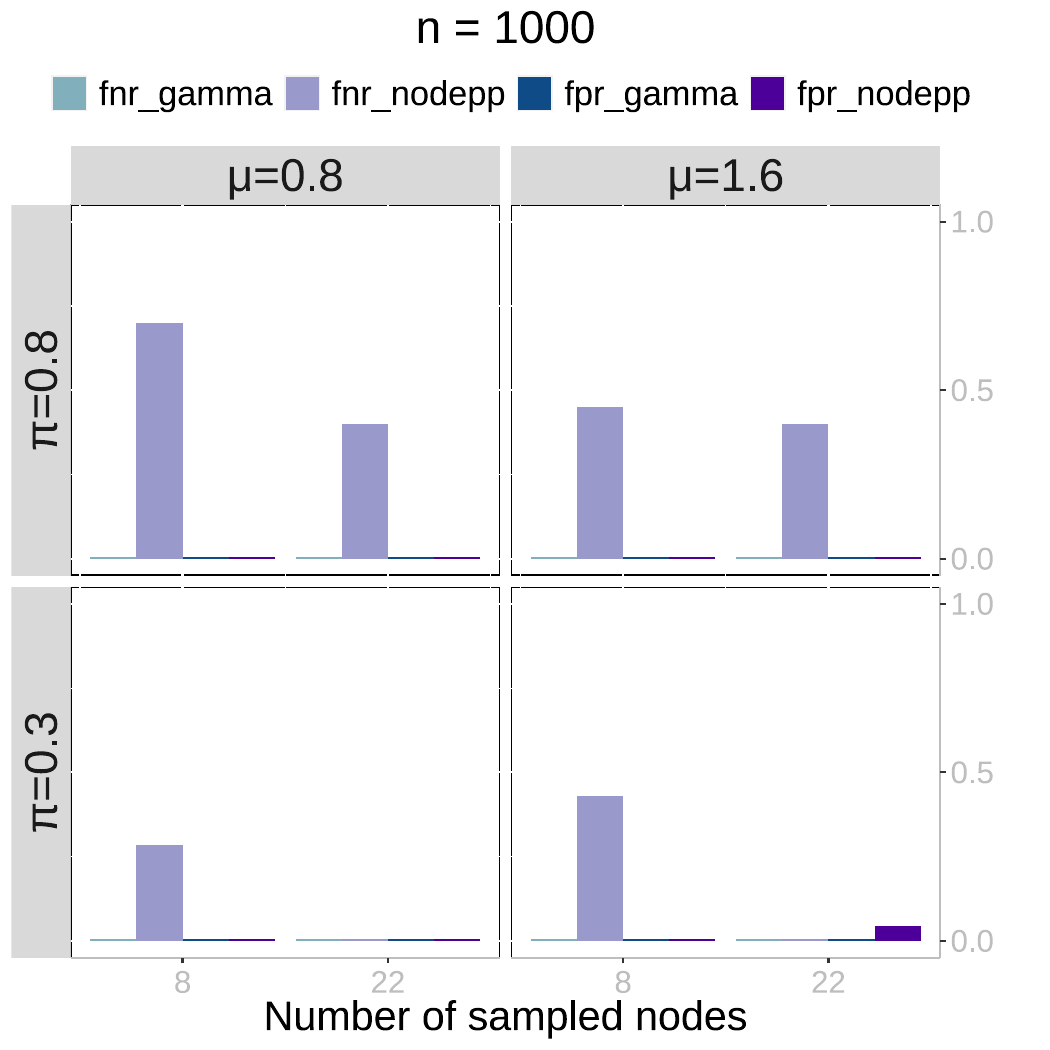}\\
\includegraphics[scale=0.34]{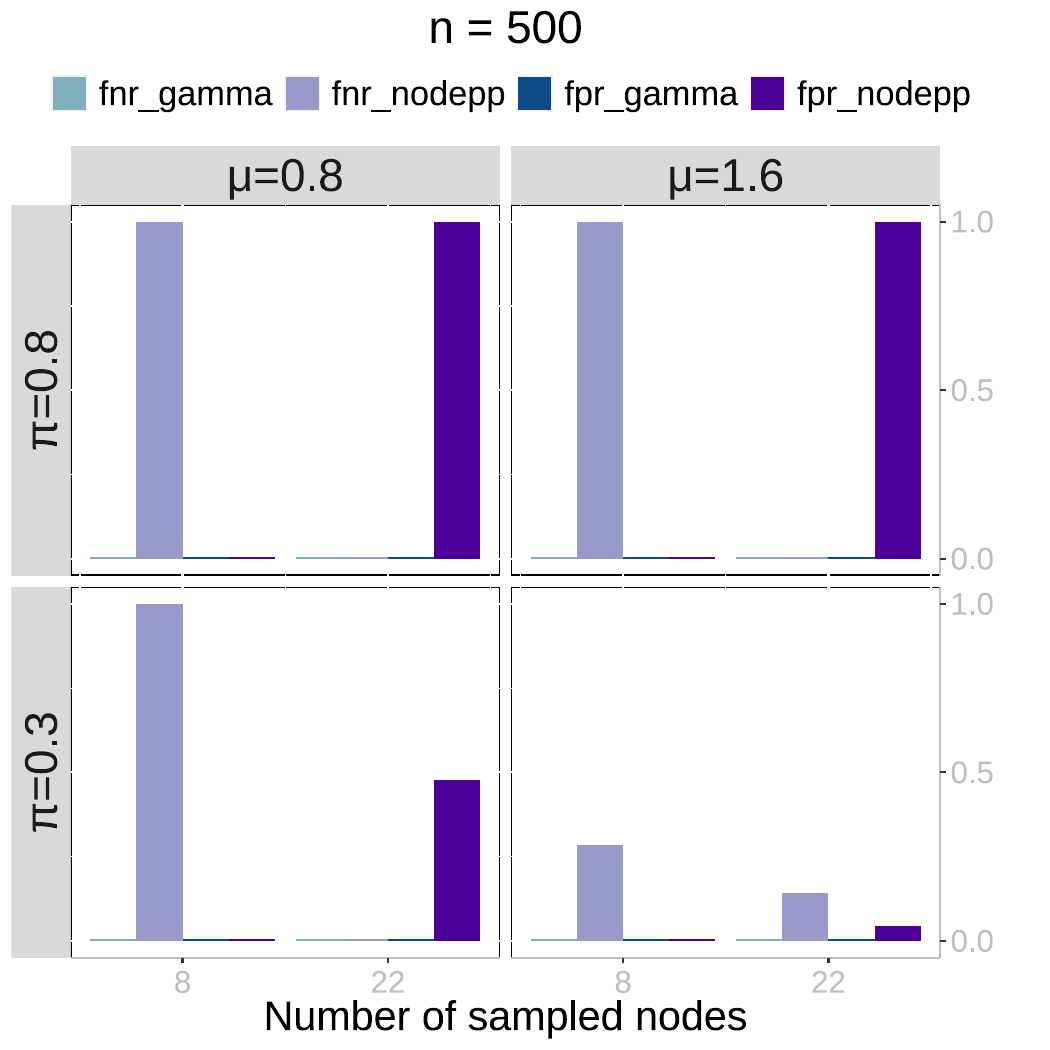}
\includegraphics[scale=0.34]{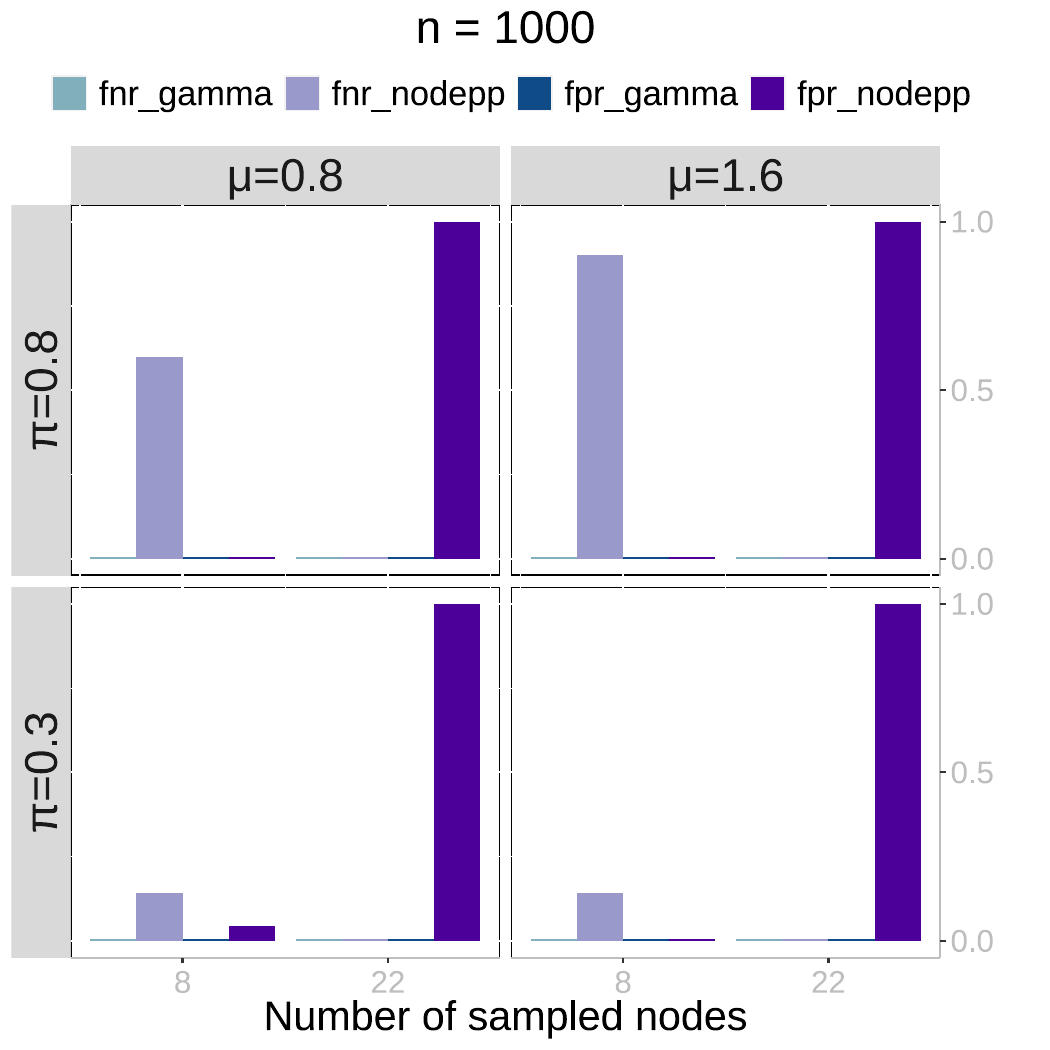}
\caption[False positive and false negative rates for influential edges and nodes for additive (top), interaction (middle) and functional redundancy (bottom) models with phylogenetic coefficients]{{\bf False positive and false negative rates for influential edges and nodes for additive (top), interaction (middle) and functional redundancy (bottom) models with phylogenetic coefficients.} X axis corresponds to the number of sampled nodes (microbes) which relates to the sparsity of the adjacency matrix $\mathbf X$. Decisions to reject for edges are based on 95\% posterior credible intervals, and for nodes are based on whether the posterior probability of influence is greater than 0.5. Within each panel, we have four plots corresponding to the two values of edge effect size ($\mu=0.8, 1.6$) and two values of probability of influential edge ($\pi=0.3, 0.8$) which controls the sparsity of the regression coefficient matrix ($\mathbf B$). Each bar corresponds to one rate: false positive rate and false negative rate for edges (dark and light blue) and false positive rate and false negative rate for nodes (dark and light purple).}
\label{fig:rates-phylo}
\end{figure}

\begin{figure}[!ht]
    \centering
    \begin{subfigure}[t]{\textwidth}
        \centering
        \includegraphics[scale=0.4]{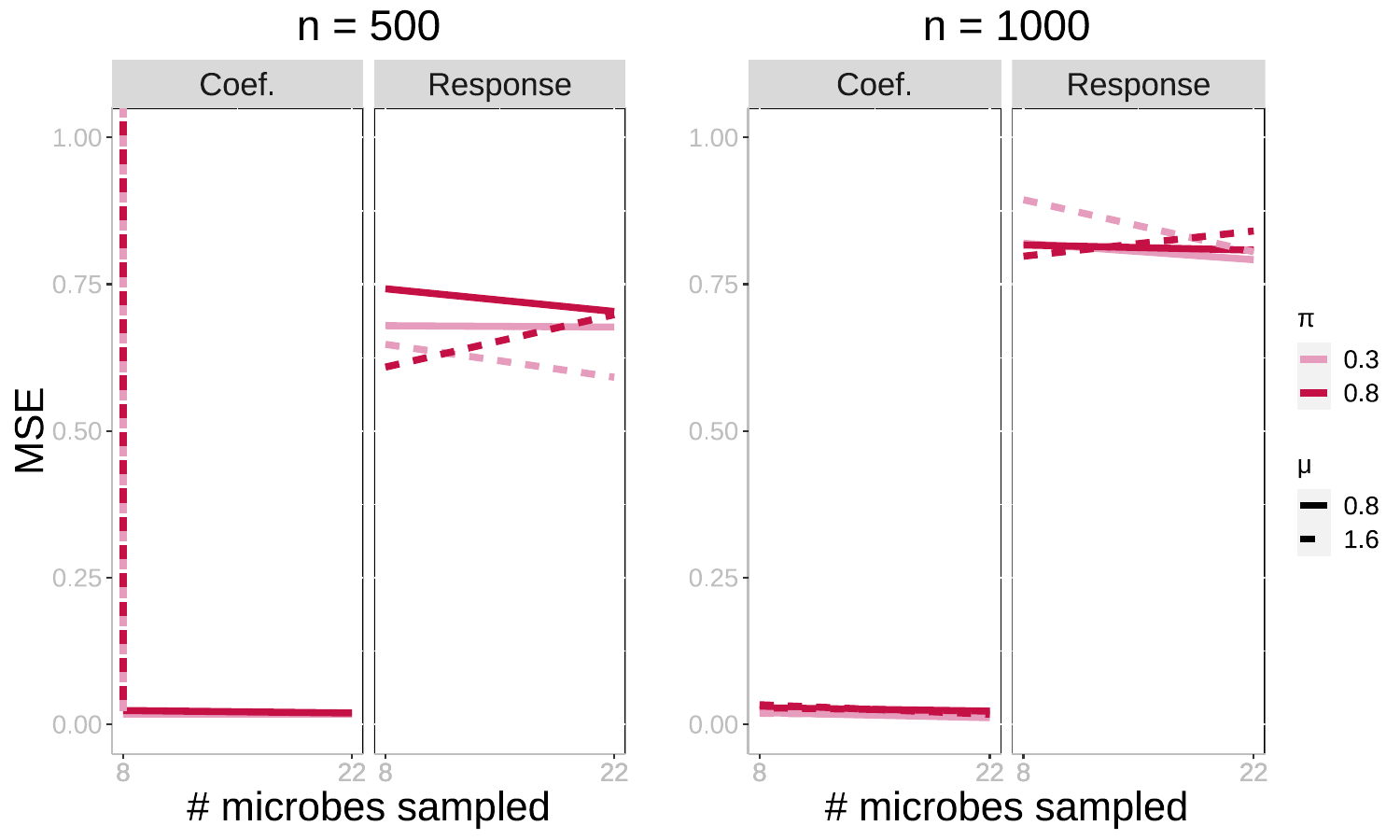}
        \caption{\textbf{Additive}}
    \end{subfigure}
    \begin{subfigure}[t]{\textwidth}
        \centering
        \includegraphics[scale=0.4]{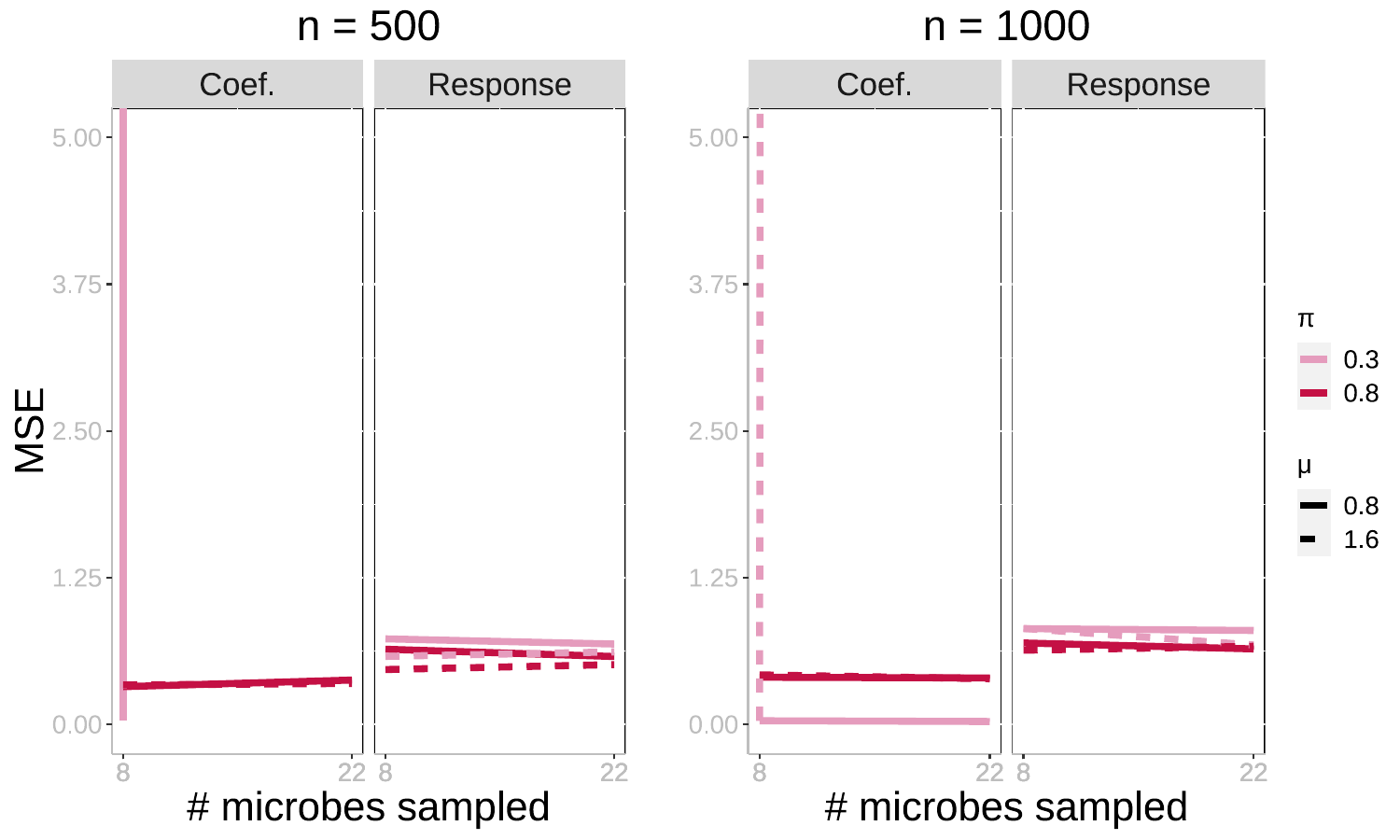}
        \caption{\textbf{Interaction}}
    \end{subfigure}
    \begin{subfigure}[t]{\textwidth}
        \centering
        \includegraphics[scale=0.4]{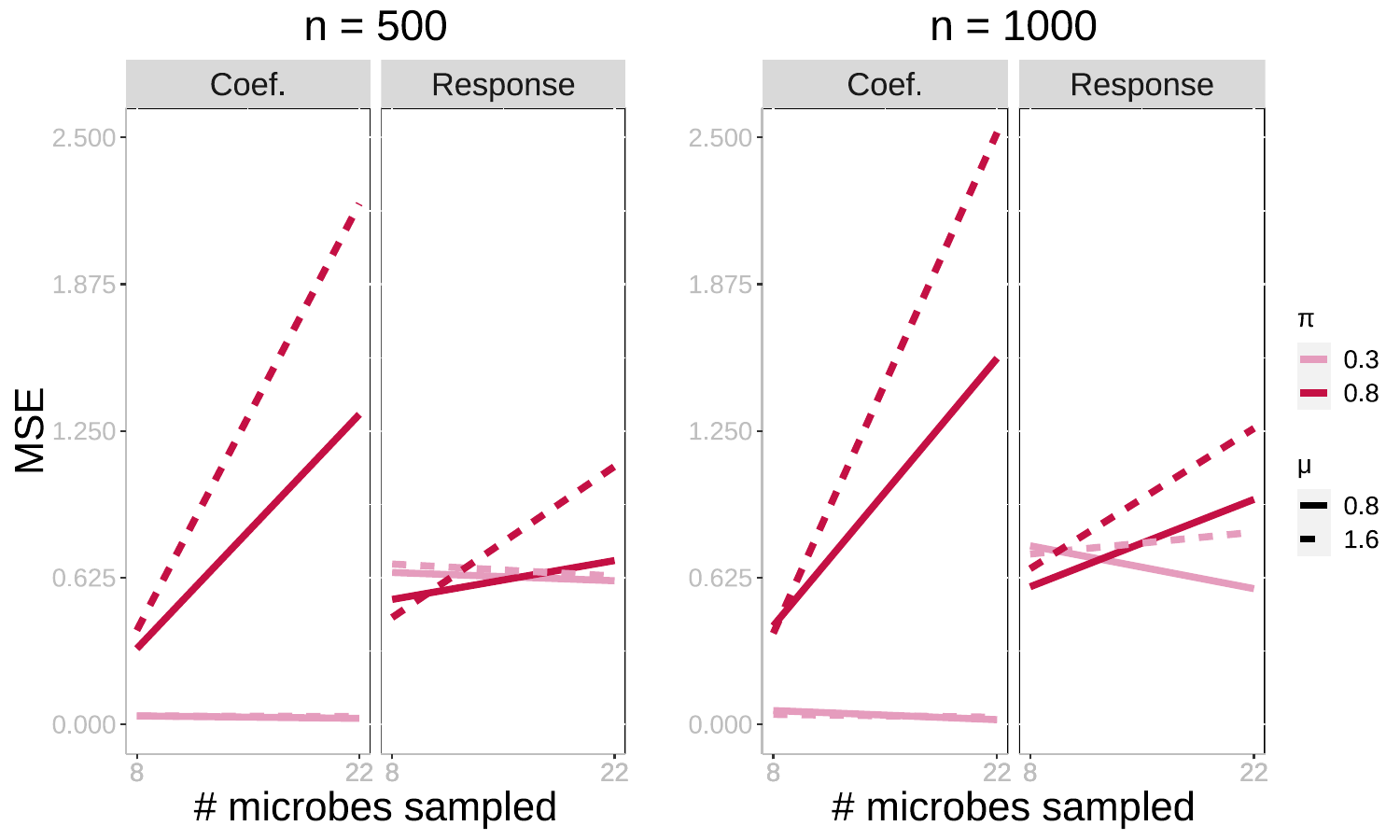}
        \caption{\textbf{Funcional redundant}}
    \end{subfigure}
    \caption[Mean Square Error of coefficients and response for additive (top), interaction (middle) and functional redundancy (bottom) models with random coefficients]{\revision{{\bf Mean Square Error of coefficients and response for additive (top), interaction (middle) and functional redundancy (bottom) models with random coefficients for $R=7$.} X axis corresponds to the number of sampled nodes (microbes) which relates to the sparsity of the adjacency matrix $\mathbf X$. Dashed lines correspond to different values of the true mean for edge effects ($\mu=0.8, 1.6$) and different colors correspond to different sparsity levels on the regression coefficient matrix $\mathbf B$ ($\pi=0.3,0.8$).}}
    \label{fig:mse-rand-adx}
\end{figure}

\begin{figure}[!ht]
    \centering
    \begin{subfigure}[t]{\textwidth}
        \centering
        \includegraphics[scale=0.4]{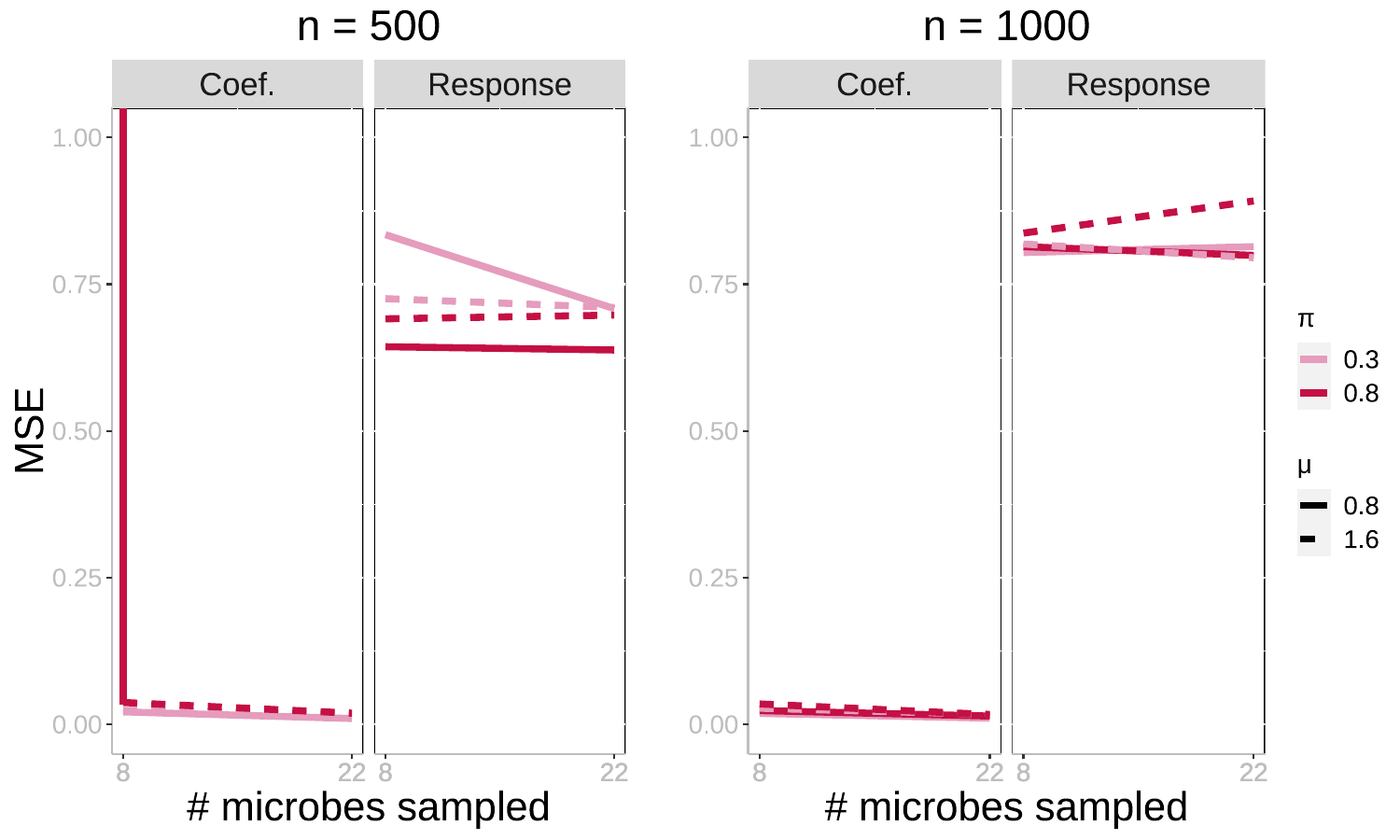}
        \caption{\textbf{Additive}}
    \end{subfigure}
    \begin{subfigure}[t]{\textwidth}
        \centering
        \includegraphics[scale=0.4]{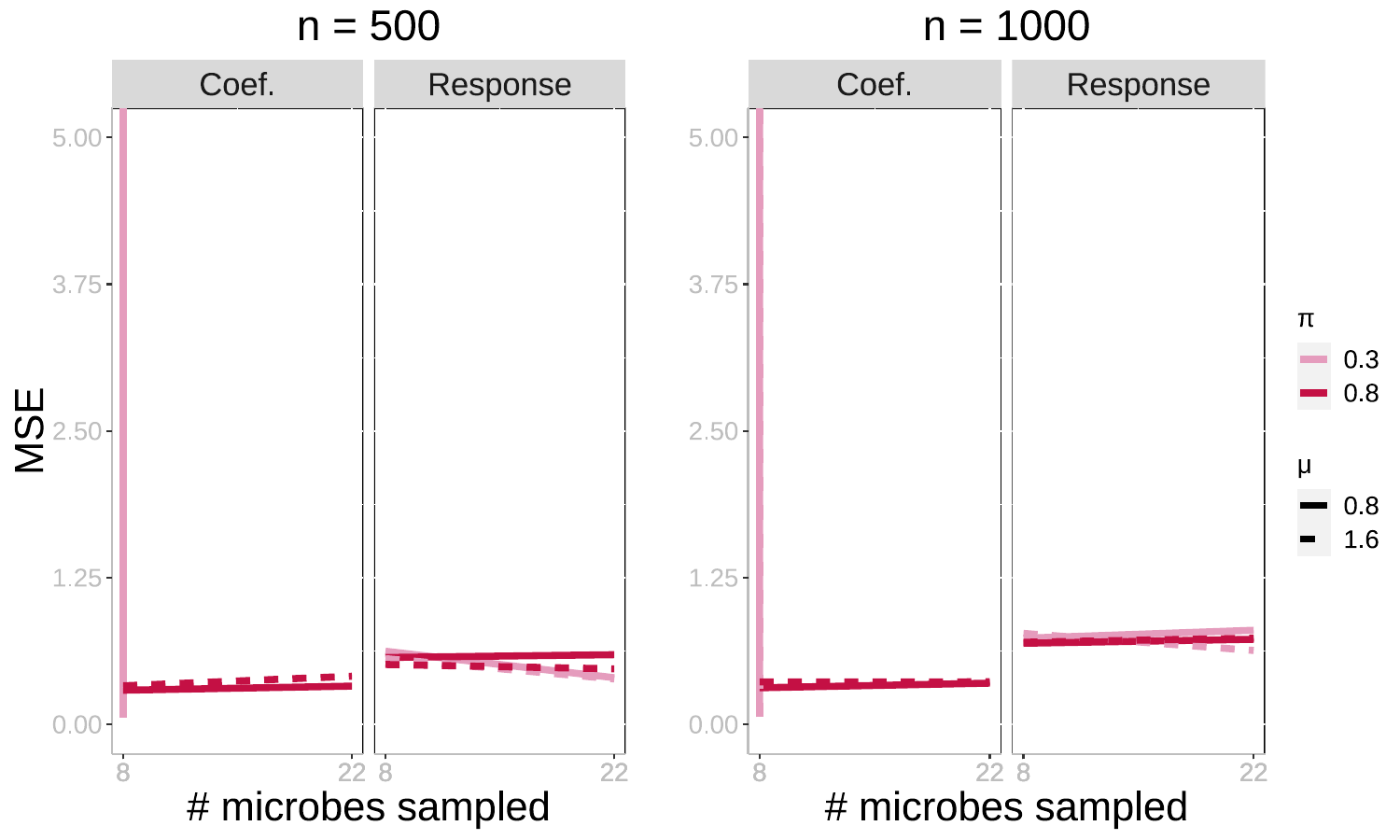}
        \caption{\textbf{Interaction}}
    \end{subfigure}
    \begin{subfigure}[t]{\textwidth}
        \centering
        \includegraphics[scale=0.4]{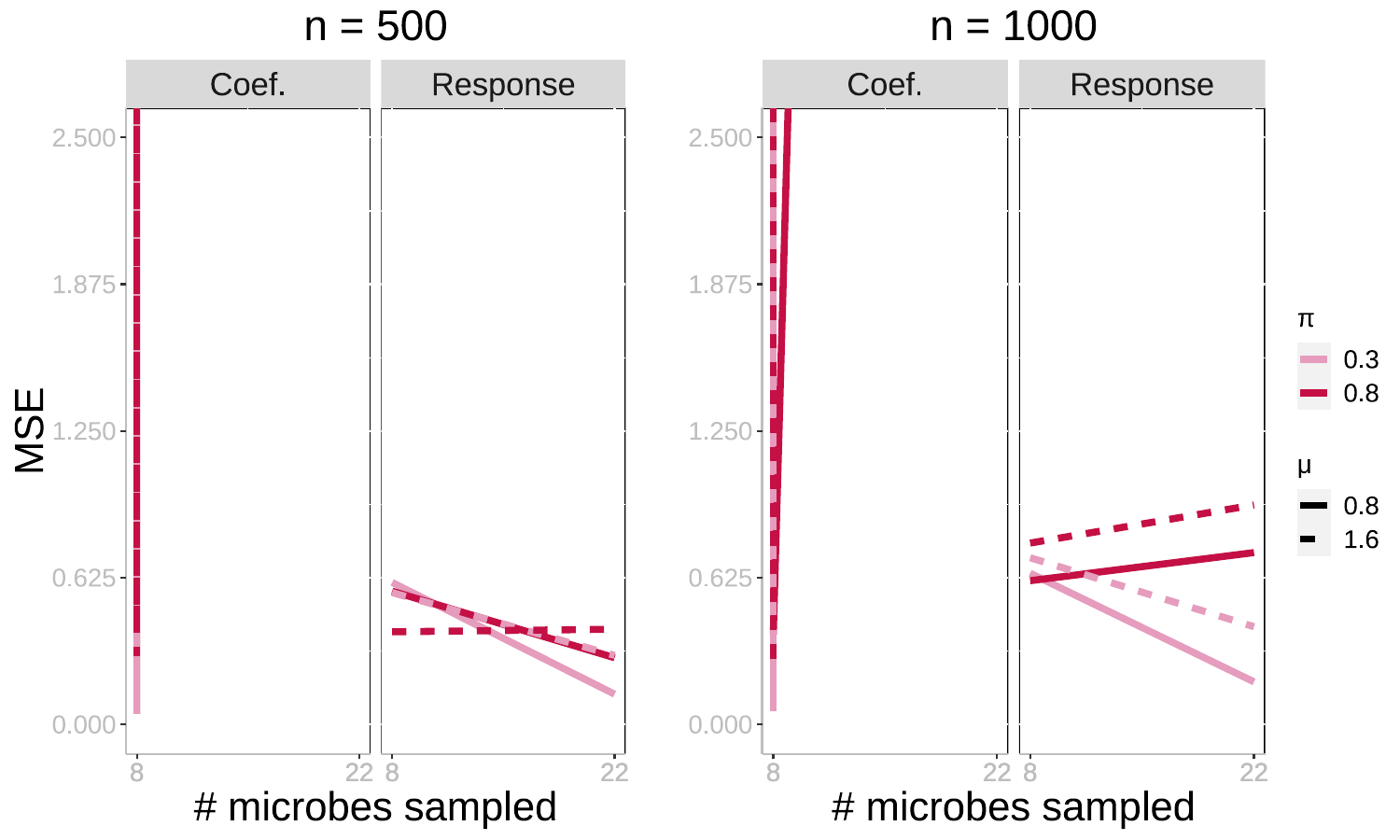}
        \caption{\textbf{Functional redundant}}
    \end{subfigure}
    \caption[Mean Square Error of coefficients and response for additive (top), interaction (middle) and functional redundancy (bottom) models with phylogenetic coefficients]{\revision{{\bf Mean Square Error of coefficients and response for additive (top), interaction (middle) and functional redundancy (bottom) models with phylogenetic coefficients} X axis corresponds to the number of sampled nodes (microbes) which relates to the sparsity of the adjacency matrix $\mathbf X$. Dashed lines correspond to different values of the true mean for edge effects ($\mu=0.8, 1.6$) and different colors correspond to different sparsity levels on the regression coefficient matrix $\mathbf B$ ($\pi=0.3,0.8$).}}
    \label{fig:mse-phylo-adx}
\end{figure}

\FloatBarrier
\subsection{Fungal and bacterial drivers of Phosphorous Leaching in Soil}
\FloatBarrier

\begin{figure}[!ht]
    \centering
    \includegraphics[scale=0.5]{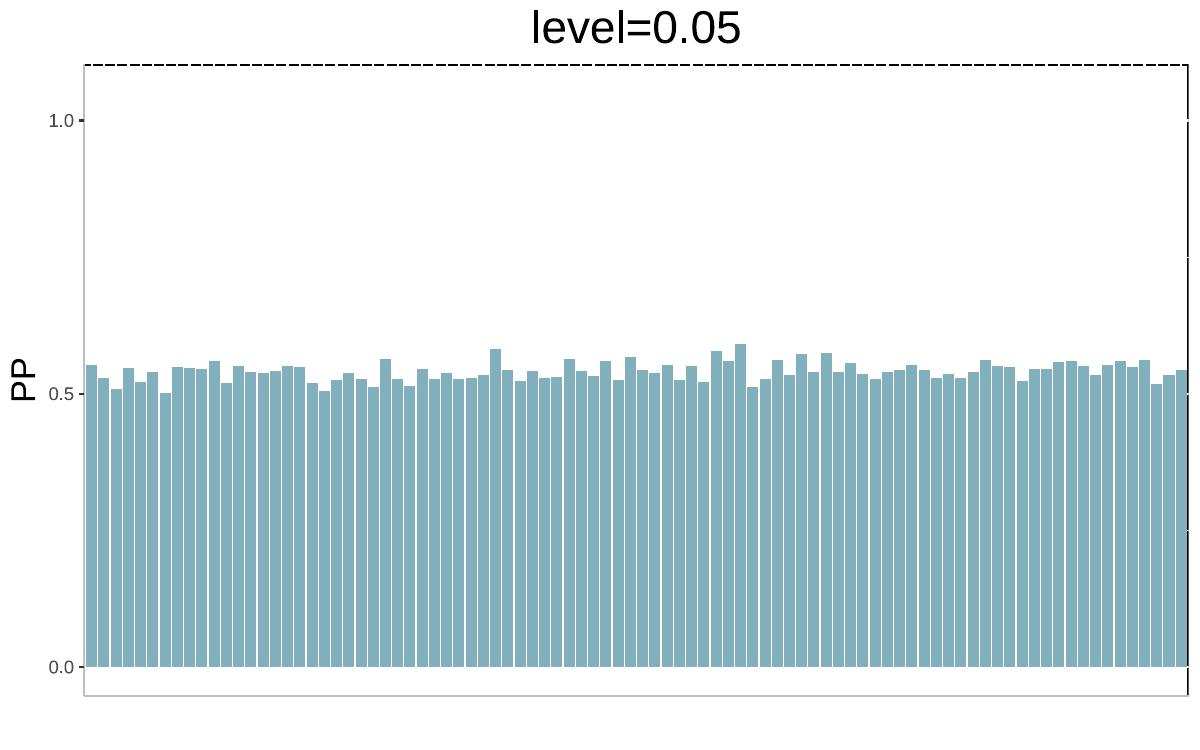}
    \caption[Posterior probability of influence for microbes (bacterial microbiome data)]{{\bf Posterior probability of influence for microbes (bacterial microbiome data).}
    \revision{Posterior probability of influence for each microbe, calculated as the average of the $\xi$ value for that node across retained Gibbs samples.}}
    \label{fig:nodes-wagg-adx}
\end{figure}

\begin{figure}[!ht]
    \centering
    \includegraphics[scale=0.6]{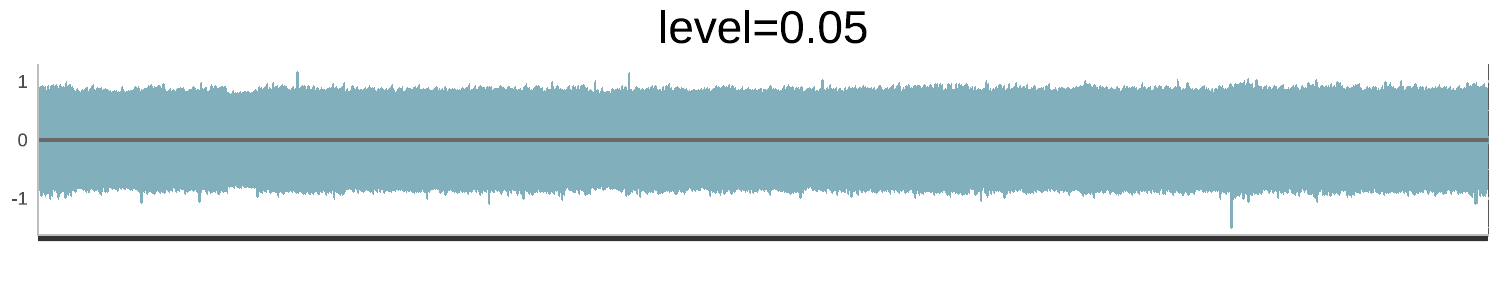}
    \caption[Posterior 95\% credible intervals for edge effects (real-world data)]{{\bf {Posterior 95\% credible intervals for edge effects (real-world data)}}. The color of the intervals depends on whether it intersects zero (light) and hence estimated to be non-influential or does not intersect zero (dark) and hence estimated to be influential by the model. Note that there are no dark intervals: no edges are found to be influential.}
    \label{fig:edges-wagg}
\end{figure}



\begin{figure}[!ht]
\centering
\centering
\includegraphics[scale=0.5]{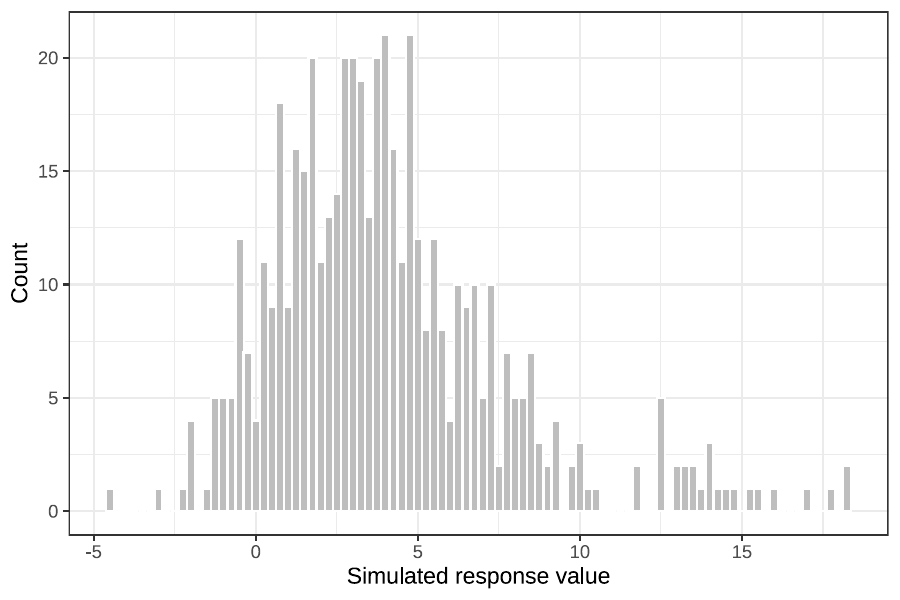}
\centering
\includegraphics[scale=0.5]{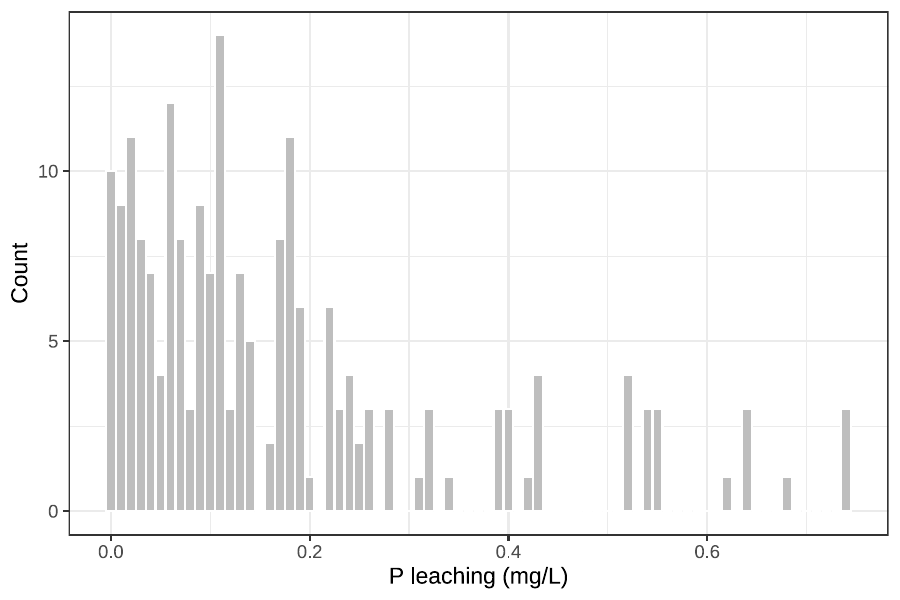}
\caption[Histogram of response values for simulated "functional redundancy" data]{Histogram of response values for simulated "functional redundancy" data with $\pi=0.8$, $\mu=0.8$, $n=500$ samples, $k=8$ sampled nodes (left) and for the phosphorous leaching data, including augmented samples (right).}
\label{fig:response_histograms}
\end{figure}

\end{document}